\documentclass[%
 reprint,
showpacs,preprintnumbers,
 nofootinbib,
 amsmath,amssymb,
 aps,
rmp,
floatfix,
eqsecnum,
]{myrevtex4-1}

\usepackage{graphicx}
\usepackage{dcolumn}

\usepackage{bm}
\usepackage{bbm}
\usepackage{mathrsfs}
\usepackage[bbgreekl]{mathbbol}

\usepackage{url}
\usepackage{hyperref}

\usepackage{feynmp}
\usepackage{subfigure}
\usepackage{multirow}
\usepackage{stackrel}

\usepackage[ddmmyy,24hr]{datetime}

\usepackage{slashed}

\usepackage{esvect}
\renewcommand{\vec}[1]{\vv{#1}}
\newcommand{\vet}[1]{\vv{#1}}

\newcommand{\nua}[1]{\ensuremath{\rlap
           {\kern-2.5pt\ensuremath
           {\overset{\scriptscriptstyle(-)}{\phantom{\nu}}}}
           {\ensuremath{{\nu}_{#1}}}}}

\newcommand{\fermion}{\ensuremath{\mathfrak{f}}}

\newcommand{\ham}{\ensuremath{\mathbbm{H}}}
\newcommand{\nff}{\ensuremath{\mathbbm{f}}}

\newcommand{\chg}{\ensuremath{\mathbbm{q}}}
\newcommand{\mgm}{\ensuremath{\bbmu}}
\newcommand{\elm}{\ensuremath{\bbespilon}}
\newcommand{\anm}{\ensuremath{\mathbbm{a}}}

\newcommand{\elechg}{\ensuremath{e}}

\newcommand{\act}{\ensuremath{a}}

\newcommand{\afl}{\ensuremath{\ell}}
\newcommand{\bfl}{\ensuremath{\ell'}}

\newcommand{\light}{\text{l}}
\newcommand{\heavy}{\text{h}}

\newcommand{\bmag}{\ensuremath{\mu_{\text{B}}}}

\begin{document}

\title{Neutrino electromagnetic interactions: a window to new physics}

\author{Carlo Giunti}
\altaffiliation[Also at the ]{Department of Physics, Torino University, Via P. Giuria 1, I-10125 Torino, Italy}
\email{giunti@to.infn.it}
\affiliation{INFN, Torino Section, Via P. Giuria 1, I-10125 Torino, Italy}

\author{Alexander Studenikin}
\email{studenik@srd.sinp.msu.ru}
\affiliation{Department of Theoretical Physics, Faculty of Physics, Moscow State University
\\
and
\\
Joint Institute for Nuclear Research, Dubna, Russia}


\begin{abstract}
We review the theory and phenomenology of neutrino electromagnetic interactions,
which give us powerful tools to probe the physics beyond the Standard Model.
After a derivation of the general structure of the electromagnetic interactions
of Dirac and Majorana neutrinos in the one-photon approximation,
we discuss the effects of neutrino electromagnetic interactions
in terrestrial experiments and in astrophysical environments.
We present the experimental bounds on neutrino electromagnetic properties and
we confront them with the predictions of theories beyond the Standard Model.
\end{abstract}

\pacs{14.60.St, 13.15.+g, 13.35.Hb, 14.60.Lm, 14.60.Pq, 26.65.+t
\\[0.1cm]
Published in: \textbf{C. Giunti and A. Studenikin, Rev. Mod. Phys. 87 (2015) 531}}

\maketitle

\tableofcontents

\section{Introduction}
\label{A001}

The theoretical and experimental investigation of neutrino properties and interactions is one of the most active
fields of research in current high-energy physics.
It brings us precious information on the physics of the Standard Model
and provides a powerful window on the physics beyond the Standard Model.

The possibility that a neutrino has a magnetic moment was considered by Pauli in his famous
1930 letter addressed to ``Dear Radioactive Ladies and Gentlemen''
(see \textcite{Pauli:1992mu}),
in which he proposed the existence of the neutrino
and he supposed that its mass could be of the same order of magnitude as the electron mass.
Neutrinos remained elusive until the detection of reactor neutrinos
by Reines and Cowan around 1956
\cite{Reines:1960pr}.
However,
there was no sign of a neutrino mass.
After the discovery of parity violation in 1957,
\textcite{Landau:1957tp,Lee:1957qr,Salam:1957st}
proposed the two-component theory of massless neutrinos,
in which a neutrino is described by a Weyl spinor
and there are only left-handed neutrinos and right-handed antineutrinos.
It was however clear
\cite{Touschek:1957,Case:1957zza,Mclennan:1957}
that
two-component neutrinos could be massive Majorana fermions
and that
the two-component theory of a massless neutrino
is equivalent to the Majorana theory in the limit of zero neutrino mass.

The two-component theory of massless neutrinos
was later incorporated in the Standard Model of
\textcite{Glashow:1961tr,Weinberg:1967tq,Salam:1968rm},
in which neutrinos are massless and have only weak interactions.
In the Standard Model
Majorana neutrino masses are forbidden by the
$\text{SU}(2)_{L} \times \text{U}(1)_{Y}$ symmetry.
Although in the Standard Model
neutrinos are electrically neutral
and do not possess electric or magnetic dipole moments,
they have a charge radius which is generated by radiative corrections.

We now know that neutrinos are massive,
because many experiments observed neutrino oscillations
(see the reviews by
\textcite{Giunti-Kim-2007,Bilenky:2010zza,Xing:2011zza,GonzalezGarcia:2012sz,Bellini:2013wra,PDG-2012}),
which are generated by neutrino masses and mixing
\cite{Pontecorvo:1957cp,Pontecorvo:1957qd,Maki:1962mu,Pontecorvo:1968fh}.
Therefore,
the Standard Model must be extended to account for the neutrino masses.
There are many possible extensions of the Standard Model
which predict different properties for neutrinos
(see \textcite{Ramond:1999vh,Mohapatra:2004,Xing:2011zza}).
Among them, most important is their fundamental Dirac or Majorana character.
In many extensions of the Standard Model neutrinos acquire also electromagnetic properties through quantum loops effects
which allow direct interactions of neutrinos with electromagnetic fields
and electromagnetic interactions of neutrinos with charged particles.

Hence,
the theoretical and experimental study of neutrino electromagnetic interactions
is a powerful tool in the search for the fundamental theory beyond the Standard Model.
Moreover,
the electromagnetic interactions of neutrinos can generate important effects,
especially in astrophysical environments,
where neutrinos
propagate over long distances in magnetic fields in vacuum and in matter.

Unfortunately,
in spite of many efforts in the search of neutrino electromagnetic interactions,
up to now there is no positive experimental indication in favor of their existence.
However,
it is expected that the Standard Model neutrino charge radii
should be measured in the near future.
This will be a test of the Standard Model
and of the physics beyond the Standard Model which contributes to
the neutrino charge radii.
Moreover,
the existence of neutrino masses and mixing implies that
neutrinos have magnetic moments.
Since their values
depend on the specific theory
which extends the Standard Model
in order to accommodate neutrino masses and mixing,
experimentalists and theorists are eagerly looking for them.

The structure of this review is as follows.
In Section~\ref{B001} we summarize the
basic theory of neutrino masses and mixing and the
phenomenology of neutrino oscillations,
which are important for the following discussion of
theoretical models and for understanding
the connection between neutrino masses and mixing and neutrino electromagnetic properties.
In Section~\ref{C001}
we derive the general form of the electromagnetic interactions of Dirac and Majorana neutrinos
in the one-photon approximation,
which are expressed in terms of electromagnetic form factors.
In Section~\ref{D001}
we discuss the phenomenology of the neutrino magnetic and electric dipole moments in laboratory experiments.
These are the most studied electromagnetic properties of neutrinos,
both experimentally and theoretically.
In Section~\ref{E001}
we discuss neutrino radiative decay in vacuum and in matter
and related processes which are induced by the neutrino magnetic and electric dipole moments.
These processes could have observable effects in
astrophysical environments and could be detected on Earth
by astronomical photon detectors.
In Section~\ref{F001}
we discuss some important effects due to the interaction
of neutrino magnetic moments with classical electromagnetic fields.
In particular,
we derive the effective potential in a magnetic field
and we discuss the corresponding
spin and spin-flavor transitions
in astrophysical environments.
In Section~\ref{G001}
we review the theory and experimental constraints on the neutrino electric charge (millicharge),
the charge radius and the anapole moment.
In conclusion,
in Section~\ref{H001}
we summarize the status of our knowledge of
neutrino electromagnetic properties and we discuss the prospects for future research.
This review has also several appendices.
We highlight here Appendix~\ref{I001},
in which we clarify the conventions and notation used in the paper and
we list some useful physical constants and formulae.

Let us also remind that
neutrino electromagnetic properties and interactions are discussed in the books by
\textcite{Bahcall:1989ks,Boehm:1992nn,CWKim-book,Raffelt:1996wa,Fukugita:2003en,Zuber:2003,Mohapatra:2004,Xing:2011zza,Barger:2012pxa,Lesgourgues-Mangano-Miele-Pastor-2013},
and in the previous reviews by
\textcite{Bilenky:1987ty,Dolgov:1981hv,Raffelt:1990yz,Salati:1993tf,Raffelt:1999gv,Raffelt:1999tx,Raffelt:2000kp,Pulido:1991fb,Dolgov:2002wy,Nowakowski:2004cv,Wong:2005pa,Studenikin:2008bd,Giunti:2008ve,Broggini:2012df,Akhmedov:2014kxa}.
In this review we improved and extended the discussion presented in our previous reviews in order to cover
in details the most important aspects of neutrino electromagnetic interactions.
\section{Neutrino masses and mixing}
\label{B001}

In the Standard Model of electroweak interactions
\cite{Glashow:1961tr,Weinberg:1967tq,Salam:1968rm},
neutrinos are described by two-component massless left-handed Weyl spinors
(see \textcite{Giunti-Kim-2007}).
The masslessness of neutrinos is due to the absence of
right-handed neutrino fields, without which it is not possible to have Dirac mass
terms, and to the absence of Higgs triplets, without which it is not possible
to have Majorana mass terms.
In the following we consider the extension of
the Standard Model with the introduction of three right-handed neutrinos.
We
will see that this seemingly innocent addition has the very powerful effect of
introducing not only Dirac mass terms, but also Majorana mass terms for the
right-handed neutrinos, which can induce Majorana masses for the observable light neutrinos through the see-saw mechanism.

Table~\ref{B012} shows the values of the weak isospin, hypercharge, and electric
charge of the lepton and Higgs doublets and singlets in the extended Standard
Model under consideration.
We work in the flavor basis in which
the mass matrix of the charged leptons is diagonal.
Hence, $ e $, $ \mu $, $
\tau $ are the physical charged leptons with definite masses.

In the following Subsections
we briefly review the theory of masses and mixing
of Dirac
(\ref{B002})
and Majorana
(\ref{B013})
neutrinos,
the standard framework of three-neutrino mixing
(\ref{B029}),
neutrino oscillations in vacuum and in matter
(\ref{B035}),
the current phenomenological status of three-neutrino mixing
(\ref{B079}),
and
the possibility of additional sterile neutrinos
(\ref{B094}.

\subsection{Dirac neutrinos}
\label{B002}

The fields in Tab.~\ref{B012} allow us to construct the Yukawa Lagrangian term
\begin{equation}
\mathscr{L}_{\text{Y}} = - \sum_{\afl,\bfl=e,\mu,\tau} Y_{\afl \bfl}
\overline{L_{{\afl}L}} \, \widetilde{\Phi} \, \nu_{\bfl R} + \text{H.c.}
,
\label{B003}
\end{equation}
where $Y$ is a matrix of Yukawa couplings and
$\widetilde{\Phi} \equiv i \sigma_{2} \Phi^{*}$.
In the Standard Model, a nonzero vacuum expectation
value of the Higgs doublet,
\begin{equation}
\langle \Phi \rangle = \frac{ 1 }{ \sqrt{2} }
\begin{pmatrix}
0 \\ v
\end{pmatrix}
, \label{B004}
\end{equation}
induces the spontaneous symmetry breaking of the Standard Model symmetries
$\text{SU}(2)_{L} \times \text{U}(1)_{Y} \to \text{U}(1)_{Q}$.
From the Yukawa Lagrangian term in
Eq.~(\ref{B003}), we obtain the neutrino Dirac mass term
\begin{equation}
\mathscr{L}_{\text{D}}
=
-
\sum_{\afl,\bfl=e,\mu,\tau}
\overline{\nu_{{\afl}L}}
\,
M^{\text{D}}_{\afl\bfl}
\,
\nu_{\bfl R} +
\text{H.c.}
,
\label{B005}
\end{equation}
with the complex $3\times3$ Dirac mass matrix
\begin{equation}
M^{\text{D}} = \frac{v}{\sqrt{2}} \, Y
.
\label{B006}
\end{equation}

If the total lepton number is conserved,
$\mathscr{L}_{\text{D}}$
is the only neutrino mass term
and the three massive neutrinos obtained through the diagonalization of
$\mathscr{L}_{\text{D}}$
are Dirac particles.
The diagonalization of
$\mathscr{L}_{\text{D}}$
is achieved through the transformations
\begin{align}
\null & \null
\nu_{{\afl}L} = \sum_{k=1}^{3} U_{\afl k} \, \nu_{kL}
,
\label{B007}
\\
\null & \null
\nu_{\bfl R} = \sum_{k=1}^{3} V_{\bfl k} \, \nu_{kR}
,
\label{B008}
\end{align}
with unitary $3\times3$ matrices $U$ and $V$ such that
\begin{equation}
\left( U^{\dagger} \, M^{\text{D}} \, V \right)_{kj} = m_k \,
\delta_{kj} , \label{B009}
\end{equation}
with real and positive masses $m_k$
(see
\textcite{Bilenky:1987ty,Giunti-Kim-2007}).
The resulting diagonal Dirac mass term is
\begin{equation}
\mathscr{L}_{\text{D}} = - \sum_{k=1}^{3} m_{k} \, \overline{\nu_{kL}} \,
\nu_{kR} + \text{H.c.} = - \sum_{k=1}^{3} m_{k} \, \overline{\nu_{k}} \,
\nu_{k} , \label{B010}
\end{equation}
with the Dirac fields of massive neutrinos
\begin{equation}
\nu_{k} = \nu_{kL} + \nu_{kR}
.
\label{B011}
\end{equation}

\begin{table*}
\begin{center}
\renewcommand{\arraystretch}{1.45}
\setlength{\tabcolsep}{0.3cm}
\begin{tabular}{cccccc}
\hline ($ \afl = e, \mu, \tau $) & & $I$ & $I_{3}$ & $Y$ & $Q$
\\
\cline{3-6} left-handed lepton doublets & $ L_{\afl L} \equiv
\begin{pmatrix}
\nu_{\afl L} \\ \afl_{L}
\end{pmatrix}
$ & $1/2$ & $
\begin{matrix}
1/2 \\ -1/2
\end{matrix}
$ & $-1$ & $
\begin{matrix}
0 \\ -1
\end{matrix}
$
\\
right-handed charged-lepton singlets & $\afl_{R}$ & $0$ & $0$ & $-2$ & $-1$
\\
right-handed neutrino singlets & $\nu_{\afl R}$ & $0$ & $0$ & $0$ & $0$
\\
Higgs doublet & $ \Phi \equiv
\begin{pmatrix}
\phi^{+} \\ \phi^{0}
\end{pmatrix}
$ & $1/2$ & $
\begin{matrix}
1/2 \\ -1/2
\end{matrix}
$ & $+1$ & $
\begin{matrix}
1 \\ 0
\end{matrix}
$
\\
\hline
\end{tabular}
\end{center}
\caption{ \label{B012}
Eigenvalues of the weak isospin $I$, of its
third component $I_{3}$, of the hypercharge $Y$, and of the charge
$Q=I_{3}+Y/2$ of the lepton and Higgs doublets and singlets in the
extension of the Standard Model with the introduction of right-handed
neutrinos.
}
\end{table*}

\subsection{Majorana neutrinos}
\label{B013}

In the above derivation of Dirac neutrino masses we have assumed
that the total lepton number is conserved.
However,
since there is not any compelling argument which imposes the conservation of the total lepton number,
it is plausible that
the right-handed singlet neutrinos have the Majorana mass term
\begin{equation}
\mathscr{L}_{R}
=
\frac{1}{2}
\sum_{\afl,\bfl=e,\mu,\tau} \nu_{\afl R}^{T} \,
\mathcal{C}^{\dagger} \, M^{R}_{\afl\bfl} \, \nu_{\bfl R} + \text{H.c.} ,
\label{B014}
\end{equation}
which violates the total lepton number by two units.
In Eq.~(\ref{B014}),
$\mathcal{C}$ is the charge-conjugation matrix
defined by Eqs.~(\ref{I035})--(\ref{I037})
and the mass matrix $M^{R}$ is complex and symmetric.

The Majorana mass term in Eq.~(\ref{B014}) is allowed by the symmetries of the
Standard Model, since right-handed neutrino fields are invariant.
On the other
hand, an analogous Majorana mass term of the left-handed neutrinos,
\begin{equation}
\mathscr{L}_{L} = \frac{1}{2} \sum_{\afl,\bfl=e,\mu,\tau}
\nu^{T}_{{\afl}L} \, \mathcal{C}^{\dagger} \, M^{L}_{\afl\bfl} \,
\nu_{{\bfl}L} + \text{H.c.} , \label{B015}
\end{equation}
is forbidden, since it has $I_{3}=1$ and $Y=-2$,
as one can find easily using Tab.~\ref{B012}.
There is no Higgs triplet in
the Standard Model to compensate these quantum numbers.

In the extension of the Standard Model with the introduction of right-handed
neutrinos, the neutrino masses and mixing are given by the Dirac--Majorana mass
term
\begin{equation}
\mathscr{L}_{\text{D+M}} = \mathscr{L}_{\text{D}} + \mathscr{L}_{R} .
\label{B016}
\end{equation}
The neutrino fields with definite masses are obtained through the
diagonalization of $\mathscr{L}_{\text{D+M}}$. It is convenient to define the
vector $N_{L}$ of six left-handed fields
\begin{equation}
N^{T}_{L}
\equiv
\left(
\nu_{eL}, \nu_{\mu L}, \nu_{\tau L},
\nu_{eR}^{c},
\nu_{\mu R}^{c},
\nu_{\tau R}^{c}
\right)
,
\label{B017}
\end{equation}
with the charge-conjugated fields
\begin{equation}
\nu_{\afl R}^{c}
=
\mathcal{C} \, \overline{\nu_{\afl R}}^{T}
.
\label{B018}
\end{equation}
The Dirac--Majorana mass term in
Eq.~(\ref{B016}) can be written in the compact form
\begin{equation}
\mathscr{L}_{\text{D+M}} = \frac{1}{2} \, N^{T}_{L} \, \mathcal{C}^{\dagger} \,
M^{\text{D+M}} \, N_{L} + \text{H.c.} , \label{B019}
\end{equation}
with the $6\times6$ symmetric mass matrix
\begin{equation}
M^{\text{D+M}} \equiv
\begin{pmatrix}
0 & {M^{\text{D}}}^{T}
\\
M^{\text{D}} & M^{R}
\end{pmatrix}
. \label{B020}
\end{equation}

The order of magnitude of the elements of the Dirac mass matrix $M^{\text{D}}$
in Eq.~(\ref{B006}) is smaller than
$ v \sim 10^{2} \, \text{GeV} $,
since the Dirac mass term (\ref{B005}) is forbidden by the symmetries of the Standard
Model and can be generated only as a consequence of symmetry breaking
below the electroweak scale $v$.
On the other hand, since the Majorana mass term in
Eq.~(\ref{B014}) is a Standard Model singlet, the elements of the Majorana mass
matrix $M_{R}$ are not related to the electroweak scale.
It is
plausible that the Majorana mass term $\mathscr{L}_{R}$ is generated by new
physics beyond the Standard Model and the right-handed chiral neutrino fields
$\nu_{\afl R}$ belong to nontrivial multiplets of the symmetries of the high-energy
theory.
The corresponding order of magnitude
of the elements of the mass matrix $M_{R}$ is given by the symmetry-breaking scale of
the high-energy physics beyond the Standard Model, which may be as large as
the grand unification scale, of the order of $10^{14}$--$10^{16} \,
\text{GeV}$.
In this case, the mass matrix can be diagonalized by blocks,
up to corrections
of the order $ \epsilon = (M^{R})^{-1} M^{\text{D}} $:
\begin{equation}
W^{T} \, M^{\text{D+M}} \, W \simeq
\begin{pmatrix}
M^{\text{M}}_{\light} & 0
\\
0 & M^{\text{M}}_{\heavy}
\end{pmatrix}
, \label{B021}
\end{equation}
with
\begin{equation}
W \simeq 1 - \frac{1}{2}
\begin{pmatrix}
\epsilon^{\dagger} \epsilon & 2 \epsilon^{\dagger}
\\
- 2 \epsilon & \epsilon \epsilon^{\dagger}
\end{pmatrix}
. \label{B022}
\end{equation}
The light symmetric $3\times3$ Majorana mass matrix $M^{\text{M}}_{\light}$
and the heavy symmetric $3\times3$
Majorana mass matrix $M^{\text{M}}_{\heavy}$ are given by
\begin{equation}
M^{\text{M}}_{\light}
\simeq
- {M^{\text{D}}}^{T} \, ( M^{R} )^{-1} \, M^{\text{D}}
,
\qquad
M^{\text{M}}_{\heavy} \simeq M^{R}
.
\label{B023}
\end{equation}
There are three heavy masses given by the eigenvalues of $M^{\text{M}}_{\heavy}$
and three light
masses given by the eigenvalues of $M^{\text{M}}_{\light}$, whose elements are
suppressed with respect to the elements of the Dirac mass matrix $M^{\text{D}}$
by the very small matrix factor $ {M^{\text{D}}}^{T} ( M^{R} )^{-1} $.
This is
the celebrated \emph{see-saw mechanism}
\cite{Minkowski:1977sc,GellMann:1980vs,Ramond:1979py,Yanagida:1979as,Mohapatra:1980ia},
which explains naturally the smallness
of light neutrino masses.
Notice, however, that the values
of the light neutrino masses and their relative sizes can vary over wide
ranges, depending on the specific values of the elements of $M^{\text{D}}$ and
$M^{R}$.

Since the off-diagonal block elements of $W$ are very small, the three flavor
neutrinos are mainly composed by the three light neutrinos.
Therefore, the
see-saw mechanism implies the effective low-energy Majorana mass term
\begin{equation}
\mathscr{L}_{\text{M}}^{\text{eff}}
=
\frac{1}{2}
\sum_{\afl,\bfl=e,\mu,\tau}
\nu^{T}_{{\afl}L}
\,
\mathcal{C}^{\dagger}
\,
(M^{\text{M}}_{\light})_{\afl\bfl}
\,
\nu_{{\bfl}L}
+
\text{H.c.}
,
\label{B024}
\end{equation}
which involves only the three active left-handed flavor neutrino fields.
The symmetric $3\times3$ Majorana mass matrix $M^{\text{M}}_{\light}$
is diagonalized by the transformation in Eq.~(\ref{B007}) with
a $3\times3$ unitary mixing matrix $U$ such that
\begin{equation}
(U^{T} M^{\text{M}}_{\light} U)_{kj} = m_{k} \delta_{kj}
,
\label{B025}
\end{equation}
with real and positive masses $m_{k}$
(see
\textcite{Bilenky:1987ty,Giunti-Kim-2007}).
In this way, the effective Majorana mass
term in Eq.~(\ref{B024}) can be written in terms of the massive fields as
\begin{align}
\mathscr{L}_{\text{M}}^{\text{eff}}
=
\null & \null
\frac{1}{2} \sum_{k=1}^{3} m_{k} \, \nu_{kL}^{T} \,
\mathcal{C}^{\dagger} \, \nu_{kL} + \text{H.c.}
\nonumber
\\
=
\null & \null
\frac{1}{2} \sum_{k=1}^{3}
m_{k} \, \nu_{k}^{T} \, \mathcal{C}^{\dagger} \, \nu_{k}
,
\label{B026}
\end{align}
with the massive Majorana fields
\begin{equation}
\nu_{k}
=
\nu_{kL} + \nu_{kL}^{c}
=
\nu_{kL} + \mathcal{C} \, \overline{\nu_{kL}}^{T}
,
\label{B027}
\end{equation}
which satisfy
the Majorana constraint
\begin{equation}
\nu_{k}
=
\nu_{k}^{c}
=
\mathcal{C} \, \overline{\nu_{k}}^{T}
.
\label{B028}
\end{equation}
Hence, a general result of the see-saw mechanism is an effective
low-energy mixing of three massive Majorana neutrinos.

\subsection{Three-neutrino mixing}
\label{B029}

In the previous two Sections we have seen that
an effective mixing of three light neutrinos is obtained in the Dirac case
assuming the conservation of the total lepton number
and in the Majorana case through the see-saw mechanism.
In both cases the mixing relation between
the three left-handed flavor neutrino
fields $\nu_{e L}$, $\nu_{\mu L}$, $\nu_{\tau L}$ which partake in weak
interactions and the three left-handed massive
neutrino fields $\nu_{1L}$, $\nu_{2L}$, $\nu_{3L}$
is given by Eq.~(\ref{B007}),
which depends on a unitary $3\times3$ mixing matrix $U$.

The mixing matrix $U$ is observable through its effects in
charged-current weak interaction processes
in which leptons are described by the current
\begin{equation}
j_{\text{CC}}^{\rho}
=
2
\sum_{\afl=e,\mu,\tau}
\overline{\nu_{{\afl}L}}
\,
\gamma^{\rho} \afl_{L}
=
2
\sum_{\afl=e,\mu,\tau}
\sum_{k=1}^{3}
U_{\afl k}^{*}
\overline{\nu_{kL}} \gamma^{\rho} \afl_{L}
.
\label{B030}
\end{equation}
A unitary $3\times3$ matrix can be parameterized in terms of three mixing angles and six
phases.
However, in the mixing matrix three phases are unphysical, because they can be eliminated by
rephasing the three charged lepton fields in $j_{\text{CC}}^{\rho}$. In the
case of Majorana massive neutrinos, no additional phase can be eliminated,
because the Majorana mass term in Eq.~(\ref{B026})
is not invariant under rephasing of $\nu_{kL}$. On the other hand, in the case
of Dirac massive neutrinos, two additional phases can be eliminated by
rephasing the massive neutrino fields.
Hence, the mixing matrix has three physical
phases in the case of Majorana massive neutrinos or one physical phase in the
case of Dirac massive neutrinos.
In general, in the case of Majorana massive
neutrinos $U$ can be written as
\begin{equation}
U = U^{\text{D}} \, D^{\text{M}} , \label{B031}
\end{equation}
where $U^{\text{D}}$ is a Dirac unitary mixing matrix which can be
parameterized in terms of three mixing angles and one physical phase, called
\emph{Dirac phase}, and $D^{\text{M}}$ is a diagonal unitary matrix with two
physical phases, usually called \emph{Majorana phases}.
In the case of Dirac neutrinos $U = U^{\text{D}}$.

The standard parameterization of $U^{\text{D}}$ is
\begin{widetext}
\begin{equation}
U^{\text{D}} =
\begin{pmatrix}
c_{12} c_{13} & s_{12} c_{13} & s_{13} e^{-i\delta_{13}}
\\
- s_{12} c_{23} - c_{12} s_{23} s_{13} e^{i\delta_{13}} & c_{12} c_{23} -
s_{12} s_{23} s_{13} e^{i\delta_{13}} & s_{23} c_{13}
\\
s_{12} s_{23} - c_{12} c_{23} s_{13} e^{i\delta_{13}} & - c_{12} s_{23} -
s_{12} c_{23} s_{13} e^{i\delta_{13}} & c_{23} c_{13}
\end{pmatrix}
, \label{B032}
\end{equation}
\end{widetext}
where $ c_{ab} \equiv \cos\vartheta_{ab} $ and $ s_{ab} \equiv
\sin\vartheta_{ab} $. $\vartheta_{12}$, $\vartheta_{13}$, $\vartheta_{23}$ are
the three mixing angles ($ 0 \leq \vartheta_{ab} \leq \pi/2 $) and
$\delta_{13}$ is the Dirac phase ($ 0 \leq \delta_{13} < 2 \pi $).
The diagonal unitary matrix $D^{\text{M}}$ can be written as
\begin{equation}
D^{\text{M}}
=
\text{diag}\!\left( 1 \, , \, e^{i\lambda_{21}} \, , \, e^{i\lambda_{31}} \right)
,
\label{B033}
\end{equation}
in terms of the two Majorana phases $\lambda_{21}$ and $\lambda_{31}$,

All the phases in the mixing matrix violate the CP symmetry
(see \textcite{Giunti-Kim-2007,Branco:2011zb}).

Let us also note that
in the leptonic weak neutral current,
\begin{equation}
j_{\text{NC}}^{\rho}
=
\sum_{\afl=e,\mu,\tau}
\overline{\nu_{{\afl}L}} \, \gamma^{\rho} \, \nu_{{\afl}L}
=
\sum_{k=1}^{3}
\overline{\nu_{kL}} \, \gamma^{\rho} \, \nu_{kL}
,
\label{B034}
\end{equation}
the unitarity of $U$ implies the absence of neutral-current transitions
among different massive neutrinos (GIM mechanism; \textcite{Glashow:1970gm}).

\subsection{Neutrino oscillations}
\label{B035}

Flavor neutrinos are produced and detected
in charged-current weak interaction processes
described by the leptonic current in Eq.~(\ref{B030}).
Hence,
a neutrino with flavor $ \afl = e, \, \mu, \, \tau $
created in a charged-current
weak interaction process from a charged lepton $\afl^{-}$ or together with a
charged antilepton $\afl^{+}$
is described by the state
\begin{equation}
| \nu_{\afl} \rangle = \sum_{k} U_{\afl k}^{*} \, | \nu_{k} \rangle
.
\label{B036}
\end{equation}
Since the mixing matrix is unitary,
we have the inverted relation
\begin{equation}
| \nu_{k} \rangle = \sum_{\afl} U_{\afl k} \, | \nu_{\afl} \rangle
.
\label{B037}
\end{equation}

The massive neutrino states $| \nu_{k} \rangle$ are eigenstates of the free
Hamiltonian with energy eigenvalues
\begin{equation}
E_{k} = \sqrt{ |\vet{p}_{k}|^{2} + m_{k}^{2} }
,
\label{B038}
\end{equation}
where $\vet{p}_{k}$ are the respective momenta.
In the plane-wave approximation
(see \textcite{Giunti-Kim-2007}),
the space-time evolution of a massive neutrino is given by
\begin{equation}
| \nu_{k} (\vec{L},T) \rangle
=
e^{ - i E_{k} T + i \vet{p}_{k} \cdot \vec{L} }
\,
| \nu_{k} \rangle
,
\label{B039}
\end{equation}
where $(\vec{L},T)$ is the space-time distance from the production point.
Inserting this equation in Eq.~(\ref{B036}) and using Eq.~(\ref{B037}),
we obtain
\begin{align}
| \nu_{\afl} (\vec{L},T) \rangle
=
\null & \null
\sum_{k} U_{\afl k}^{*} \,
e^{ - i E_{k} T + i \vet{p}_{k} \cdot \vec{L} }
\,
| \nu_{k} \rangle
\nonumber
\\
=
\null & \null
\sum_{\bfl=e,\mu,\tau}
\left(
\sum_{k}
U_{\afl k}^{*}
\,
e^{ - i E_{k} T + i \vet{p}_{k} \cdot \vec{L} }
\,
U_{\bfl k} \right)
| \nu_{\bfl} \rangle
.
\label{B040}
\end{align}
Then, the phase differences of
different massive neutrinos generate flavor transitions with probability
\begin{align}
P_{\nu_{\afl}\to\nu_{\bfl}}(\vec{L},T)
=
\null & \null
| \langle \nu_{\bfl} | \nu_{\afl} (\vec{L},T) \rangle |^{2}
\nonumber
\\
=
\null & \null
\left|
\sum_{k}
U_{\afl k}^{*}
\,
e^{- i E_{k} T + i \vet{p}_{k} \cdot \vec{L}}
\,
U_{\bfl k} \right|^{2}
.
\label{B041}
\end{align}

Since the source-detector distance $ L \equiv |\vec{L}| $ is macroscopic, we
can consider all massive neutrino momenta $\vet{p}_{k}$ aligned along $\vec{L}$. Moreover, taking into account the smallness of neutrino masses, in
oscillation experiments in which the neutrino propagation time $T$ is not
measured it is possible to approximate $T = L$
(see \textcite{Giunti-Kim-2007}). With these approximations, the
phases in Eq.~(\ref{B041}) reduce to
\begin{align}
- E_{k} T + p_{k} L =
\null & \null
- \left( E_{k} - p_{k} \right) L = - \frac{ E_{k}^{2} -
p_{k}^{2} }{ E_{k} + p_{k} } \, L
\nonumber
\\
=
\null & \null
- \frac{ m_{k}^{2} }{ E_{k} + p_{k} } \, L
\simeq - \frac{ m_{k}^{2} }{ 2 E_{\nu} } \, L
,
\label{B042}
\end{align}
at lowest order in the neutrino masses.
Here, $ p_{k} \equiv |\vet{p}_{k}| $ and $E_{\nu}$ is the neutrino energy neglecting mass
contributions.
Equation~(\ref{B042}) shows that the phases of massive
neutrinos relevant for oscillations are independent of the values of
the energies and momenta of different massive neutrinos, because of
the relativistic dispersion relation in Eq.~(\ref{B038}). The flavor
transition probabilities are
\begin{align}
\null & \null
P_{\nu_{\afl}\to\nu_{\bfl}}(L,E_{\nu})
=
\delta_{\afl\bfl}
\nonumber
\\
\null & \null
- 4 \sum_{k>j} \operatorname{Re}\!\left( U_{{\afl}k}^{*} \,
U_{{\bfl}k} \, U_{{\afl}j} \, U_{{\bfl}j}^{*} \right)
\sin^{2}\left( \frac{\Delta{m}^{2}_{kj} L}{4E_{\nu}} \right)
\nonumber
\\
\null & \null
-
2 \sum_{k>j} \operatorname{Im}\!\left( U_{\afl k} U_{\afl j}^{*} U_{\bfl k}^{*} U_{\bfl j} \right)
\,
\sin\!\left( \frac{\Delta{m}^{2}_{kj} L}{2E_{\nu}} \right)
.
\label{B043}
\end{align}
where
$\Delta{m}^{2}_{kj}=m_{k}^{2}-m_{j}^{2}$.

In the approximation of two-neutrino mixing, in which one of the three massive
neutrino components of two flavor neutrinos is neglected, the mixing matrix
reduces to
\begin{equation}
U =
\begin{pmatrix}
\cos\vartheta & \sin\vartheta
\\
-\sin\vartheta & \cos\vartheta
\end{pmatrix}
, \label{B044}
\end{equation}
where $\vartheta$ is the mixing angle ($0 \leq \vartheta \leq \pi/2$). In this
approximation, there is only one squared-mass difference $\Delta{m}^{2}$ and the
transition probability is given by
\begin{equation}
P_{\nu_{\afl}\to\nu_{\bfl}}^{2\nu}(L,E_{\nu}) = \sin^{2} 2\vartheta \, \sin^{2}\!\left(
\frac{\Delta{m}^{2} L}{4E_{\nu}} \right) \qquad (\afl\neq\bfl) . \label{B045}
\end{equation}
The corresponding survival probabilities are given by
\begin{equation}
P_{\nu_{\afl}\to\nu_{\afl}}^{2\nu}(L,E_{\nu}) = 1 - \sin^{2} 2\vartheta \,
\sin^{2}\!\left( \frac{\Delta{m}^{2} L}{4E_{\nu}} \right) . \label{B046}
\end{equation}
These simple expressions are often used in the analysis of experimental data.

When neutrinos propagate in matter, the potential generated by the coherent
forward elastic scattering with the particles in the medium (electrons and
nucleons) modifies mixing and oscillations \cite{Wolfenstein:1978ue}. In a
medium with varying density it is possible to have resonant flavor transitions
\cite{Mikheev:1986gs}. This is the famous MSW effect.

\begin{figure}
\begin{center}
\begin{minipage}[r]{0.43\linewidth}
\begin{center}
\subfigure[]{\label{B047}
\includegraphics*[bb=250 641 340 711, width=0.8\linewidth]{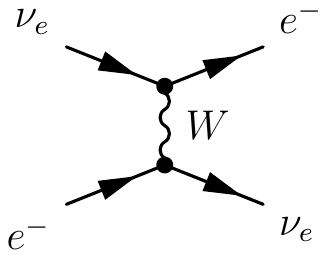}
}
\end{center}
\end{minipage}
\hfill
\begin{minipage}[l]{0.53\linewidth}
\begin{center}
\subfigure[]{\label{B048}
\includegraphics*[bb=244 665 347 717, width=0.8\linewidth]{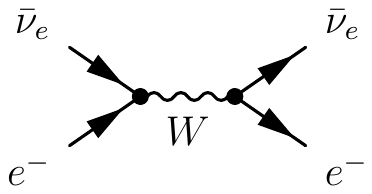}
}
\end{center}
\end{minipage}
\end{center}
\caption{ \label{B049}
Feynman diagrams of the coherent charged-current forward elastic
scattering processes that generate the potentials
$V_{\text{CC}}$ \subref{B047} and $\overline{V}_{\text{CC}}$ \subref{B048}.
}
\end{figure}

\begin{figure}
\null
\hfill
\begin{minipage}[r]{0.43\linewidth}
\begin{center}
\subfigure[]{\label{B056}
\includegraphics*[bb=251 645 340 711, width=0.8\linewidth]{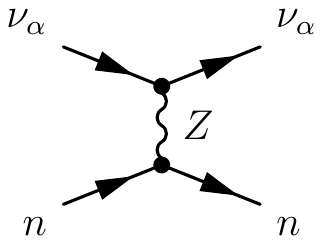}
}
\end{center}
\end{minipage}
\hfill
\begin{minipage}[l]{0.43\linewidth}
\begin{center}
\subfigure[]{\label{B057}
\includegraphics*[bb=251 645 340 713, width=0.8\linewidth]{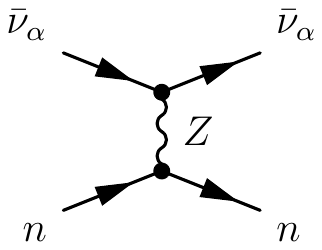}
}
\end{center}
\end{minipage}
\hfill
\null
\caption{ \label{B058}
Feynman diagram of the coherent neutral-current forward elastic
scattering processes that generate the potentials
$V_{\text{NC}}$ \subref{B056} and $\overline{V}_{\text{NC}}$ \subref{B057}.
}
\end{figure}

The effective potentials for $\nu_{\afl}$ and $\bar\nu_{\afl}$ are,
respectively,
\begin{equation}
V_{\afl} = V_{\text{CC}} \, \delta_{\afl e} + V_{\text{NC}} , \qquad
\overline{V}_{\afl} = - V_{\afl} , \label{B050}
\end{equation}
with the charged-current and neutral-current potentials
\begin{equation}
V_{\text{CC}} = \sqrt{2} \, G_{\text{F}} \, N_{e} , \qquad V_{\text{NC}} = -
\frac{1}{2} \, \sqrt{2} \, G_{\text{F}} \, N_{n} , \label{B051}
\end{equation}
generated, respectively, by the Feynman diagrams in Fig.~\ref{B047} and \ref{B056}. Here $N_{e}$ and $N_{n}$
are the electron and neutron number densities in the medium (in an electrically
neutral medium the neutral-current potentials of protons and electrons cancel
each other). In normal matter, these potentials are very small, because
\begin{equation}
\sqrt{2} \, G_{\text{F}} \simeq 7.63 \times 10^{-14} \, \frac{ \text{eV} \,
\text{cm}^{3} }{ N_{\text{A}} } , \label{B052}
\end{equation}
where $N_{\text{A}}$ is Avogadro's number given in Eq.~(\ref{I002}).

Let us consider, for simplicity, two-neutrino $\nu_{e}$--$\nu_{\act}$ mixing,
where $\nu_{\act}$
is a linear combination of
$\nu_{\mu}$ and $\nu_{\tau}$
(which can be pure $\nu_{\mu}$ or $\nu_{\tau}$ as special cases).
This is a good approximation for solar neutrinos.
In general, a neutrino produced at $x=0$ is described at a distance $x$ by a state
\begin{equation}
| \nu(x) \rangle = \varphi_{e}(x) \, | \nu_{e} \rangle + \varphi_{\act}(x) \, |
\nu_{\act} \rangle . \label{B053}
\end{equation}
Taking into account that for ultrarelativistic neutrinos the distance $x$
is approximately equal to the propagation time $t$,
the evolution of the flavor amplitudes $\varphi_{e}(x)$ and $\varphi_{\act}(x)$
with the distance $x$ is given by the Schr\"odinger equation
\cite{Wolfenstein:1978ue}
\begin{equation}
i \frac{ \text{d} }{ \text{d}x }
\begin{pmatrix}
\varphi_{e}(x)
\\
\varphi_{\act}(x)
\end{pmatrix}
=
\mathrm{H}
\begin{pmatrix}
\varphi_{e}(x)
\\
\varphi_{\act}(x)
\end{pmatrix}
,
\label{B054}
\end{equation}
with the effective Hamiltonian matrix
\begin{equation}
\mathrm{H}
=
\frac{1}{4E_{\nu}}
\begin{pmatrix}
- \Delta{m}^{2} \cos{2\vartheta} + A_{\text{CC}}
&
\Delta{m}^{2} \sin{2\vartheta}
\\
\Delta{m}^{2} \sin{2\vartheta}
&
\Delta{m}^{2} \cos{2\vartheta} - A_{\text{CC}}
\end{pmatrix}
,
\label{B055}
\end{equation}
where
$A_{\text{CC}} = 2 E_{\nu} V_{\text{CC}}$.
In Eq.~(\ref{B055}) we took into account only the difference $V_{\text{CC}}$ of the potentials of
$\nu_{e}$ and $\nu_{\act}$,
which affects neutrino oscillations.
In the framework of three-neutrino mixing
the neutral-current potential $V_{\text{NC}}$,
which is common to the three neutrino flavors, does not have any effect.
However,
one must be aware that the neutral-current potential $V_{\text{NC}}$
must be taken into account in extensions of three-neutrino mixing involving sterile states
(see Subsection~\ref{B094})
and/or spin-flavor transitions
(see Subsection~\ref{F023}).

For an initial $\nu_{e}$, as in the case of solar neutrinos,
the boundary condition for the solution of the
differential equation is
\begin{equation}
\begin{pmatrix}
\varphi_{e}(0)
\\
\varphi_{\act}(0)
\end{pmatrix}
=
\begin{pmatrix}
1
\\
0
\end{pmatrix}
, \label{B059}
\end{equation}
and the probabilities of $\nu_{e}\to\nu_{\act}$ transitions and $\nu_{e}$
survival are, respectively,
\begin{align}
\null & \null
P_{\nu_{e}\to\nu_{\act}}(x) = |\varphi_{\act}(x)|^{2}
,
\label{B060}
\\
\null & \null
P_{\nu_{e}\to\nu_{e}}(x) = |\varphi_{e}(x)|^{2} = 1 -
P_{\nu_{e}\to\nu_{\act}}(x)
.
\label{B061}
\end{align}

The effective Hamiltonian matrix in Eq.~(\ref{B055})
can be diagonalized with the transformation
\begin{equation}
\begin{pmatrix}
\varphi_{e}(x)
\\
\varphi_{\act}(x)
\end{pmatrix}
=
U_{\text{M}}
\begin{pmatrix}
\varphi^{\text{M}}_{1}(x)
\\
\varphi^{\text{M}}_{2}(x)
\end{pmatrix}
,
\label{B062}
\end{equation}
with the effective orthogonal
($U_{\text{M}}^{T}=U_{\text{M}}^{-1}$)
mixing matrix in matter
\begin{equation}
U_{\text{M}} =
\begin{pmatrix}
\cos\vartheta_{\text{M}} & \sin\vartheta_{\text{M}}
\\
-\sin\vartheta_{\text{M}} & \cos\vartheta_{\text{M}}
\end{pmatrix}
,
\label{B063}
\end{equation}
such that
\begin{equation}
U_{\text{M}}^{T} \, \mathrm{H} \, U_{\text{M}}
=
\frac{\text{diag}(-\Delta{m}^{2}_{\text{M}},\Delta{m}^{2}_{\text{M}})}{4E_{\nu}} \,
.
\label{B064}
\end{equation}
The amplitudes
$\varphi^{\text{M}}_{1}(x)$
and
$\varphi^{\text{M}}_{2}(x)$
correspond to the effective massive neutrinos in matter
$\nu^{\text{M}}_{1}(x)$
and
$\nu^{\text{M}}_{2}(x)$,
which have the effective squared-mass difference
\begin{equation}
\Delta{m}^{2}_{\text{M}} = \sqrt{ \left( \Delta{m}^{2}\cos2\vartheta - 2 E_{\nu}
V_{\text{CC}} \right)^{2} + \left( \Delta{m}^{2}\sin2\vartheta \right)^{2} }
.
\label{B065}
\end{equation}
The effective mixing angle in matter
$\vartheta_{\text{M}}$ is given by
\begin{equation}
\tan2\vartheta_{\text{M}} = \dfrac{\tan2\vartheta}{1-\dfrac{2 E_{\nu}
V_{\text{CC}}}{\Delta{m}^{2}\cos2\vartheta}} . \label{B066}
\end{equation}
The most interesting characteristic of this expression is that there is a
resonance \cite{Mikheev:1986gs} when
\begin{equation}
V_{\text{CC}} = \frac{ \Delta{m}^{2} }{ 2 E_{\nu} } \, \cos2\vartheta ,
\label{B067}
\end{equation}
which corresponds to the electron number density
\begin{equation}
N_{e}^{\text{R}} = \frac{ \Delta{m}^{2} \cos2\vartheta }{ 2 \sqrt{2} E_{\nu}
G_{\text{F}} } . \label{B068}
\end{equation}
At the resonance the effective mixing angle is equal to $\pi/4$, i.e.\ the
mixing is maximal, leading to the possibility of total transitions between the
two flavors if the resonance region is wide enough.

In general, the evolution equation~(\ref{B054}) must be solved numerically or
with appropriate approximations.
In a constant matter density, it is easy to
derive an analytic solution, leading to the transition probability
\begin{equation}
P_{\nu_{e}\to\nu_{\act}}^{2\nu}(x) = \sin^{2} 2\vartheta_{\text{M}} \, \sin^{2}\left(
\frac{\Delta{m}^{2}_{\text{M}} x}{4E_{\nu}} \right) . \label{B069}
\end{equation}
This expression has
the same structure as the two-neutrino transition probability in
vacuum in Eq.~(\ref{B045}), with the mixing angle and the squared-mass
difference replaced by their effective values in matter.

\begin{figure*}
\begin{center}
\begin{minipage}[r]{0.49\linewidth}
\begin{center}
\subfigure[]{
\includegraphics*[bb=15 22 558 558, width=0.9\linewidth]{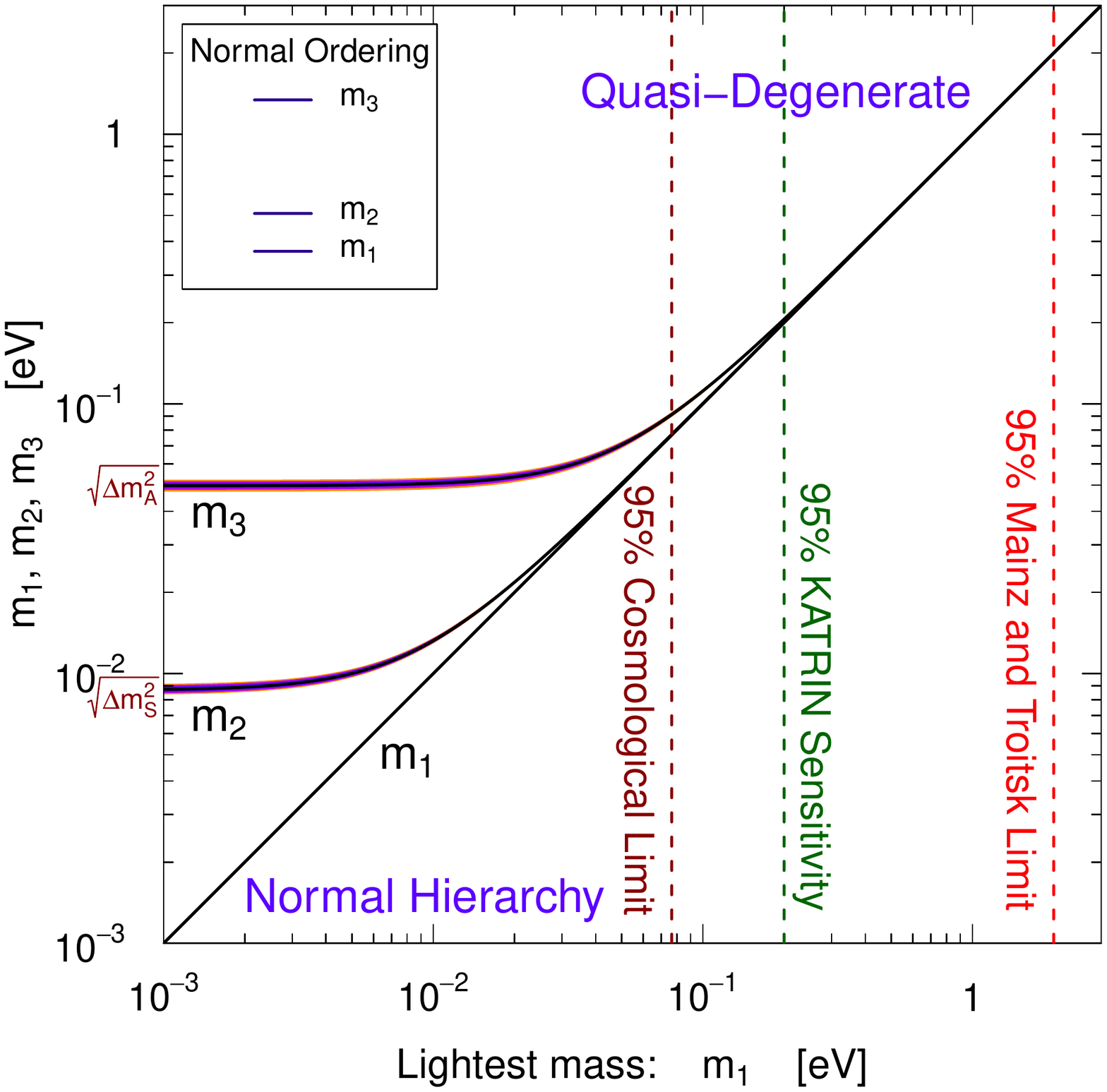}
\label{B070}
}
\end{center}
\end{minipage}
\hfill
\begin{minipage}[l]{0.49\linewidth}
\begin{center}
\subfigure[]{
\includegraphics*[bb=15 22 558 558, width=0.9\linewidth]{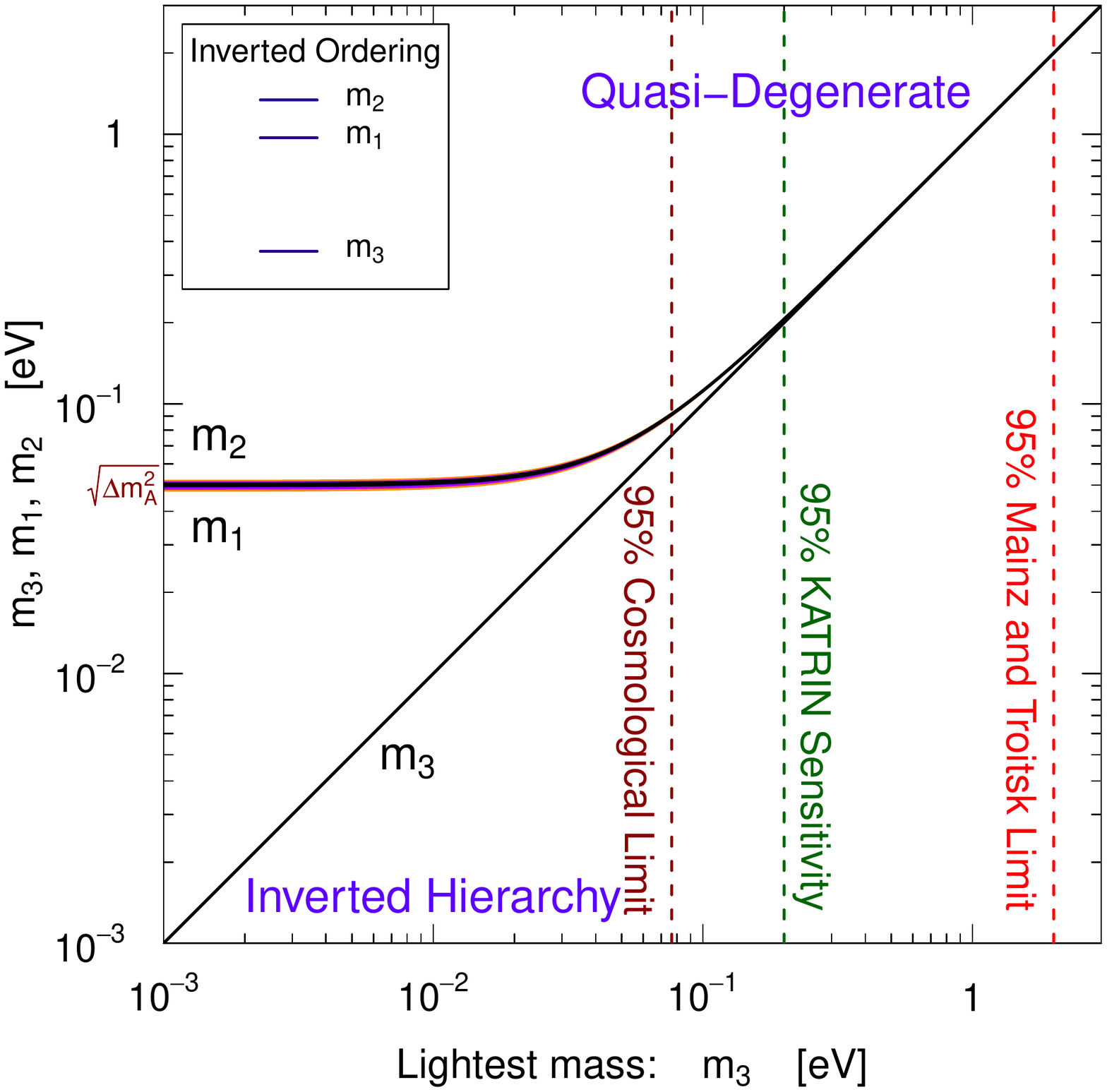}
\label{B071}
}
\end{center}
\end{minipage}
\end{center}
\caption{\label{B072}
Values of the neutrino masses as functions of the lightest mass in the two possible cases of
normal ordering \subref{B070}
and
inverted ordering \subref{B071}.
They have been obtained using the squared-mass differences in Tab.~\ref{B082}.
}
\end{figure*}

The matter effect
is especially important for solar neutrinos,
which are created
as electron neutrinos by thermonuclear reactions in the center of the Sun,
where the electron number density
$N_{e}$
is of the order of $10^2 \, N_{\text{A}} \, \text{cm}^{-3}$,
and propagate out of the Sun through an electron density
which decreases approximately in an exponential way
(see \textcite{Giunti-Kim-2007}).
In a first approximation which neglects the small effects due to $\vartheta_{13}$,
$\nu_{e}$ is mixed only with $\nu_{1}$ and $\nu_{2}$,
which are almost equally mixed with $\nu_{\mu}$ and $\nu_{\tau}$
(see Subsection~\ref{B079}).
In this approximation,
the oscillations of solar neutrinos are well described
by the two-neutrino $\nu_{e}$--$\nu_{\act}$ mixing formalism
with
$\vartheta = \vartheta_{12}$.
The oscillations are generated by the solar squared-mass difference
\begin{equation}
\Delta{m}^{2}_{\text{S}} \approx 8 \times 10^{-5} \, \text{eV}^{2}
,
\label{B073}
\end{equation}
and
\begin{align}
|\nu_{\act}\rangle
\simeq
\null & \null
\cos\vartheta_{23} \, |\nu_{\mu}\rangle
-
\sin\vartheta_{23} \, |\nu_{\tau}\rangle
\nonumber
\\
\approx
\null & \null
(
|\nu_{\mu}\rangle
-
|\nu_{\tau}\rangle
)
/
\sqrt{2}
.
\label{B074}
\end{align}

An electron neutrino created in the center of the Sun
is the linear combination of effective massive neutrinos
\begin{equation}
|\nu_{e}^{0}\rangle
=
\cos\vartheta_{\text{M}}^{0}
|\nu^{0}_{1}\rangle
+
\sin\vartheta_{\text{M}}^{0}
|\nu^{0}_{2}\rangle
,
\label{B075}
\end{equation}
where
$\nu^{0}_{1}$
and
$\nu^{0}_{2}$
are the effective massive neutrinos
at the point of neutrino production near the center of the Sun
and $\vartheta_{\text{M}}^{0}$
is the corresponding effective mixing angle.
Since the resonance is crossed adiabatically,
there are no transitions between
the effective massive neutrinos during propagation
and the state which emerges from the Sun is
\begin{equation}
|\nu_{\text{S}}\rangle
=
\cos\vartheta_{\text{M}}^{0}
|\nu_{1}\rangle
+
\sin\vartheta_{\text{M}}^{0}
|\nu_{2}\rangle
,
\label{B076}
\end{equation}
where
$\nu_{1}$
and
$\nu_{2}$
are the massive neutrinos in vacuum.
Since the two massive neutrinos lose coherence
during the long propagation from the Sun to the Earth
\cite{Dighe:1999id},
experiments on Earth measure the average electron neutrino survival probability
\cite{Parke:1986jy}
\begin{align}
\overline{P}_{\nu_{e}\to\nu_{e}}^{\text{S},2\nu}
=
\null & \null
\cos^2\vartheta_{\text{M}}^{0}
|\langle\nu_{e}|\nu_{1}\rangle|^2
+
\sin^2\vartheta_{\text{M}}^{0}
|\langle\nu_{e}|\nu_{2}\rangle|^2
\nonumber
\\
=
\null & \null
\frac{1}{2}
+
\frac{1}{2}
\cos2\vartheta_{\text{M}}^{0}
\cos2\vartheta_{12}
.
\label{B077}
\end{align}
This is a surprisingly simple expression, which depends only
on the mixing angle in vacuum $\vartheta_{12}$
and on the effective mixing angle in the center of the Sun
$\vartheta_{\text{M}}^{0}$,
which can be easily calculated using Eq.~(\ref{B066}).
Notice that $\vartheta_{\text{M}}^{0}$ depends on the neutrino energy.
With the value of $\Delta{m}^{2}_{\text{S}}$
in Eq.~(\ref{B073}),
$\vartheta_{\text{M}}^{0}\simeq\vartheta_{12}$ for $E_{\nu} \lesssim 1 \, \text{MeV}$
and
$\vartheta_{\text{M}}^{0}\simeq\pi/2$ for $E_{\nu} \gtrsim 5 \, \text{MeV}$
(see \textcite{Giunti-Kim-2007}).
Therefore,
\begin{equation}
\overline{P}_{\nu_{e}\to\nu_{e}}^{\text{S},2\nu}
\simeq
\left\{
\begin{array}{lll} \displaystyle
1
-
0.5 \sin^2 2 \vartheta_{12}
&
\quad \text{for} \quad
&
E_{\nu} \lesssim 1 \, \text{MeV}
,
\\ \displaystyle
\sin^2 \vartheta_{12}
&
\quad \text{for} \quad
&
E_{\nu} \gtrsim 5 \, \text{MeV}
.
\end{array}
\right.
\label{B078}
\end{equation}

\subsection{Status of three-neutrino mixing}
\label{B079}

The results of several solar, atmospheric and long-baseline
neutrino oscillation experiments have proved that
neutrinos are massive and mixed particles
(see
\textcite{Giunti-Kim-2007,Bilenky:2010zza,Xing:2011zza,GonzalezGarcia:2012sz,Bellini:2013wra,NuFIT-2014,Capozzi:2013csa}).
There are two groups of experiments which measured
two types of flavor transition generated by
two independent squared-mass differences ($\Delta{m}^{2}$):
the solar squared-mass difference in Eq.~(\ref{B073})
and the atmospheric squared-mass difference
\begin{equation}
\Delta{m}^{2}_{\text{A}} \approx 2 \times 10^{-3} \, \text{eV}^{2}
.
\label{B080}
\end{equation}
Since in the framework of
three-neutrino mixing described in Subsection~\ref{B029}
there are just two independent squared-mass differences,
solar, atmospheric and long-baseline data have led us to the current
three-neutrino mixing paradigm,
with the standard assignments
\begin{equation}
\Delta{m}^{2}_{\text{S}}
=
\Delta{m}^{2}_{21}
\ll
\Delta{m}^{2}_{\text{A}}
=
\frac{1}{2}
\left|
\Delta{m}^{2}_{31}
+
\Delta{m}^{2}_{32}
\right|
.
\label{B081}
\end{equation}
The absolute value in the definition of $\Delta{m}^{2}_{\text{A}}$
is necessary,
because
there are the two possible orderings of the neutrino masses
illustrated schematically in the insets of the two corresponding
panels in Fig.~\ref{B072}:
the normal ordering
(NO)
with
$m_{1}<m_{2}<m_{3}$
and
$\Delta{m}^{2}_{13}, \, \Delta{m}^{2}_{23} > 0$;
the inverted ordering
(IO)
with
$m_{3}<m_{1}<m_{2}$
and
$\Delta{m}^{2}_{13}, \, \Delta{m}^{2}_{23} < 0$.

The three-neutrino mixing parameters can be determined with good precision
with a global fit of neutrino oscillation data.
In Tab.~\ref{B082}
we report the results of the latest global fit presented in \textcite{Capozzi:2013csa},
which agree, within the uncertainties,
with the NuFIT-v1.2
\cite{NuFIT-2014}
update of the global analysis presented in \textcite{GonzalezGarcia:2012sz}.
One can see that all the oscillation parameters are determined with precision between
about 3\% and 10\%.
The largest uncertainty is that of $\vartheta_{23}$,
which is known to be close to maximal ($\pi/4$),
but it is not known if it is smaller or larger than $\pi/4$.
For the Dirac CP-violating phase $\delta$,
there is an indication in favor of $\delta \approx 3\pi/2$,
which would give maximal CP violation,
but at $3\sigma$ all the values of $\delta$ are allowed,
including the CP-conserving values $\delta=0,\pi$.

An open problem in the framework of three-neutrino mixing
is the determination of the absolute scale of neutrino masses,
which cannot be determined with neutrino oscillation experiments,
because oscillations depend only on the differences of neutrino masses.
However,
the measurement in neutrino oscillation experiments of the
neutrino squared-mass differences allows us to constrain the allowed patterns of neutrino masses.
A convenient way to see the allowed patterns of neutrino masses
is to plot the values of the masses as functions of the unknown lightest mass,
which is
$m_{1}$ in the normal ordering
and
$m_{3}$ in the inverted ordering,
as shown in Figs.~\ref{B072}.
We used the squared-mass differences in Tab.~\ref{B082}.
Figure~\ref{B072} shows that there are three extreme possibilities:

\begin{table*}
\begin{tabular}{lcccccc}
Parameter & Ordering & Best fit & $1\sigma$ range & $2\sigma$ range & $3\sigma$ range & rel. unc. \\
\hline
$\Delta{m}^2_{\text{S}}/10^{-5}\,\text{eV}^2 $ & & 7.54 & 7.32 -- 7.80 & 7.15 -- 8.00 & 6.99 -- 8.18 & 3\% \\
\hline
$\sin^2 \vartheta_{12}/10^{-1}$ & & 3.08 & 2.91 -- 3.25 & 2.75 -- 3.42 & 2.59 -- 3.59 & 5\% \\
\hline
\multirow{2}{*}{$\Delta{m}^2_{\text{A}}/10^{-3}\,\text{eV}^2$}
& NO & 2.43 & 2.37 -- 2.49 & 2.30 -- 2.55 & 2.23 -- 2.61 & 3\% \\
& IO & 2.38 & 2.32 -- 2.44 & 2.25 -- 2.50 & 2.19 -- 2.56 & 3\% \\
\hline
\multirow{2}{*}{$\sin^2 \vartheta_{23}/10^{-1}$}
& NO & 4.37 & 4.14 -- 4.70 & 3.93 -- 5.52 & 3.74 -- 6.26 & 10\% \\
& IO & 4.55 & 4.24 -- 5.94 & 4.00 -- 6.20 & 3.80 -- 6.41 & 10\% \\
\hline
\multirow{2}{*}{$\sin^2 \vartheta_{13}/10^{-2}$}
& NO & 2.34 & 2.15 -- 2.54 & 1.95 -- 2.74 & 1.76 -- 2.95 & 8\% \\
& IO & 2.40 & 2.18 -- 2.59 & 1.98 -- 2.79 & 1.78 -- 2.98 & 8\% \\
\hline
\end{tabular}
\caption{\label{B082}
Values of the neutrino mixing parameters obtained with a
global analysis of neutrino oscillation data
presented in \textcite{Capozzi:2013csa}
in the framework of three-neutrino mixing
with the normal ordering (NO) and the inverted ordering (IO).
The relative uncertainty (rel. unc.) has been obtained from the
$3\sigma$ range divided by 6.}
\end{table*}

\begin{description}

\item[A normal hierarchy]
$m_{1} \ll m_{2} \ll m_{3}$.
In this case
\begin{align}
\null & \null
m_{2} \simeq \sqrt{\Delta{m}^{2}_{\text{S}}} \approx 9 \times 10^{-3} \, \text{eV}
,
\label{B083}
\\
\null & \null
m_{3} \simeq \sqrt{\Delta{m}^{2}_{\text{A}}} \approx 5 \times 10^{-2} \, \text{eV}
.
\label{B084}
\end{align}

\item[An inverted hierarchy]
$m_{3} \ll m_{1} \lesssim m_{2}$
In this case
\begin{equation}
m_{1}
\lesssim
m_{2}
\simeq \sqrt{\Delta{m}^{2}_{\text{A}}} \approx 5 \times 10^{-2} \, \text{eV}
.
\label{B085}
\end{equation}

\item[Quasi-degenerate masses]
$m_{1} \lesssim m_{2} \lesssim m_{3} \simeq m_{\nu}$ in the normal scheme
and
$m_{3} \lesssim m_{1} \lesssim m_{2} \simeq m_{\nu}$ in the inverted scheme,
with
\begin{equation}
m_{\nu}
\gg
\sqrt{\Delta{m}^{2}_{\text{A}}} \approx 5 \times 10^{-2} \, \text{eV}
.
\label{B086}
\end{equation}

\end{description}

There are three main sources of information
on the absolute scale of neutrino masses:

\begin{description}

\item[Beta decay]
The most robust information on neutrino masses
can be obtained in $\beta$-decay experiments
which measure the kinematical effect of neutrino masses
on the energy spectrum of the emitted electron.
Tritium $\beta$-decay experiments obtained the most stringent bounds
on the neutrino masses by limiting the effective electron neutrino mass
$m_{\beta}$ given by
(see \textcite{Giunti-Kim-2007,Bilenky:2010zza,Xing:2011zza})
\begin{equation}
m_{\beta}^2
=
\sum_{k=1}^{3} |U_{ek}|^{2} m^{2}_{k}
.
\label{B087}
\end{equation}
The most stringent 95\% CL limits obtained
in the Mainz \cite{hep-ex/0412056} and Troitsk \cite{1108.5034} experiments,
\begin{align}
\null & \null
m_{\beta} \leq 2.3 \, \text{eV}
\quad
(\text{Mainz})
,
\label{B088}
\\
\null & \null
m_{\beta} \leq 2.1 \, \text{eV}
\quad
(\text{Troitsk})
,
\label{B089}
\end{align}
are shown in Fig.~\ref{B072}.
The KATRIN experiment \cite{Fraenkle:2011uu},
which is scheduled to start data taking in 2016,
is expected to have a sensitivity to
$m_{\beta}$ of about 0.2 eV
(also shown in Fig.~\ref{B072}).

\item[Neutrinoless double-beta decay]
This process occurs only if massive neutrinos are Majorana fermions
and depends on the effective Majorana mass
(see
\textcite{Giunti-Kim-2007,Bilenky:2010zza,Xing:2011zza,Bilenky:2014uka})
\begin{equation}
m_{\beta\beta} = \left| \sum_{k=1}^{3} U^{2}_{ek} m_{k} \right|
.
\label{B090}
\end{equation}
The most stringent 90\%CL limits, have been obtained
combining the results of
EXO \cite{Auger:2012ar}
and
KamLAND-Zen \cite{Gando:2012zm}
experiments with
$^{136}\text{Xe}$,
\begin{equation}
m_{\beta\beta}
\lesssim
0.12 - 0.25 \, \text{eV}
\,
\text{[\protect\textcite{Gando:2012zm}]}
,
\label{EXO+KZ}
\end{equation}
and
combining the results of
Heidelberg-Moscow \cite{Klapdor-Kleingrothaus:2001yx},
IGEX \cite{Aalseth:2002rf}
and
GERDA \cite{Agostini:2013mzu}
with
$^{76}\text{Ge}$\footnote{
The claim of observation of
neutrinoless double-beta decay of $^{76}\text{Ge}$
presented by \textcite{KlapdorKleingrothaus:2004wj}
is strongly disfavored by the recent results of the GERDA experiment
\cite{Agostini:2013mzu}
and by the combined bound in Eq.~(\ref{HM+IGEX+GERDA}).
See also the discussions in \textcite{Elliott:2004hr,Aalseth:2004hb,Strumia:2006db,Schwingenheuer:2012zs,Bilenky:2014uka}.
},
\begin{equation}
m_{\beta\beta}
\lesssim
0.2 - 0.4 \, \text{eV}
\,
\text{[\protect\textcite{Agostini:2013mzu}]}
.
\label{HM+IGEX+GERDA}
\end{equation}
The intervals are caused by nuclear physics uncertainties
(see \textcite{1205.0649}).

\item[Cosmology]
Since light massive neutrinos are hot dark matter,
cosmological data give information on the sum of neutrino masses
(see
\textcite{Giunti-Kim-2007,Bilenky:2010zza,Xing:2011zza,Lesgourgues-Mangano-Miele-Pastor-2013}).
The analysis of cosmological data in the framework of the standard
Cold Dark Matter model with a cosmological constant ($\Lambda$CDM)
disfavors neutrino masses larger than some fraction of eV,
but the value of the upper bound on the sum of neutrino masses
depends on model assumptions and on the considered data set
(see \textcite{1111.1436}).
Figure~\ref{B072}
shows the 95\% limit
\begin{equation}
\sum_{k=1}^{3} m_{k} < 0.32 \, \text{eV}
,
\label{B093}
\end{equation}
obtained recently by the Planck collaboration
\cite{Ade:2013zuv}.
See
\textcite{Archidiacono:2013fha,Abazajian:2013oma,Lesgourgues:2014zoa}
for recent reviews of the implications of cosmological data for
neutrino physics.

\end{description}

\subsection{Sterile neutrinos}
\label{B094}

In the previous Subsections we have considered the standard framework of three-neutrino mixing
which can explain the numerous existing measurements of neutrino oscillations as explained in Subsection~\ref{B079}.
However, it is possible that there are additional massive neutrinos,
such as those at the eV scale suggested by anomalies found in short-baseline oscillation experiments
(see \textcite{Aguilar:2001ty,Abdurashitov:2005tb,Giunti:2010zu,Mention:2011rk,Kopp:2011qd,Conrad:2012qt,Giunti:2012tn,Kopp:2013vaa,Giunti:2013aea})
or those at the keV scale,
which could constitute warm dark matter according to the Neutrino Minimal Standard Model ($\nu$MSM)
\cite{Asaka:2005an,Asaka:2005pn,Asaka:2006ek,Asaka:2006rw,Asaka:2006nq}
(see also the reviews in \textcite{Boyarsky:2009ix,Kusenko:2009up,Drewes:2013gca,Boyarsky:2012rt}).
In the flavor basis,
which describes the interacting neutrino states,
the additional neutrinos are sterile,
because we know
from the measurement of the invisible width of the $Z$ boson in the LEP experiments
that the number of light active neutrinos is three
\cite{ALEPH:2005ab},
and
the existence of a heavy fourth generation of active fermions with an active neutrino heavier than $m_{Z}/2$
is disfavored by the experimental data
\cite{Vysotsky:2013gfa,Lenz:2013iha}.
From the theoretical point of view,
it is likely that if there are sterile neutrinos,
all neutrinos are Majorana particles,
but the Dirac case is not excluded.

Let us consider the general case of $N_{s}$ sterile neutrinos
$\nu_{s_{1}}, \ldots, \nu_{s_{N_{s}}}$.
In the mass basis there are $N=3+N_{s}$ massive neutrino fields
$\nu_{1}, \ldots, \nu_{N}$
and
the mixing of the left-handed neutrino fields is given by
\begin{equation}
\nu_{\afl L}
=
\sum_{k=1}^{N}
U_{\afl k} \nu_{k L}
\quad
(\afl=e,\mu,\tau,s_{1},\ldots,s_{N_{s}})
,
\label{B095}
\end{equation}
where $U$ is a $N{\times}N$ unitary mixing matrix.
The three massive neutrinos
$\nu_{1}$,
$\nu_{2}$,
$\nu_{3}$
coincide with those in the standard three-neutrino mixing framework discussed in Subsection~\ref{B029},
and
$\nu_{4}, \ldots, \nu_{N}$
are the additional nonstandard $N_{s}$ massive neutrinos.
In order to preserve approximately the three-neutrino mixing explanation of oscillation data described in Subsection~\ref{B079},
the mixing of the three active neutrinos
$\nu_{e}$,
$\nu_{\mu}$,
$\nu_{\tau}$
with the nonstandard massive neutrinos
$\nu_{4}, \ldots, \nu_{N}$
must be very small:
\begin{equation}
|U_{\afl k}| \ll 1
\quad
\text{for}
\quad
\afl=e,\mu,\tau
\quad
\text{and}
\quad
k \geq 4
,
\label{B096}
\end{equation}
which implies that
\begin{equation}
|U_{s_{n}k}| \ll 1
\quad
\text{for}
\quad
n=1,\ldots,N_{s}
\quad
\text{and}
\quad
k \leq 3
.
\label{B097}
\end{equation}
Since the mixing in the sterile sector is arbitrary,
it is convenient to choose
\begin{equation}
U_{s_{n}k} = 0
\quad
\text{for}
\quad
n \neq k - 3
\quad
\text{and}
\quad
k \geq 4
.
\label{B098}
\end{equation}
Then, from Eq.~(\ref{B097}) we have
\begin{equation}
1 - |U_{s_{k - 3}k}|^2 \ll 1
,
\quad
\text{for}
\quad
k \geq 4
.
\label{B099}
\end{equation}

The numerical values of the inequalities (\ref{B096})--(\ref{B099})
depend on the model and on the experimental data under consideration.
In this review we consider only these generic inequalities in order to present
general results on the neutrino dipole moments in Subsections~\ref{D009} and \ref{D024}
and
on neutrino radiative decay in Subsection~\ref{E005}.
\section{Electromagnetic form factors}
\label{C001}

The importance of neutrino electromagnetic properties
was first mentioned by Pauli in 1930, when he postulated the
existence of this particle and discussed the possibility that the
neutrino might have a magnetic moment
\cite{Pauli:1992mu}.
Systematic theoretical
studies of neutrino electromagnetic properties started after
it was shown that in the extended Standard Model with
right-handed neutrinos the magnetic moment of a massive neutrino
is, in general, nonvanishing and that its value is determined by
the neutrino mass
\cite{Marciano:1977wx,Lee:1977tib,Fujikawa:1980yx,Petcov:1976ff,Pal:1981rm,Shrock:1982sc,Bilenky:1987ty}.

Neutrino electromagnetic properties are important
because they are directly connected to fundamentals of particle
physics.
For example, neutrino electromagnetic properties can be
used to distinguish Dirac and Majorana neutrinos,
because Dirac neutrinos can have both diagonal and off-diagonal
magnetic and electric dipole moments,
whereas only the off-diagonal ones
are allowed for Majorana neutrinos
(see
\textcite{Schechter:1981hw,Shrock:1982sc,Pal:1981rm,Nieves:1981zt,Kayser:1982br,Kayser:1984ge}).
This is shown in details in the following Subsections.
Another important case in which
Dirac and Majorana neutrinos
have quite different observable effects is
the spin-flavor precession in an external magnetic field
discussed in Subsection~\ref{F023}.
Neutrino electromagnetic properties are also probes of new physics beyond the Standard Model,
because in the Standard Model neutrinos can have only a charge radius
(see Subsection~\ref{C089} and Subsection~\ref{G043}).
The discovery of other neutrino electromagnetic properties
would be a signal of new physics beyond the Standard Model (see
\textcite{Bell:2005kz,Bell:2006wi,Bell:2007nu,NovalesSanchez:2008tn}).

In this Section we discuss the general form of the electromagnetic interactions of Dirac and Majorana neutrinos
in the one-photon approximation.
In Subsection~\ref{C002} we derive the general expression of the effective electromagnetic coupling of Dirac neutrinos
in terms of electromagnetic form factors
and we discuss the properties of the form factors under CP and CPT transformations.
In Subsection~\ref{C065} we consider Majorana neutrinos
and
in Subsection~\ref{C089} we consider the Standard Model case of massless Weyl neutrinos.

\subsection{Dirac neutrinos}
\label{C002}

\begin{figure}
\begin{center}
\begin{minipage}[r]{0.48\linewidth}
\begin{center}
\subfigure[]{\label{C003}
\includegraphics*[bb=240 681 352 767, width=0.8\linewidth]{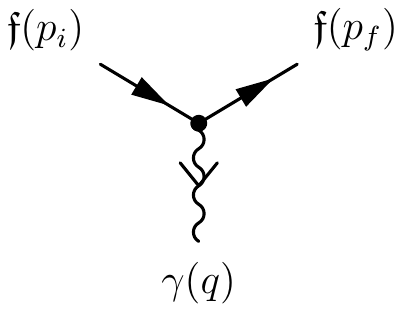}
}
\end{center}
\end{minipage}
\hfill
\begin{minipage}[l]{0.48\linewidth}
\begin{center}
\subfigure[]{\label{C004}
\includegraphics*[bb=238 681 355 767, width=0.8\linewidth]{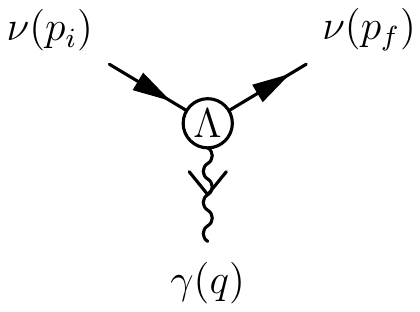}
}
\end{center}
\end{minipage}
\end{center}
\caption{ \label{C005}
Tree-level coupling of a
charged fermion $\fermion$ with a photon $\gamma$
\subref{C003}
and
effective one-photon coupling of a
neutrino with a photon
\subref{C004}.
}
\end{figure}

In the Standard Model,
the interaction of a fermionic field $\fermion(x)$
with the electromagnetic field $A^{\mu}(x)$ is given by the interaction Hamiltonian
\begin{equation}
\mathcal{H}_{\text{em}}^{(\fermion)}(x)
=
j_{\mu}^{(\fermion)}(x) A^{\mu}(x)
=
\chg_{\fermion} \overline{\fermion}(x) \gamma_{\mu} \fermion(x) A^{\mu}(x)
,
\label{C006}
\end{equation}
where
$\chg_{\fermion}$ is the charge of the fermion $\fermion$.
Figure~\ref{C003} shows the corresponding
tree-level Feynman diagram
(the photon $\gamma$ is the quantum of the electromagnetic field $A^{\mu}(x)$).

For neutrinos the electric charge is zero and there are no electromagnetic interactions at tree-level\footnote{
However, in some theories beyond the Standard Model
neutrinos can be millicharged particles
(see Subsection~\ref{G012}).
}.
However, such interactions can arise at the quantum level from loop diagrams at higher order of the perturbative expansion of the interaction.
In the one-photon approximation\footnote{
Some cases in which the one-photon approximation breaks down are discussed in
Subsection~\ref{G012}.
},
the electromagnetic interactions of a neutrino field $\nu(x)$
can be described
by the effective interaction Hamiltonian
\begin{equation}
\mathcal{H}_{\text{em}}^{(\nu)}(x)
=
j_{\mu}^{(\nu)}(x) A^{\mu}(x)
=
\overline{\nu}(x) \Lambda_{\mu} \nu(x) A^{\mu}(x)
,
\label{C007}
\end{equation}
where,
$j_{\mu}^{(\nu)}(x)$
is the neutrino effective electromagnetic current four-vector
and
$\Lambda_{\mu}$
is a $4\times4$ matrix in spinor space which can contain space-time derivatives,
such that
$j_{\mu}^{(\nu)}(x)$
transforms as a four-vector.
Since radiative corrections are generated by weak interactions which are not invariant under a parity transformation,
$j_{\mu}^{(\nu)}(x)$ can be a sum of
polar and axial parts.
The corresponding
diagram for the interaction of a
neutrino with a photon is shown in Fig.~\ref{C004},
where the blob represents the quantum loop contributions.

As we will see in the following,
the neutrino electromagnetic properties corresponding to the diagram in Fig.~\ref{C004}
include charge and magnetic form factors.
Let us emphasize that these neutrino electromagnetic properties
can exist even if neutrinos are elementary particles,
without an internal structure,
because they are generated by quantum loop effects.
Thus,
the neutrino charge and magnetic form factors
have a different origin from the neutron charge and magnetic form factors
(also called Dirac and Pauli form factors),
which are mainly due to its internal quark structure.
For example, the neutrino magnetic moment
(which is the magnetic form factor for interactions with real photons,
i.e. $q^{2}=0$ in Fig.~\ref{C004})
have the same quantum origin as the
anomalous magnetic moment of the electron
(see \textcite{Greiner:1992bv}).

We are interested in the neutrino part of the amplitude corresponding to the diagram in Fig.~\ref{C004},
which is given by the
matrix element
\begin{equation}
\langle \nu(p_{f},h_{f}) |
j_{\mu}^{(\nu)}(x)
| \nu(p_{i},h_{i}) \rangle
,
\label{C008}
\end{equation}
where
$p_{i}$ ($p_{f}$)
and
$h_{i}$ ($h_{f}$)
are the four-momentum and helicity of the initial (final) neutrino.
Taking into account that
\begin{equation}
\partial^{\mu} j_{\mu}^{(\nu)}(x)
=
i \left[ \mathcal{P}^{\mu}, j_{\mu}^{(\nu)}(x) \right]
,
\label{C009}
\end{equation}
where
$\mathcal{P}^{\mu}$
is the four-momentum operator which generate translations,
the effective current can be written as
\begin{equation}
j_{\mu}^{(\nu)}(x)
=
e^{i \mathcal{P} \cdot x}
j_{\mu}^{(\nu)}(0)
e^{- i \mathcal{P} \cdot x}
.
\label{C010}
\end{equation}
Since
$
\mathcal{P}^{\mu} | \nu(p) \rangle
=
p^{\mu} | \nu(p) \rangle
$,
we have
\begin{equation}
\langle \nu(p_{f}) |
j_{\mu}^{(\nu)}(x)
| \nu(p_{i}) \rangle
=
e^{i (p_{f}-p_{i}) \cdot x}
\langle \nu(p_{f}) |
j_{\mu}^{(\nu)}(0)
| \nu(p_{i}) \rangle
,
\label{C011}
\end{equation}
where we suppressed for simplicity the helicity labels
which are not of immediate relevance.
Here we see that the unknown quantity which determines the neutrino-photon interaction is
$
\langle \nu(p_{f}) |
j_{\mu}^{(\nu)}(0)
| \nu(p_{i}) \rangle
$.
Considering that the incoming and outgoing
neutrinos are free particles which are described by
free Dirac fields with the Fourier expansion in Eq.~(\ref{I067}),
we have
\begin{equation}
\langle \nu(p_{f}) |
j_{\mu}^{(\nu)}(0)
| \nu(p_{i}) \rangle
=
\overline{u}(p_{f})
\Lambda_{\mu}(p_{f},p_{i})
u(p_{i})
.
\label{C012}
\end{equation}
The electromagnetic properties of neutrinos are embodied by the vertex function
$\Lambda_{\mu}(p_{f},p_{i})$,
which is a matrix in spinor space and can be decomposed
in terms of linearly independent products
of Dirac $\gamma$ matrices and the available kinematical four-vectors
$p_{i}$
and
$p_{f}$.
As shown in Appendix~\ref{J001},
the most general decomposition can be written as
\begin{align}
\null & \null
\Lambda_{\mu}(p_{f},p_{i})
=
\nff_{1}(q^{2}) q_{\mu}
+
\nff_{2}(q^{2}) q_{\mu} \gamma_{5}
+
\nff_{3}(q^{2}) \gamma_{\mu}
\nonumber
\\
\null & \null
+
\nff_{4}(q^{2}) \gamma_{\mu} \gamma_{5}
+
\nff_{5}(q^{2}) \sigma_{\mu\nu} q^{\nu}
+
\nff_{6}(q^{2}) \epsilon_{\mu\nu\rho\gamma} q^{\nu} \sigma^{\rho\gamma}
,
\label{C013}
\end{align}
where
$\nff_{k}(q^{2})$
are six Lorentz-invariant form factors
($k=1,\ldots,6$)
and
$q$ is the four-momentum of the photon,
which is given by
\begin{equation}
q = p_{i} - p_{f}
,
\label{C014}
\end{equation}
from energy-momentum conservation.
Notice that the form factors depend only on $q^{2}$,
which is the only available Lorentz-invariant
kinematical quantity,
since
$(p_{i}+p_{f})^{2} = 4 m^{2} - q^{2}$.
Therefore,
$\Lambda_{\mu}(p_{f},p_{i})$
depends only on
$q$
and from now on we will denote it as
$\Lambda_{\mu}(q)$.

Since the Hamiltonian and the electromagnetic field are Hermitian
($\mathcal{H}_{\text{em}}^{(\nu)\dagger}=\mathcal{H}_{\text{em}}^{(\nu)}$
and
$A^{\mu\dagger}=A^{\mu}$),
the effective current must be Hermitian,
$j_{\mu}^{(\nu)\dagger}=j_{\mu}^{(\nu)}$.
Hence,
we have
\begin{equation}
\langle \nu(p_{f}) |
j_{\mu}^{(\nu)}(0)
| \nu(p_{i}) \rangle
=
\langle \nu(p_{i}) |
j_{\mu}^{(\nu)}(0)
| \nu(p_{f}) \rangle^{*}
,
\label{C015}
\end{equation}
which leads to
\begin{equation}
\Lambda_{\mu}(q)
=
\gamma^{0} \Lambda_{\mu}^{\dagger}(-q) \gamma^{0}
.
\label{C016}
\end{equation}
Using the properties of the Dirac matrices
(see Appendix~\ref{I001}),
one can find that this constraint implies that
\begin{equation}
\nff_{2}
,
\quad
\nff_{3}
,
\quad
\nff_{4}
\quad
\text{are real}
,
\label{C017}
\end{equation}
and
\begin{equation}
\nff_{1}
,
\quad
\nff_{5}
,
\quad
\nff_{6}
\quad
\text{are imaginary}
.
\label{C018}
\end{equation}

The number of independent form factors can be reduced by imposing current conservation,
$\partial^{\mu} j_{\mu}^{(\nu)}(x) = 0$,
which is required by gauge invariance
(i.e.
invariance of $\mathcal{H}_{\text{em}}^{(\nu)}(x)$ under the transformation
$A^{\mu}(x) \to A^{\mu}(x) + \partial^{\mu} \varphi(x)$
for any $\varphi(x)$,
which leaves invariant the electromagnetic tensor
$F^{\mu\nu} = \partial^{\mu} A^{\nu} - \partial^{\nu} A^{\mu}$).
Using Eq.~(\ref{C009}),
current conservation implies that
\begin{equation}
\langle \nu(p_{f}) |
\left[ \mathcal{P}^{\mu}, j_{\mu}^{(\nu)}(0) \right]
| \nu(p_{i}) \rangle
=
0
.
\label{C019}
\end{equation}
Hence,
in momentum space we have the constraint
\begin{equation}
q^{\mu}
\,
\overline{u}(p_{f})
\Lambda_{\mu}(q)
u(p_{i})
=
0
,
\label{C020}
\end{equation}
which implies that
\begin{equation}
\nff_{1}(q^{2}) q^{2}
+
\nff_{2}(q^{2}) q^{2} \gamma_{5}
+
2 m \nff_{4}(q^{2}) \gamma_{5}
=
0
.
\label{C021}
\end{equation}
Since $\gamma_{5}$ and the unity matrix are linearly independent,
we obtain the constraints
\begin{equation}
\nff_{1}(q^{2}) = 0
,
\quad
\nff_{4}(q^{2})
=
- \nff_{2}(q^{2}) q^{2} / 2 m
.
\label{C022}
\end{equation}
Therefore, in the most general case consistent with Lorentz and
electromagnetic gauge invariance, the vertex function
$\Lambda_{\mu}(q)$
is defined in terms of
four form factors \cite{Nieves:1981zt,Kayser:1982br,Kayser:1984ge},
\begin{align}
\Lambda_{\mu}(q)
=
\null & \null
\nff_{Q}(q^{2}) \gamma_{\mu}
-
\nff_{M}(q^{2}) i \sigma_{\mu\nu} q^{\nu}
+
\nff_{E}(q^{2}) \sigma_{\mu\nu} q^{\nu} \gamma_{5}
\nonumber
\\
\null & \null
+
\nff_{A}(q^{2}) (q^{2} \gamma_{\mu} - q_{\mu} \slashed{q}) \gamma_{5}
,
\label{C023}
\end{align}
where
$\nff_{Q} = \nff_{3}$,
$\nff_{M} = i \nff_{5}$,
$\nff_{E} = - 2 i \nff_{6}$ and
$\nff_{A} = - \nff_{2} / 2m$
are the real
charge, dipole magnetic and electric, and anapole neutrino form factors.
The term involving the electric form factor corresponds to the last term in Eq.~(\ref{C013}),
in which we took into account the identity in Eq.~(\ref{I027}).
In the term involving the anapole form factor we used the identity
$
\overline{u}(p_{f})
\slashed{q} \gamma^{5}
u(p_{i})
=
2 m
\,
\overline{u}(p_{f})
\gamma^{5}
u(p_{i})
$,
which is easily obtained from Eqs.~(\ref{I018}) and (\ref{I043}).

The physical meaning of the
dipole magnetic and electric neutrino form factors
is discussed in Section~\ref{D001}
and that of the charge and anapole in Section~\ref{G001}.
Here we only remark that for the coupling with a real photon ($q^{2}=0$)
\begin{equation}
\nff_{Q}(0) = \chg
,
\quad
\nff_{M}(0) = \mgm
,
\quad
\nff_{E}(0) = \elm
,
\quad
\nff_{A}(0) = \anm
,
\label{C024}
\end{equation}
where
$\chg$,
$\mgm$,
$\elm$ and
$\anm$
are, respectively,
the neutrino charge, magnetic moment, electric moment and anapole moment.
Although above we stated that $\chg=0$,
here we did not enforce this equality
because
in some theories beyond the Standard Model
neutrinos can be millicharged particles,
as explained in Subsection~\ref{G012}.

Now it is interesting to study the properties of $\mathcal{H}_{\text{em}}^{(\nu)}(x)$ under a CP transformation,
in order to find which of the terms in Eq.~(\ref{C023}) violate CP.
The reason is that,
whereas it is well known that weak interactions violate maximally C and P,
the violation of CP is a more exotic phenomenon,
which has been observed so far only in the hadron sector
(see \textcite{Bilenky:2007ne}).

Using the transformation (\ref{I078}) of a fermion field under an active CP transformation
one can find that for the Standard Model electric current $j_{\mu}(x)$ in Eq.~(\ref{C006}) we have
\begin{equation}
j_{\mu}(x)
\xrightarrow{\;\text{CP}\;}
\mathsf{U}_{\text{CP}}
j_{\mu}(x)
\mathsf{U}_{\text{CP}}^{\dagger}
=
-
j^{\mu}(x_{\text{P}})
.
\label{C025}
\end{equation}
Hence,
the Standard Model electromagnetic interaction Hamiltonian $\mathcal{H}_{\text{em}}^{(\nu)}(x)$
is left invariant by\footnote{
The transformation
$x \to x_{\text{P}}$
is irrelevant since all amplitudes are obtained by integrating over
$d^4x$, as in Eq.~(\ref{E007}).
}
\begin{equation}
A_{\mu}(x)
\xrightarrow{\;\text{CP}\;}
- A^{\mu}(x_{\text{P}})
.
\label{C026}
\end{equation}
CP is conserved in neutrino electromagnetic interactions
(in the one-photon approximation)
if $j_{\mu}^{(\nu)}(x)$
transforms as $j_{\mu}(x)$:
\begin{equation}
\text{CP}
\quad
\Longleftrightarrow
\quad
\mathsf{U}_{\text{CP}}
j_{\mu}^{(\nu)}(x)
\mathsf{U}_{\text{CP}}^{\dagger}
=
-
j^{\mu}_{(\nu)}(x_{\text{P}})
.
\label{C027}
\end{equation}
For the matrix element (\ref{C012})
we obtain
\begin{equation}
\text{CP}
\quad
\Longleftrightarrow
\quad
\Lambda_{\mu}(q)
\xrightarrow{\;\text{CP}\;}
- \Lambda^{\mu}(q)
.
\label{C028}
\end{equation}
Using the formulae in Appendix~\ref{I001},
one can find that under a CP transformation we have\footnote{
The operators in $j_{\mu}^{(\nu)}(x)$
are implicitly assumed to be normally ordered
(see \textcite{Giunti-Kim-2007}).
}
\begin{equation}
\Lambda_{\mu}(q)
\xrightarrow{\;\text{CP}\;}
\gamma^{0}
\mathcal{C}
\Lambda_{\mu}^{T}(q_{\text{P}})
\mathcal{C}^{\dagger}
\gamma^{0}
,
\label{C029}
\end{equation}
with
$q^{\mu}_{\text{P}}=q_{\mu}$.
Using the form-factor expansion in Eq.~(\ref{C023}),
we obtain
\begin{align}
\null & \null
\Lambda_{\mu}(q)
\xrightarrow{\;\text{CP}\;}
-
\big[
\nff_{Q}(q^{2}) \gamma^{\mu}
-
\nff_{M}(q^{2}) i \sigma^{\mu\nu} q_{\nu}
\nonumber
\\
\null & \null
-
\nff_{E}(q^{2}) \sigma^{\mu\nu} q_{\nu} \gamma_{5}
+
\nff_{A}(q^{2}) (q^{2} \gamma^{\mu} - q^{\mu} \slashed{q}) \gamma_{5}
\big]
.
\label{C030}
\end{align}
Therefore,
only the electric dipole form factor violates CP:
\begin{equation}
\text{CP}
\quad
\Longleftrightarrow
\quad
\nff_{E}(q^{2}) = 0
.
\label{C031}
\end{equation}

\begin{figure}
\begin{center}
\begin{minipage}[l]{0.48\linewidth}
\begin{center}
\includegraphics*[bb=235 681 360 767, width=0.8\linewidth]{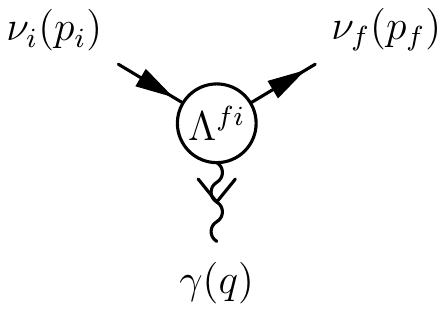}
\end{center}
\end{minipage}
\end{center}
\caption{ \label{C032}
Effective one-photon coupling of
neutrinos with the electromagnetic field, taking into account possible transitions
between two different initial and final massive neutrinos $\nu_{i}$ and $\nu_{f}$.
}
\end{figure}

So far,
in this Section we have considered only one massive neutrino field $\nu(x)$,
but from the discussion of neutrino mixing in Section~\ref{B001}
we know that there are at least three massive neutrino fields in nature.
Therefore, we must generalize the discussion to the case of $N$ massive neutrino fields
$\nu_{k}(x)$ with respective masses $m_{k}$ ($k=1,\ldots,N$).
The effective electromagnetic interaction Hamiltonian in Eq.~(\ref{C007})
is generalized to
\begin{equation}
\mathcal{H}_{\text{em}}^{(\nu)}(x)
=
j_{\mu}^{(\nu)}(x) A^{\mu}(x)
=
\sum_{k,j=1}^{N}
\overline{\nu_{k}}(x) \Lambda^{kj}_{\mu} \nu_{j}(x) A^{\mu}(x)
,
\label{C033}
\end{equation}
where we take into account possible transitions between different massive neutrinos.
The physical effect of $\mathcal{H}_{\text{em}}^{(\nu)}$
is described by the effective electromagnetic vertex in Fig.~\ref{C032},
with the neutrino matrix element
\begin{equation}
\langle \nu_{f}(p_{f}) |
j_{\mu}^{(\nu)}(0)
| \nu_{i}(p_{i}) \rangle
=
\overline{u_{f}}(p_{f})
\Lambda^{fi}_{\mu}(p_{f},p_{i})
u_{i}(p_{i})
.
\label{C034}
\end{equation}
As in the case of one massive neutrino field (see Appendix~\ref{J001}),
$\Lambda^{fi}_{\mu}(p_{f},p_{i})$
depends only on the four-momentum $q$ transferred to the photon
and can be expressed in terms of six Lorentz-invariant form factors:
\begin{align}
\null & \null
\Lambda^{fi}_{\mu}(q)
=
\nff_{1}^{fi}(q^{2}) q_{\mu}
+
\nff_{2}^{fi}(q^{2}) q_{\mu} \gamma_{5}
+
\nff_{3}^{fi}(q^{2}) \gamma_{\mu}
\nonumber
\\
\null & \null
+
\nff_{4}^{fi}(q^{2}) \gamma_{\mu} \gamma_{5}
+
\nff_{5}^{fi}(q^{2}) \sigma_{\mu\nu} q^{\nu}
+
\nff_{6}^{fi}(q^{2}) \epsilon_{\mu\nu\rho\gamma} q^{\nu} \sigma^{\rho\gamma}
.
\label{C035}
\end{align}
The Hermitian nature of $j_{\mu}^{(\nu)}$
implies that
$
\langle \nu_{f}(p_{f}) |
j_{\mu}^{(\nu)}(0)
| \nu_{i}(p_{i}) \rangle
=
\langle \nu_{i}(p_{i}) |
j_{\mu}^{(\nu)}(0)
| \nu_{f}(p_{f}) \rangle^{*}
$,
leading to the constraint
\begin{equation}
\Lambda^{fi}_{\mu}(q)
=
\gamma^{0} [\Lambda^{if}_{\mu}(-q)]^{\dagger} \gamma^{0}
.
\label{C036}
\end{equation}
Considering the $N{\times}N$ form-factor matrices
$\nff_{k}$
in the space of massive neutrinos
with components
$\nff_{k}^{fi}$
for
$k=1,\ldots,6$,
we find that
\begin{equation}
\nff_{2}
,
\quad
\nff_{3}
,
\quad
\nff_{4}
\quad
\text{are Hermitian}
,
\label{C037}
\end{equation}
and
\begin{equation}
\nff_{1}
,
\quad
\nff_{5}
,
\quad
\nff_{6}
\quad
\text{are antihermitian}
.
\label{C038}
\end{equation}

Following the same method used in Eqs.~(\ref{C009})--(\ref{C021}),
one can find that current conservation
implies the constraints
\begin{align}
\null & \null
\nff_{1}^{fi}(q^{2}) q^{2}
+
\nff_{3}^{fi}(q^{2}) (m_{f}-m_{i})
=
0
,
\label{C039}
\\
\null & \null
\nff_{2}^{fi}(q^{2}) q^{2}
+
\nff_{4}^{fi}(q^{2}) (m_{f}+m_{i})
=
0
.
\label{C040}
\end{align}
Therefore,
we obtain
\begin{align}
\Lambda^{fi}_{\mu}(q)
=
\null & \null
\left( \gamma_{\mu} - q_{\mu} \slashed{q}/q^{2} \right)
\left[
\nff_{Q}^{fi}(q^{2})
+
\nff_{A}^{fi}(q^{2}) q^{2} \gamma_{5}
\right]
\nonumber
\\
\null & \null
-
i \sigma_{\mu\nu} q^{\nu}
\left[
\nff_{M}^{fi}(q^{2})
+ i
\nff_{E}^{fi}(q^{2}) \gamma_{5}
\right]
,
\label{C041}
\end{align}
where
$\nff_{Q}^{fi} = \nff_{3}^{fi}$,
$\nff_{M}^{fi} = i \nff_{5}^{fi}$,
$\nff_{E}^{fi} = - 2 i \nff_{6}^{fi}$ and
$\nff_{A}^{fi} = - \nff_{2}^{fi} / (m_{f}+m_i)$,
with
\begin{equation}
\nff_{\Omega}^{fi} = (\nff_{\Omega}^{if})^{*}
\qquad
(\Omega=Q,M,E,A)
.
\label{C042}
\end{equation}
Note that since
$
\overline{u_{f}}(p_{f})
\slashed{q}
u_{i}(p_{i})
=
(m_{f} - m_{i})
\,
\overline{u_{f}}(p_{f})
u_{i}(p_{i})
$,
if $f=i$
Eq.~(\ref{C041}) correctly reduces to Eq.~(\ref{C023}).

The form-factors with $f=i$ are called ``diagonal'',
whereas those with $f{\neq}i$ are called ``off-diagonal'' or
``transition form-factors''.
This terminology follows from the expression
\begin{align}
\Lambda_{\mu}(q)
=
\null & \null
\left( \gamma_{\mu} - q_{\mu} \slashed{q}/q^{2} \right)
\left[
\nff_{Q}(q^{2})
+
\nff_{A}(q^{2}) q^{2} \gamma_{5}
\right]
\nonumber
\\
\null & \null
-
i \sigma_{\mu\nu} q^{\nu}
\left[
\nff_{M}(q^{2})
+ i
\nff_{E}(q^{2}) \gamma_{5}
\right]
,
\label{C043}
\end{align}
in which $\Lambda_{\mu}(q)$
is a $N{\times}N$ matrix
in the space of massive neutrinos expressed
in terms of the four Hermitian $N{\times}N$ matrices of form factors
\begin{equation}
\nff_{\Omega} = \nff_{\Omega}^{\dagger}
\qquad
(\Omega=Q,M,E,A)
.
\label{C044}
\end{equation}

For the coupling with a real photon ($q^{2}=0$) we have
\begin{equation}
\nff_{Q}^{fi}(0) = \chg_{fi}
,
\
\nff_{M}^{fi}(0) = \mgm_{fi}
,
\
\nff_{E}^{fi}(0) = \elm_{fi}
,
\
\nff_{A}^{fi}(0) = \anm_{fi}
,
\label{C045}
\end{equation}
where
$\chg_{fi}$,
$\mgm_{fi}$,
$\elm_{fi}$ and
$\anm_{fi}$
are, respectively,
the neutrino charge, magnetic moment, electric moment and anapole moment
of diagonal ($f=i$) and transition ($f{\neq}i$) types.

Considering now CP invariance,
the transformation (\ref{C027}) of $j_{\mu}^{(\nu)}(x)$
implies the constraint in Eq.~(\ref{C028})
for the $N{\times}N$ matrix $\Lambda_{\mu}(q)$
in the space of massive neutrinos.
Using the formulae in Appendix~\ref{I001},
we obtain
\begin{equation}
\Lambda_{\mu}^{fi}(q)
\xrightarrow{\;\text{CP}\;}
\xi^{\text{CP}}_{f} {\xi^{\text{CP}}_{i}}^{*}
\gamma^{0}
\mathcal{C}
[\Lambda_{\mu}^{if}(q_{\text{P}})]^{T}
\mathcal{C}^{\dagger}
\gamma^{0}
,
\label{C046}
\end{equation}
where
$\xi^{\text{CP}}_{k}$
is the CP phase of $\nu_{k}$.
Since the massive neutrinos take part to standard charged-current weak
interactions\footnote{
Here we consider massive neutrinos which are mixed with the three active flavor neutrinos
$\nu_{e}$,
$\nu_{\mu}$,
$\nu_{\tau}$.
This is the case in standard three-neutrino mixing (see Section~\ref{B001})
and in its extensions with Dirac sterile neutrinos which mix with the active ones.
If there are Dirac sterile neutrinos
which are not mixed with the active ones
and have nonstandard interactions,
the CP phases of the corresponding massive neutrinos
could be different from that of the standard massive neutrinos.
However,
since the production and detection of such sterile neutrinos would be very problematic,
this case is not interesting in practice.
},
their CP phases are equal if CP is conserved
(see \textcite{Giunti-Kim-2007}).
Hence, we have
\begin{equation}
\Lambda_{\mu}^{fi}(q)
\xrightarrow{\;\text{CP}\;}
\gamma^{0}
\mathcal{C}
[\Lambda_{\mu}^{if}(q_{\text{P}})]^{T}
\mathcal{C}^{\dagger}
\gamma^{0}
.
\label{C047}
\end{equation}
Using the form-factor expansion in Eq.~(\ref{C041}),
we obtain
\begin{align}
\Lambda_{\mu}^{fi}(q)
\xrightarrow{\;\text{CP}\;}
\null & \null
-
\Big\{
\left( \gamma^{\mu} - q^{\mu} \slashed{q}/q^{2} \right)
\left[
\nff_{Q}^{if}(q^{2})
+
\nff_{A}^{if}(q^{2}) q^{2} \gamma_{5}
\right]
\nonumber
\\
\null & \null
-
i \sigma^{\mu\nu} q_{\nu}
\left[
\nff_{M}^{if}(q^{2})
- i
\nff_{E}^{if}(q^{2}) \gamma_{5}
\right]
\Big\}
.
\label{C048}
\end{align}
Imposing the constraint in Eq.~(\ref{C028}),
for the form factors we obtain
\begin{equation}
\text{CP}
\quad
\Longleftrightarrow
\quad
\left\{
\begin{array}{l} \displaystyle
\nff_{\Omega}^{fi} = \nff_{\Omega}^{if} = (\nff_{\Omega}^{fi})^{*}
\qquad
(\Omega=Q,M,A)
,
\\ \displaystyle
\nff_{E}^{fi} = - \nff_{E}^{if} = - (\nff_{E}^{fi})^{*}
,
\end{array}
\right.
\label{C049}
\end{equation}
where, in the last equalities, we took into account the constraints (\ref{C042}).
Therefore,
diagonal electric form factors violate CP,
in agreement with the one-generation constraint in Eq.~(\ref{C031}).
For the Hermitian $N{\times}N$ form-factor matrices, we obtain that
if CP is conserved
$\nff_{Q}$,
$\nff_{M}$ and
$\nff_{A}$
are real and symmetric
and
$\nff_{E}$ is imaginary and antisymmetric:
\begin{equation}
\text{CP}
\quad
\Longleftrightarrow
\quad
\left\{
\begin{array}{l} \displaystyle
\nff_{\Omega} = \nff_{\Omega}^{\text{T}} = \nff_{\Omega}^{*}
\qquad
(\Omega=Q,M,A)
,
\\ \displaystyle
\nff_{E} = - \nff_{E}^{\text{T}} = - \nff_{E}^{*}
.
\end{array}
\right.
\label{C050}
\end{equation}

Let us now consider antineutrinos.
Using for the massive neutrino fields the Fourier expansion in Eq.~(\ref{I067}),
the effective antineutrino matrix element
for
$\bar\nu_{i}(p_{i}) \to \bar\nu_{f}(p_{f})$
transitions is given by
\begin{equation}
\langle \bar\nu_{f}(p_{f}) |
j_{\mu}^{(\nu)}(0)
| \bar\nu_{i}(p_{i}) \rangle
=
-
\overline{v_{i}}(p_{i})
\Lambda^{if}_{\mu}(q)
v_{f}(p_{f})
.
\label{C051}
\end{equation}
Using the relation (\ref{I048}) we can write it as
\begin{equation}
\langle \bar\nu_{f}(p_{f}) |
j_{\mu}^{(\nu)}(0)
| \bar\nu_{i}(p_{i}) \rangle
=
\overline{u_{f}}(p_{f})
\mathcal{C}
[\Lambda^{if}_{\mu}(q)]^{T}
\mathcal{C}^{\dagger}
u_{i}(p_{i})
,
\label{C052}
\end{equation}
where transposition operates in spinor space.
Therefore,
the effective form-factor matrix in spinor space for antineutrinos is given by
\begin{equation}
\overline{\Lambda}^{fi}_{\mu}(q)
=
\mathcal{C}
[\Lambda^{if}_{\mu}(q)]^{T}
\mathcal{C}^{\dagger}
.
\label{C053}
\end{equation}
Using the properties of the charge-conjugation matrix,
the expression (\ref{C041}) for $\Lambda^{if}_{\mu}(q)$,
and the hermiticity in Eq.~(\ref{C042}),
we obtain
the antineutrino form factors
\begin{align}
\null & \null
\overline{\nff}_{\Omega}^{fi} = - \nff_{\Omega}^{if} = - (\nff_{\Omega}^{fi})^{*}
\qquad
(\Omega=Q,M,E)
,
\label{C054}
\\
\null & \null
\overline{\nff}_{A}^{fi} = \nff_{A}^{if} = (\nff_{A}^{fi})^{*}
.
\label{C055}
\end{align}
Therefore,
in particular
the diagonal magnetic and electric moments of
neutrinos and antineutrinos,
which are real,
have the same size with opposite signs,
as the charge, if it exists.
On the other hand,
the real diagonal neutrino and antineutrino anapole moments are equal.

It is interesting to note that the relations in Eqs.~(\ref{C054}) and (\ref{C055})
between neutrino and antineutrino form factors are a consequence of
CPT symmetry,
which is a fundamental symmetry of local relativistic Quantum Field Theory
(see \textcite{hep-ph/0309309}).
In order to prove this statement,
let us first consider the CPT transformation of the Standard Model electric current $j_{\mu}(x)$ in Eq.~(\ref{C006}):
using Eq.~(\ref{I080}) we have
\begin{equation}
j_{\mu}(x)
\xrightarrow{\;\text{CPT}\;}
\mathsf{U}_{\text{CPT}}
j_{\mu}(x)
\mathsf{U}_{\text{CPT}}^{\dagger}
=
-
j_{\mu}(-x)
.
\label{C056}
\end{equation}
Therefore,
the Standard Model electromagnetic interaction Hamiltonian $\mathcal{H}_{\text{em}}^{(\nu)}(x)$
is left invariant by
\begin{equation}
A_{\mu}(x)
\xrightarrow{\;\text{CPT}\;}
- A_{\mu}(-x)
.
\label{C057}
\end{equation}
CPT is conserved by the neutrino effective electromagnetic interaction Hamiltonian in Eq.~(\ref{C033}) if
$j_{\mu}^{(\nu)}(x)$
transforms as $j_{\mu}(x)$:
\begin{equation}
\text{CPT}
\quad
\Longleftrightarrow
\quad
\mathsf{U}_{\text{CPT}}
j_{\mu}^{(\nu)}(x)
\mathsf{U}_{\text{CPT}}^{\dagger}
=
-
j_{\mu}^{(\nu)}(-x)
.
\label{C058}
\end{equation}
In order to find the implications of this relation for the antineutrino matrix element in Eq.~(\ref{C051}),
we need to consider it taking into account the helicities of the initial and final neutrinos,
because CPT reverses helicities.
Thus,
assuming CPT and inserting
$
\mathsf{U}_{\text{CPT}}^{\dagger}
\mathsf{U}_{\text{CPT}}
=
1
$
on both sides of $j_{\mu}^{(\nu)}(0)$,
we obtain
\begin{align}
\null & \null
\overline{M}_{fi}
=
\langle \bar\nu_{f}(p_{f},h_{f}) |
j_{\mu}^{(\nu)}(0)
| \bar\nu_{i}(p_{i},h_{i}) \rangle
\nonumber
\\
\null & \null
=
-
\langle \bar\nu_{f}(p_{f},h_{f}) |
\mathsf{U}_{\text{CPT}}^{\dagger}
j_{\mu}^{(\nu)}(0)
\mathsf{U}_{\text{CPT}}
| \bar\nu_{i}(p_{i},h_{i}) \rangle
.
\label{C059}
\end{align}
Now we take into account that the application of $\mathsf{U}_{\text{CPT}}$
to a neutrino state transforms it into an antineutrino state.
Using the notation and conventions of \textcite{Giunti-Kim-2007}
we have
\begin{equation}
\mathsf{U}_{\text{CPT}}
| \bar\nu_{k}(p_{k},h_{k}) \rangle
=
- \zeta(h) \, {\xi^{\text{CPT}}_{k}}^{*} | \nu_{k}(p_{k},-h_{k}) \rangle
,
\label{C060}
\end{equation}
where
$\zeta(h)$
is a phase coming from the relation
\begin{equation}
\gamma^{5} \, v^{(-h)}(p)
=
\zeta(h) \, u^{(h)}(p)
,
\label{C061}
\end{equation}
and
$\zeta(-h)=-\zeta(h)$.
For the CPT phases $\xi^{\text{CPT}}_{k}$, we assume that they are all equal,
as we have done for the CP phases in Eq.~(\ref{C046}).
Then,
using Eq.~(\ref{C060}) and taking into account the antiunitarity of $\mathsf{U}_{\text{CPT}}$,
Eq.~(\ref{C059}) becomes
\begin{equation}
\overline{M}_{fi}
=
- \zeta(h_{f}) \zeta^{*}(h_{i})
\langle \nu_{i}(p_{i},-h_{i}) |
j_{\mu}^{(\nu)}(0)
| \nu_{f}(p_{f},-h_{f}) \rangle
.
\label{C062}
\end{equation}
This is the crucial relation between the neutrino and antineutrino matrix elements
which follows from CPT invariance.
Using for the neutrino matrix element the expression (\ref{C034})
and the relation (\ref{C061}),
we obtain
\begin{equation}
\overline{M}_{fi}
=
\overline{v_{i}^{(h_{i})}}(p_{i})
\gamma^{5}
\Lambda^{if}_{\mu}(-q)
\gamma^{5}
v_{f}^{(h_{f})}(p_{f})
.
\label{C063}
\end{equation}
Taking into account the form-factor expression of
$\Lambda^{fi}_{\mu}(q)$
in Eq.~(\ref{C041}),
we have
$
\gamma^{5}
\Lambda^{if}_{\mu}(-q)
\gamma^{5}
=
- \Lambda^{if}_{\mu}(q)
$,
which leads to
\begin{equation}
\overline{M}_{fi}
=
-
\overline{v_{i}^{(h_{i})}}(p_{i})
\Lambda^{if}_{\mu}(q)
v_{f}^{(h_{f})}(p_{f})
.
\label{C064}
\end{equation}
This expression for the antineutrino matrix element
coincides with Eq.~(\ref{C051})
and implies the relations
(\ref{C054}) and (\ref{C055})
for the form factors.

Thus, we obtained the expression (\ref{C051}) for the antineutrino matrix element
in a complicated way,
assuming only
CPT invariance and
the expression (\ref{C034}) for the neutrino matrix element.
This result is a tautology in the theoretical framework in which we are working,
because CPT is a fundamental symmetry of any local relativistic Quantum Field Theory
(see \textcite{hep-ph/0309309}).
However,
in some theories beyond the Standard Model
small CPT violations can exist
(see \textcite{1006.4989}),
which may be revealed by finding violations
of the equalities in Eqs.~(\ref{C054}) and (\ref{C055}).

\subsection{Majorana neutrinos}
\label{C065}

A Majorana neutrino is a neutral spin 1/2 particle which coincides with its antiparticle.
The four degrees of freedom of a Dirac field
(two helicities and two particle-antiparticle)
are reduced to two
(two helicities) by the Majorana constraint in Eq.~(\ref{B028}).
Since a Majorana field has half the degrees of freedom of a Dirac field,
it is possible that its electromagnetic properties are reduced.
From the relations (\ref{C054}) and (\ref{C055})
between neutrino and antineutrino form factors in the Dirac case,
we can infer that in the Majorana case
the charge, magnetic and electric form-factor matrices are antisymmetric and
the anapole form-factor matrix is symmetric.
In order to confirm this deduction, let us calculate the neutrino matrix element
corresponding to
the effective electromagnetic vertex in Fig.~\ref{C032},
with the effective interaction Hamiltonian in Eq.~(\ref{C033}),
which takes into account possible transitions
between two different initial and final massive Majorana neutrinos $\nu_{i}$ and $\nu_{f}$.
Using the Fourier expansion (\ref{I071}) for the neutrino Majorana fields we obtain
\begin{align}
\langle \nu_{f}(p_{f}) |
j_{\mu}^{(\nu)}(0)
| \nu_{i}(p_{i}) \rangle
\null & \null
=
\overline{u_{f}}(p_{f})
\Lambda^{fi}_{\mu}(p_{f},p_{i})
u_{i}(p_{i})
\nonumber
\\
\null & \null
-
\overline{v_{i}}(p_{i})
\Lambda^{if}_{\mu}(p_{f},p_{i})
v_{f}(p_{f})
.
\label{C066}
\end{align}
Using Eq.~(\ref{I048}),
we can write it as
\begin{equation}
\overline{u_{f}}(p_{f})
\left\{
\Lambda^{fi}_{\mu}(p_{f},p_{i})
+
\mathcal{C}
[\Lambda^{if}_{\mu}(p_{f},p_{i})]^{T}
\mathcal{C}^{\dagger}
\right\}
u_{i}(p_{i})
,
\label{C067}
\end{equation}
where transposition operates in spinor space.
Therefore the effective
form-factor matrix in spinor space for Majorana neutrinos is given by
\begin{equation}
\Lambda^{\text{M}fi}_{\mu}(p_{f},p_{i})
=
\Lambda^{fi}_{\mu}(p_{f},p_{i})
+
\mathcal{C}
[\Lambda^{if}_{\mu}(p_{f},p_{i})]^{T}
\mathcal{C}^{\dagger}
.
\label{C068}
\end{equation}
As in the case of Dirac neutrinos,
$\Lambda^{fi}_{\mu}(p_{f},p_{i})$
depends only on $q=p_{f}-p_{i}$
and can be expressed in terms of six Lorentz-invariant form factors
according to Eq.~(\ref{C035}).
Hence, we can write
the $N{\times}N$ matrix $\Lambda^{\text{M}}_{\mu}(p_{f},p_{i})$
in the space of massive Majorana neutrinos
as
\begin{align}
\Lambda^{\text{M}}_{\mu}(q)
=
\null & \null
\nff^{\text{M}}_{1}(q^{2}) q_{\mu}
+
\nff^{\text{M}}_{2}(q^{2}) q_{\mu} \gamma_{5}
+
\nff^{\text{M}}_{3}(q^{2}) \gamma_{\mu}
\nonumber
\\
\null & \null
+
\nff^{\text{M}}_{4}(q^{2}) \gamma_{\mu} \gamma_{5}
+
\nff^{\text{M}}_{5}(q^{2}) \sigma_{\mu\nu} q^{\nu}
\nonumber
\\
\null & \null
+
\nff^{\text{M}}_{6}(q^{2}) \epsilon_{\mu\nu\rho\gamma} q^{\nu} \sigma^{\rho\gamma}
,
\label{C069}
\end{align}
with
\begin{align}
\null & \null
\nff^{\text{M}}_{k} = \nff_{k} + \nff_{k}^{T}
\Rightarrow
\nff^{\text{M}}_{k} = (\nff^{\text{M}}_{k})^{T}
\null & \null
\text{for}
\,
k=1,2,4
,
\label{C070}
\\
\null & \null
\nff^{\text{M}}_{k} = \nff_{k} - \nff_{k}^{T}
\Rightarrow
\nff^{\text{M}}_{k} = - (\nff^{\text{M}}_{k})^{T}
\null & \null
\text{for}
\,
k=3,5,6
.
\label{C071}
\end{align}
Now we can follow the discussion in Subsection~\ref{C002}
for Dirac neutrinos taking into account the additional constraints
(\ref{C070}) and (\ref{C071})
for Majorana neutrinos.
The hermiticity of $j_{\mu}^{(\nu)}$
and current conservation lead to an expression similar to that
in Eq.~(\ref{C043}):
\begin{align}
\Lambda^{\text{M}}_{\mu}(q)
=
\null & \null
\left( \gamma_{\mu} - q_{\mu} \slashed{q}/q^{2} \right)
\left[
\nff^{\text{M}}_{Q}(q^{2})
+
\nff^{\text{M}}_{A}(q^{2}) q^{2} \gamma_{5}
\right]
\nonumber
\\
\null & \null
-
i \sigma_{\mu\nu} q^{\nu}
\left[
\nff^{\text{M}}_{M}(q^{2})
+ i
\nff^{\text{M}}_{E}(q^{2}) \gamma_{5}
\right]
,
\label{C072}
\end{align}
with
$\nff^{\text{M}}_{Q} = \nff^{\text{M}}_{3}$,
$\nff^{\text{M}}_{M} = i \nff^{\text{M}}_{5}$,
$\nff^{\text{M}}_{E} = - 2 i \nff^{\text{M}}_{6}$ and
$\nff^{\text{M}}_{A} = - \nff^{\text{M}}_{2} / (m_{f}+m_i)$.
For the Hermitian $N{\times}N$ form-factor matrices
in the space of massive neutrinos,
\begin{equation}
\nff^{\text{M}}_{\Omega} = (\nff^{\text{M}}_{\Omega})^{\dagger}
\qquad
(\Omega=Q,M,E,A)
,
\label{C073}
\end{equation}
the Majorana constraints (\ref{C070}) and (\ref{C071})
imply that
\begin{align}
\null & \null
\nff^{\text{M}}_{\Omega} = - (\nff^{\text{M}}_{\Omega})^{T}
\quad
(\Omega=Q,M,E)
,
\label{C074}
\\
\null & \null
\nff^{\text{M}}_{A} = (\nff^{\text{M}}_{A})^{T}
.
\label{C075}
\end{align}
These relations
confirm the expectation discussed above that
for Majorana neutrinos
the charge, magnetic and electric form-factor matrices are antisymmetric and
the anapole form-factor matrix is symmetric.

Since
$\nff^{\text{M}}_{Q}$,
$\nff^{\text{M}}_{M}$ and
$\nff^{\text{M}}_{E}$
are antisymmetric, a Majorana neutrino does not have
diagonal charge and dipole magnetic and electric form factors
\cite{Touschek:1957,Case:1957zza}.
It can only have a diagonal anapole form factor.
On the other hand,
Majorana neutrinos can have as many off-diagonal (transition) form-factors as Dirac neutrinos.

Since the form-factor matrices are Hermitian as in the Dirac case,
$\nff^{\text{M}}_{Q}$,
$\nff^{\text{M}}_{M}$ and
$\nff^{\text{M}}_{E}$
are imaginary,
whereas
$\nff^{\text{M}}_{A}$ is real:
\begin{align}
\null & \null
\nff^{\text{M}}_{\Omega} = - (\nff^{\text{M}}_{\Omega})^{*}
\qquad
(\Omega=Q,M,E)
,
\label{C076}
\\
\null & \null
\nff^{\text{M}}_{A} = (\nff^{\text{M}}_{A})^{*}
.
\label{C077}
\end{align}
Taking into account these properties,
in the standard case of three-neutrino mixing the
charge, magnetic and electric Majorana form factors can be written as
\begin{equation}
\nff^{\text{M}fi}_{\Omega}(q^2) = i \sum_{j=1}^{3} \epsilon^{fij} \, \tilde{\nff}^{\text{M}j}_{\Omega}(q^2)
,
\label{C078}
\end{equation}
for
$\Omega=Q,M,E$,
in terms of three vectors of real form factors
\begin{equation}
(
\tilde{\nff}^{\text{M}1}_{\Omega},
\tilde{\nff}^{\text{M}2}_{\Omega},
\tilde{\nff}^{\text{M}3}_{\Omega}
)
=
- i
(
\tilde{\nff}^{\text{M}23}_{\Omega},
\tilde{\nff}^{\text{M}31}_{\Omega},
\tilde{\nff}^{\text{M}12}_{\Omega}
)
.
\label{C079}
\end{equation}

Considering now CP invariance,
the case of Majorana neutrinos is rather different from that of Dirac neutrinos,
because the CP phases of Majorana neutrinos are constrained by the CP invariance of the Majorana mass term.
In order to prove this statement,
let us first notice that since a massive Majorana neutrino field
$\nu_{k}$
is constrained by the Majorana relation in Eq.~(\ref{B028}),
only the parity transformation part is effective in a CP transformation.
Indeed,
from Eqs.~(\ref{B028}) and (\ref{I078}) we obtain
\begin{equation}
\mathsf{U}_{\text{CP}}
\nu_{k}(x)
\mathsf{U}_{\text{CP}}^{\dagger}
=
\xi^{\text{CP}}_{k} \gamma^{0} \nu_{k}(x_{\text{P}})
.
\label{C080}
\end{equation}
Considering the Majorana mass term in Eq.~(\ref{B026}),
we have
\begin{equation}
\mathsf{U}_{\text{CP}}
\nu_{k}^{T} \, \mathcal{C}^{\dagger} \, \nu_{k}
\mathsf{U}_{\text{CP}}^{\dagger}
=
- {\xi^{\text{CP}}_{k}}^{2}
\,
\nu_{k}^{T} \, \mathcal{C}^{\dagger} \, \nu_{k}
.
\label{C081}
\end{equation}
Therefore,
\begin{equation}
\text{CP}
\quad
\Longleftrightarrow
\quad
\xi^{\text{CP}}_{k}
=
\eta_{k} \, i
,
\label{C082}
\end{equation}
with
$\eta_{k} = \pm 1$.
These CP signs can be different for the different massive neutrinos,
even if they all take part to the standard charged-current weak interactions
through neutrino mixing,
because they can be compensated by the Majorana CP phases in the mixing matrix
(see \textcite{Giunti-Kim-2007}).
Therefore,
from Eq.~(\ref{C046}) we have
\begin{equation}
\Lambda^{\text{M}fi}_{\mu}(q)
\xrightarrow{\;\text{CP}\;}
\eta_{f} \eta_{i}
\gamma^{0}
\mathcal{C}
[\Lambda^{\text{M}if}_{\mu}(q_{\text{P}})]^{T}
\mathcal{C}^{\dagger}
\gamma^{0}
.
\label{C083}
\end{equation}
Imposing a CP constraint analogous to that in Eq.~(\ref{C028}),
we obtain
\begin{equation}
\text{CP}
\quad
\Longleftrightarrow
\quad
\left\{
\begin{array}{l} \displaystyle
\nff^{\text{M}fi}_{\Omega} = \eta_{f} \eta_{i} \nff^{\text{M}fi}_{\Omega} = \eta_{f} \eta_{i} (\nff^{\text{M}fi}_{\Omega})^{*}
,
\\ \displaystyle
\nff^{\text{M}fi}_{E} = - \eta_{f} \eta_{i} \nff^{\text{M}fi}_{E} = - \eta_{f} \eta_{i} (\nff^{\text{M}fi}_{E})^{*}
,
\end{array}
\right.
\label{C084}
\end{equation}
with
$\Omega=Q,M,A$.
Taking into account the constraints (\ref{C076}) and (\ref{C077}),
we have two cases:
\begin{equation}
\text{CP}
\quad
\text{and}
\quad
\eta_{f} = \eta_{i}
\quad
\Longleftrightarrow
\quad
\nff^{\text{M}fi}_{Q} = \nff^{\text{M}fi}_{M} = 0
,
\label{C085}
\end{equation}
and
\begin{equation}
\text{CP}
\quad
\text{and}
\quad
\eta_{f} = - \eta_{i}
\quad
\Longleftrightarrow
\quad
\nff^{\text{M}fi}_{E} = \nff^{\text{M}fi}_{A} = 0
.
\label{C086}
\end{equation}
Therefore,
if CP is conserved
two massive Majorana neutrinos can have either
a transition electric form factor
or
a transition magnetic form factor,
but not both,
and
the transition electric form factor can exist only together with a transition anapole form factor,
whereas
the transition magnetic form factor can exist only together with a transition charge form factor.
In the diagonal case $f=i$,
Eq.~(\ref{C085}) does not give any constraint,
because only diagonal
anapole form factors are allowed for Majorana neutrinos.

We consider now the CPT symmetry.
Following the method used at the end of the previous Subsection~\ref{C002} for Dirac neutrinos
and taking into account the particle-antiparticle equality of Majorana neutrinos,
one can show that the relations
(\ref{C074}) and (\ref{C075})
are a consequence of CPT symmetry
\cite{Nieves:1981zt,Kayser:1982br,Kayser:1984ge}.
Therefore,
in particular the existence of diagonal magnetic or electric moments
of Majorana neutrinos
would be a signal of CPT violation.

Let us finally note that the determination of which are the allowed form factors for Majorana neutrinos
can be also performed at the field level considering the
neutrino electromagnetic current
$j_{\mu}^{(\nu)}$ in Eq.~(\ref{C033})
and taking into account the chiral decomposition (\ref{B027})
of a Majorana field.
For example,
the magnetic dipole moment $\mgm^{\text{M}}_{kj}$ is generated by
\begin{equation}
\overline{\nu_{k}} \sigma^{\mu\nu} \nu_{j}
=
\overline{\nu_{kL}} \sigma^{\mu\nu} \nu_{jL}^{c}
+
\overline{\nu_{kL}^{c}} \sigma^{\mu\nu} \nu_{jL}
.
\label{C087}
\end{equation}
Taking into account the antisymmetry of fermion fields and the
properties of the charge-conjugation matrix,
one can find that
\begin{equation}
\overline{\nu_{k}} \sigma^{\mu\nu} \nu_{j}
=
-
\overline{\nu_{j}} \sigma^{\mu\nu} \nu_{k}
.
\label{C088}
\end{equation}
Therefore,
Majorana neutrinos can have only
off-diagonal (transition) magnetic dipole moments.

\subsection{Massless Weyl neutrinos}
\label{C089}

In Section~\ref{B001}
we have seen that neutrinos are known to be massive and mixed.
However,
it is interesting to study the electromagnetic properties of neutrinos
in the Standard Model,
where they are described by the two-component massless left-handed Weyl spinors
$\nu_{\afl L}(x)$,
with
$\afl=e,\mu,\tau$.
In this case,
taking into account that there is no mixing,
the neutrino effective electromagnetic current is
\begin{equation}
j_{\mu}^{(\nu)}(x)
=
\sum_{\afl,\bfl=e,\mu,\tau}
\overline{\nu_{\afl L}}(x) \, \Lambda^{\afl\bfl}_{\mu} \, \nu_{\bfl L}(x)
.
\label{C090}
\end{equation}
Since neutrinos are strictly left-handed,
the effective electromagnetic vertex in Fig.~\ref{C032}
is given by the matrix element
\begin{equation}
\langle \nu_{\afl}(p_{\afl},-) |
j_{\mu}^{(\nu)}(0)
| \nu_{\bfl}(p_{\bfl},-) \rangle
=
\overline{u_{\afl}^{(-)}}(p_{\afl})
\Lambda^{\afl\bfl}_{\mu}(q)
u_{\bfl}^{(-)}(p_{\bfl})
,
\label{C091}
\end{equation}
with
$q = p_{\bfl} - p_{\afl}$.
Since for massless neutrinos Eq.~(\ref{K007}) leads to the equality
\begin{equation}
\gamma^{5} u^{(-)}(p)
=
- u^{(-)}(p)
,
\label{C092}
\end{equation}
we can reduce the general expression of $\Lambda_{\mu}$
in Eq.~(\ref{C043}) to
\cite{Bernstein:1963qh}
\begin{equation}
\Lambda_{\mu}(q)
=
\left( \gamma_{\mu} - q_{\mu} \slashed{q}/q^{2} \right)
\nff(q^{2})
,
\label{C093}
\end{equation}
with
\begin{equation}
\nff(q^{2})
=
\nff_{Q}(q^{2})
-
\nff_{A}(q^{2}) q^{2}
.
\label{C094}
\end{equation}
Therefore,
massless left-handed Weyl neutrinos
have only one type of form factor given by the difference of
the charge form factor and the anapole form factor multiplied by $q^2$.

It is important that
massless left-handed Weyl neutrinos cannot have
diagonal or off-diagonal
electric or magnetic dipole moments,
because
\begin{equation}
\overline{\nu_{\afl L}} \sigma^{\mu\nu} \nu_{\bfl L}
=
\overline{\nu_{\afl L}} \sigma^{\mu\nu} \gamma_{5} \nu_{\bfl L}
=
0
.
\label{C095}
\end{equation}
The physical reason is that
in the case of massless neutrinos
the interactions generated by
electric and magnetic dipole moments flip helicity, as explained in Appendix~\ref{K001},
but the helicity flip of a massless left-handed Weyl neutrino
is not possible if the corresponding right-handed state does not exist.

In the Standard Model neutrinos are electrically neutral and
$
\nff(0)
=
\nff_{Q}(0)
=
0
$.
However,
radiative corrections
generate a finite
$\nff(q^{2})$
for
$q^{2} \neq 0$,
as explained in Subsection~\ref{G043},
where
$
\left.
d\nff_{Q}(q^{2}) / dq^{2}
\right|_{q^2=0}
$
is interpreted as the neutrino charge radius.
The equivalence between the
charge radius and anapole moment interpretations of
$\nff(q^{2})$
is explained in Subsection~\ref{G060}.

Let us also note that the Lorentz symmetry allows to write
an effective current of the type
\begin{equation}
\tilde{j}_{\mu}^{(\nu)}(x)
=
\sum_{\afl,\bfl=e,\mu,\tau}
\overline{\nu_{\afl L}}(x) \, \widetilde\Lambda^{\afl\bfl}_{\mu} \, \nu^{c}_{\bfl L}(x)
+
\text{H.c.}
.
\label{C096}
\end{equation}
However,
this current violates the total lepton number by two units
and cannot be generated in the framework of the Standard Model
where the total lepton number is conserved.
In theories beyond the Standard Model in which the total lepton number is violated,
neutrinos are Majorana particles and the discussion in Subsection~\ref{C065} applies.
For example, the magnetic moment terms in Eq.~(\ref{C096}) are of the form in Eq.~(\ref{C087}).
\section{Magnetic and electric dipole moments}
\label{D001}

The magnetic and electric dipole moments are theoretically
the most well-studied electromagnetic properties of neutrinos.
They also attract the interest of experimentalists, although
the magnetic moments of Dirac neutrinos in the simplest extension
of the Standard Model with the addition of right-handed neutrinos are
proportional to the corresponding neutrino mass and therefore they are many orders of
magnitude smaller than the present experimental limits.
However,
if there is new physics beyond the minimally extended Standard Model with right-handed neutrinos,
the magnetic and electric dipole moments of neutrinos
can be much larger
and observable by future experiments.

In Subsection~\ref{D009} we discuss this prediction for Dirac neutrinos
and in Subsection~\ref{D024} we present the predictions for the transition magnetic moments of Majorana neutrinos
in minimal extensions of the Standard Model.
In Subsection~\ref{D034}
we discuss the observable effects of electric and magnetic dipole moments in neutrino-electron elastic scattering
and
in Subsection~\ref{D043} we review the derivation of the effective dipole moments
in scattering experiments.
In Subsection~\ref{D064} we present the most relevant experimental limits on the values
of the effective dipole moments
and in Subsection~\ref{D068}
we conclude with some considerations
on the theoretical possibilities to have large magnetic moments.

\begin{figure}
\begin{center}
\begin{tabular}{cc}
\subfigure[]{\label{D002}
\includegraphics*[bb=248 628 343 720, width=0.3\linewidth]{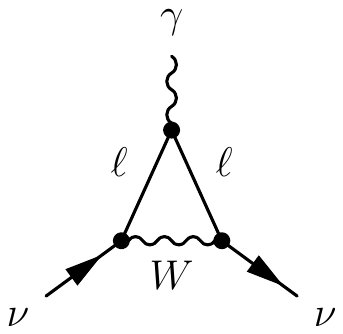}
}
&
\subfigure[]{\label{D003}
\includegraphics*[bb=248 628 343 720, width=0.3\linewidth]{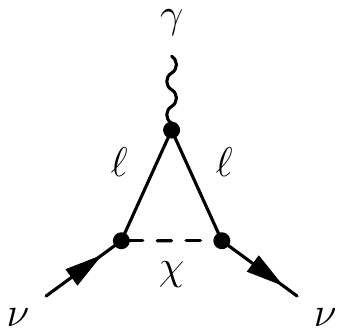}
}
\\
\subfigure[]
{\label{D004}
\includegraphics*[bb=248 628 343 720, width=0.3\linewidth]{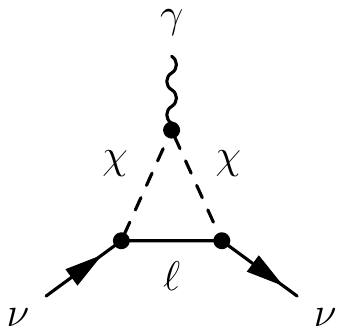}
}
&
\subfigure[]
{\label{D005}
\includegraphics*[bb=248 628 343 720, width=0.3\linewidth]{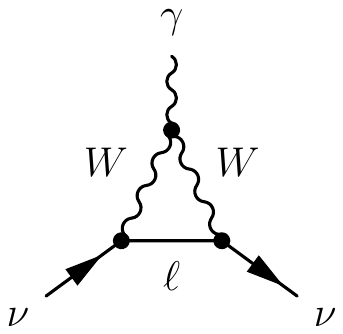}
}
\\
\subfigure[]
{\label{D006}
\includegraphics*[bb=248 628 343 720, width=0.3\linewidth]{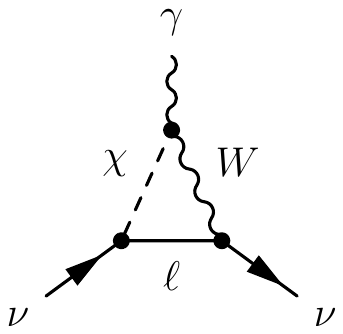}
}
&
\subfigure[]
{\label{D007}
\includegraphics*[bb=248 628 343 720, width=0.3\linewidth]{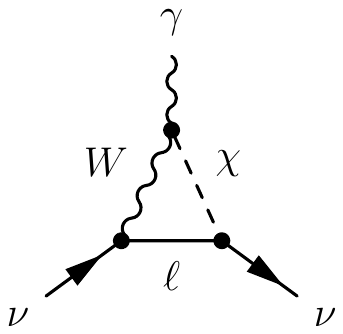}
}
\end{tabular}
\end{center}
\caption{\label{D008}
Feynman diagrams of proper vertices contributing to the neutrino vertex function
at one loop
in the extended Standard Model with
right-handed neutrinos
\cite{Dvornikov:2003js,Dvornikov:2004sj}.
$\chi$ is the unphysical would-be charged scalar boson.
}
\end{figure}

\subsection{Theoretical predictions for Dirac neutrinos}
\label{D009}

The first calculations of the one-loop
electromagnetic vertex of an initial fermion $\fermion$,
a final fermion $\fermion'$ (with $\fermion'=\fermion$ or $\fermion'\neq\fermion$)
and a photon $\gamma$
in the minimal extension of the
Standard Model with right-handed neutrinos
were presented in
\textcite{Petcov:1976ff,Marciano:1977wx,Lee:1977tib},
with applications to
$\mu \to e \gamma$
and
$\mu \to e e \bar{e}$
decays and to the radiative neutrino decay process discussed in Subsection~\ref{E005},
which depends on the transition electric and magnetic moments of the corresponding neutrinos.
The electric and magnetic moments of neutrinos have been explicitly calculated in
\textcite{Fujikawa:1980yx,Pal:1981rm,Shrock:1982sc,Dvornikov:2003js,Dvornikov:2004sj}
by evaluating the one-loop radiative diagrams
shown in Fig.~\ref{D008}.
The result is
\cite{Shrock:1982sc}
\begin{equation}\label{D010}
\left.
\setlength{\arraycolsep}{0pt}
\begin{array}{r}
\mgm^{\text{D}}_{kj}
\\
i
\elm^{\text{D}}_{kj}
\end{array}
\right\}
=
\frac{e G_{\text{F}}}{8\sqrt{2}\pi^{2}}
\left( m_{k} \pm m_{j} \right)
\sum_{\afl=e,\mu,\tau} f(a_{\afl}) U^{*}_{\afl k} U_{\afl j}
,
\end{equation}
where
the superscript ``D`` indicate Dirac neutrinos,
\begin{equation}
f(a_{\afl})
=
\frac{3}{4}
\left[1+\frac{1}{1-a_{\afl}}-\frac{2a_{\afl}}{(1-a_{\afl})^{2}}-\frac{2a_{\afl}^{2}\ln a_{\afl}}{(1-a_{\afl})^3}\right]
,
\label{D011}
\end{equation}
and
\begin{equation}
a_{\afl}
=
\frac{m_{\afl}^{2}}{m_{W}^{2}}
\leq
\frac{m_{\tau}^{2}}{m_{W}^{2}}
\simeq
5 \times 10^{-4}
,
\label{D012}
\end{equation}
for $\afl=e,\mu,\tau$.
Since all the $a_{\afl}$'s are very small, we can approximate
\begin{equation}\label{D013}
f(a_{\afl})
\simeq
\frac{3}{2}
\left(1 - \frac{a_{\afl}}{2}\right)
,
\end{equation}
and we obtain
\begin{align}
\left.
\setlength{\arraycolsep}{0pt}
\begin{array}{r}
\mgm^{\text{D}}_{kj}
\\
i
\elm^{\text{D}}_{kj}
\end{array}
\right\}
\simeq
\null & \null
\frac{3 e G_{\text{F}}}{16\sqrt{2}\pi^{2}}
\left( m_{k} \pm m_{j} \right)
\nonumber
\\
\null & \null
\times
\left(
\delta_{kj}
-
\frac{1}{2} \sum_{\afl=e,\mu,\tau}
U^{*}_{\afl k} U_{\afl j}
\frac{m_{\afl}^{2}}{m_{W}^{2}}
\right)
.
\label{D014}
\end{align}

It is clear that in this model there are no diagonal electric dipole moments ($\elm^{\text{D}}_{kk}=0$).
The diagonal magnetic moments are given by
\begin{equation}\label{D015}
\mgm^{\text{D}}_{kk}
\simeq
\frac{3e G_{\text{F}} m_{k}}{8\sqrt{2} \pi^{2}}
.
\end{equation}
Here we neglected the corrections due to the very small $a_{\afl}$'s in Eq.~(\ref{D012}).
Note also that higher-order electromagnetic corrections,
which have been neglected in Eq.~(\ref{D010}),
can be of the same order of magnitude or larger
(for example, the ratio of the contributions of two-loop and one-loop diagrams can be of the order of
$\alpha/\pi \simeq 2 \times 10^{-3}$).

The expression (\ref{D015}) exhibits the following important features.
Each diagonal magnetic moments is proportional to the corresponding neutrino
mass and vanishes in the massless limit,
even if in the extension of the Standard Model under consideration there are right-handed neutrinos.
This case is different from that of massless Weyl neutrinos
discussed in Subsection~\ref{C089},
in which all electric and magnetic, diagonal and off-diagonal
dipole moments are forbidden by the absence of right-handed states.
In this case we have
both spinors $u^{(-)}(p)$ and $u^{(+)}(p)$.
As shown in Appendix~\ref{K001},
in the massless limit helicity equals chirality,
because
$
\gamma^{5} u^{(\pm)}(p)
=
\pm u^{(\pm)}(p)
$.
Since
$\overline{u^{(\pm)}}(p) \sigma^{\mu\nu} u^{(\pm)}(p) = 0$
and
$\overline{u^{(\pm)}}(p) \sigma^{\mu\nu} u^{(\mp)}(p) \neq 0$,
the existence of a magnetic moment corresponds to the existence of
an helicity and chirality flipping interaction with the electromagnetic field.
However,
in the minimal extension of the
Standard Model with right-handed neutrinos
a magnetic moment is generated by the radiative diagrams in Fig.~\ref{D008},
which cannot flip chirality,
because the weak interaction vertices in the diagrams in Fig.~\ref{D008}
involve only left-handed neutrinos.

At the leading order in the small ratios $m_{\afl}^{2} / m_{W}^{2}$,
the diagonal magnetic moments
are independent of the neutrino mixing matrix and
of the values of the charged lepton masses.
Their numerical values are given by
\begin{equation}
\mgm^{\text{D}}_{kk}
\simeq
3.2 \times 10^{-19}
\left( \frac{m_{k}}{\text{eV}} \right) \bmag
.
\label{D016}
\end{equation}
Taking into account the existing constraint of the order of 1 eV on the neutrino masses
(see Subsection~\ref{B079}),
these values are several orders of magnitude smaller than the present experimental limits,
which are discussed in Subsection~\ref{D064}.

Let us consider now the neutrino transition dipole moments, which are
given by Eqs.~(\ref{D010}) and (\ref{D014}) for $k \neq j$.
Considering only
the leading term
$f(a_{\afl}) \simeq 3/2$
in the expansion (\ref{D013}),
one gets vanishing transition dipole moments,
because of the unitarity relation
\begin{equation}
\sum_{\afl=e,\mu,\tau} U^{*}_{\afl k} U_{\afl j}
=
\delta_{kj}
.
\label{D017}
\end{equation}
Therefore,
the first nonvanishing contribution comes from the second term in the expansion
(\ref{D013})
of $f(a_{\afl})$, which contains the additional small factor
$a_{\afl}=m_{\afl}^{2}/m_{W}^{2}$:
\begin{equation}
\left.
\setlength{\arraycolsep}{0pt}
\begin{array}{r}
\mgm^{\text{D}}_{kj}
\\
i
\elm^{\text{D}}_{kj}
\end{array}
\right\}
\simeq
-
\frac{3e G_{\text{F}}}{32\sqrt{2}\pi^{2}}
\left( m_{k} \pm m_{j} \right)
\sum_{\afl=e,\mu,\tau}
U^{*}_{\afl k} U_{\afl j}
\,
\frac{m_{\afl}^{2}}{m_{W}^{2}}
,
\label{D018}
\end{equation}
for $k \neq j$.
Thus, the transition magnetic moment
$\mgm^{\text{D}}_{kj}$
is suppressed with respect to the largest of the diagonal
magnetic moments of $\nu_{k}$ and $\nu_{j}$,
which are given by Eq.~(\ref{D015}).
This suppression is called
``GIM mechanism'',
in analogy with the suppression of flavor-changing neutral currents
in hadronic processes discovered by
\textcite{Glashow:1970gm}.
Numerically, the transition dipole moments
are given by
\begin{align}
\left.
\setlength{\arraycolsep}{0pt}
\begin{array}{r}
\mgm^{\text{D}}_{kj}
\\
i
\elm^{\text{D}}_{kj}
\end{array}
\right\}
\simeq
\null & \null
-
3.9 \times 10^{-23} \bmag
\left(\frac{m_{k} \pm m_j}{\text{eV}}\right)
\nonumber
\\
\null & \null
\times
\sum_{\afl=e,\mu,\tau}
U^{*}_{\afl k} U_{\afl j}
\left(\frac{m_{\afl}}{m_{\tau}}\right)^{2}
.
\label{D019}
\end{align}
Hence, the suppression of $\mgm^{\text{D}}_{kj}$ with respect to the numerical values
of the largest of the diagonal
magnetic moments of $\nu_{k}$ and $\nu_{j}$,
which are given by Eq.~(\ref{D016}),
is at least a factor of the order of $10^{-4}$.
The transition electric moments are even smaller than the transition magnetic moment
because of the mass difference,
and they are the only electric moments in the extension of the Standard Model under consideration.

So far in this Subsection we considered the standard framework of three-neutrino mixing
in which the unitarity relation (\ref{D017}) applies.
However, it is possible that there are additional nonstandard sterile neutrinos,
as discussed in Subsection~\ref{B094}.
In this case,
the unitarity relation (\ref{D017}) becomes
\begin{equation}
\sum_{\afl=e,\mu,\tau} U^{*}_{\afl k} U_{\afl j}
=
\delta_{kj}
-
\sum_{n=1}^{N_{s}}
U^{*}_{s_{n}k} U_{s_{n}j}
,
\label{D020}
\end{equation}
where $N_{s}$ is the number of sterile neutrinos,
which correspond in the mass basis to
$N_{s}$ nonstandard massive neutrinos.
From Eqs.~(\ref{D010}) and (\ref{D013}),
the diagonal magnetic moments are given by
\begin{equation}
\mgm^{\text{D}}_{kk}
\simeq
\frac{3e G_{\text{F}} m_{k}}{8\sqrt{2} \pi^{2}}
\left(
1
-
\sum_{n=1}^{N_{s}}
|U_{s_{n}k}|^2
\right)
.
\label{D021}
\end{equation}
From the inequality (\ref{B097}) it follows that the diagonal magnetic moments of the three standard massive neutrinos
($k=1,2,3$)
are practically the same as those in Eq.~(\ref{D015}).
On the other hand,
for the nonstandard massive neutrinos
Eq.~(\ref{B098}) implies that
\begin{equation}
\mgm^{\text{D}}_{kk}
\simeq
\frac{3e G_{\text{F}} m_{k}}{8\sqrt{2} \pi^{2}}
\left(
1
-
|U_{s_{k - 3}k}|^2
\right)
\quad
\text{for}
\quad
k \geq 4
.
\label{D022}
\end{equation}
Hence,
the diagonal magnetic moments of the nonstandard massive neutrinos are suppressed by the inequality (\ref{B099}).

The GIM mechanism does not operate for the transition dipole moments,
which are given by
\begin{align}
\null & \null
\left.
\setlength{\arraycolsep}{0pt}
\begin{array}{r}
\mgm^{\text{D}}_{kj}
\\
i
\elm^{\text{D}}_{kj}
\end{array}
\right\}
\simeq
-
\frac{3 e G_{\text{F}}}{16\sqrt{2}\pi^{2}}
\left( m_{k} \pm m_{j} \right)
\nonumber
\\
\null & \null
\times
\left(
\sum_{n=1}^{N_{s}}
U^{*}_{s_{n} k} U_{s_{n} j}
+
\frac{1}{2} \sum_{\afl=e,\mu,\tau}
U^{*}_{\afl k} U_{\afl j}
\frac{m_{\afl}^{2}}{m_{W}^{2}}
\right)
,
\label{D023}
\end{align}
for $k{\neq}j$.
However, the inequality (\ref{B097}) suppresses quadratically the additional contribution
$
\sum_{n}
U^{*}_{s_{n} k} U_{s_{n} j}
$
to the transition dipole moments
between two standard massive neutrinos
($k,j \leq 3$).
From Eqs.~(\ref{B096}) and (\ref{B098}),
the transition dipole moments
between two nonstandard massive neutrinos
($k,j \geq 4$)
are strongly suppressed.
On the other hand,
the transition dipole moments
between a standard massive neutrino and a nonstandard massive neutrino
($k \leq 3$ and $j \geq 4$ or vice versa)
are suppressed only linearly by the inequality (\ref{B097}).

\subsection{Theoretical predictions for Majorana neutrinos}
\label{D024}

Majorana neutrinos can have only transition magnetic and electric moments,
as discussed in Subsection~\ref{C065}.
The simplest models with Majorana neutrinos
can be obtained by extending the Standard Model with the addition of a
$\text{SU}(2)_L$ Higgs triplet \cite{Gelmini:1980re}
or with the addition of right-handed neutrinos
and a $\text{SU}(2)_L$ Higgs singlet
\cite{Chikashige:1980qk}
(see \textcite{Mohapatra:2004}).
Neglecting the model-dependent Feynman diagrams which depend on the details of the scalar sector,
the Majorana magnetic and electric transition moments are given by
\cite{Shrock:1982sc}
\begin{align}
\mgm^{\text{M}}_{kj}
\simeq
\null & \null
-
\frac{3 i e G_{\text{F}}}{16\sqrt{2}\pi^{2}}
\left( m_{k} + m_{j} \right)
\sum_{\afl=e,\mu,\tau}
\operatorname{Im}\left[U^{*}_{\afl k} U_{\afl j}\right]
\frac{m_{\afl}^{2}}{m_{W}^{2}}
,
\label{D025}
\\
\elm^{\text{M}}_{kj}
\simeq
\null & \null
\frac{3 i e G_{\text{F}}}{16\sqrt{2}\pi^{2}}
\left( m_{k} - m_{j} \right)
\sum_{\afl=e,\mu,\tau}
\operatorname{Re}\left[U^{*}_{\afl k} U_{\afl j}\right]
\frac{m_{\afl}^{2}}{m_{W}^{2}}
.
\label{D026}
\end{align}
Apart from the increase by a factor of 2 of the first coefficient
with respect to the Dirac case in Eq.~(\ref{D018}),
it is difficult to compare the expressions of the Dirac and Majorana
dipole moments,
because the mixing matrices are different in the two cases,
due to the possible presence of additional phases in the Majorana case
(see Eq.~(\ref{B031})).
In any case,
it is clear that
also the Majorana transition dipole moments are suppressed by the GIM mechanism
and they are expected to have the same order of magnitude (\ref{D019}) of the Dirac transition dipole moments.
However,
the model-dependent contributions of the scalar sector
can enhance the Majorana transition dipole moments
(see \textcite{Pal:1981rm,Barr:1990um,Pal:1991qr}).

If CP is conserved,
we must distinguish the two cases in which
$\nu_{k}$ and $\nu_{j}$
have the same or opposite CP phases,
as explained in Subsection~\ref{C065}.
It can be shown
(see \textcite{Giunti-Kim-2007})
that if CP is conserved the elements of the mixing matrix can be written as
\begin{equation}
U_{\afl k} = \mathcal{O}_{\afl k} \, e^{i\lambda_{k}}
,
\label{D027}
\end{equation}
where $\mathcal{O}$ is a real orthogonal matrix (e.g. $U^{\text{D}}$ in Eq.~(\ref{B032}) with $\delta_{13}=0,\pi$)
and the Majorana CP phases $\lambda_{k}$ such that
\begin{equation}
e^{-2i(\lambda_{k}-\lambda_{j})} = \eta_{k} / \eta_{j}
.
\label{D028}
\end{equation}
Here $\eta_{k}=\pm1$ is the sign of the CP phase in Eq.~(\ref{C082}) of the massive Majorana neutrino $\nu_{k}$.
Then, we have
\begin{equation}
U^{*}_{\afl k} U_{\afl j}
=
\mathcal{O}_{\afl k} \, \mathcal{O}_{\afl j} \, e^{-i(\lambda_{k}-\lambda_{j})}
=
\mathcal{O}_{\afl k} \, \mathcal{O}_{\afl j} \, \sqrt{\eta_{k} / \eta_{j}}
.
\label{D029}
\end{equation}
Then,
if $\nu_{k}$ and $\nu_{j}$
have the same CP phase
($\eta_{k}=\eta_{j}$),
the products $U^{*}_{\afl k} U_{\afl j} = \mathcal{O}_{\afl k} \mathcal{O}_{\afl j}$ are real
and the dipole moments are given by
\cite{Schechter:1981hw,Pal:1981rm}
\begin{equation}
\mgm^{\text{M}}_{kj} = 0
\qquad
\text{and}
\qquad
\elm^{\text{M}}_{kj}
=
2 \elm^{\text{D}}_{kj}
,
\label{D030}
\end{equation}
with
$\elm^{\text{D}}_{kj}$
and
$\mgm^{\text{D}}_{kj}$
given by Eq.~(\ref{D010}).
On the other hand,
if $\nu_{k}$ and $\nu_{j}$
have opposite CP phases
($\eta_{k}=-\eta_{j}$),
the products $U^{*}_{\afl k} U_{\afl j} = i \mathcal{O}_{\afl k} \mathcal{O}_{\afl j}$ are imaginary
and the dipole moments are given by
\cite{Schechter:1981hw,Pal:1981rm}
\begin{equation}
\mgm^{\text{M}}_{kj}
=
2 \mgm^{\text{D}}_{kj}
\qquad
\text{and}
\qquad
\elm^{\text{D}}_{kj} = 0
.
\label{D031}
\end{equation}
The vanishing of $\mgm^{\text{M}}_{kj}$ in the first case
and the vanishing of $\elm^{\text{D}}_{kj}$ in the second case are consistent with the
general results in Eqs.~(\ref{C085}) and (\ref{C086}).

Let us consider now the case of additional sterile neutrinos discussed in Subsection~\ref{B094}.
Taking into account the unitarity relation (\ref{D020}),
in the Majorana case one can infer from \textcite{Shrock:1982sc}
that the transition dipole moments are given by
\begin{align}
\null & \null
\mgm^{\text{M}}_{kj}
\simeq
-
\frac{3 i e G_{\text{F}}}{16\sqrt{2}\pi^{2}}
\left( m_{k} + m_{j} \right)
\nonumber
\\
\null & \null
\times
\operatorname{Im}\left[
\sum_{n=1}^{N_{s}}
U^{*}_{s_{n} k} U_{s_{n} j}
+
\sum_{\afl=e,\mu,\tau}
U^{*}_{\afl k} U_{\afl j}
\,
\frac{m_{\afl}^{2}}{m_{W}^{2}}
\right]
,
\label{D032}
\\
\null & \null
\elm^{\text{M}}_{kj}
\simeq
\frac{3 i e G_{\text{F}}}{16\sqrt{2}\pi^{2}}
\left( m_{k} - m_{j} \right)
\nonumber
\\
\null & \null
\times
\operatorname{Re}\left[
\sum_{n=1}^{N_{s}}
U^{*}_{s_{n} k} U_{s_{n} j}
+
\sum_{\afl=e,\mu,\tau}
U^{*}_{\afl k} U_{\afl j}
\,
\frac{m_{\afl}^{2}}{m_{W}^{2}}
\right]
.
\label{D033}
\end{align}
Here the situation is similar to the case of Dirac neutrinos discussed at the end of Subsection~\ref{D009}:
the additional contribution
$
\sum_{n}
U^{*}_{s_{n} k} U_{s_{n} j}
$
to the transition dipole moments
between two standard massive neutrinos
($k,j \leq 3$)
is suppressed quadratically by the inequality (\ref{B097});
the transition dipole moments
between two nonstandard massive neutrinos
($k,j \geq 4$)
are strongly suppressed by Eqs.~(\ref{B096}) and (\ref{B098});
the transition dipole moments
between a standard massive neutrino and a nonstandard massive neutrino
($k \leq 3$ and $j \geq 4$ or vice versa)
are suppressed only linearly by the inequality (\ref{B097}).

\subsection{Neutrino-electron elastic scattering}
\label{D034}

The most sensitive and widely used method for the experimental
investigation of the neutrino magnetic moment is provided by
direct laboratory measurements of low-energy elastic scattering of neutrinos and antineutrinos
with electrons in
reactor, accelerator and solar experiments\footnote{
The effects of a neutrino magnetic moment in other processes which can be observed in laboratory experiments
have been discussed in
\textcite{Kim:1974xx,Kim:1978xk,Dicus:1978rz,Rosado:1982fr}.
}.
Detailed descriptions
of several experiments can be found in
\textcite{Wong:2005pa,Beda:2007hf}.

Extensive experimental studies of the neutrino magnetic moment,
performed during many years, are stimulated by the hope to observe
a value much larger than the prediction in Eq.~(\ref{D016}) of the
minimally extended Standard Model with right-handed neutrinos.
It would be a clear indication of new physics beyond the extended
Standard Model.
For example,
the effective magnetic moment in
$\bar\nu_{e}$-$e$ elastic scattering
in a class of extra-dimension models
can be as large as about
$10^{-10} \bmag$
\cite{Mohapatra:2004ce}.
Future higher precision
reactor experiments can therefore be used to provide new
constraints on large extra-dimensions.

The possibility for neutrino-electron elastic scattering due to neutrino
magnetic moment was first considered in \textcite{Carlson:1932rk}
and the cross section of this process was calculated in
\textcite{Bethe:1935cp}
(for related short historical notes see
\textcite{Kyuldjiev:1984kz}).
Here we would like to recall the
paper by \textcite{Domogatsky:1971tu},
where the cross section of \textcite{Bethe:1935cp}
was corrected and the antineutrino-electron
cross section was considered in the context of the earlier experiments
with reactor antineutrinos of
\textcite{Cowan:1954pq,Cowan:1957pp},
which were aimed to reveal the effects of the neutrino
magnetic moment.
Discussions on the derivation of the cross section
and on the optimal conditions for bounding the neutrino magnetic
moment, as well as a collection of cross section formulae for
elastic scattering of neutrinos (antineutrinos) on electrons, nucleons,
and nuclei can be found in \textcite{Kyuldjiev:1984kz,Vogel:1989iv}.

Let us consider the elastic scattering
\begin{equation}
\nua{\afl} + e^{-} \to \nua{\afl} + e^{-}
\label{D035}
\end{equation}
of a neutrino or antineutrino with flavor $\afl=e,\mu,\tau$ and energy $E_{\nu}$
with an electron at rest in the laboratory frame.
There are two observables:
the kinetic
energy $T_{e}$ of the recoil electron
and the recoil angle $\chi$
with respect to the neutrino beam,
which are related by
\begin{equation}
\cos\chi = \frac{E_{\nu}+m_{e}}{E_{\nu}}\Big[\frac{T_{e}}{T_{e}+2m_{e}}\Big]^{1/2}
\label{D036}
.
\end{equation}
The electron kinetic energy is constrained from the energy-momentum
conservation by
\begin{equation}
T_{e} \leq \frac {2E_{\nu}^{2}}{2E_{\nu} + m_{e}}
.
\end{equation}

Since,
in the ultrarelativistic limit,
the neutrino
magnetic moment interaction changes the neutrino helicity
and the Standard Model weak interaction conserves the neutrino helicity
(see Appendix~\ref{K001}),
the two contributions add incoherently in the cross section\footnote{
The small interference term
due to neutrino masses has been derived by \textcite{Grimus:1997aa}.
}
which can be written as
\cite{Vogel:1989iv},
\begin{equation}\label{D037}
\frac{d\sigma_{\nu_{\afl}e^{-}}}{dT_{e}}
=
\left(\frac{d\sigma_{\nu_{\afl}e^{-}}}{dT_{e}}\right)_{\text{SM}}
+
\left(\frac{d\sigma_{\nu_{\afl}e^{-}}}{dT_{e}}\right)_{\text{mag}}
.
\end{equation}

The weak-interaction cross section is given by
\begin{align}
\left(\frac{d\sigma_{\nu_{\afl}e^{-}}}{dT_{e}}\right)_{\text{SM}}
=
\null & \null
\frac{G_{\text{F}}^2 m_{e}}{2\pi}
\bigg\{
(g_{V}^{\nu_{\afl}} + g_{A}^{\nu_{\afl}})^{2}
\nonumber
\\
\null & \null
+
(g_{V}^{\nu_{\afl}} - g_{A}^{\nu_{\afl}})^{2}
\left(1-\frac{T_{e}}{E_{\nu}}\right)^{2}
\nonumber
\\
\null & \null
+
\left[ (g_{A}^{\nu_{\afl}})^{2} - (g_{V}^{\nu_{\afl}})^{2} \right]
\frac{m_{e}T_{e}}{E_{\nu}^{2}}
\bigg\}
,
\label{D038}
\end{align}
with the standard coupling constants $g_{V}$ and $g_{A}$ given by
\begin{align}
\null & \null
g_{V}^{\nu_{e}}
=
2\sin^{2} \theta_{W} + 1/2
,
\quad
\null && \null
g_{A}^{\nu_{e}}
=
1/2
,
\label{D039}
\\
\null & \null
g_{V}^{\nu_{\mu,\tau}}
=
2\sin^{2} \theta_{W} - 1/2
,
\quad
\null && \null
g_{A}^{\nu_{\mu,\tau}}
=
- 1/2
.
\label{D040}
\end{align}
For antineutrinos one must substitute
$g_A \to -g_A$.

The neutrino magnetic-moment contribution to the cross section is given by
\cite{Vogel:1989iv}
\begin{equation}
\left(\frac{d\sigma_{\nu_{\afl}e^{-}}}{dT_{e}}\right)_{\text{mag}}
=
\frac{\pi\alpha^{2}}{m_{e}^{2}}
\left(\frac{1}{T_{e}}-\frac{1}{E_{\nu}}\right)
\left(\frac{\mgm_{\nu_{\afl}}}{\bmag}\right)^{2}
,
\label{D041}
\end{equation}
where $\mgm_{\nu_{\afl}}$ is the effective magnetic moment discussed in the following Subsection~\ref{D043}.
It is called traditionally ``magnetic moment'',
but it receives equal contributions from both the electric and magnetic dipole moments.

\begin{figure}
\begin{center}
\includegraphics*[bb=60 104 2249 1719, width=\linewidth]{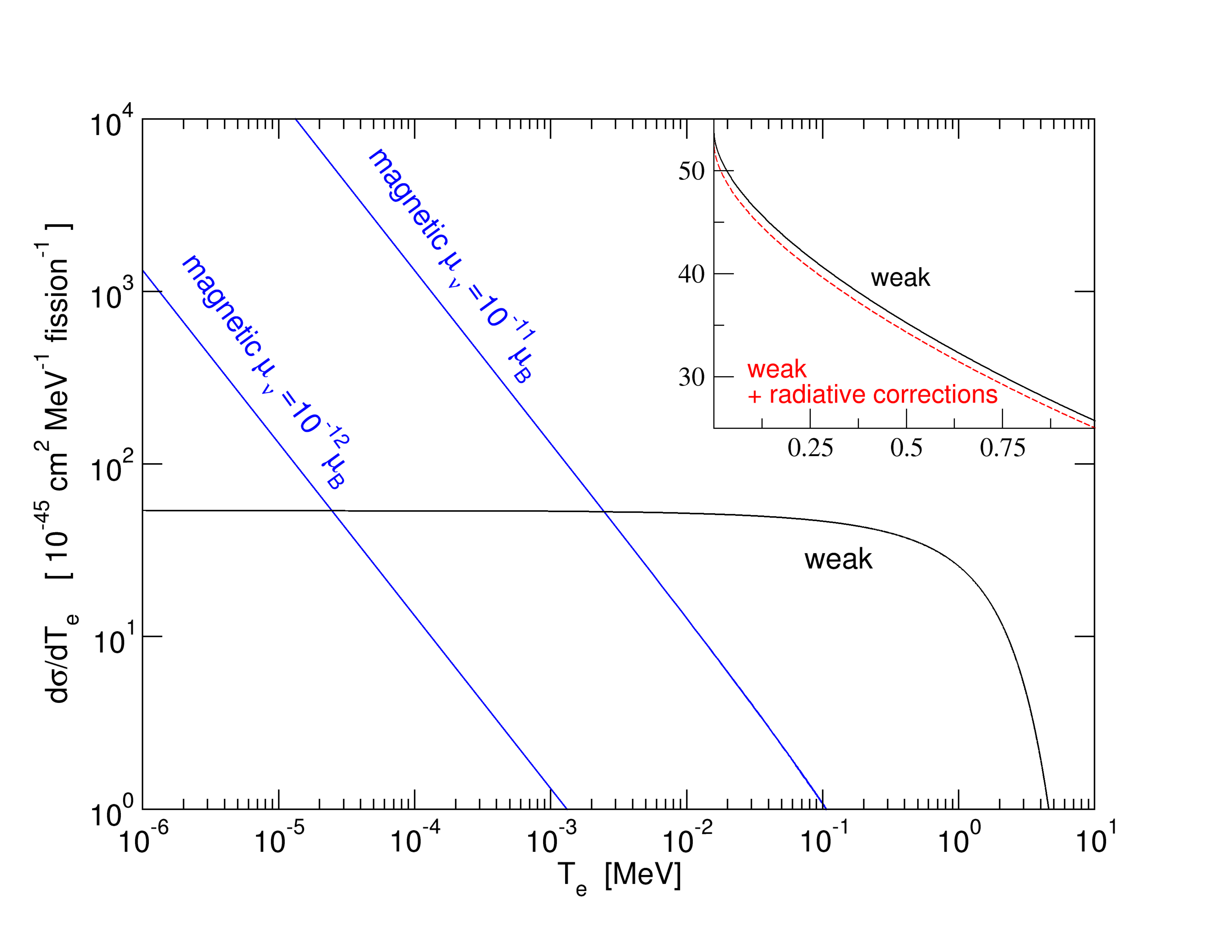}
\end{center}
\caption{\label{D042}
Standard Model weak and magnetic moment electromagnetic
contributions to the differential cross section of neutrino-electron scattering averaged over the antineutrino spectrum of fissioning $^{235}$U.
The inset plot is the weak correction on the linear scale both with (dashed line) and without (solid line) radiative corrections \cite{Sarantakos:1982bp}.
Figure from \textcite{Balantekin:2013sda}.
}
\end{figure}

The two terms $(d\sigma_{\nu_{\afl}e^{-}}/dT_{e})_{\text{SM}}$ and $(d\sigma_{\nu_{\afl}e^{-}}/dT_{e})_{\text{mag}}$ exhibit quite different
dependencies on the experimentally observable electron kinetic energy $T_{e}$,
as illustrated in Fig.~\ref{D042}
taken from \textcite{Balantekin:2013sda}
(see also
\textcite{Vogel:1989iv,Beda:2007hf}).
One can see that small values of the neutrino magnetic moment
can be probed by lowering the electron recoil energy threshold.
In fact,
considering
$T_{e} \ll E_{\nu}$ in Eq.~(\ref{D041})
and neglecting the coefficients due to
$g_{V}^{\nu_{\afl}}$ and $g_{A}^{\nu_{\afl}}$
in Eq.~(\ref{D038}),
one can find that
$(d\sigma/dT_{e})_{\text{mag}}$ exceeds
$(d\sigma/dT_{e})_{\text{SM}}$ for
\begin{equation}
T_{e}
\lesssim
\frac{\pi^{2}\alpha^{2}}{G_{\text{F}}^{2}m_{e}^3}
\left(\frac {\mgm_{\nu}}{\bmag}\right)^{2}
.
\end{equation}

\subsection{Effective magnetic moment}
\label{D043}

In scattering experiments
the neutrino is created at some distance
from the detector as a flavor neutrino,
which is a superposition of massive neutrinos.
Therefore,
the magnetic moment that is measured in these experiment is not that of a massive neutrino,
but it is an effective magnetic moment
which takes into account neutrino mixing and the oscillations
during the propagation between source and detector
\cite{Grimus:1997aa,Beacom:1999wx}.

Let us consider
an initial neutrino with flavor $ \afl = e, \, \mu, \, \tau $,
which is described by the flavor state in Eq.~(\ref{B036}).
The state of the neutrino which is
detected through a scattering process at
a space-time distance $(\vec{L},T)$
from the source is given by the superposition of massive neutrinos
in the first line of Eq.~(\ref{B040}).
Considering an incoming left-handed neutrino,
the amplitude of $\nu_{j}$ production in low-$q^{2}$ electromagnetic scattering
of a neutrino which has traveled a space-time distance $(\vec{L},T)$
from a source of $\nu_{\afl}$ is
\begin{align}
A_{\afl j}(\vec{L},T)
\null & \null
\propto
\sum_{k}
U_{\afl k}^{*}
e^{ - i E_{k} T + i \vet{p}_{k} \cdot \vec{L} }
\nonumber
\\
\null & \null
\times
\sum_{h_{j}}
\overline{u_{j}^{(h_{j})}}
\sigma_{\mu\nu}
q^{\nu}
\left(
\mgm_{jk} + i \elm_{jk} \gamma_{5}
\right)
u_{k}^{(-)}
.
\label{D044}
\end{align}
Since for an incoming ultrarelativistic left-handed neutrino
the additional $\gamma^{5}$ in the electric dipole term
has only the effect of changing a sign
(see Eq.~(\ref{K007})),
the amplitude of $\nu_{k}\to\nu_{j}$ transitions
is proportional to
$\mgm_{jk} - i \elm_{jk}$,
leading to
\begin{equation}
A_{\afl j}(\vec{L},T)
\propto
\sum_{k}
U_{\afl k}^{*}
e^{ - i E_{k} T + i \vet{p}_{k} \cdot \vec{L} }
\left(
\mgm_{jk} - i \elm_{jk}
\right)
.
\label{D045}
\end{equation}
The total cross section of electromagnetic scattering with an electron or a nucleon is given by
\begin{equation}
\sigma_{\nu_{\afl}e^{-}}(\vec{L},T)
\propto
\sum_{j}
|A_{\afl j}(\vec{L},T)|^2
.
\label{D046}
\end{equation}
Taking into account that for ultrarelativistic neutrinos $T=L$,
from the approximation in Eq.~(\ref{B042})
we obtain that
the cross section is proportional to the squared effective magnetic moment
\begin{equation}
\mgm_{\nu_{\afl}}^{2}(L,E_{\nu})
=
\sum_{j}
\left|
\sum_{k}
U_{\afl k}^{*}
e^{- i \Delta{m}^{2}_{kj} L / 2 E_{\nu} }
\left(
\mgm_{jk} - i \elm_{jk}
\right)
\right|^{2}
.
\label{D047}
\end{equation}
In this expression of the effective $\mgm_{\nu_{\afl}}$
one can see that in general both the magnetic and electric dipole moments
contribute to the elastic scattering.
Note also that,
as neutrino oscillations discussed in Section~\ref{B001},
the effective magnetic moment
$\mgm_{\nu_{\afl}}(L,E_{\nu})$ depends on the neutrino squared-mass differences,
not on the absolute values of neutrino masses.

Considering antineutrinos,
the mixing of antineutrinos is obtained from that of neutrinos in Eq.~(\ref{B036})
with the substitution $U \to U^{*}$.
From Eq.~(\ref{C054}) it follows that the
electric and magnetic moments of antineutrinos are obtained with the substitutions
$\mgm_{jk} \to - \mgm_{jk}^{*}$
and
$\elm_{jk} \to - \elm_{jk}^{*}$.
Moreover,
we must take into account that incoming antineutrinos are right-handed.
Hence,
for antineutrinos we have
\begin{align}
\overline{A}_{\afl j}(\vec{L},T)
\null & \null
\propto
\sum_{k}
U_{\afl k}
e^{ - i E_{k} T + i \vet{p}_{k} \cdot \vec{L} }
\nonumber
\\
\null & \null
\times
\sum_{h_{j}}
\overline{u_{j}^{(h_{j})}}
\sigma_{\mu\nu}
q^{\nu}
\left(
\mgm_{jk}^{*} + i \elm_{jk}^{*} \gamma_{5}
\right)
u_{k}^{(+)}
.
\label{D048}
\end{align}
For an incoming ultrarelativistic right-handed neutrino
the additional $\gamma^{5}$ in the electric dipole term
has no effect
(see Eq.~(\ref{K007}))
and we obtain
\begin{align}
\mgm_{\bar\nu_{\afl}}^{2}(L,E_{\nu})
=
\null & \null
\sum_{j}
\left|
\sum_{k}
U_{\afl k}
e^{- i \Delta{m}^{2}_{kj} L / 2 E_{\nu} }
\left(
\mgm_{jk}^{*} + i \elm_{jk}^{*}
\right)
\right|^{2}
\nonumber
\\
=
\null & \null
\sum_{j}
\left|
\sum_{k}
U_{\afl k}^{*}
e^{i \Delta{m}^{2}_{kj} L / 2 E_{\nu} }
\left(
\mgm_{jk} - i \elm_{jk}
\right)
\right|^{2}
.
\label{D049}
\end{align}
Therefore,
there can be only a phase difference between the terms contributing to
$\mgm_{\nu_{\afl}}^{2}(L,E_{\nu})$
and
$\mgm_{\bar\nu_{\afl}}^{2}(L,E_{\nu})$,
which is induced by neutrino oscillations.

As discussed in the following Subsection~\ref{D064},
the laboratory experiments which are most sensitive to small values of
the effective magnetic moment
are reactor and accelerator experiments
which detect the elastic scattering of flavor neutrinos on electrons
at a short distance from the neutrino source.
In this case,
the value in Eq.~(\ref{B080}) of the largest squared-mass difference
$\Delta{m}^{2}_{\text{A}}$
in the standard case of three-neutrino mixing
is such that
$\Delta{m}^{2}_{\text{A}} L / 2 E_{\nu} \ll 1$.
Therefore,
it is possible to approximate all the exponentials in Eqs.~(\ref{D047}) and (\ref{D049}) with unity
and obtain the effective short-baseline magnetic moment of flavor neutrinos and antineutrinos
\begin{align}
\mgm_{\nu_{\afl}}^{2}
\simeq
\null & \null
\mgm_{\bar\nu_{\afl}}^{2}
\simeq
\sum_{j}
\left|
\sum_{k}
U_{\afl k}^{*}
\left(
\mgm_{jk} - i \elm_{jk}
\right)
\right|^2
\nonumber
\\
=
\null & \null
\left[
U
\left( \mgm^2 + \elm^2 \right)
U^{\dagger}
+
2 \operatorname{Im}\!\left( U \, \mgm \, \elm \, U^{\dagger} \right)
\right]_{\afl\afl}
,
\label{D050}
\end{align}
where we took into account that
$\mgm=\mgm^{\dagger}$
and
$\elm=\elm^{\dagger}$.
In this approximation the effective magnetic moment is independent of the neutrino energy
and from the source-detector distance.

In the following,
when we refer to an effective magnetic moment of a flavor neutrino without indication of
a source-detector distance $L$
it is implicitly understood that $L$ is small
and the effective magnetic moment is given by
Eq.~(\ref{D050}).

It is interesting to note that
flavor neutrinos can have effective magnetic moments
even if massive neutrinos are Majorana particles.
In this case,
since massive Majorana neutrinos do not have diagonal magnetic and electric dipole moments,
the effective magnetic moments of flavor neutrinos
receive contributions only from the transition dipole moments.
For example,
in the three-generation case,
following Eq.~(\ref{C078}), we can write $\mgm_{jk}$ and $\elm_{jk}$
as
\begin{equation}
\mgm_{jk}
=
i
\sum_{m=1}^{3}
\epsilon_{jkm}
\tilde{\mgm}_{m}
,
\quad
\elm_{jk}
=
i
\sum_{m=1}^{3}
\epsilon_{jkm}
\tilde{\elm}_{m}
,
\label{D051}
\end{equation}
with real $\tilde{\mgm}_{m}$ and $\tilde{\elm}_{m}$.
Thus, we obtain
\begin{equation}
\mgm_{\nu_{\afl}}^{2}
\simeq
\sum_{k=1}^{3} \left( \tilde{\mgm}_{k}^2 + \tilde{\elm}_{k}^2 \right)
-
\left|
\sum_{k=1}^{3}
U_{\afl k}
\left(
\tilde{\mgm}_{k} - i \tilde{\elm}_{k}
\right)
\right|^2
.
\label{D052}
\end{equation}

Another case in which the effective magnetic moment does not
depend on the neutrino energy
and on the source-detector distance
is when the source-detector distance is much larger than all the oscillation lengths
$L_{kj} = 4 \pi E_{\nu} / |\Delta{m}^{2}_{kj}|$.
In this case the interference terms in Eqs.~(\ref{D047}) and (\ref{D049})
are washed out by the finite energy resolution of the detector,
leading to
\begin{align}
\mgm_{\nu_{\afl}}^{2}
\null & \null
(\infty)
\simeq
\mgm_{\bar\nu_{\afl}}^{2}(\infty)
\simeq
\sum_{k}
|U_{\afl k}|^2
\sum_{j}
\left|
\mgm_{jk} - i \elm_{jk}
\right|^2
\nonumber
\\
\null & \null
=
\sum_{k}
|U_{\afl k}|^2
\left[
(\mgm^2)_{kk} + (\elm^2)_{kk} + 2 \operatorname{Im}(\mgm\elm)_{kk}
\right]
.
\label{D053}
\end{align}
For three-generations of Majorana neutrinos,
from Eq.~(\ref{D051})
we obtain
\begin{equation}
\mgm_{\nu_{\afl}}^{2}(\infty)
\simeq
\mgm_{\bar\nu_{\afl}}^{2}(\infty)
\simeq
\sum_{k=1}^{3}
\left( 1 - |U_{\afl k}|^2 \right)
\left(
\tilde{\mgm}_{k}^2 + \tilde{\elm}_{k}^2
\right)
.
\label{D054}
\end{equation}

\begin{table*}
\renewcommand{\arraystretch}{1.2}
\begin{tabular}{lllll}
Method & Experiment & Limit & CL & Reference\\
\hline
\multirow{5}{*}{Reactor $\bar\nu_e$-$e^-$}
&Krasnoyarsk		&$\mgm_{\nu_e} < 2.4 \times 10^{-10}\,\bmag$	&90\%	&\textcite{Vidyakin:1992nf}		\\
&Rovno			&$\mgm_{\nu_e} < 1.9 \times 10^{-10}\,\bmag$	&95\%	&\textcite{Derbin:1993wy}		\\
&MUNU			&$\mgm_{\nu_e} < 9   \times 10^{-11}\,\bmag$	&90\%	&\textcite{Daraktchieva:2005kn}	\\
&TEXONO			&$\mgm_{\nu_e} < 7.4 \times 10^{-11}\,\bmag$	&90\%	&\textcite{Wong:2006nx}		\\
&GEMMA			&$\mgm_{\nu_e} < 2.9 \times 10^{-11}\,\bmag$	&90\%	&\textcite{Beda:2012zz}		\\
\hline
\multirow{1}{*}{Accelerator $\nu_e$-$e^-$}
&LAMPF			&$\mgm_{\nu_e} < 1.1 \times 10^{-9}\,\bmag$		&90\%	&\textcite{Allen:1992qe}		\\
\hline
\multirow{1}{*}{Accelerator ($\nu_{\mu},\bar\nu_{\mu}$)-$e^-$}
&BNL-E734		&$\mgm_{\nu_{\mu}} < 8.5 \times 10^{-10}\,\bmag$	&90\%	&\textcite{Ahrens:1990fp}		\\
&LAMPF			&$\mgm_{\nu_{\mu}} < 7.4 \times 10^{-10}\,\bmag$	&90\%	&\textcite{Allen:1992qe}		\\
&LSND			&$\mgm_{\nu_{\mu}} < 6.8 \times 10^{-10}\,\bmag$	&90\%	&\textcite{Auerbach:2001wg}		\\
\hline
\multirow{1}{*}{Accelerator ($\nu_{\tau},\bar\nu_{\tau}$)-$e^-$}
&DONUT			&$\mgm_{\nu_{\tau}} < 3.9 \times 10^{-7}\,\bmag$	&90\%	&\textcite{Schwienhorst:2001sj}	\\
\hline
\multirow{2}{*}{Solar $\nu_e$-$e^-$}
&Super-Kamiokande	&$\mgm_{\text{S}}(E_{\nu} \gtrsim 5 \, \text{MeV}) < 1.1 \times 10^{-10}\,\bmag$	&90\%	&\textcite{Liu:2004ny}		\\
&Borexino		&$\mgm_{\text{S}}(E_{\nu} \lesssim 1 \, \text{MeV}) < 5.4 \times 10^{-11}\,\bmag$	&90\%	&\textcite{Arpesella:2008mt}	\\
\hline
\end{tabular}
\caption{\label{D055}
Experimental limits for different neutrino effective magnetic moments.
}
\end{table*}

So far, in this Subsection we have considered the effects of neutrino mixing
and oscillations on the effective magnetic moment for neutrinos propagating in vacuum.
In the case of solar neutrinos,
which have been used by the
Super-Kamiokande
\cite{Liu:2004ny}
and Borexino
\cite{Arpesella:2008mt}
experiments to search for neutrino magnetic moment effects,
one must take into account
the matter effects discussed in Subsection~\ref{B035}.
The state which describes the neutrinos emerging
from the Sun is the following generalization of the state in Eq.~(\ref{B076})
which takes into account three-neutrino mixing and
the squared-mass hierarchy in Eq.~(\ref{B081}):
\begin{equation}
|\nu_{\text{S}}\rangle
=
\sum_{k=1}^{3}
(U_{ek}^{\text{M}})^{*}
|\nu_{k}\rangle
,
\label{D056}
\end{equation}
with
\begin{align}
\null & \null
U_{e1}^{\text{M}}
=
\cos\vartheta_{13}
\cos\vartheta_{\text{M}}^{0}
,
\label{D057}
\\
\null & \null
U_{e2}^{\text{M}}
=
\cos\vartheta_{13}
\sin\vartheta_{\text{M}}^{0}
,
\label{D058}
\\
\null & \null
U_{e3}^{\text{M}}
=
U_{e3}
=
\sin\vartheta_{13} e^{-i\delta_{13}}
,
\label{D059}
\end{align}
where
$\vartheta_{\text{M}}^{0}$
is the effective mixing angle at the point of neutrino production inside the Sun.
Following the same reasoning that led to Eq.~(\ref{D047}),
we obtain that the effective magnetic moment measured by an experiment on Earth is
\begin{equation}
\mgm_{\text{S}}^{2}(L,E_{\nu})
=
\sum_{j}
\left|
\sum_{k}
(U_{ek}^{\text{M}})^{*}
e^{- i \Delta{m}^{2}_{kj} L / 2 E_{\nu} }
\left(
\mgm_{jk} - i \elm_{jk}
\right)
\right|^{2}
,
\label{D060}
\end{equation}
where
$L$ is the Sun-Earth distance.
Since the Sun-Earth distance is much larger than the oscillation lengths,
the interference terms in Eqs.~(\ref{D060})
are washed out by the finite energy resolution of the detector
and we obtain the effective magnetic moment
\begin{equation}
\mgm_{\text{S}}^{2}(E_{\nu})
=
\sum_{k}
|U_{ek}^{\text{M}}|^2
\sum_{j}
\left|
\mgm_{jk} - i \elm_{jk}
\right|^2
.
\label{D061}
\end{equation}
This expression is similar to that in Eq.~(\ref{D053}),
but takes into account the effective mixing at the point of neutrino production inside the Sun.
Note that $\mgm_{\text{S}}$
depends on the neutrino energy
through the dependence of $\vartheta_{\text{M}}^{0}$
on $E_{\nu}$
(see Eq.~(\ref{B066})).
As remarked before Eq.~(\ref{B078}),
in practice we have
$\vartheta_{\text{M}}^{0}\simeq\vartheta_{12}$ for $E_{\nu} \lesssim 1 \, \text{MeV}$
and
$\vartheta_{\text{M}}^{0}\simeq\pi/2$ for $E_{\nu} \gtrsim 5 \, \text{MeV}$.
Therefore,
\begin{equation}
\mgm_{\text{S}}(E_{\nu} \lesssim 1 \, \text{MeV})
\simeq
\mgm_{\nu_{e}}(\infty)
,
\label{D062}
\end{equation}
and
\begin{align}
\mgm_{\text{S}}^{2}(E_{\nu} \gtrsim
\null & \null
5 \, \text{MeV})
\simeq
\cos^2\vartheta_{13}
\sum_{j}
\left|
\mgm_{j2} - i \elm_{j2}
\right|^2
\nonumber
\\
\null & \null
+
\sin^2\vartheta_{13}
\sum_{j}
\left|
\mgm_{j3} - i \elm_{j3}
\right|^2
.
\label{D063}
\end{align}

\subsection{Experimental limits}
\label{D064}

The constraints on the neutrino magnetic moment in direct
laboratory experiments have been obtained so far from the lack of
any observable distortion of the recoil electron energy spectrum.
Experiments of this type have started in the 50's at
the Savannah River Laboratory where the ${\bar\nu_{e}}$-$e^{-}$
elastic scattering process was studied
\cite{Cowan:1954pq,Cowan:1957pp,Reines:1976pv} with somewhat controversial results,
as discussed by \textcite{Vogel:1989iv}.
The most significant experimental limits on the effective magnetic moment
$\mgm_{\nu_e}$
which have been obtained in reactor ${\bar\nu_{e}}$-$e^{-}$ after about 1990
are listed in Tab.~\ref{D055}
(some details of the different experimental setups are reviewed in \textcite{Broggini:2012df})\footnote{
An attempt to improve the experimental bound on
$\mgm_{\nu_e}$
in reactor experiments
was undertaken in \textcite{Wong:2010pb},
where it was suggested that
in $\bar\nu_{e}$ interactions on an atomic target
the atomic electron binding
(``atomic-ionization effect'')
can significantly increase the electromagnetic
contribution to the differential cross section with respect to the
free-electron approximation.
However,
as explained in Appendix~\ref{L001},
the dipole approximation used to derive the atomic-ionization effect
is not valid for the electron antineutrino cross
section in reactor neutrino magnetic moment experiments.
Instead, the free
electron approximation is appropriate for the interpretation of
the data of reactor neutrino experiments and the current constraints
in Tab.~\ref{D055}
cannot be improved by considering the atomic electron binding
\cite{Voloshin:2010vm,Kouzakov:2010tx,Kouzakov:2011ig,Kouzakov:2011ka,Kouzakov:2011vx,Kouzakov:2011uq,Chen:2013lba}.
The history and present status of the theory of
neutrino-atom collisions is reviewed in \textcite{Kouzakov:2014lka}.
}.

The current best limit on $\mgm_{\nu_e}$
has been obtained in 2012 in the GEMMA experiment at the Kalinin Nuclear Power Plant
(Russia)
with a 1.5 kg highly pure germanium detector
exposed at a $\bar\nu_{e}$ flux of $2.7 \times 10^{13} \, \text{cm}^{-2} \, \text{s}^{-1}$
at a distance of 13.9 m from the core of a $3 \, \text{GW}_{\text{th}}$
commercial water-moderated reactor
\cite{Beda:2012zz}.
The competitive
TEXONO experiment is based at the Kuo-Sheng Reactor Neutrino Laboratory (Taiwan),
where a 1.06 kg highly pure germanium detector
was exposed to the flux of $\bar\nu_{e}$ at a distance of 28 m
from the core of a $2.9 \, \text{GW}_{\text{th}}$
commercial reactor
\cite{Wong:2006nx}\footnote{
The TEXONO and GEMMA data
have been also used by
\textcite{Barranco:2011wx,Healey:2013vka}
to constrain neutrino nonstandard interactions.
}.

Searches for effects of neutrino magnetic moments
have been performed also in accelerator experiments.
The LAMPF bounds on
$\mgm_{\nu_{e}}$
in Tab.~\ref{D055}
has been obtained with $\nu_{e}$
from $\mu^+$ decay
\cite{Allen:1992qe}.
The LAMPF and LSND bounds on $\mgm_{\nu_{\mu}}$
in Tab.~\ref{D055}
has been obtained with $\nu_{\mu}$ and $\bar\nu_{\mu}$
from $\pi^+$ and $\mu^+$ decay
\cite{Allen:1992qe,Auerbach:2001wg}.
The DONUT collaboration \cite{Schwienhorst:2001sj} investigated
$\nu_{\tau}$-$e^-$
and
$\bar\nu_{\tau}$-$e^-$
elastic scattering,
finding the limit on $\mgm_{\nu_{\tau}}$ in Tab.~\ref{D055}.

Solar neutrino experiments can also search for a neutrino magnetic moment signal by
studying the shape of the electron spectrum
\cite{Beacom:1999wx}.
The effective magnetic moment $\mgm_{\text{S}}$ in solar $\nu_e$-$e^-$ scattering experiments
is given in Eq.~(\ref{D061}).
Table~\ref{D055}
gives the limits on obtained in the
Super-Kamiokande experiment \cite{Liu:2004ny}
for
$\mgm_{\text{S}}(E_{\nu} \gtrsim 5 \, \text{MeV})$
and that obtained in the
Borexino experiment \cite{Arpesella:2008mt}
for
$\mgm_{\text{S}}(E_{\nu} \lesssim 1 \, \text{MeV})$
(see Eqs.~(\ref{D063}) and (\ref{D062})).

Information on neutrino magnetic moments has been obtained also with
global fits of solar neutrino data
\cite{Joshipura:2002bp,Grimus:2002vb,Tortola:2004vh}.
Considering Majorana three-neutrino mixing,
\textcite{Tortola:2004vh} obtained, at 90\% CL,
\begin{equation}
\sqrt{ |\mgm_{12}|^2 + |\mgm_{23}|^2 + |\mgm_{31}|^2 } < 4.0 \times 10^{-10} \, \bmag
,
\label{D065}
\end{equation}
from the analysis of solar and KamLAND,
and
\begin{equation}
\sqrt{ |\mgm_{12}|^2 + |\mgm_{23}|^2 + |\mgm_{31}|^2 } < 1.8 \times 10^{-10} \, \bmag
,
\label{D066}
\end{equation}
adding the
Rovno \cite{Derbin:1993wy},
TEXONO \cite{Li:2002pn}
and
MUNU \cite{Daraktchieva:2003dr}
constraints.

As we have seen in Subsection~\ref{D034}
the neutrino magnetic moment contribution to the
$\nua{\afl}$--$e^{-}$
elastic scattering process flips the neutrino helicity.
If neutrinos are Dirac particles,
this process transforms active left-handed neutrinos into sterile right-handed neutrinos,
leading to dramatic effects on the explosion of a core-collapse supernova
\cite{Dar:1987yv,Nussinov:1987zr,Goldman:1987fg,Lattimer:1988mf,Barbieri:1988nh,Notzold:1988kz,Voloshin:1988xu,Ayala:1998qz,Ayala:1999xn,Balantekin:2007xq},
where there are also contributions from
$\nua{\afl}$--$p$
and
$\nua{\afl}$--$n$
elastic scattering.
Requiring that the entire energy in a supernova collapse is not carried away by the
escaping sterile right-handed neutrinos created in the supernova core,
\textcite{Ayala:1998qz,Ayala:1999xn}
obtained the following upper limit on a generic neutrino magnetic moment:
\begin{equation}
\mgm_{\nu}
\lesssim
\left( 0.1 - 0.4 \right) \times 10^{-11} \, \bmag
,
\label{D067}
\end{equation}
which is slightly more stringent than the bound
$
\mgm_{\nu}
\lesssim
\left( 0.2 - 0.8 \right) \times 10^{-11} \, \bmag
$
obtained by
\textcite{Barbieri:1988nh}.

\subsection{Theoretical considerations}
\label{D068}

There is a gap of many orders
of magnitude between the present experimental limits on neutrino magnetic moments of the order of
$10^{-11} \, \bmag$
(discussed in Subsection~\ref{D064})
and the prediction smaller than about
$10^{-19} \, \bmag$
in Eq.~(\ref{D016}) of the
minimal extension of the Standard Model with right-handed neutrinos.
The hope to reach in the near future an experimental sensitivity of this order of magnitude
is very weak,
taking into account that the
experimental sensitivity
of reactor ${\bar\nu_{e}}$-$e$
elastic scattering experiments
have improved by only one order of
magnitude during a period of about twenty years
(see \textcite{Vogel:1989iv}, where a sensitivity of the order of $10^{-10}\bmag$ is discussed).
However,
the experimental studies of neutrino magnetic moments
are stimulated by the hope that
new physics beyond the minimally extended Standard Model with right-handed neutrinos might
give much stronger contributions.

One of the
examples in which it is possible to avoid the neutrino magnetic
moment being proportional to a (small) neutrino mass, that would
in principle make a neutrino magnetic moment accessible for
experimental observations, is realized in
the left-right symmetric model with
direct right-handed neutrino interactions
\cite{Shrock:1974nd,Kim:1976gk,Marciano:1977wx,Beg:1977xz,Shrock:1982sc,Duncan:1987ki,Liu:1987nf,Rajpoot:1990hj,Czakon:1998rf,Nemevsek:2012iq,Boyarkin:2014oza}.
In this model there is a new charged boson $W_{R}$ which mediates right-handed charged-current weak interactions
and mixes with the Standard Model $W_{L}$ boson which mediates left-handed charged-current weak interactions.
The massive gauge bosons states $W_1$ and $W_2$ are given by
\begin{align}
\null & \null
W_1 = W_{L} \cos\xi - W_R e^{i\varphi} \sin\xi
,
\label{D069}
\\
\null & \null
W_2 = W_{L} e^{-i\varphi} \sin\xi + W_R \cos\xi
,
\label{D070}
\end{align}
where $\xi$ is a small mixing angle
and
$\varphi$ is a possible CP-violating phase.
Neglecting the contributions of neutrino masses and
the terms suppressed by the small ratio
$m_{W_1}/m_{W_2}$,
the magnetic moments of Dirac neutrinos
are given by
\cite{Shrock:1982sc,Fukugita:2003en}
\begin{align}
\mgm_{kj}
=
\null & \null
\frac{eG_{\text{F}}}{4\sqrt{2}\pi^{2}}
\sin2\xi
\nonumber
\\
\null & \null
\times
\sum_{\afl=e,\mu,\tau}
m_{\afl}
\left[
e^{i\varphi}
U^{*}_{\afl k}
V_{\afl j}
+
e^{-i\varphi}
V^{*}_{\afl k}
U_{\afl j}
\right]
,
\label{D071}
\end{align}
where
$U$ is the standard mixing matrix of left-handed neutrinos
and
$V$
is the mixing matrix of right-handed neutrinos.
Hence,
in this case the neutrino magnetic moments depend on
the values of the charged lepton masses.
However,
one must take into account the coefficient
$\sin2\xi$,
which must be very small in order to have small Dirac neutrino masses
\cite{Czakon:1998rf}.
For example,
in the model of
\textcite{Chang:1986bp}
$\sin\xi \lesssim 10^{-7}$
for
$m_{W_2} \gtrsim 2.5 \, \text{TeV}$
\cite{Beall:1981ze,Ecker:1985vv,Maiezza:2010ic},
which implies that
$ \mgm_{kj} \lesssim 10^{-16} \bmag$.
However,
larger values of the magnetic moments have been obtained by
\textcite{Rajpoot:1990hj}
by adding to the left-right symmetric model
a charged scalar singlet,
following the idea of
\textcite{Fukugita:1987ti}.

Other interesting possibilities of obtaining neutrino magnetic
moments larger than the prediction in Eq.~(\ref{D016}) of the
minimal extension the Standard Model with right-handed neutrinos have been considered
in the literature.
For example,
the analysis performed by \textcite{Aboubrahim:2013yfa} of the Dirac
neutrino magnetic moment in the framework of a Minimal Supersymmetric Standard Model\footnote{
Other Supersymmetric models have been considered in
\textcite{Biswas:1983iy,Frank:1999nb,Fukuyama:2003uz,Gozdz:2009zz}.
}
extension with a vectorlike lepton generation
showed that a neutrino magnetic moment
as large as $\left( 10^{-12} - 10^{-14} \right) \bmag$ can be obtained.
These values lie within reach of improved laboratory experiments in the future.

\textcite{Gozdz:2006iz}
obtained Majorana transition magnetic moments as large as about $10^{-17} \, \bmag$,
significantly larger than those in Eq.~(\ref{D019}),
in the framework of the
Minimal Supersymmetric Standard Model
with $R$-parity violating interactions,
constrained by grand unification.

It is possible to estimate a generic
relation between the size of a neutrino magnetic moment $\mgm_{\nu}$ and the
corresponding neutrino mass
$m_{\nu}$
\cite{Voloshin:1987qy,Barr:1990um,Pal:1991pm,Davidson:2005cs,Bell:2006wi,Bell:2007nu}.
Suppose that a large neutrino magnetic moment is
generated by physics beyond a minimal extension of the Standard
Model at an energy scale characterized by $\Lambda$.
This contribution to $\mgm_{\nu}$ is described
by the Feynman diagram in Fig.~\ref{C004},
with the blob representing the effects of new physics beyond the
Standard Model.
The contribution of this diagram to the magnetic moment is
\begin{equation}\label{D072}
\mgm_{\nu} \sim \frac{eG}{\Lambda},
\end{equation}
where $e$ is the electric charge and $G$ is a combination of coupling
constants and loop factors
\cite{Bell:2006wi,Bell:2007nu}.
The diagram of Fig.~\ref{C004} without the photon line gives a new physics contribution to the
neutrino mass of the order
\begin{equation}\label{D073}
\delta{m}_{\nu} \sim G\Lambda.
\end{equation}
Combining the estimates (\ref{D072}) and (\ref{D073}),
one can get the relation
\begin{equation}\label{D074}
\delta{m}_{\nu}
\sim
\frac{\Lambda^{2}}{2m_{e}}
\frac{\mgm_{\nu}}{\bmag}
=
\frac{\mgm_{\nu}}{10^{-18}\bmag}
\left(
\frac{\Lambda}{\text{TeV}}
\right)^{2}
\,
\text{eV}
\end{equation}
between the new physics contribution to the neutrino mass and the
neutrino magnetic moment.

It follows that, generally, in theoretical models that predict
large values for the neutrino magnetic moment, simultaneously
large contributions to the neutrino mass arise.
Therefore, a particular fine tuning is needed to get a large value for the
neutrino magnetic moment while keeping the neutrino mass within
experimental bounds,
unless the ratio $m_{\nu}/\mgm_{\nu}$ is suppressed by a symmetry.
\textcite{Voloshin:1987qy} proposed a $\text{SU}(2)_{\nu}$
under which the neutrino and antineutrino fields, $\nu$ and $\nu^{c}$,
transform as a doublet.
Taking into account that fermion fields anticommute, a Dirac mass term can be written as
\begin{equation}
\overline{\nu} \nu
=
- \nu^{T} \overline{\nu}^{T}
=
- \nu^{T} \mathcal{C}^{\dagger} \mathcal{C} \overline{\nu}^{T}
=
- \nu^{T} \mathcal{C}^{\dagger} \nu^{c}
,
\label{D075}
\end{equation}
and a magnetic moment term can be written as
\begin{equation}
\overline{\nu} \sigma_{\alpha\beta} \nu
=
- \nu^{T} \sigma_{\alpha\beta}^{T} \overline{\nu}^{T}
=
\nu^{T} \mathcal{C}^{\dagger} \sigma_{\alpha\beta} \nu^{c}
.
\label{D076}
\end{equation}
One can see that
the mass term is invariant under the change
$\nu \leftrightarrows \nu^{c}$,
whereas the magnetic moment term changes sign.
Therefore,
the magnetic moment term is a singlet under the $\text{SU}(2)_{\nu}$ symmetry,
whereas the mass term transforms as a triplet and is forbidden\footnote{
Denoting the doublet as
$
\psi^{T}
=
\begin{pmatrix}
\nu & \nu^{c}
\end{pmatrix}^{T}
$
and the Pauli matrices acting in the $\text{SU}(2)_{\nu}$ space as
$\vet{\tau} = (\tau^{1},\tau^{2},\tau^{3})$,
we have
$
\overline{\nu} \nu
=
- \frac{1}{2} \psi^{T} \mathcal{C}^{\dagger} \tau^{1} \psi
$
and
$
\overline{\nu} \sigma_{\alpha\beta} \nu
=
\frac{1}{2} \psi^{T} \mathcal{C}^{\dagger} \sigma_{\alpha\beta} i\tau^{2} \psi
$.
One can verify that the magnetic moment is invariant under a
$\text{SU}(2)_{\nu}$
transformation
$
\psi \to \psi' = e^{i \vet{\lambda} \cdot \vet{\tau}} \psi
$,
whereas the mass term is not invariant.
}.
If, as it happens in a realistic model, the $\text{SU}(2)_{\nu}$
symmetry is broken and if this breaking is small, the ratio
$m_{\nu}/\mgm_{\nu}$ is also small, giving a natural way to obtain
a magnetic moment of the order of $\sim 10^{-11}\bmag$ without
contradictions with the neutrino mass experimental constraints.
Several possibilities based on the general idea of
\textcite{Voloshin:1987qy}
were considered by
\textcite{Leurer:1989hx,Babu:1990wv,Georgi:1990se,Ecker:1989ph,Chang:1991ri,Barbieri:1988fh}.

Another idea of neutrino mass suppression without suppression of
the neutrino magnetic moment was discussed by \textcite{Barr:1990um}
within the Zee model \cite{Zee:1980ai}, which is based on the
Standard Model gauge group $\text{SU}(2)_{L}\times \text{U}(1)_{Y}$ and
contains at least three Higgs doublets and a charged field which
is a singlet of $\text{SU}(2)_{L}$. For this kind of models there is a
suppression of the neutrino mass diagram, while the magnetic
moment diagram is not suppressed.

\textcite{Bell:2005kz,Davidson:2005cs,Bell:2006wi,Bell:2007nu}
derived ``natural'' upper bounds
for the magnetic moments of Dirac and Majorana neutrinos
generated by new physics above the electroweak scale.
They considered an effective low-energy theory
in which the effects of the new physics above the electroweak scale
are described by high-dimension nonrenormalizable operators
whose coefficients are not fine-tuned.
The low-energy effective Lagrangian
must respect the Standard Model $\text{SU}(2)_L \times \text{U}(1)_Y$
symmetry and is constructed with Standard Model
fields plus right-handed neutrino fields $\nu_{R}$
(with implicit flavor indices),
in order to have Dirac neutrino masses.
This low-energy effective Lagrangian can be written as
\begin{equation}
\mathcal{L}_{\text{eff}}
=
\sum_{n,j} \frac{\mathcal{C}^{n}_{j}(\mu)}{\Lambda^{n-4}}\mathcal{O}_{j}^{(n)}(\mu)+ \text{H.c.}
,
\label{D077}
\end{equation}
where $\mu$ is the renormalization scale, $n\geq 4$ denotes the
operator dimension and $j$ runs over independent operators of a
given dimension.
For $n=4$, a Dirac neutrino mass arises from the operator
$\mathcal{O}^{(4)}_{1}=\overline{L}\tilde{\Phi}\nu_{R}$, where
${\tilde \Phi}=i\sigma_2 \Phi^{*}$.
For $n=6$ there are two operators which generate, after electroweak symmetry breaking,
the magnetic moment operator
$\overline{\nu} \sigma_{\alpha\beta} \nu F^{\alpha\beta}$.
These operators can generate a contribution to the neutrino mass operator
$\mathcal{O}^{(4)}_{1}$
through loop diagrams.
Using dimensional analysis,
\textcite{Bell:2005kz}
estimated that the corresponding contribution $\delta{m}_{\nu}^{(4)}$ to the Dirac neutrino mass
is given by
\begin{equation}
\delta{m}_{\nu}^{(4)}
\sim
\frac{\alpha}{16\pi}
\frac{\Lambda^{2}}{m_{e}}
\frac{\mgm_{\nu}^{\text{D}}}{\bmag}
.
\label{D078}
\end{equation}
Apart from the different coefficient,
the dependence on $\Lambda$ and $\bmag$ is the same as in Eq.~(\ref{D074}).
The ${\Lambda}^{2}$ dependence is due to the
quadratic divergence in the renormalization of the dimension-four
neutrino mass operator.
Imposing that $\delta{m}_{\nu}$ is smaller than the neutrino mass $m_{\nu}$, we obtain
\begin{equation}
\mgm_{\nu}^{\text{D}}
\lesssim
3 \times 10^{-15} \, \bmag
\left(
\frac{m_{\nu}}{\text{eV}}
\right)
\left(
\frac{\text{TeV}}{\Lambda}
\right)^{2}
.
\label{D079}
\end{equation}
For $\Lambda \sim 1 \, \text{TeV}$
and
$m_{\nu} \lesssim 1 \, \text{eV}$,
one obtains
$\mgm_{\nu}^{\text{D}} \lesssim 3 \times 10^{-15} \, \bmag$,
which is some orders of magnitude stronger than the
experimental constraints in Tab.~\ref{D055}.

\textcite{Bell:2005kz}
noted that if the scale $\Lambda$
is close to the electroweak scale,
an important contribution to the neutrino mass can arise also from
an $n=6$ operator.
In order to obtain a natural upper bound on $\mgm_{\nu}^{\text{D}}$
they assumed that at the scale $\Lambda$ the coefficient of the $n=6$ mass operator
is zero,
so that the contribution to the neutrino mass is generated entirely
by radiative corrections involving insertions of the
$n=6$ magnetic moment operators.
Solving the renormalization group equations
from the scale $\Lambda$ to the electroweak scale,
they found the following relation between
the contribution $\delta{m}_{\nu}^{(6)}$ neutrino mass
and the neutrino magnetic moment:
\begin{equation}\label{D080}
\frac{\mgm_{\nu}^{\text{D}}}{\bmag}
=
\frac{16\sqrt{2} G_{\text{F}} m_{e} \sin^{4}\theta_{W}}{9\alpha^{2} |f| \ln\left( \Lambda / v \right)}
\,
\delta{m}_{\nu}^{(6)}
,
\end{equation}
where $\alpha$ is the fine structure constant,
$v$ is the vacuum expectation value of the Higgs doublet,
\begin{equation}\label{D081}
f=1-r-\frac{2}{3}\tan^{2} \theta_{W} -\frac{1}{3}(1+r)\tan^4 \theta_{W},
\end{equation}
and $r$ is a ratio of effective operator coefficients defined at the scale
$\Lambda$ which is of order unity without fine-tuning.
If the neutrino magnetic
moment is generated by new physics at a scale $\Lambda \sim 1 \, \text{TeV}$ and
the corresponding contribution to the neutrino mass is
$\delta{m}_{\nu}^{(6)} \lesssim 1 \, \text{eV}$,
then $\mgm_{\nu} \lesssim 8 \times 10^{-15} \bmag$.
Also this bound is some orders of magnitude stronger than the
experimental constraints in Tab.~\ref{D055}.

Following a similar method,
\textcite{Bell:2006wi} calculated natural upper bounds for
the transition magnetic moments of Majorana neutrinos
(see also \textcite{Davidson:2005cs}).
They found that the most general naturalness upper bounds for the
Majorana transition magnetic moments in the flavor basis
are given by
\begin{equation}
\mu_{\afl\bfl}^{\text{M}}
\lesssim
4 \times 10^{-9} \, \bmag
\left(
\frac{M_{\afl\bfl}^{\text{M}}}{\text{eV}}
\right)
\left(
\frac{\text{TeV}}{\Lambda}
\right)^{2}
\left|
\frac{m_{\tau}^2}{m_{\afl}^2-m_{\bfl}^2}
\right|
,
\label{D082}
\end{equation}
where
$M_{\afl\bfl}^{\text{M}}$
is the Majorana neutrino mass matrix in the flavor basis.
For Majorana neutrinos the flavor and mass bases are related by a transformation similar to that in Eq.~(\ref{B025}):
$(U^{T} M^{\text{M}} U)_{kj} = m_{k} \delta_{kj}$,
where $U$ is the neutrino mixing matrix.
For the magnetic moments we have
\begin{equation}
\mu_{kj}^{\text{M}}
=
\sum_{\afl,\bfl}
U_{\afl k} \mu_{\afl\bfl}^{\text{M}} U_{\bfl j}
.
\label{D083}
\end{equation}
The limits (\ref{D082}) are much weaker than those in the Dirac case,
because for a Majorana neutrino the magnetic moment contribution
to the mass is Yukawa suppressed\footnote{
Since in the Majorana case the magnetic moment matrix is antisymmetric,
it is generated by an antisymmetric magnetic moment operator.
On the other hand,
the mass matrix and the corresponding mass operator of Majorana neutrinos are diagonal in the mass basis
and symmetric in the flavor basis.
Therefore,
with respect to the Dirac case in which there are no such constraints,
additional Yukawa couplings
are needed to convert an antisymmetric magnetic moment operator into a
symmetric mass operator
\cite{Davidson:2005cs,Bell:2006wi,Bell:2007nu}.
}.
Hence,
if a neutrino transition magnetic moment larger than about $10^{-14} \bmag$ is observed in an experiment,
it would indicate that it is plausible that neutrinos are Majorana rather than Dirac
particles.

Let us emphasize that the natural upper bounds on neutrino magnetic moments
derived by \textcite{Bell:2005kz,Davidson:2005cs,Bell:2006wi,Bell:2007nu}
apply in models with new physics well above the electroweak scale,
for which only the first terms of the effective Lagrangian expansion in Eq.~(\ref{D077})
are not negligible.
This is not the case,
for example,
in the model discussed by
\textcite{Aboubrahim:2013yfa},
in which there is new physics at the electroweak scale.

An unusual case of a large observable effect of the small magnetic moments in Eq.~(\ref{D016})
is that of $\bar\nu_{e}$--$e^{-}$ elastic scattering in large extra dimension
brane-bulk models with three bulk neutrinos
discussed by
\textcite{Mohapatra:2004ce}.
They showed that the magnetic moment contribution to $\bar\nu_{e}$--$e^{-}$ elastic scattering
due to the tower of Kaluza-Klein right-handed neutrino states,
each contribution with a magnetic moment given by Eq.~(\ref{D016}),
can be comparable with that of a single neutrino in four-dimensional space-time with magnetic moment of the order of
$10^{-10} \, \bmag$
and the different shapes of the spectra can distinguish the two cases.
Hence,
$\bar\nu_{e}$--$e^{-}$ elastic scattering experiments searching for the effects of neutrino magnetic moments
can probe the existence of large extra dimensions.
\section{Radiative decay and related processes}
\label{E001}

\begin{figure}
\null
\hfill
\subfigure[]{\label{E002}
\includegraphics*[bb=238 657 344 768, width=0.3\linewidth]{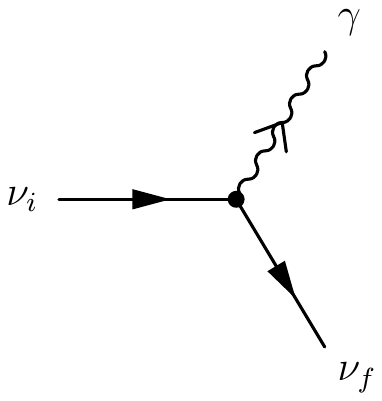}
}
\hfill
\subfigure[]{\label{E003}
\includegraphics*[bb=240 658 344 770, width=0.3\linewidth]{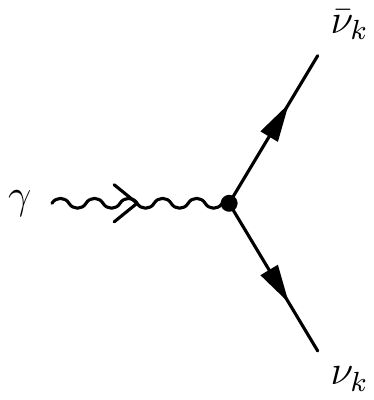}
}
\hfill
\null
\caption{\label{E004}
Feynman diagrams for
neutrino radiative decay and Cherenkov radiation \subref{E002}
and
plasmon decay \subref{E003}.}
\end{figure}

The magnetic and electric (transition) dipole moments of neutrinos,
as well as possible very small electric charges (millicharges),
describe direct couplings of neutrinos with photons
which induce several observable decay processes.
In this Section we discuss the decay processes
generated by the diagrams in Fig.~\ref{E004}:
the diagram in Fig.~\ref{E002} generates
neutrino radiative decay $\nu_{i}\to\nu_{f}+\gamma$
and the processes of
neutrino Cherenkov radiation
and
spin light ($SL\nu$)
of a neutrino propagating in a medium;
the diagram in Fig.~\ref{E003} generates
photon (plasmon) decay to an
neutrino-antineutrino pair in a plasma ($\gamma^{*} \to \nu \bar\nu$).

In Subsections~\ref{E005} and \ref{E024}
we review neutrino radiative decay in vacuum and in matter, respectively.
In Subsection~\ref{E036} we discuss neutrino Cherenkov radiation.
In Subsection~\ref{E043} we consider the process of
plasmon decay into a neutrino-antineutrino pair,
which can be important in dense astrophysical environments
as the interior of stars.
In Subsection~\ref{E049} we review the spin light process
of a neutrino propagating in a medium.

\subsection{Radiative decay}
\label{E005}

If the masses of neutrinos are nondegenerate, the radiative decay of
a heavier neutrino $\nu_{i}$ into a lighter neutrino $\nu_{f}$ (with $m_{i}>m_{f}$ ) with
emission of a photon,
\begin{equation}
\nu_{i} \to \nu_{f} + \gamma
,
\label{E006}
\end{equation}
may proceed in vacuum
\cite{Shrock:1974nd,Marciano:1977wx,Lee:1977tib,Petcov:1976ff,Goldman:1977jx,Zatsepin:1978iy,Pal:1981rm,Shrock:1982sc}.
Early discussions of the possible role of neutrino
radiative decay in different astrophysical and cosmological
settings can be found in
\textcite{Dicus:1977nn,Sato:1977ye,Stecker:1980bu,Kimble:1980vz,Melott:1981iw,De_Rujula:1980qd}.

The neutrino radiative decay process is generated by the interaction in Fig.~\ref{C032} with a real photon.
The decay amplitude is given by
\begin{align}
\null & \null
\langle \nu_{f}(p_{f},h_{f}), \gamma(q,\varepsilon) |
\int d^4x
\mathcal{H}_{\text{em}}^{(\nu)}(x)
| \nu_{i}(p_{i},h_{i}) \rangle
\nonumber
\\
=
\null & \null
(2\pi)^4
\delta^{4}(q - p_{i} + p_{f})
\overline{u^{(h_{f})}}(p_{f})
\Lambda^{fi}_{\mu}(q)
u^{(h_{i})}(p_{i})
\varepsilon^{\mu}
,
\label{E007}
\end{align}
where
$p_{i}$ ($p_{f}$)
and
$h_{i}$ ($h_{f}$)
are the four-momentum and helicity of the initial (final) neutrino
and
$q$ and $\varepsilon$
are the four-momentum and polarization four-vectors of the photon.
The Dirac $\delta$-function implements energy-momentum conservation.

Taking into account that for a real photon
\begin{equation}
q^{2}=0
\quad
\text{and}
\quad
\varepsilon^{\mu} q_{\mu}
=
0
,
\label{E008}
\end{equation}
from the general expression of $\Lambda_{\mu}(q)$ for Dirac neutrinos in Eq.~(\ref{C041})
and from Eq.~(\ref{C045}),
we obtain
\begin{equation}
\Lambda^{fi}_{\mu}(q)
\varepsilon^{\mu}
=
\chg_{fi} \slashed{\epsilon}
-
i \sigma_{\mu\nu} \varepsilon^{\mu} q^{\nu}
\left(
\mgm_{fi}
+ i
\elm_{fi} \gamma_{5}
\right)
,
\label{E009}
\end{equation}
where
$\chg_{fi} \neq 0$
only if neutrinos are millicharged particles
(see Subsection~\ref{G012}).
Therefore,
the radiative decay of a neutrino $\nu_{i}$ into a lighter neutrino $\nu_{f}$
depends on the corresponding
transition
charge, magnetic moment and electric moment.
Assuming
$\chg_{fi} = 0$,
the decay rate in the rest frame (rf) of the decaying neutrino $\nu_{i}$ is given by
(see \textcite{Raffelt:1996wa,Raffelt:1999gv,Raffelt:1999tx})
\begin{equation}
\Gamma_{\nu_{i}\to\nu_{f}+\gamma}^{\text{rf}}
=
\frac{1}{8\pi}
\left(
\frac{m_{i}^{2}-m_{f}^{2}}{m_{i}}
\right)^3
\left(
|\mgm_{fi}|^{2}
+
|\elm_{fi}|^{2}
\right)
.
\label{E010}
\end{equation}
This expression is valid for both Dirac and Majorana neutrinos,
because both can have transition magnetic and electric moments
and the corresponding expression (\ref{C069}) for $\Lambda_{\mu}(q)$ in the Majorana case
is equivalent to that in Eq.~(\ref{C041}) for Dirac neutrinos.

The transition magnetic and electric dipole moments
of Dirac neutrinos in the minimal extension of the
Standard Model with right-handed neutrinos
are given approximately by
Eq.~(\ref{D018}).
In this case,
the radiative decay rate is given by
\cite{Shrock:1982sc}
\begin{align}
\Gamma_{\nu^{\text{D}}_{i}\to\nu^{\text{D}}_{f}+\gamma}^{\text{rf}}
\simeq
\null & \null
\frac{\alpha}{2}
\left(
\frac{3 G_{\text{F}}}{32 \pi^2}
\right)^2
\left(\frac{m^{2}_{i}-m^{2}_{f}}{m_{i}}\right)^3
\left(m^{2}_{i}+m^{2}_{f}\right)
\nonumber
\\
\null & \null
\times
\left|
\sum_{\afl=e,\mu,\tau}
U^{*}_{\afl i} U_{\afl f}
\,
\frac{m_{\afl}^2}{m_{W}^2}
\right|^{2}
.
\label{E011}
\end{align}
The radiative decay rate is suppressed by
the small phase space due to the smallness of neutrino masses,
by the proportionality of the magnetic (electric) transition moment
to the sum (difference) of the masses of the two neutrinos involved in the decay
and by a coefficient which is smaller than
$(m_{\tau}/m_{W})^4 \simeq 2 \times 10^{-7}$.
Note, however, that there are models
(see, for instance,
\textcite{Petcov:1982en,Aboubrahim:2013gfa,Aboubrahim:2013yfa})
in which the neutrino radiative decay rate
(as well as the magnetic moment discussed in Section~\ref{D001})
of a Dirac neutrino are much larger than those predicted in the minimally
extended Standard Model.

The expression of the decay rate for Majorana neutrinos
in the simplest extensions of the Standard Model
(without taking into account model-dependent contributions of the scalar sector)
can be derived
from the expressions in Eqs.~(\ref{D025}) and (\ref{D026})
of the Majorana magnetic and electric transition moments
\cite{Shrock:1982sc}:
\begin{align}
\Gamma_{\nu^{\text{M}}_{i}\to\nu^{\text{M}}_{f}+\gamma}^{\text{rf}}
\simeq
\null & \null
\alpha
\left(
\frac{3 G_{\text{F}}}{32 \pi^2}
\right)^2
\left(\frac{m^{2}_{i}-m^{2}_{f}}{m_{i}}\right)^3
\nonumber
\\
\null & \null
\times
\left\{
\left(m_{i}+m_{f}\right)^2
\left|
\sum_{\afl=e,\mu,\tau}
U^{*}_{\afl i} U_{\afl f}
\,
\frac{m_{\afl}^2}{m_{W}^2}
\right|^{2}
\right.
\nonumber
\\
\null & \null
\hspace{-2cm}
\left.
- 4 m_{i} m_{f}
\left(
\operatorname{Re}\left[
\sum_{\afl=e,\mu,\tau}
U^{*}_{\afl i} U_{\afl f}
\,
\frac{m_{\afl}^2}{m_{W}^2}
\right]
\right)^2
\right\}
.
\label{E012}
\end{align}
In the case of CP conservation,
from Eqs.~(\ref{D030}) and (\ref{D031})
it follows that the decay process is induced purely by the neutrino
electric or magnetic transition dipole moment if the CP phases of
$\nu_{i}$ and $\nu_{f} $ are,
respectively, equal or opposite.

For numerical estimations it is convenient to express the lifetime
$\tau_{\nu_{i}\to\nu_{f}+\gamma} = \Gamma_{\nu_{i}\to\nu_{f}+\gamma}^{-1}$
in the following form:
\begin{equation}
\tau_{\nu_{i}\to\nu_{f}+\gamma}^{\text{rf}}
\simeq
0.19
\left(\frac{m_{i}^{2}}{m^{2}_{i}-m^{2}_{f}}\right)^3
\left(\frac{\text{eV}}{m_{i}}\right)^3
\left(\frac{\bmag}{\mgm_{fi}^{\text{eff}}}\right)^{2}
\text{s}
,
\label{E013}
\end{equation}
with the neutrino effective magnetic moment
\begin{equation}
\mgm_{fi}^{\text{eff}}
=
\sqrt{|\mgm_{fi}|^{2}+|\elm_{fi}|^{2}}
.
\label{E014}
\end{equation}
Since $\mgm_{fi}^{\text{eff}}$ is very small,
the lifetime in Eq.~\ref{E013} is very long.
Indeed,
in the case of Dirac neutrinos in the minimal extension of the
Standard Model with right-handed neutrinos,
considering only the dominant $\tau$ contribution in Eq.~(\ref{E011})
and neglecting $m_{f}$,
we obtain
\begin{equation}
\tau_{\nu^{\text{D}}_{i}\to\nu^{\text{D}}_{f}+\gamma}^{\text{rf}}
\simeq
\frac{6.2 \times 10^{43} \, \text{s}}{|U_{\tau i}|^2 |U_{\tau f}|^2}
\left(\frac{\text{eV}}{m_{i}}\right)^5
.
\label{E015}
\end{equation}
For
$m_{i} \lesssim 1 \, \text{eV}$,
this lifetime is much larger than the age of the Universe,
which is about
$4.3 \times 10^{17} \, \text{s}$
\cite{PDG-2012}.

\begin{figure}
\begin{center}
\includegraphics*[bb=5 368 576 806, width=\linewidth]{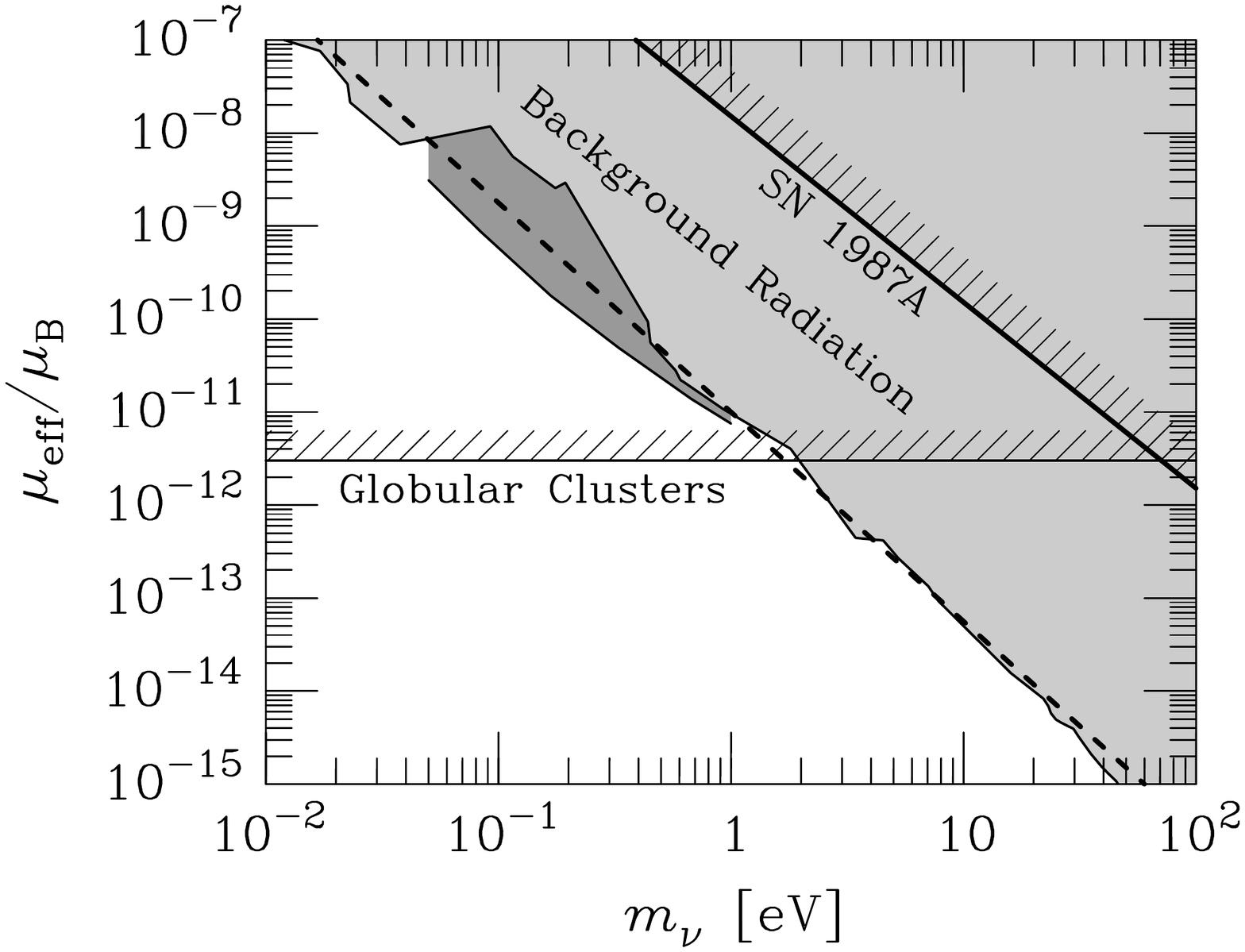}
\end{center}
\caption{\label{E016}
Astrophysical limits on neutrino transition moments.
Figure from \textcite{Raffelt:1999tx}.
}
\end{figure}

The neutrino radiative decay can be constrained by the absence of
decay photons in reactor $\bar\nu_{e}$ and solar $\nu_{e}$ fluxes.
The limits on $\mgm_{fi}^{\text{eff}}$ that are obtained from these
considerations are much weaker than those obtained from neutrino
scattering terrestrial experiments.
Stronger constraints on
$\mgm_{fi}^{\text{eff}}$ (though still weaker than those obtained in terrestrial experiments) are
obtained from the neutrino decay limit set by SN 1987A
\cite{Kolb:1988pe,Jaffe:1995sw}
and
from the measurements of the diffuse cosmic infrared background
and those of the cosmic microwave background
\cite{Cowsik:1977vz,Sato:1977ye,Dicus:1977av,DeRujula:1980qd,Stecker:1980bu,Kimble:1980vz,Dolgov:1981hv,Ressell:1989rz,Biller:1998nc,Raffelt:1998xu,Masso:1999wj,Mirizzi:2007jd}.
These limits,
shown in Fig.~\ref{E016},
can be expressed as (see \textcite{Raffelt:1996wa,Raffelt:1999gv,Raffelt:1999tx})
\begin{equation}
\frac{\mgm_{fi}^{\text{eff}}}{\bmag}
<
\left\{\begin{array}{ll}
0.9{\times}10^{-1}\left(\text{eV}/m_{\nu}\right)^{2}	& \text{Reactor ($\bar\nu_{e}$)},\\
0.5{\times}10^{-5}\left(\text{eV}/m_{\nu}\right)^{2}	& \text{Sun ($\nu_{e}$)}, \\
1.5{\times}10^{-8}\left(\text{eV}/m_{\nu}\right)^{2}	& \text{SN 1987A (all flavors)}, \\
1.0{\times}10^{-11}\left(\text{eV}/m_{\nu}\right)^{9/4}	& \begin{array}{l}\text{Cosmic background}\\\text{(all flavors)}.\end{array}
\end{array}
\right.
\end{equation}

Let us also recall the studies of
the effect of neutrino radiative decay on primordial Big-Bang Nucleosynthesis in
\textcite{Sato:1977ye,Dicus:1977av,Miyama:1978mn,Audouze:1985be,Terasawa:1988my}
(see also the review by \textcite{Dolgov:2002wy}).

Until now in this Subsection we considered the standard framework of three-neutrino mixing
in which there are three massive neutrinos,
but it is possible that additional massive neutrinos which are mainly sterile exist,
as explained in Subsection~\ref{B094}.
The radiative decay of heavy massive neutrinos
is a topic of current interest in view of the
recent indication\footnote{
See, however, also the negative result of the searches in \textcite{Jeltema:2014qfa,Malyshev:2014xqa,Anderson:2014tza,Carlson:2014lla,1412.1869}.
}
of an astrophysical monochromatic X-ray line at and energy of about 3.5 keV
\cite{Bulbul:2014sua,Boyarsky:2014jta},
which could be due to the radiative decay of a heavy neutrino
with a mass of about 7 keV
\cite{Abazajian:2014gza,Vincent:2014rja,Harada:2014lma}
in agreement with the prediction of the $\nu$MSM
(see \textcite{Boyarsky:2009ix,Kusenko:2009up,Drewes:2013gca,Boyarsky:2012rt}).
In fact,
from energy-momentum conservation in the two-body decay (\ref{E006})
the energy of the emitted photon in the rest frame of the decaying neutrino $\nu_{i}$ is given by
\begin{equation}
E_{\gamma}
=
\frac{m_{i}^2 - m_{f}^2}{2 m_{i}}
\simeq
\frac{m_{i}}{2}
\quad
\text{for}
\quad
m_{i} \gg m_{f}
.
\label{E017}
\end{equation}

Let us first consider the radiative decay of Dirac neutrinos.
Using Eq.~(\ref{E010}),
the transition dipole moments in Eq.~(\ref{D023}) imply the
decay rates
\begin{align}
\null & \null
\Gamma_{\nu^{\text{D}}_{i}\to\nu^{\text{D}}_{f}+\gamma}^{\text{rf}}
\simeq
\frac{\alpha}{2}
\left(
\frac{3 G_{\text{F}}}{16 \pi^2}
\right)^2
\left(\frac{m^{2}_{i}-m^{2}_{f}}{m_{i}}\right)^3
\left(m^{2}_{i}+m^{2}_{f}\right)
\nonumber
\\
\null & \null
\hspace{1cm}
\times
\left|
\sum_{n=1}^{N_{s}}
U^{*}_{s_{n} i} U_{s_{n} f}
+
\frac{1}{2} \sum_{\afl=e,\mu,\tau}
U^{*}_{\afl i} U_{\afl f}
\frac{m_{\afl}^{2}}{m_{W}^{2}}
\right|^{2}
.
\label{E018}
\end{align}
The inequality (\ref{B097}) suppresses quadratically the sterile contribution
to the decays between two standard massive neutrinos
($k,j \leq 3$)
and
the decays between two nonstandard massive neutrinos
are strongly suppressed by Eqs.~(\ref{B096}) and (\ref{B098}).
On the other hand,
the decay of a nonstandard heavy massive neutrino $\nu_{h}$ with $h\geq4$ into a lighter standard massive neutrino $\nu_{l}$ with $l\leq3$
can be significant if $|U_{s_{h-3} l}|$
is not too small.
Neglecting the small contributions due to the charged lepton masses
and considering $m_{h} \gg m_{l}$,
we have
\begin{equation}
\Gamma_{\nu^{\text{D}}_{h}\to\nu^{\text{D}}_{l}+\gamma}^{\text{rf}}
\simeq
\frac{\alpha}{2}
\left(
\frac{3 G_{\text{F}}}{16 \pi^2}
\right)^2
m_{h}^5
\,
|U_{s_{h-3} h}|^2
|U_{s_{h-3} l}|^2
.
\label{E019}
\end{equation}
If the mixing of $\nu_{s_{h-3}}$ with the three light neutrinos is dominated by $|U_{s_{h-3} l}|^2$,
we can define an approximate effective mixing angle $\vartheta_{hl}$ such that
\begin{equation}
\cos^2\vartheta_{hl} \simeq |U_{s_{h-3} h}|^2
,
\quad
\sin^2\vartheta_{hl} \simeq |U_{s_{h-3} l}|^2
,
\label{E020}
\end{equation}
and we can write the decay rate as
\begin{equation}
\Gamma_{\nu^{\text{D}}_{h}\to\nu^{\text{D}}_{l}+\gamma}^{\text{rf}}
\simeq
\frac{\alpha}{2}
\left(
\frac{3 G_{\text{F}}}{32 \pi^2}
\right)^2
m_{h}^5
\,
\sin^2 2 \vartheta_{hl}
.
\label{E021}
\end{equation}
This approximation is convenient for the analysis of experimental data,
because the decay rate depends on only two unknown parameters,
the heavy neutrino mass $m_{h}$ and the effective mixing angle $\vartheta_{hl}$.

Let us consider now the decay of heavy nonstandard massive neutrinos in the Majorana framework,
which applies to the $\nu$MSM explanation of the
astrophysical 3.5 keV X-ray line mentioned above
(see \textcite{Boyarsky:2009ix,Kusenko:2009up,Drewes:2013gca,Boyarsky:2012rt}).
The decay rates are generalizations of those in Eq.~(\ref{E012})
taking into account the transition magnetic and electric moments in Eqs.~(\ref{D032}) and (\ref{D033}).
For simplicity,
let us consider only
the decay of a heavy neutrino $\nu_{h}$ with $h\geq4$ into a light neutrino $\nu_{l}$ with $l\leq3$.
Neglecting the small contributions due to the charged lepton masses
and considering $m_{h} \gg m_{l}$
we obtain
\begin{equation}
\Gamma_{\nu^{\text{M}}_{h}\to\nu^{\text{M}}_{l}+\gamma}^{\text{rf}}
\simeq
\alpha
\left(
\frac{3 G_{\text{F}}}{16 \pi^2}
\right)^2
m_{h}^5
\,
|U_{s_{h-3} h}|^2
|U_{s_{h-3} l}|^2
.
\label{E022}
\end{equation}
This expression is twice of that in Eq.~(\ref{E019}) in the Dirac case.
Under the approximation (\ref{E020})
we obtain
\begin{equation}
\Gamma_{\nu^{\text{M}}_{h}\to\nu^{\text{M}}_{l}+\gamma}^{\text{rf}}
\simeq
\alpha
\left(
\frac{3 G_{\text{F}}}{32 \pi^2}
\right)^2
m_{h}^5
\,
\sin^2 2 \vartheta_{hl}
.
\label{E023}
\end{equation}
This expression is typically used in the phenomenological studies of
heavy neutrino radiative decay in the $\nu$MSM model
(see \textcite{Boyarsky:2009ix,Kusenko:2009up,Drewes:2013gca,Boyarsky:2012rt}).

Let us finally mention that the radiative decay of heavy neutrinos
may be observable also in hadron collider experiments
\cite{Boyarkin:2014hva}.

\subsection{Radiative decay in matter}
\label{E024}

\begin{figure}
\begin{center}
\begin{minipage}[r]{0.49\linewidth}
\begin{center}
\subfigure[]{\label{E028}
\includegraphics*[bb=234 628 370 768, width=0.8\linewidth]{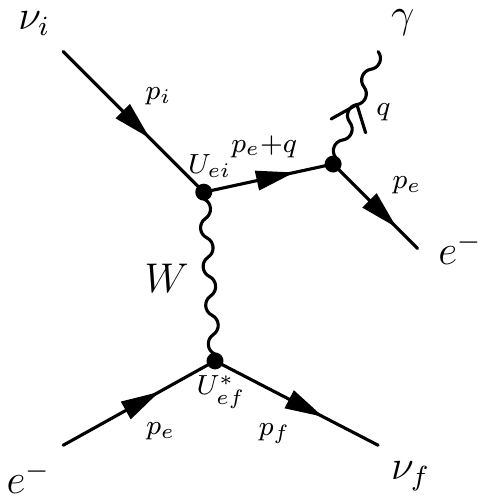}
}
\end{center}
\end{minipage}
\hfill
\begin{minipage}[l]{0.49\linewidth}
\begin{center}
\subfigure[]{\label{E029}
\includegraphics*[bb=234 628 369 768, width=0.8\linewidth]{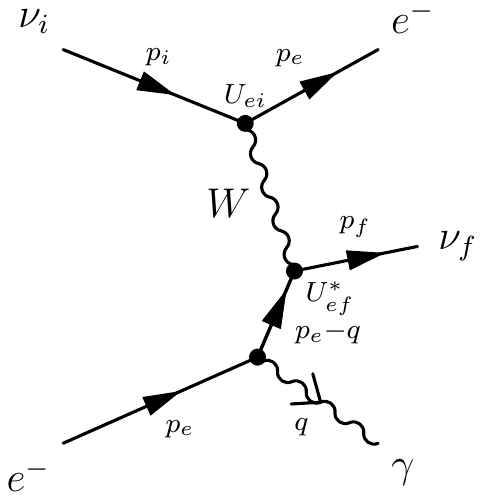}
}
\end{center}
\end{minipage}
\end{center}
\caption{ \label{E030}
Feynman diagrams of the coherent contribution of background electrons
to the radiative decay
$\nu_{i}(p_{i}) \to \nu_{f}(p_{f}) + \gamma(q)$
in matter.
}
\end{figure}

As explained in Subsection~\ref{B035},
the evolution of neutrinos propagating in matter is affected
by the potential generated by the coherent
forward elastic scattering with the particles in the medium.
It turns out that the coherent interaction with an electron background
induces the radiative decay in Eq.~(\ref{E006})
with a rate that is not suppressed by the GIM mechanism as the decay rate in vacuum in Eq.~(\ref{E011})
\cite{D'Olivo:1989un}.
Following the approach of \textcite{Giunti:1992sy},
the process of radiative decay in an electron background can be represented by the two Feynman diagrams
in Fig.~\ref{E030}
which are obtained from the CC potential diagram in Fig.~\ref{B047}
attaching a final photon line at the initial or final electron line.
As in the case of the calculation of the potential
(see \textcite{Giunti-Kim-2007}),
the coherent contribution of the electron background
is obtained by considering equal initial and final four-momenta of the electron.
The resulting decay rate in the rest frame of the electron background is
\begin{equation}
\Gamma_{\nu_{i}\to\nu_{f}+\gamma}^{\text{mat}}
=
\frac{\alpha G_{\text{F}}^2 N_{e}^2}{2 m_{e}^2}
\left( \frac{m_{i}^2 - m_{f}^2}{m_{i}} \right)
|U_{ei}|^2 |U_{ef}|^2 F(v_{i})
,
\label{E025}
\end{equation}
where
$N_{e}$ is the electron number density,
$v_{i}=|\vec{p}_{i}|/E_{i}$ is the velocity of the initial neutrino,
and
\begin{equation}
F(v_{i})
=
\sqrt{1-v_{i}^2}
\left[
\frac{2}{v_{i}} \ln\left( \frac{1+v_{i}}{1-v_{i}} \right)
- 3 + \frac{m_{f}^2}{m_{i}^2}
\right]
.
\label{E026}
\end{equation}
In the realistic case of ultrarelativistic initial neutrinos,
we have
\begin{equation}
F(v_{i})
\xrightarrow[v_{i}\to1]{}
4 \, m_{i} / E_i
.
\label{E027}
\end{equation}
Note that the matter-induced radiative decay is independent of the Dirac or Majorana
nature of neutrinos,
because it is generated by the coherent weak interactions with matter,
which are the same for left-handed neutrinos.

Neglecting the final neutrino mass in Eq.~(\ref{E025}),
the numerical value of the lifetime
$\tau_{\nu_{i}\to\nu_{f}+\gamma}^{\text{mat}} = (\Gamma_{\nu_{i}\to\nu_{f}+\gamma}^{\text{mat}})^{-1}$
for ultrarelativistic initial neutrinos
is given by
\begin{equation}
\tau_{\nu_{i}\to\nu_{f}+\gamma}^{\text{mat}}
\simeq
\frac{4.0 \times 10^{30} \, \text{s}}{|U_{ei}|^2 |U_{ef}|^2}
\left(\frac{\text{eV}}{m_{i}}\right)^2
\left(\frac{E_{i}}{\text{MeV}}\right)
\left(\frac{N_{\text{A}}\,\text{cm}^{-3}}{N_{e}}\right)^2
.
\label{E031}
\end{equation}
In order to compare the radiative lifetime in matter
in Eq.~(\ref{E031})
with the radiative lifetime in vacuum
in Eq.~(\ref{E015}),
obtained in the minimal extension of the
Standard Model with right-handed neutrinos,
we must take into account the Lorentz boost factor
$\gamma_{i} = E_{i} / m_{i}$
from the rest frame of the
decaying neutrino to the rest frame of the electron background:
\begin{align}
\frac
{\tau_{\nu_{i}\to\nu_{f}+\gamma}^{\text{mat}}}
{\gamma_{i}\tau_{\nu^{\text{D}}_{i}\to\nu^{\text{D}}_{f}+\gamma}^{\text{rf}}}
\simeq
\null & \null
1.1 \times 10^{-19}
\left(\frac{|U_{\tau i}|^2 |U_{\tau f}|^2}{|U_{ei}|^2 |U_{ef}|^2}\right)
\nonumber
\\
\null & \null
\times
\left(\frac{m_{i}}{\text{eV}}\right)^4
\left(\frac{N_{\text{A}}\,\text{cm}^{-3}}{N_{e}}\right)^2
.
\label{E032}
\end{align}
Therefore,
the radiative decay rate in an electron background
is many orders of magnitude larger than the radiative decay rate in vacuum
in the minimal extension of the
Standard Model with right-handed neutrinos.
However, the large value of the lifetime in Eq.~(\ref{E031})
indicate that it is very difficult, if not impossible,
to find a realistic application of this effect.

So far we have considered the radiative decay rate in a background of electrons,
assuming that the temperature is not very high.
For a temperature $T \gg m_{e}$
both electrons and positrons are present in the background
and the radiative decay rate is given by
\textcite{D'Olivo:1989un}
\begin{equation}
\Gamma_{\nu_{i}\to\nu_{f}+\gamma}^{(T \gg m_{e})}
=
\frac{\alpha G_{\text{F}}^2 T^4}{72}
\left( \frac{m_{i}^2 - m_{f}^2}{m_{i}} \right)
|U_{ei}|^2 |U_{ef}|^2 F(v_{i})
.
\label{E033}
\end{equation}
Neglecting the final neutrino mass,
for ultrarelativistic initial neutrinos we have
\begin{equation}
\tau_{\nu_{i}\to\nu_{f}+\gamma}^{(T \gg m_{e})}
\simeq
\frac{1.2 \times 10^{16} \, \text{s}}{|U_{ei}|^2 |U_{ef}|^2}
\left(\frac{\text{eV}}{m_{i}}\right)^2
\left(\frac{E_{i}}{\text{MeV}}\right)
\left(\frac{\text{MeV}}{T}\right)^4
.
\label{E034}
\end{equation}
Therefore, in this case the radiative decay in matter is enormously faster than that
in vacuum in the minimal extension of the
Standard Model with right-handed neutrinos:
\begin{align}
\frac
{\tau_{\nu_{i}\to\nu_{f}+\gamma}^{(T \gg m_{e})}}
{\gamma_{i}\tau_{\nu^{\text{D}}_{i}\to\nu^{\text{D}}_{f}+\gamma}^{\text{rf}}}
\simeq
\null & \null
3.3 \times 10^{-34}
\left(\frac{|U_{\tau i}|^2 |U_{\tau f}|^2}{|U_{ei}|^2 |U_{ef}|^2}\right)
\nonumber
\\
\null & \null
\times
\left(\frac{m_{i}}{\text{eV}}\right)^4
\left(\frac{\text{MeV}}{T}\right)^4
.
\label{E035}
\end{align}

Let us finally mention that:
\textcite{Nieves:1997md}
calculated the radiative decay rate of neutrinos propagating in a thermal background of electrons and photons, taking into account the effect of the stimulated emission of photons in the thermal bath;
\textcite{Grasso:1998td}
calculated the decay rate of a neutrino induced by the emission or absorption of a photon in a plasma
taking into account the effective mass of the photons (plasmons);
\textcite{Skobelev:1995pf,Zhukovsky:1996bi,Kachelriess:1996up}
calculated the radiative decay rate of neutrinos propagating in magnetic fields;
\textcite{Ternov:2003yi,Ternov:2013ana}
calculated the radiative decay rate of neutrinos propagating in a magnetized plasma.

\subsection{Cherenkov radiation}
\label{E036}

It is well known that a charged particle moving through a medium at a velocity greater than the speed of light in the medium,
$v>c/n$ ($n$ is the medium refractive index), can emit Cherenkov radiation.
In the same way, neutrinos with a magnetic moment
(and/or an electric dipole moment) propagating in a medium with a velocity larger
than the velocity of light in the medium can emit Cherenkov radiation.
This possibility was first discussed by \textcite{Radomski:1975re},
who studied a solution of the solar neutrino problem
in which the rate of solar neutrino detection is lowered
by the loss of energy of the neutrinos due to the emission of Cherenkov radiation in the solar matter.
However, the effect was found to be too
small to reduce significantly the solar neutrino flux.

The Cerenkov radiation is the helicity flip process
\begin{equation}\label{E037}
\nu_{L}(p)\rightarrow \nu_{R}(p') + \gamma(k)
,
\end{equation}
where
$\nu_{L}(p)$ and $\nu_{R}(p')$
denote the initial and final states of the same neutrino with
negative and positive helicities, respectively.
The amplitude of the transition due to a neutrino magnetic moment $\mgm$
is given by
\begin{equation}\label{E038}
M
=
\frac{\mgm}{n}
\,
\overline{u^{(+)}}(p')
\,
\sigma_{\mu\nu} k^{\mu}
\,
u^{(-)}(p)
\,
\varepsilon^{\nu}(k,\lambda)
,
\end{equation}
where $p=(E, \vec{p})$ and $p'=(E', \vec{p}')$ are
the four-momenta of the initial and final neutrinos and $k=(\omega, \vec{k})$ and
$\varepsilon^{\nu}(k,\lambda)$ are the four-momenta and polarization four-vector of the emitted photon
($\lambda$ denotes the photon helicity),
with
$|\vec{k}| = n \omega$
and
$n > 1$
in matter.
The rate of the Cherenkov process is given by
\cite{Grimus:1993np,Mohanty:1995bx}
\begin{equation}
\Gamma
=
\frac{1}{{2(2\pi)}^2 E}
\int
\frac{d^3 p'}{2E'}
\,
\frac{d^3 k}{2\omega}
\left| M \right|^2
\delta^{4}(p-p'-k)
.
\label{E039}
\end{equation}
After integration with use of the $\delta$-function, we obtain
\begin{align}
\Gamma
=
\null & \null
\frac{1}{16 \pi E^2 v}
\int
n^2
\,
d\omega
\,
d(\cos\theta)
\left| M \right|^2
\nonumber
\\
\null & \null
\times
\delta\left(\cos\theta -\frac{2 \omega E + \left( n^2 -1 \right) \omega^2}{2 n \omega E v}\right)
,
\label{E040}
\end{align}
where
$v=|\vec p|$ / ${E}$ is the initial neutrino velocity
and
$\theta$ is the angle between the emitted photon and the direction of propagation of the
initial neutrino.
The remaining $\delta$-function constrains the photon emission angle
to have the value
\begin{equation}
\cos\theta
=
\frac{1}{n v}
\left( 1 + (n^2-1) \frac{\omega}{2E} \right)
.
\label{E041}
\end{equation}
After performing the analytic integrals and taking into account
Eq.~(\ref{E041}),
we obtain
\begin{align}
\Gamma =
\null & \null
\frac{\mgm^2}{4 \pi E^2 v}
\int_{\omega_{\text{min}}}^{\omega_{\text{max}}}
\nonumber
\left\{
\left[
\frac{(n^2 -1)^2}{n^2}E^2 + (n^2-1)m_{\nu}^2
\right]
\omega^2
\right.
\\
\null & \null
\left.
-
\frac{(n^2 -1)^2}{n^2}E\omega^3 - \frac{(n^2 -1)^3}{4 n^2}\omega^4
\right\}
d\omega
.
\label{E042}
\end{align}
The range of integration from $\omega_{\text{min}}$ to $\omega_{\text{max}}$
corresponds to the range of frequencies of the emitted photon
which is allowed by the kinematical condition
$|\cos\theta| \leq 1$.
The determination of this range is nontrivial,
because in general
the refractive index $n$ depends on $\omega$.

The general expression (\ref{E042}) of the rate of the Cherenkov process
can be used for analyses of possible phenomenological consequences of the
neutrino magnetic moment Cherenkov radiation in different environments.
For example, \textcite{Grimus:1993np} estimated that if
solar neutrinos have an effective magnetic moment of about $3 \times 10^{-11} \, \bmag$,
they emit about 5 Cherenkov photons per day in a terrestrial $ 1 \, \text{km}^3$ water detector.

The Cherenkov mechanism is of interest for astrophysical applications also
because it flips the neutrino helicity.
If neutrinos are Dirac particles,
this helicity flip transforms
active left-handed neutrinos into sterile right-handed neutrinos.
This mechanism can have important consequences, for instance,
for the evolution of a supernova core.
Imposing that the luminosity
of the sterile neutrinos is less then the total energy
$10^{53} \, \text{ergs} \, \text{sec}^{-1}$
emitted by a typical core-collapse supernova,
\textcite{Mohanty:1995bx}
found an upper bound on the neutrino effective magnetic moment of about
$3 \times 10^{-14} \, \bmag$.

Neutrinos can emit Cherenkov radiation
also when they propagate in vacuum in the presence of a magnetic field.
This can occur even if neutrinos are massless with only standard-model couplings,
because the magnetic field induces an effective neutrino-photon vertex and modifies the photon dispersion relation
in such a way that the Cherenkov condition is fulfilled
(see \textcite{Ioannisian:1996pn}).
This mechanism was discussed by
\textcite{Galtsov:1972xp,Skobelev:1976at,Gvozdev:1992np,Gvozdev:1996kx,Skobelev:1995pf,Kachelriess:1996up,Ioannisian:1996pn}.
However,
in order to produce appreciable effects
this mechanism requires an extremely strong magnetic field.
The strongest magnetic fields known in nature
are those near pulsars.
Even considering a magnetic field as strong as the critical field
$B_{\text{cr}} = m_e^2/e = 4.41 \times 10^{13} \, \text{G}$,
since its spatial extensions near a pulsar is only of some tens of kilometers,
the Cherenkov radiation emitted by the neutrinos escaping from the pulsar
is too small to be of practical importance for pulsar physics
\cite{Ioannisian:1996pn}.

There is also another possible mechanism of electromagnetic radiation of neutrinos in a medium,
also called ``Cherenkov radiation''
\cite{Sawyer:1992aj,D'Olivo:1995gy}.
This mechanism is based on the expectation that neutrinos moving in a medium
acquire an electric charge as a consequence of their weak interaction with the particles of the background
\cite{Oraevsky:1986dt}.
Note that this effect exists even for massless neutrinos and no physics beyond the
Standard Model is needed.
The magnetic moment Cherenkov radiation estimated by
\textcite{Grimus:1993np}
in the optical range is much larger than the Cherenkov radiation
due to the induced charge.
However, the Cherenkov radiation due to the induced neutrino charge
becomes important for photons with higher energies
and might be of interest for applications in astrophysics.

Let us finally mention the studies of
photon emission of a massive neutrino with a
magnetic moment in magnetic fields and in electromagnetic waves
\cite{Borisov:1988wy,Borisov:1989yw,Skobelev:1976at,Skobelev:1991pv,Chistyakov:1999ii}
and
that of a neutrino with a magnetic moment which crosses
the interface of two media with different refractive indices
\cite{Sakuda:1993aq,Sakuda:1994zq,Grimus:1994ug,Ioannisian:2011mf,D'Olivo:2012zz}.

\subsection{Plasmon decay into a neutrino-antineutrino pair}
\label{E043}

The most interesting process, for the purpose of constraining
neutrino electromagnetic properties, is the photon (plasmon) decay
into a neutrino-antineutrino pair,
$\gamma^{*} \to \nu + \bar\nu$
\cite{Bernstein:1963qh,Sutherland:1975dr}.
This plasmon process becomes kinematically allowed
in media, because a photon with the dispersion relation
$\omega_{\gamma}^{2} - \vec{k}_{\gamma}^{2} >0$ roughly behaves as
a particle with an effective mass.
For example,
photons in a nonrelativistic plasma have the dispersion relation
$\omega_{\gamma}^{2} - \vec{k}_{\gamma}^{2} = \omega_{\text{P}}^2$,
where
$\omega_{\text{P}} = 4 \pi \alpha N_{e} / m_{e}$ is the plasma frequency
\cite{Raffelt:1996wa}.
For
$\omega_{\text{P}} > 2 m_{\nu}$
the plasmon decay
$\gamma^{*} \to\nu + {\bar\nu }$
is kinematically possible.

The plasmon decay rate is
\cite{Sutherland:1975dr,Raffelt:1996wa}
\begin{equation}
\Gamma_{\gamma^{*}\to\nu {\bar\nu}}
=
\frac{\mgm_{\nu}^{2}}{24\pi}
\,
Z
\,
\frac{(\omega_{\gamma}^{2} - k_{\gamma}^{2})^{2}}{\omega_ \gamma}
,
\label{E044}
\end{equation}
where $\mgm_{\nu}$ is the effective magnetic moment
\begin{equation}
\mgm_{\nu}
=
\sum_{k,j}
\left(
|\mgm_{kj}|^2 + |\elm_{kj}|^2
\right)
.
\label{E045}
\end{equation}
The quantity $Z$ is a renormalization factor which depends on the polarization of the plasmon.
For transverse plasmons
$\omega_{\gamma}^{2} - k_{\gamma}^{2} = \omega_{\text{P}}^2$
and
$Z=1$,
whereas
for longitudinal plasmons
$\omega_{\gamma} \simeq \omega_{\text{P}}$
and
$Z \simeq ( 1 - k_{\gamma}^{2} / \omega_{\text{P}}^2 )^{-1}$
\cite{Raffelt:1996wa}.

The process of plasmon decay into a neutrino-antineutrino pair
was first considered
by \textcite{Bernstein:1963qh} as a new energy-loss channel for the Sun.
In general,
a plasmon decay in a star
liberates the energy
$\omega_{\gamma}$ in the form of neutrinos that freely escape the stellar
environment.
The corresponding energy-loss rate per unit volume is
\begin{equation}
Q_{\gamma^{*}\to\nu {\bar\nu}}=\frac{g}{(2
\pi)^3}\int\omega_{\gamma} f_{k_{\gamma}}\Gamma_{\gamma \to\nu {\bar
\nu}}d^3k_{\gamma},
\end{equation}
where $f_{k_\gamma}$ is the photon Bose-Einstein distribution function and $g=2$
is the number of polarization states.

The requirement that the plasmon-decay energy-loss channel does not
exceed the standard solar model luminosity leads to the constraint
\cite{Raffelt:1996wa,Raffelt:1999gv,Raffelt:1999tx}
\begin{equation}
\mgm_{\nu} \lesssim 4 \times 10^{-10} \bmag
.
\label{E046}
\end{equation}

However,
the tightest astrophysical bound on $\mgm_{\nu}$
comes from the constraints on the possible delay of helium ignition of red giant star in globular clusters
due to the cooling induced the plasmon-decay energy loss.
From the lack of observational evidence of this effect,
the following limit has been found \cite{Raffelt:1989xu,Raffelt:1990pj,Raffelt:1992pi}:
\begin{equation}
\mgm_{\nu} \lesssim 3 \times 10^{-12} \bmag.
\label{E047}
\end{equation}
See also \textcite{Castellani:1993hs,Catelan:1995ba}.
Recently the limit has been updated by
\textcite{Viaux:2013hca}
using state-of-the-art astronomical observations and stellar evolution codes,
with the results
\begin{equation}
\mgm_{\nu}
<
\left\{
\begin{array}{l} \displaystyle
2.6 \times 10^{-12} \bmag
\quad
\text{(68\% CL)}
,
\\ \displaystyle
4.5 \times 10^{-12} \bmag
\quad
\text{(95\% CL)}
.
\end{array}
\right.
\label{E048}
\end{equation}
This astrophysical constraint on a neutrino
magnetic moment is applicable to both Dirac and Majorana neutrinos
and
constrains all diagonal and transition
dipole moments
according to Eq.~(\ref{E045}).

It has also been shown by \cite{Heger:2008er} that
the additional cooling due to processes induced by neutrino magnetic moments
(plasmon decay
$\gamma^{*} \to \nu \bar\nu$,
photo processes
$\gamma e^{-} \to e^{-} \nu \bar\nu$,
pair processes
$e^{+} e^{-} \to \nu \bar\nu$,
bremsstrahlung
$e^{-} (Z e) \to (Z e) e^{-} \nu \bar\nu$)
generate qualitative changes to the structure and evolution of stars
with masses between 7 and 18 solar masses,
rather than simply changing the time scales of their burning.
The resulting sensitivity to the neutrino magnetic moment
has been estimated by
\textcite{Heger:2008er}
to be at the level of $(2-4) \times 10^{-11} \, \bmag$.

\subsection{Spin light}
\label{E049}

It is known from
classical electrodynamics that a system with zero electric charge but nonzero magnetic (or electric) moment
can produce electromagnetic radiation which is called ``magnetic (or electric) dipole radiation''.
It is due to the rotation of the magnetic (or electric) moment.

A similar mechanism of radiation exists
in the case of a neutrino
with a magnetic (or electric) moment propagating in matter
\cite{Lobanov:2002ur}.
This phenomenon,
called ``spin light of neutrino''
($SL\nu$),
is different from
the neutrino Cherenkov radiation in matter discussed in Subsection~\ref{E036},
because it can exist even when the emitted photon refractive index
is equal to unity.
The $SL\nu$ is a radiation produced by the neutrino on its own,
rather than a radiation of the background particles.
Since the $SL\nu$ process is a transition between neutrino states with equal masses,
it can only become possible because of an external environment
influence on the neutrino states.

\begin{figure}
\begin{center}
\includegraphics*[bb=236 604 346 719, width=0.3\linewidth]{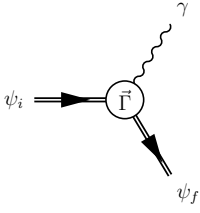}
\end{center}
\caption{\label{E050}
The spin light of neutrino ($SL\nu$) radiation diagram.}
\end{figure}

The $SL\nu$ was first studied by
\textcite{Lobanov:2002ur,Lobanov:2004um,Grigoriev:2004bm}
with a quasi-classical treatment based on a Lorentz-invariant approach to the neutrino spin
evolution that implies the use of the generalized Bargmann-Michel-Telegdi equation
\cite{Egorov:1999ah,Lobanov:2001ar}
(for further details see Appendix~\ref{N001}).
The full quantum theory of the $SL\nu$ has been elaborated by
\textcite{Studenikin:2004dx,Grigorev:2005sw,Grigoriev:2005bc,Grigoriev:2005gq,Studenikin:2007zz,Grigoriev:2012pw}
(see also \textcite{Lobanov:2004wa,Lobanov:2005zn}).
The method is based on the exact solution of the modified Dirac
equation for the neutrino wave function in matter
\cite{StuDeBroglei:2006zz,Studenikin:2005bq,Studenikin:2008qk,Grigoriev:2009zz}.

The Feynman diagram of the $SL\nu$ process is shown in
Fig.~\ref{E050},
where the neutrino initial ($\psi_{i}$) and final ($\psi_{f}$) states
(indicated by broad lines) are exact solutions of the corresponding
Dirac equations accounting for the interactions with matter.
The neutrino wave functions and energy spectrum are given by Eqs.(\ref{P013}) and (\ref{P014}) of Appendix~\ref{N001}.
Here we consider a generic flavor neutrino with an effective magnetic moment $\mgm_{\nu}$ and
effective mass $m_{\nu}$.
The $SL\nu$ process for a relativistic neutrino
is a transition from an initial neutrino state to a less energetic final neutrino state with the emission of a photon
and a neutrino helicity flip \cite{Studenikin:2004dx,Grigorev:2005sw}.

The amplitude of the $SL\nu$ process is given by \cite{Studenikin:2004dx}
\begin{equation}\label{E051}
S_{fi}
=
- \mgm_{\nu} \sqrt{4\pi}
\int d^{4} x
{\bar\psi}_{f}(x)(\vec\Gamma \cdot \vec\varepsilon^{*})\frac{e^{ikx}}{\sqrt{2\omega L^{3}}}\psi_{i}(x)
,
\end{equation}
where
$L^{3}$
is the normalization volume and
\begin{equation}
\vec{\Gamma}=i\omega\big\{\big[\vec{\Sigma} \times
\vec{\varkappa}\big]+i\gamma^{5}\vec{\Sigma}\big\}
,
\quad
\text{with}
\quad
\vec\Sigma
=
\begin{pmatrix}
\vec\sigma & 0
\\
0 & \vec\sigma
\end{pmatrix}
.
\label{E052}
\end{equation}
Here,
$k^{\mu}=(\omega,\vec{k})$ and $\vec\varepsilon$
are the photon momentum and polarization vectors,
and
$\vec{\varkappa}=\vec{k}/{\omega}$ is the unit vector
pointing in the direction of propagation of the emitted photon.
$\psi_{i}(x)$ and $\psi_{f}(x)$ are the
initial and final neutrino wave functions in presence of matter
obtained as the exact solutions of the effective Dirac equation
\begin{equation}\label{E053}
\Big\{ i\gamma_{\mu}\partial^{\mu}-\frac{1}{2}
\gamma_{\mu}(1+\gamma_{5})\widetilde{f}^{\mu}-m_{\nu}
\Big\}\psi_{i,f}(x)=0,
\end{equation}
(see Eqs. ({\ref{P013}) and ({\ref{P014}) of Appendix~\ref{P001}).
From the energy-momentum conservation relations
\begin{equation}\label{E054}
p_0=p'_0+\omega,
\quad
\vec{p}=\vec{p}'+\vec{\varkappa},
\end{equation}
where $(p_0, \vec p)$ and $(p'_0, \vec{p}')$ are the initial and final neutrino
energy and momenta, it follows that the photon energy is given by
\begin{equation}\label{E055}
\omega =\frac{2\widetilde{\alpha} m_{\nu} p\left[ (p_0-\widetilde{\alpha} m_{\nu} )-\left( p+\widetilde{\alpha}
m_{\nu} \right) \cos\theta \right] }{\left( p_0-\widetilde{\alpha} m_{\nu} -p\cos\theta
\right)^{2}-\left( \widetilde{\alpha} m_{\nu} \right)^{2}},
\end{equation}
where
$p=|\vec{p}|$
and
$\theta$
is the angle between $\vec{\varkappa}$ and the initial neutrino propagation.
For an electron neutrino propagating in a medium composed of electrons, protons and neutrons,
the matter density parameter $\widetilde{\alpha}$ is given by
\begin{equation}
\widetilde{\alpha} = \frac{G_{\text{F}}}{2{\sqrt 2}m_{\nu}}
\Big[n_{e} (1+4\sin^2\theta_W )+ n_{p}(1-4\sin^2\theta_W )-n_{n}\Big],
\end{equation}
where $n_{e}, n_{p}$ and $n_{n}$ are the number densities of the background electrons, protons and neutrons, respectively.
From the amplitude
(\ref{E051}) and the photon energy
(\ref{E055}) the $SL\nu$ transition rate and
total radiation power can be obtained:
\begin{eqnarray}\label{E056}
\Gamma &=&\mgm_{\nu}^{2}
\int_{0}^{\pi}\frac{\omega^{3}}{1+\tilde\beta'y}S\sin \theta d\theta,
\end{eqnarray}
\begin{equation}\label{E057}
I=\mgm_{\nu}^{2}\int_{0}^{\pi}\frac{\omega^{4}}{1+\tilde\beta'y}S\sin \theta d\theta,
\end{equation}
where
\begin{equation}\label{E058}
S=(\tilde\beta \tilde\beta'+1)(1-y\cos\theta)-(\tilde\beta +\tilde\beta') (\cos\theta -y),
\end{equation}
and
\begin{equation}\label{E059}
\tilde \beta =\frac{p+\widetilde{\alpha} m_{\nu} }{p_0-\widetilde{\alpha}m_{\nu} }, \ \tilde \beta'=\frac{p'-\widetilde{\alpha}m_{\nu} }{p'_0-\widetilde{\alpha}m_{\nu} },
\ y=\frac{\omega -p\cos\theta }{p'},
\end{equation}
where $p'=|\vec{p}'|$.
For the case of a relativistic neutrino with $p\gg m_{\nu}$,
the total rate and power are given by
\begin{equation}\label{E060}
\Gamma = \left\{
\begin{array}{lcl}
\frac{64}{3} \mgm_{\nu}^{2} \widetilde{\alpha}^3 p^2 m_{\nu}
&\text{for}&
\widetilde{\alpha} \ll \frac{m_{\nu}}{p},
\\
4 \mgm_{\nu}^{2} \widetilde{\alpha}^2 m_{\nu}^2 p
&\text{for}&
\frac{m_{\nu}}{p} \ll \widetilde{\alpha} \ll \frac{p}{m_{\nu}},
\\
4 \mgm_{\nu}^{2} \widetilde{\alpha}^3 m_{\nu}^3
&\text{for}&
\widetilde{\alpha} \gg \frac{p}{m_{\nu}},
\end{array}
\right.
\end{equation}
\begin{equation}\label{E061}
I = \left\{
\begin{array}{lcl}
\frac{128}{3} \mgm_{\nu}^{2}\widetilde{\alpha}^{4}p^{4}
&\text{for}&
\widetilde{\alpha} \ll \frac{m_{\nu}}{p},
\\
\frac{4}{3} \mgm_{\nu}^{2} \widetilde{\alpha}^2 m_{\nu}^2 p^2
&\text{for}&
\frac{m_{\nu}}{p} \ll \widetilde{\alpha} \ll \frac{p}{m_{\nu}},
\\
4 \mgm_{\nu}^{2} \widetilde{\alpha}^4 m_{\nu}^4
&\text{for}&
\widetilde{\alpha} \gg \frac{p}{m_{\nu}}.
\end{array}
\right.
\end{equation}

Since the rate and power of $SL\nu$
are proportional to $\mgm_{\nu}^2$,
they are in general very small.
However, some specific features of the $SL\nu$ might be phenomenologically interesting for astrophysics
\cite{Lobanov:2002ur,Studenikin:2004bu,Studenikin:2007zz,StuDeBroglei:2006zz,Lobanov:2004um,Grigoriev:2004bm,Studenikin:2004dx,Grigorev:2005sw,Grigoriev:2005bc,Grigoriev:2005gq,Studenikin:2007zz,Lobanov:2004wa,Lobanov:2005zn,Studenikin:2005bq,Studenikin:2008qk,Grigoriev:2009zz,Grigoriev:2006dx,Grigoriev:2008zq,Grigoriev:2012pw}.

As it can be seen from Eqs.~(\ref{E060}) and (\ref{E061}),
for a wide range of matter densities
the $SL\nu$ rate and power increase with the neutrino momentum.
For ultrahigh energy neutrinos ($p_0\sim 10^{18} \, \text{eV}$) propagating through a dense matter characterized by the value of the density parameter
$\widetilde{\alpha}m_{\nu} \sim 10 \, \text{eV}$
(this value is typical for a neutron star with $n_{e,p,n}$ of the order of $10^{38} \, \text{cm}^{-3}$),
the rate of the $SL\nu$ process is about $0.7 \, \text{s}^{-1}$.

For the average emitted photon energy
\begin{equation}\label{E062}
\left\langle \omega\right\rangle = I / \Gamma
,
\end{equation}
we obtain
\begin{equation}\label{E063}
\left\langle\omega\right\rangle \simeq \left\{
\begin{array}{lcl}
2\widetilde{\alpha} p^{2} / m_{\nu}
&\text{for}&
\widetilde{\alpha} \ll \frac{m_{\nu}}{p},
\\
\frac{1}{3}p
&\text{for}&
\frac{m_{\nu}}{p} \ll \widetilde{\alpha} \ll \frac{p}{m_{\nu}},
\\
\widetilde{\alpha} m_{\nu}
&\text{for}&
\widetilde{\alpha} \gg \frac{p}{m_{\nu}}.
\end{array}
\right.
\end{equation}
Therefore, in the most interesting case of astrophysical ultra-high energy
neutrinos, the average energy of the $SL\nu$ photons is one third
of the neutrino momentum and the $SL\nu$
spectrum spans the energy range of gamma-rays.

Another interesting property of the $SL\nu$ is its spatial distribution.
As it follows from Eqs.~(\ref{E056}) and (\ref{E057}) the radiation is collimated along the direction of
neutrino propagation.
In the case of relativistic neutrinos
($p\gg m_\nu$)
we have
$1 \ll \widetilde{\alpha} \ll p / m_\nu$
for a wide range of matter densities
and the radiation power is emitted in a narrow cone
with thickness
$\delta\theta \simeq m_\nu / p$
around a very small angle
$\theta_{\text{max}}$
given by
$\cos \theta_{\text{max}} \simeq 1 - \frac{2}{3} \widetilde\alpha m_\nu / p$.
The image drawn by the $SL\nu$ radiation in the
plane per\-pen\-dicular to
the neutrino direction of motion
in dense matter is a narrow ring with a very small radius
centered on the neutrino path.

When neutrinos propagate in a plasma,
the $SL\nu$ radiation is affected by the influence of the background plasma on the propagation of the emitted photons.
This effect has been first discussed by
\textcite{Studenikin:2004dx,Grigorev:2005sw,Grigoriev:2005bc,Grigoriev:2005gq}
and
was further studied in
\textcite{Kuznetsov:2006ci,Kuznetsov:2007ar},
where the role of the $SL\nu$ plasmon mass was taken into account.
In the case of ultra-high energy neutrino, the $SL\nu$
rate of \textcite{Kuznetsov:2006ci,Kuznetsov:2007ar}
reproduces exactly the result obtained in
\textcite{Studenikin:2004dx,Grigorev:2005sw,Grigoriev:2005bc,Grigoriev:2005gq}.
For a more detailed discussion on the historical aspects of this
issue see \textcite{Studenikin:2008qk,Grigoriev:2006in,
Grigoriev:2006dx, Grigoriev:2008zq,Grigoriev:2009zz}.
The most complete and consistent study of the $SL\nu$ accounting for the plasma effects
can be found in \textcite{Grigoriev:2012pw}.

The $SL\nu$ process with transitions between neutrinos with
different masses was considered in \textcite{Grigoriev:2010ni}
and
the $SL\nu$ mechanism taking into account possible effects
of Lorentz invariance violation was discussed in \textcite{Kruglov:2013oia}.
\section{Interactions with electromagnetic fields}
\label{F001}

If neutrinos have nontrivial electromagnetic properties,
they can interact with classical electromagnetic fields.
Significant effects can occur, in particular,
in neutrino astrophysics,
since neutrinos can propagate over very long distances in astrophysical environments
with magnetic fields.
In this case even a very weak interaction can have large cumulative effects.

In Subsection~\ref{F002}
we derive the effective potential of a neutrino propagating in a classical electromagnetic field.
This potential can generate spin and spin flavor transitions,
which are discussed in Subsection~\ref{F023}.
We also review the limits on the neutrino effective magnetic moment
obtained from analyses of solar neutrino data.
In Subsection~\ref{F087}
we discuss the modifications of neutrino magnetic moments
in very strong magnetic fields.
In Subsection~\ref{F097} we review the effects of a strong magnetic field on neutron decay.
In Subsection~\ref{F104} we review neutrino-antineutrino pair production in a magnetic field
and
in Subsection~\ref{F111} we discuss neutrino-antineutrino pair production due to vacuum instability in a very strong magnetic field.
In Subsection~\ref{F115} we review the energy quantization of neutrinos propagating in rotating media.

\subsection{Effective potential}
\label{F002}

The coherent interactions of neutrinos with classical electromagnetic fields
generate potentials which are similar to the matter potentials in Eq.~(\ref{B051})
and must be taken into account in the study of flavor and spin evolution
with an equation analogous to the MSW equation (\ref{B054}).
This evolution in a magnetic field is discussed in detail in Subsection~\ref{F023}.
Here we discuss the derivation of the neutrino effective potential in a classical electromagnetic field,
which corresponds to the amplitude of coherent
forward elastic scattering:
\begin{equation}
V_{h_{i} \to h_{f}}
=
\lim_{q\to0}
\frac
{
\langle \nu(p_{f},h_{f}) |
\int d^3x
\mathcal{H}_{\text{em}}^{(\nu)}(x)
| \nu(p_{i},h_{i}) \rangle
}{
\langle \nu(p,h) | \nu(p,h) \rangle
}
,
\label{F003}
\end{equation}
where $q=p_{i}-p_{f}$ as above and the denominator enforces the correct normalization
($p=p_{i}=p_{f}$
in the limit $q\to0$
and $h$ is arbitrary).
The interaction Hamiltonian
$\mathcal{H}_{\text{em}}^{(\nu)}(x)$
is that in Eq.~(\ref{C007}).
Here we consider for simplicity
only one neutrino
(the generalization to more than one neutrino,
with
the possibility of coherent transitions between different massive neutrinos
generated by transition form factors,
is discussed later),
allowing for possible helicity transitions ($h_{f} \neq h_{i}$),
which are important in magnetic fields (see Subsection~\ref{F023}).
Note that the hermiticity of $\mathcal{H}_{\text{em}}^{(\nu)}(x)$ implies that
\begin{equation}
V_{h_{f} \to h_{i}}
=
V_{h_{i} \to h_{f}}^{*}
.
\label{F004}
\end{equation}

From the normalization of states in Eq.~(\ref{I070}) and Eqs.~(\ref{C007})--(\ref{C013}),
we obtain
\begin{equation}
V_{h_{i} \to h_{f}}
=
\frac{1}{2E_{\nu}VT}
\lim_{q\to0}
\overline{u^{(h_{f})}}(p_{f})
\Lambda_{\mu}(q)
u^{(h_{i})}(p_{i})
\widetilde{A}^{\mu}(q)
,
\label{F005}
\end{equation}
where
$T$ is the normalization time,
$E_{\nu}=E_{i}=E_{f}$
in the limit $q\to0$, and
\begin{equation}
\widetilde{A}^{\mu}(q)
=
\int d^4x
e^{- i q \cdot x}
A^{\mu}(x)
\label{F006}
\end{equation}
is the Fourier transform of $A^{\mu}(x)$.
Integrating by parts and neglecting an irrelevant surface term
(which vanishes for well-behaved physical fields which vanish at infinity),
we have
\begin{equation}
q_{\alpha} \widetilde{A}^{\mu}(q)
=
- i
\int d^4x
e^{- i q \cdot x}
\partial_{\alpha}
A^{\mu}(x)
.
\label{F007}
\end{equation}
Using the expression (\ref{C023}) for $\Lambda_{\mu}(q)$,
and the Gordon identity (\ref{I072}) for the $\gamma^{\mu}$ term,
we obtain,
in the limit $q\to0$,
\begin{align}
\null & \null
V_{h_{i} \to h_{f}}
=
\frac{1}{VT}
\int d^4x
\Big[
\chg
\frac{p_{\mu}}{E_{\nu}} A^{\mu}(x) \delta_{h_{f}h_{i}}
\nonumber
\\
\null & \null
+
\frac{1}{4E_{\nu}}
\overline{u^{(h_{f})}}(p)
\sigma_{\mu\nu} F^{\mu\nu}(x)
\left(
\frac{\chg}{2m}
+
\mgm
+ i
\elm
\gamma_{5}
\right)
u^{(h_{i})}(p)
\nonumber
\\
\null & \null
-
\frac{\anm}{2E_{\nu}}
\overline{u^{(h_{f})}}(p)
j^{\mu}(x)
\gamma_{\mu}
\gamma_{5}
u^{(h_{i})}(p)
\Big]
,
\label{F008}
\end{align}
where
$p^{\mu}=p_{i}^{\mu}=p_{f}^{\mu}$.
The electromagnetic tensor
$F^{\mu\nu}(x)$
defined in Eq.~(\ref{I084})
contains the physical
electric field $\vec{E}(x)$ and magnetic field $\vec{B}(x)$
(see Eq.~(\ref{I086})).

Now, we take into account that propagating neutrinos are described by wave packets
whose size is limited
(see \textcite{Giunti-Kim-2007}).
Considering fields which are approximately constant over the extension of the neutrino wave packet,
we can extract them from the integral in Eq.~(\ref{F008}).
Then, the integral simplifies with $VT$,
leading to
\begin{align}
\null & \null
V_{h_{i} \to h_{f}}
=
\chg
\frac{p_{\mu}}{E_{\nu}} A^{\mu} \delta_{h_{f}h_{i}}
\nonumber
\\
\null & \null
+
\frac{1}{4E_{\nu}}
\overline{u^{(h_{f})}}(p)
\sigma_{\mu\nu} F^{\mu\nu}
\left(
\frac{\chg}{2m}
+
\mgm
+ i
\elm
\gamma_{5}
\right)
u^{(h_{i})}(p)
\nonumber
\\
\null & \null
-
\frac{\anm}{2E_{\nu}}
\overline{u^{(h_{f})}}(p)
j^{\mu}
\gamma_{\mu}
\gamma_{5}
u^{(h_{i})}(p)
.
\label{F009}
\end{align}
From Eq.~(\ref{F009}) one can see that
$V_{h_{i} \to h_{f}}$ depends on the four neutrino electromagnetic form factors
at $q^2=0$,
but the anapole moment contributes only in very special environments
in which the medium is charged.
Since
we will discuss this special case in Subsection~\ref{G060}
devoted to the anapole moment,
in the following part of this Section we do not consider
the anapole moment, assuming
$j^{\mu}(x)=0$.

Let us consider the first term in Eq.~(\ref{F009}).
In an electrostatic field
$A^{\mu}=(A^{0},0,0,0)$,
we have
$
V_{h_{i} \to h_{f}}^{(1)}
=
\chg
A^{0}
\delta_{h_{f}h_{i}}
$.
This is the expected result,
taking into account that $A^{0}$ is the electric potential.
Of course this term can contribute to the neutrino potential only if
neutrinos are millicharged particles
(see Subsection~\ref{G012}).

Let us now consider the more interesting contribution of
the second term in Eq.~(\ref{F009}),
which depends on the dipole magnetic and electric moments.
Note that the charge generates a magnetic moment
\begin{equation}
\mgm_{\chg}
=
\frac{\chg}{2m}
=
g \, \mgm_{\text{cl}}^{(1/2)}
,
\quad
\text{with}
\quad
g=2
,
\label{F010}
\end{equation}
where
$
\mgm_{\text{cl}}^{(s)}
=
\chg s / 2 m
$
is the classical magnetic moment of a spin-$s$ particle
(see \textcite{Jackson-1999}).
This is the same magnetic moment
obtained from the Dirac equation of a charged particle,
with the well-known gyromagnetic ratio $g=2$.
For a normally-charged particle the additional contribution $\mgm$ in Eq.~(\ref{F009})
to the magnetic moment would be called ``anomalous magnetic moment'',
which is generated by an internal structure in the case of nucleons
or by quantum loop corrections in the case of leptons
(measured in the famous $g-2$ experiments).
Since neutrinos are at most millicharged particles,
the $\mgm$ in Eq.~(\ref{F009})
is traditionally called ``magnetic moment'',
and the possible contribution of $\mgm_{\chg}$ is neglected.
Moreover,
$\mgm_{\chg}$
does not contribute to helicity transitions,
because it generates a spin precession which has the same frequency as
the precession of the angular momentum generated by $\chg$
(see \textcite{Sakurai-AQM-1967}).

In the following,
we study the effects of $\mgm$ and $\elm$
assuming
$\chg = 0$.
We also wish to establish the connection of the neutrino potential with the
classical potential for a nonrelativistic particle
(see \textcite{Jackson-1999}),
\begin{equation}
V_{\text{cl}}
=
- \vec{\mgm} \cdot \vec{B}
- \vec{\elm} \cdot \vec{E}
,
\label{F011}
\end{equation}
and the torque
\begin{equation}
\vec{T}_{\text{cl}}
=
\vec{\mgm} \times \vec{B}
+
\vec{\elm} \times \vec{E}
,
\label{F012}
\end{equation}
which generates the precession of the spin $\vec{S}$
through
$d\vec{S}/dt = \vec{T}_{\text{cl}}$.

Let us first consider the helicity-conserving potential
$V_{h \to h}$.
Using the method described in Appendix~\ref{M001},
we obtain
\begin{equation}
V_{h \to h}
=
- \frac{m}{E_{\nu}}
\left(
\vec{\mgm} \cdot \vec{B}
+
\vec{\elm} \cdot \vec{E}
\right)
,
\label{F015}
\end{equation}
with
\begin{equation}
\vec{\mgm}
=
h \, \frac{\vet{p}}{|\vet{p}|} \, \mgm
,
\qquad
\vec{\elm}
=
h \, \frac{\vet{p}}{|\vet{p}|} \, \elm
.
\label{F016}
\end{equation}
Hence,
the helicity-conserving potential is proportional to the
longitudinal components of the magnetic and electric fields.
In the nonrelativistic limit ($E_{\nu} \simeq m$) we obtain a potential which correspond to the
classical one in Eq.~(\ref{F011}).
Note, however,
that this potential is strongly suppressed by the small fraction
$m/E_{\nu}$
for ultrarelativistic neutrinos in realistic experiments.

Considering now the helicity-flipping potential
$V_{-h \to h}$,
using the method described in Appendix~\ref{M001},
if there is only an electric field $\vec{E}$,
we obtain
\begin{equation}
V_{-h \to h}(\vec{E})
=
\left(
\elm
+
i h \, \frac{|\vet{p}|}{E_{\nu}} \, \mgm
\right)
E_{\perp}
,
\label{F017}
\end{equation}
where $E_{\perp}$
is the transverse component of the electric field,
i.e. that orthogonal to $\vet{p}$.
In the case of a pure magnetic field $\vec{B}$,
we have, with a similar notation,
\begin{equation}
V_{-h \to h}(\vec{B})
=
\left(
\mgm
-
i h \, \frac{|\vet{p}|}{E_{\nu}} \, \elm
\right)
B_{\perp}
.
\label{F018}
\end{equation}
where $B_{\perp}$
is the transverse component of the magnetic field.
The expression of $V_{-h \to h}$ in the general case of an electromagnetic field is given
in Eq.~(\ref{M012}),
from which one can see that in any case
the helicity-flipping potential
depends only on the transverse components of the electric and magnetic fields.

Notice that for nonrelativistic neutrinos
($|\vet{p}| \ll E_{\nu}$)
in practice
$V_{-h \to h}(\vec{E})$
depends only on $\elm$
and
$V_{-h \to h}(\vec{B})$
depends only on $\mgm$,
as one may have expected:
\begin{align}
\null & \null
V_{-h \to h}^{\text{nr}}(\vec{E})
\simeq
\elm
E_{\perp}
=
|\vec{\elm} \times \vec{E}|
,
\label{F019}
\\
\null & \null
V_{-h \to h}^{\text{nr}}(\vec{B})
\simeq
\mgm
B_{\perp}
=
|\vec{\mgm} \times \vec{B}|
.
\label{F020}
\end{align}
Hence,
in the nonrelativistic limit
the helicity-flipping potential corresponds to the classical torque
in Eq.~(\ref{F012}),
which rotates the spin of the particle,
causing periodic changes of the helicity.

The additional dependence of
$V_{-h \to h}(\vec{E})$ on $\mgm$
and that of
$V_{-h \to h}(\vec{B})$ on $\elm$
for relativistic neutrinos
are explained in Appendix~\ref{M001}
as a consequence of the relativistic transformations
of the electric and magnetic fields
and the correspondence of the
electric and magnetic dipole moments with their classical counterparts
only in the nonrelativistic limit.

Let us finally consider the
potential between different massive neutrinos,
which is generated by transition
electric and magnetic dipole moments,
\begin{equation}
V_{\nu_{i}^{(h_{i})}\to\nu_{f}^{(h_{f})}}
=
\lim_{q\to0}
\frac
{
\langle \nu_{f}(p_{f},h_{f}) |
\int d^3x
\mathcal{H}_{\text{em}}^{(\nu)}(x)
| \nu_{i}(p_{i},h_{i}) \rangle
}{
\langle \nu(p,h) | \nu(p,h) \rangle
}
,
\label{F021}
\end{equation}
which is especially interesting for Majorana neutrinos
which do not have diagonal electric and magnetic dipole moments.
Here one can notice that
it is impossible to have $p_{i}=p_{f}$ if $m_{i} \neq m_{f}$.
However,
we must remember that observable neutrinos are ultrarelativistic
and their energy-momentum uncertainty is much larger than
their mass differences
(see \textcite{Giunti-Kim-2007}).
In this case,
$\nu_{i}\to\nu_{f}$
transitions are possible in an electromagnetic field,
as well as the coherent production of
different massive neutrinos which is necessary for the oscillations
discussed in Subsection~\ref{B035}.
In practice this means that in the calculation of $V_{fi}$
we can approximate the neutrinos as massless.
Under this approximation,
the helicity-flipping potential in a transverse magnetic field
in Eq.~(\ref{F018})
can be generalized to
\begin{equation}
V_{\nu_{i}^{(-h)}\to\nu_{f}^{(h)}}(\vec{B})
=
\left(
\mgm_{fi}
-
i h \, \frac{|\vet{p}|}{E_{\nu}} \, \elm_{fi}
\right)
B_{\perp}
.
\label{F022}
\end{equation}
This potential is interesting because it determines
the neutrino spin-flavor precession in a transverse magnetic field
discussed in Subsection~\ref{F023}.

\subsection{Spin-flavor precession}
\label{F023}

If neutrinos have magnetic moments, the spin can precess in a transverse magnetic field
\cite{Cisneros:1971nq,Fujikawa:1980yx,Voloshin:1986ty,Okun:1986na}.

Let us first derive the spin precession of an ultrarelativistic
Dirac neutrino generated by its diagonal magnetic moment
$\mgm$.
We consider a neutrino
with four-momentum $p$ which at the initial time $t=0$ has a definite helicity $h_{i}$
and is described by the state
$| \nu(p,h_{i}) \rangle$.
After propagation in a magnetic field
$\vec{B}$,
at the time $t$ the neutrino is described by a superposition of both helicities:
\begin{equation}
| \nu(t) \rangle
=
\sum_{h=\pm1} \psi_{h}(t) | \nu(p,h) \rangle
.
\label{F024}
\end{equation}
The temporal evolution of $| \nu(t) \rangle$
is given by the Schr\"odinger equation
\begin{equation}
i \, \frac{\text{d}}{\text{d}t} \,
| \nu(t) \rangle
=
\ham_{\text{em}}(t) \,
| \nu(t) \rangle
,
\label{F025}
\end{equation}
where
$
| \nu(0) \rangle
=
| \nu(p,h_{i}) \rangle
$
and
$
\ham_{\text{em}}(t) = \int d^3x \, \mathcal{H}_{\text{em}}^{(\nu)}(x)
$
is the effective interaction Hamiltonian, which can depend on time if the magnetic field is not constant.
here we neglect the irrelevant contribution of the vacuum Hamiltonian,
which does not cause any change in helicity because the two helicity states
have the same mass.

Multiplying Eq.~(\ref{F025}) on the left by
$\langle \nu(p,h) |$,
we obtain the evolution equation for the helicity amplitudes
\begin{equation}
i \, \frac{\text{d}\psi_{h}(t)}{\text{d}t}
=
\sum_{h'=\pm1}
\psi_{h'}(t)
\,
V_{h' \to h}(t)
,
\label{F026}
\end{equation}
with the potential $V_{h' \to h}(t)$ given in Eq.~(\ref{F003})
and
$\psi_{h}(0) = \delta_{hh_{i}}$.

In Eq.~(\ref{F015}) we have seen that the
helicity-conserving potential,
which depends on the longitudinal component of the magnetic field,
is strongly suppressed for ultrarelativistic neutrinos.
Hence,
in practice only the transverse component of the magnetic field
contributes through the helicity-flipping potential in Eq.~(\ref{F018}).
Considering for simplicity only the contribution of the magnetic moment $\mgm$,
we have
\begin{equation}
V_{h' \to h}(t)
=
\mgm
\,
B_{\perp}(t)
\,
\delta_{-hh'}
.
\label{F027}
\end{equation}
Then,
the evolution equation (\ref{F026}) can be written in the standard matrix form
\begin{equation}
i \frac{ \text{d} }{ \text{d}x }
\begin{pmatrix}
\psi_{L}(x)
\\
\psi_{R}(x)
\end{pmatrix}
=
\begin{pmatrix}
0 & \mgm B_{\perp}(x)
\\
\mgm B_{\perp}(x) & 0
\end{pmatrix}
\begin{pmatrix}
\psi_{L}(x)
\\
\psi_{R}(x)
\end{pmatrix}
, \label{F028}
\end{equation}
where we approximated the distance $x$ along the neutrino trajectory with the time $t$
for ultrarelativistic neutrinos
and
we adopted the standard notation
$\psi_{L} \equiv \psi_{-1}$
and
$\psi_{R} \equiv \psi_{+1}$
for the negative and positive helicity amplitudes
of the left-handed and right-handed neutrinos,
which are described, respectively, by the states
$| \nu_{L} \rangle = | \nu(p,-1) \rangle$
and
$| \nu_{R} \rangle = | \nu(p,+1) \rangle$.
The differential equation (\ref{F028}) can be solved through the
transformation
\begin{equation}
\begin{pmatrix}
\psi_{L}(x)
\\
\psi_{R}(x)
\end{pmatrix}
= \frac{1}{\sqrt{2}}
\begin{pmatrix}
1 & 1
\\
-1 & 1
\end{pmatrix}
\begin{pmatrix}
\varphi_{-}(x)
\\
\varphi_{+}(x)
\end{pmatrix}
. \label{F029}
\end{equation}
The amplitudes $\varphi_{-}(x)$ and $\varphi_{+}(x)$ satisfy
decoupled differential equations, whose solutions are
\begin{equation}
\varphi_{\mp}(x) = \exp\!\left[ \pm i \int_{0}^{x} \text{d}x' \, \mgm \, B_{\perp}(x') \right] \varphi_{\mp}(0) \,.
\label{F030}
\end{equation}
If we consider an initial left-handed neutrino, we have
\begin{equation}
\begin{pmatrix}
\psi_{L}(0)
\\
\psi_{R}(0)
\end{pmatrix}
=
\begin{pmatrix}
1
\\
0
\end{pmatrix}
\quad \Rightarrow \quad
\begin{pmatrix}
\varphi_{-}(0)
\\
\varphi_{+}(0)
\end{pmatrix}
= \frac{1}{\sqrt{2}}
\begin{pmatrix}
1
\\
1
\end{pmatrix}
. \label{F031}
\end{equation}
Then, the probability of $\nu_{L}\to\nu_{R}$ transitions is given
by
\begin{equation}
P_{\nu_{L}\to\nu_{R}}(x) = |\psi_{R}(x)|^{2} = \sin^{2}\!\left( \int_{0}^{x} \text{d}x' \, \mgm \, B_{\perp}(x') \right)
. \label{F032}
\end{equation}
Note that the transition probability is independent of the
neutrino energy (contrary to the case of flavor oscillations) and
the amplitude of the oscillating probability is unity.
Hence, when
the argument of the sine is equal to $\pi/2$ there is complete
$\nu_{L}\to\nu_{R}$ conversion.

The precession $\nu_{eL}\to\nu_{eR}$ in the magnetic field of the
Sun was considered in 1971 \cite{Cisneros:1971nq} as a possible
solution of the solar neutrino problem.
If neutrinos are Dirac
particles, right-handed neutrinos are sterile and a
$\nu_{eL}\to\nu_{eR}$ conversion could explain the disappearance of
active solar $\nu_{eL}$'s.

In 1986 it was realized \cite{Voloshin:1986ty,Okun:1986na} that
the matter effect during neutrino propagation inside of the Sun
suppresses $\nu_{eL}\to\nu_{eR}$ transition by lifting the
degeneracy of $\nu_{eL}$ and $\nu_{eR}$
(see also \textcite{Barbieri:1987xm}).
Indeed, taking into
account matter effects, the evolution equation~(\ref{F028}) becomes
\begin{equation}
i \frac{ \text{d} }{ \text{d}x }
\begin{pmatrix}
\psi_{L}(x)
\\
\psi_{R}(x)
\end{pmatrix}
=
\begin{pmatrix}
V(x) & \mgm B_{\perp}(x)
\\
\mgm B_{\perp}(x) & 0
\end{pmatrix}
\begin{pmatrix}
\psi_{L}(x)
\\
\psi_{R}(x)
\end{pmatrix}
, \label{F033}
\end{equation}
with the appropriate potential $V$ which depends on the neutrino
flavor, according to Eq.~(\ref{B050}).
In the case of a
constant matter density, this differential equation can be solved
analytically with the orthogonal transformation
\begin{equation}
\begin{pmatrix}
\psi_{L}(x)
\\
\psi_{R}(x)
\end{pmatrix}
=
\begin{pmatrix}
\cos\xi & \sin\xi
\\
- \sin\xi & \cos\xi
\end{pmatrix}
\begin{pmatrix}
\varphi_{-}(x)
\\
\varphi_{+}(x)
\end{pmatrix}
. \label{F034}
\end{equation}
The angle $\xi$ is chosen in order to diagonalize the matrix
operator in Eq.~(\ref{F033}):
\begin{equation}
\sin 2 \xi = \frac{ 2 \mgm B_{\perp} }{ \Delta{E}_{\text{M}} } \,,
\label{F035}
\end{equation}
with the effective energy splitting in matter
\begin{equation}
\Delta{E}_{\text{M}} = \sqrt{ V^{2} + \left( 2 \mgm B_{\perp} \right)^{2} } \,. \label{F036}
\end{equation}
The decoupled evolution of $\varphi_{\mp}(x)$ is given by
\begin{equation}
\varphi_{\mp}(x) = \exp\!\left[ - \frac{i}{2} \left( V \mp \Delta{E}_{\text{M}} \right) \right] \varphi_{\mp}(0) \,.
\label{F037}
\end{equation}
Considering an initial left-handed neutrino,
\begin{equation}
\begin{pmatrix}
\varphi_{-}(0)
\\
\varphi_{+}(0)
\end{pmatrix}
=
\begin{pmatrix}
\cos\xi
\\
\sin\xi
\end{pmatrix}
,
\label{F038}
\end{equation}
we obtain the oscillatory transition probability
\begin{equation}
P_{\nu_{L}\to\nu_{R}}(x) = |\psi_{R}(x)|^{2} = \sin^{2} 2 \xi
\sin^{2}\!\left( \frac{1}{2} \, \Delta{E}_{\text{M}} x \right) \,.
\label{F039}
\end{equation}

Since in matter $ \Delta{E}_{\text{M}}> 2 \mgm B_{\perp}$, the
matter effect suppresses the amplitude of $\nu_{L}\to\nu_{R}$
transitions.
However, these transitions are still independent of
the neutrino energy, which does not enter in the evolution
equation (\ref{F033}).

When it was known, in 1986 \cite{Voloshin:1986ty,Okun:1986na},
that the matter potential has the effect of suppressing
$\nu_{L}\to\nu_{R}$ transitions because it breaks the degeneracy
of left-handed and right-handed states, it did not take long to
realize, in 1988
\cite{Akhmedov:1988nc,Lim:1988tk},
that the
matter potentials can cause resonant spin-flavor precession if
different flavor neutrinos have transition magnetic moments
(spin-flavor precession in vacuum was previously discussed by
\textcite{Schechter:1981hw}).
The application of this mechanism to solar neutrinos
has been discussed in the following years by many authors
\cite{Minakata:1988gm,Akhmedov:1989ds,Balantekin:1990jg,Raghavan:1991em,Akhmedov:1991nt,Akhmedov:1991uk,Pulido:1991fb,Shi:1992ek,Balantekin:1992dv,Akhmedov:1993fv,Akhmedov:1993ta,Akhmedov:1994ix,Pulido:1999xp,Akhmedov:2000fj,Dev:2000wx,Chauhan:2002jw,Barranco:2002te,Akhmedov:2002mf,Chauhan:2003wr,Miranda:2003yh,Miranda:2004nz,Balantekin:2004tk,Chauhan:2004sf,Chauhan:2005pn,Pulido:2005pt,Guzzo:2005rr,Friedland:2005xh,Chauhan:2006yd,Picariello:2007sw,Yilmaz:2008vh,Raffelt:2009mm,Das:2009kw,Guzzo:2012rf}.

Let us consider
a neutrino state which is a superposition of different massive neutrinos with
both helicities:
\begin{equation}
| \nu(t) \rangle
=
\sum_{k} \sum_{h=\pm1} \psi_{k,h}(t) | \nu_{k}(p,h) \rangle
,
\label{F040}
\end{equation}
where
$\psi_{kh}(t)$
is the amplitude of the neutrino with mass $m_{k}$ and helicity $h$.
The temporal evolution of $| \nu(t) \rangle$
is given by the Schr\"odinger equation
\begin{equation}
i \, \frac{\text{d}}{\text{d}t} \,
| \nu(t) \rangle
=
\ham(t) \,
| \nu(t) \rangle
,
\label{F041}
\end{equation}
with
the initial condition
$
| \nu(0) \rangle
=
| \nu_{i}(p,h_{i}) \rangle
$.
Multiplying the evolution equation on the left by
$\langle \nu_{k}(p,h) |$,
we obtain the evolution equation for the helicity amplitudes of the different massive neutrinos
\begin{equation}
i \, \frac{\text{d}\psi_{kh}(t)}{\text{d}t}
=
\sum_{j}
\sum_{h'=\pm1}
\frac{\langle \nu_{k}(p,h) | \ham(t) | \nu_{j}(p,h') \rangle}{\langle \nu(p,h) | \nu(p,h) \rangle}
\psi_{j,h'}(t)
,
\label{F042}
\end{equation}
with
$\psi_{kh}(0) = \delta_{hh_{i}} \delta_{ki}$.
The effective Hamiltonian
$\ham(t)$
is the sum of
a vacuum Hamiltonian
$\ham_{0}$,
a weak interaction Hamiltonian
$\ham_{\text{w}}(t)$
which generates the effective potentials (\ref{B050}) of flavor neutrinos in matter,
and the electromagnetic Hamiltonian
$\ham_{\text{em}}(t)$
already considered in Eq.~(\ref{F025}).
For ultrarelativistic neutrinos,
\begin{equation}
\frac{\langle \nu_{k}(p,h) | \ham_{0} | \nu_{j}(p,h') \rangle}{\langle \nu(p,h) | \nu(p,h) \rangle}
=
\left(
E_{\nu} + \frac{m_{k}^{2}}{2E_{\nu}}
\right)
\delta_{kj}
\delta_{hh'}
,
\label{F043}
\end{equation}
where
$E_{\nu}$ is the neutrino energy neglecting mass contributions.

In order to calculate the matrix element of $\ham_{\text{w}}(t)$,
we must take into account the mixing of neutrino states in Eq.~(\ref{B037}),
which applies to left-handed neutrinos:
\begin{equation}
| \nu_{k}(p,-) \rangle = \sum_{\afl} U_{\afl k} \, | \nu_{\afl}(p,-) \rangle
.
\label{F044}
\end{equation}
For right-handed Dirac neutrinos the mixing is arbitrary,
because right-handed Dirac neutrinos are sterile to weak interactions.
On the other hand,
since right-handed Majorana neutrinos interact as right-handed Dirac antineutrinos,
their mixing is given by
\begin{equation}
| \nu_{k}^{\text{M}}(p,+) \rangle = \sum_{\afl} U_{\afl k}^{*} \, | \nu_{\afl}(p,+) \rangle
.
\label{F045}
\end{equation}
Therefore,
we define the generalized mixing relation
\begin{equation}
| \nu_{k}(p,h) \rangle = \sum_{\afl} U^{(h)}_{\afl k} \, | \nu_{\afl}(p,h) \rangle
,
\label{F046}
\end{equation}
with $U^{(-)} = U$
and
\begin{align}
\text{Dirac}:
\quad
\null & \null
U^{(+)} = U
;
\label{F047}
\\
\text{Majorana}:
\quad
\null & \null
U^{(+)} = U^{*}
.
\label{F048}
\end{align}
The arbitrary choice for Dirac neutrinos
has been made for simple convenience.
Then,
for the matrix element of $\ham_{\text{w}}(t)$
we obtain
\begin{equation}
\frac{\langle \nu_{k}(p,h) | \ham_{\text{w}}(t) | \nu_{j}(p,h') \rangle}{\langle \nu(p,h) | \nu(p,h) \rangle}
=
\sum_{\afl}
U_{\afl k}^{(h)*}
U_{\afl j}^{(h)}
V_{\afl}^{(h)}(t)
\delta_{hh'}
,
\label{F049}
\end{equation}
where
$V_{\afl}^{(-)} = V_{\afl}$,
with the potential $V_{\afl}$ in Eq.~(\ref{B050}),
and
\begin{align}
\text{Dirac}:
\quad
\null & \null
V_{\afl}^{(+)} = 0
;
\label{F050}
\\
\text{Majorana}:
\quad
\null & \null
V_{\afl}^{(+)} = - V_{\afl}
.
\label{F051}
\end{align}

As remarked before Eq.~(\ref{F027}),
the helicity-conserving potential generated by $\ham_{\text{em}}^{(\nu)}(t)$,
which depends on the longitudinal component of the magnetic field,
is strongly suppressed for ultrarelativistic neutrinos.
Then, from Eq.~(\ref{F022}),
considering for simplicity only the contribution of the magnetic moments,
we have
\begin{equation}
\frac{\langle \nu_{k}(p,h) | \ham_{\text{em}}^{(\nu)}(t) | \nu_{j}(p,h') \rangle}{\langle \nu(p,h) | \nu(p,h) \rangle}
=
\mgm_{kj}
\,
B_{\perp}(t)
\,
\delta_{-hh'}
.
\label{F052}
\end{equation}

Plugging Eqs.~(\ref{F043}), (\ref{F049}) and (\ref{F052}) in Eq.~(\ref{F042}),
neglecting the irrelevant common energy contribution in Eq.~(\ref{F043})
and
approximating the distance $x$ along the neutrino trajectory with the time $t$
for ultrarelativistic neutrinos,
one obtains the evolution equations of the helicity amplitudes of the different massive neutrinos:
\begin{align}
i \, \frac{\text{d}\psi_{k,h}(x)}{\text{d}x}
=
\sum_{j}
\null & \null
\sum_{h'=\pm1}
\Big[
\Big(
\frac{m_{k}^{2}}{2E_{\nu}}
\delta_{kj}
\nonumber
\\
\null & \null
\phantom{\sum_{h'=\pm1}\Big[\Big(}
+
\sum_{\afl}
U_{\afl k}^{(h)*}
V_{\afl}^{(h)}(x)
U_{\afl j}^{(h)}
\Big)
\delta_{hh'}
\nonumber
\\
\null & \null
+
\mgm_{kj}
B_{\perp}(x)
\delta_{-hh'}
\Big]
\psi_{j,h'}(x)
,
\label{F053}
\end{align}
In order to study flavor and helicity transitions,
it is more convenient to work in the flavor basis.
Using the mixing of neutrino states in Eq.~(\ref{B036}),
the state (\ref{F040}) with $t=x$ can be written as
\begin{equation}
| \nu(x) \rangle
=
\sum_{\afl} \sum_{h=\pm1} \psi_{\afl,h}(x) | \nu_{\afl}(p,h) \rangle
,
\label{F054}
\end{equation}
with the flavor and helicity amplitudes
\begin{equation}
\psi_{\afl,h}(x)
=
\sum_{k}
U_{\afl k}^{(h)}
\,
\psi_{k,h}(x)
,
\label{F055}
\end{equation}
which obey the evolution equation
\begin{align}
i \, \frac{\text{d}\psi_{\afl,h}(x)}{\text{d}x}
=
\sum_{\bfl}
\null & \null
\sum_{h'=\pm1}
\Big[
\Big(
\sum_{k}
U_{\afl k}^{(h)}
\frac{m_{k}^{2}}{2E_{\nu}}
U_{\bfl k}^{(h)*}
\nonumber
\\
\null & \null
\phantom{\sum_{h'=\pm1}\Big[\Big(}
+
V_{\afl}^{(h)}(x)
\delta_{\afl\bfl}
\Big)
\delta_{hh'}
\nonumber
\\
\null & \null
+
\mgm_{\afl\bfl}^{(h,h')}
B_{\perp}(x)
\delta_{-hh'}
\Big]
\psi_{\bfl,h'}(x)
,
\label{F056}
\end{align}
with the effective magnetic moments in the flavor basis
\begin{equation}
\mgm_{\afl\bfl}^{(h,h')}
=
\sum_{k,j}
U_{\afl k}^{(h)}
\mgm_{kj}
U_{\bfl j}^{(h')*}
.
\label{F057}
\end{equation}

For Dirac neutrinos,
from Eq.~(\ref{F047}) we have
\begin{equation}
\mgm_{\afl\bfl}^{(-,+)}
=
\mgm_{\afl\bfl}^{(+,-)}
=
\sum_{k,j}
U_{\afl k}
\mgm_{kj}
U_{\bfl j}^{*}
\equiv
\mgm_{\afl\bfl}
.
\label{F058}
\end{equation}
Then, from Eq.~(\ref{C042}) we obtain
\begin{equation}
\mgm_{jk} = {\mgm_{kj}}^{*}
\quad
\Rightarrow
\quad
\mgm_{\bfl\afl}
=
\mgm_{\afl\bfl}^{*}
.
\label{F059}
\end{equation}

For Majorana neutrinos,
from Eq.~(\ref{F048}) we have
\begin{align}
\null & \null
\mgm_{\afl\bfl}^{(-,+)}
=
\sum_{k,j}
U_{\afl k}
\mgm_{kj}
U_{\bfl j}
,
\label{F060}
\\
\null & \null
\mgm_{\afl\bfl}^{(+,-)}
=
\sum_{k,j}
U_{\afl k}^{*}
\mgm_{kj}
U_{\bfl j}^{*}
.
\label{F061}
\end{align}
From Eqs.~(\ref{C074}) and (\ref{C076})
it follows that for Majorana neutrinos the matrix of magnetic moments is antisymmetric
and the transition magnetic moments are imaginary:
\begin{equation}
\mgm_{jk} = - \mgm_{kj} = \mgm_{kj}^{*}
.
\label{F062}
\end{equation}
The antisymmetric property is preserved in the flavor basis:
\begin{equation}
\mgm_{\afl\bfl}^{(-,+)}
=
-
\mgm_{\bfl\afl}^{(-,+)}
,
\quad
\mgm_{\afl\bfl}^{(+,-)}
=
-
\mgm_{\bfl\afl}^{(+,-)}
.
\label{F063}
\end{equation}
Hence,
there are no diagonal magnetic moments in the flavor basis as in the mass basis.
Moreover,
we have
\begin{equation}
\mgm_{\afl\bfl}^{(+,-)}
=
-
\mgm_{\afl\bfl}^{(-,+)*}
.
\label{F064}
\end{equation}

In the following we discuss the spin-flavor evolution equation
in the two-neutrino mixing approximation,
which is interesting for understanding the relevant features of
neutrino spin-flavor precession.
Having in mind the application to solar neutrinos,
we consider the $\nu_{e}$--$\nu_{\act}$ mixing
discussed in Subsection~\ref{B035},
where $\nu_{\act}$
is the linear combination of
$\nu_{\mu}$ and $\nu_{\tau}$ in Eq.~(\ref{B074}).
Neglecting the small effects due to $\vartheta_{13}$,
we have
\begin{equation}
\begin{pmatrix}
\psi_{e,h}(x)
\\
\psi_{\act,h}(x)
\end{pmatrix}
=
R_{12}
\begin{pmatrix}
\psi_{1,h}(x)
\\
\psi_{2,h}(x)
\end{pmatrix}
,
\label{F065}
\end{equation}
with
\begin{equation}
R_{12}
=
\begin{pmatrix}
\cos\vartheta_{12} & \sin\vartheta_{12}
\\
- \sin\vartheta_{12} & \cos\vartheta_{12}
\end{pmatrix}
.
\label{F066}
\end{equation}

Considering Dirac neutrinos,
from Eq.~(\ref{F056}) it follows that
the generalization of Eq.~(\ref{F028})
to two-neutrino $\nu_{e}$--$\nu_{\act}$ mixing
is,
using the analogous notation
$\psi_{\afl L} \equiv \psi_{\afl,-1}$
and
$\psi_{\afl R} \equiv \psi_{\afl,+1}$,
\begin{equation}
i \frac{ \text{d} }{ \text{d}x }
\begin{pmatrix}
\psi_{eL}(x)
\\
\psi_{\act L}(x)
\\
\psi_{eR}(x)
\\
\psi_{\act R}(x)
\end{pmatrix}
=
\mathrm{H}
\begin{pmatrix}
\psi_{eL}(x)
\\
\psi_{\act L}(x)
\\
\psi_{eR}(x)
\\
\psi_{\act R}(x)
\end{pmatrix}
, \label{F067}
\end{equation}
with the effective Hamiltonian matrix
\begin{widetext}
\begin{equation}
\mathrm{H} =
\begin{pmatrix}
- \frac{ \Delta{m}^{2} }{ 4 E_{\nu} } \cos{2\vartheta_{12}} + V_{e} & \frac{ \Delta{m}^{2} }{ 4 E_{\nu} } \sin{2\vartheta_{12}} & \mgm_{ee}
B_{\perp}(x) & \mgm_{e\act} B_{\perp}(x)
\\
\frac{ \Delta{m}^{2} }{ 4 E_{\nu} } \sin{2\vartheta_{12}} & \frac{ \Delta{m}^{2} }{ 4 E_{\nu} } \cos{2\vartheta_{12}} + V_{\act} & \mgm_{e\act}^{*}
B_{\perp}(x) & \mgm_{\act\act} B_{\perp}(x)
\\
\mgm_{ee} B_{\perp}(x) & \mgm_{e\act} B_{\perp}(x) & - \frac{ \Delta{m}^{2} }{ 4 E_{\nu} } \cos{2\vartheta_{12}} & \frac{
\Delta{m}^{2} }{ 4 E_{\nu} } \sin{2\vartheta_{12}}
\\
\mgm_{e\act}^{*} B_{\perp}(x) & \mgm_{\act\act} B_{\perp}(x) & \frac{ \Delta{m}^{2} }{ 4 E_{\nu} } \sin{2\vartheta_{12}} & \frac{
\Delta{m}^{2} }{ 4 E_{\nu} } \cos{2\vartheta_{12}}
\end{pmatrix}
,
\label{F068}
\end{equation}
\end{widetext}
with the effective magnetic moments in the flavor basis
given by
\begin{equation}
\begin{pmatrix}
\mgm_{ee} & \mgm_{e\act}
\\
\mgm_{e\act}^{*} & \mgm_{\act\act}
\end{pmatrix}
=
R_{12}
\begin{pmatrix}
\mgm_{11} & \mgm_{12}
\\
\mgm_{12}^{*} & \mgm_{22}
\end{pmatrix}
R_{12}^{T}
.
\label{F069}
\end{equation}

The matter potential can generate resonances,
which occur when two diagonal
elements of $\mathrm{H}$ become equal.
Besides the standard MSW resonance in the $\nu_{eL}\leftrightarrows\nu_{\act L}$ channel
discussed in Subsection~\ref{B035},
there are two possibilities:
\begin{enumerate}
\item
There is a resonance in the
$\nu_{eL}\leftrightarrows\nu_{\act R}$
channel for
\begin{equation}
V_{e} = \frac{ \Delta{m}^{2} }{ 2 E_{\nu} } \, \cos2\vartheta_{12}
.
\label{F070}
\end{equation}
The density at which this resonance occurs is not the same as that of the MSW resonance,
given by Eq.~(\ref{B067}),
because of the neutral-current contribution to $V_{e}=V_{\text{CC}}+V_{\text{NC}}$.
The location of this resonance
depends on both $N_{e}$ and $N_{n}$.
\item There is a resonance in the $\nu_{\act L}\leftrightarrows\nu_{eR}$ channel for
\begin{equation}
V_{\act} = - \frac{ \Delta{m}^{2} }{ 2 E_{\nu} } \, \cos2\vartheta_{12}
.
\label{F071}
\end{equation}
If $\cos2\vartheta_{12}>0$, this resonance is possible in normal matter, since the sign of $V_{\act}=V_{\text{NC}}$ is
negative, as one can see from Eq.~(\ref{B051}).
\end{enumerate}
In practice the effect of these resonances could be the disappearance
of active $\nu_{eL}$ or $\nu_{\act L}$ into sterile right-handed
states.

Let us consider now the case of Majorana neutrinos.
The evolution equation of the amplitudes is given by Eq.~(\ref{F067}) with the effective Hamiltonian matrix
\begin{widetext}
\begin{equation}
\mathrm{H} =
\begin{pmatrix}
- \frac{ \Delta{m}^{2} }{ 4 E_{\nu} } \cos{2\vartheta_{12}} + V_{e} & \frac{ \Delta{m}^{2} }{ 4 E_{\nu} } \sin{2\vartheta_{12}} & 0 &
\mgm_{e\act} B_{\perp}(x)
\\
\frac{ \Delta{m}^{2} }{ 4 E_{\nu} } \sin{2\vartheta_{12}} & \frac{ \Delta{m}^{2} }{ 4 E_{\nu} } \cos{2\vartheta_{12}} + V_{\act} & -
\mgm_{e\act} B_{\perp}(x) & 0
\\
0 & -\mgm_{e\act}^{*} B_{\perp}(x) & - \frac{ \Delta{m}^{2} }{ 4 E_{\nu} } \cos{2\vartheta_{12}} - V_{e} & \frac{ \Delta{m}^{2} }{ 4 E_{\nu}
} \sin{2\vartheta_{12}}
\\
\mgm_{e\act}^{*} B_{\perp}(x) & 0 & \frac{ \Delta{m}^{2} }{ 4 E_{\nu} } \sin{2\vartheta_{12}} & \frac{ \Delta{m}^{2} }{ 4 E_{\nu} }
\cos{2\vartheta_{12}} - V_{\act}
\end{pmatrix}
,
\label{F072}
\end{equation}
\end{widetext}
with
\begin{equation}
\mgm_{e\act} \equiv \mgm_{e\act}^{(-,+)} = \mgm_{12} e^{i\lambda_{12}}
,
\label{F073}
\end{equation}
where $\lambda_{12}$ is the Majorana phase in Eq.~(\ref{B033}).

As in the Dirac case,
there are two possible resonances
besides the standard MSW resonance in the $\nu_{eL}\leftrightarrows\nu_{\act L}$ channel:
\begin{enumerate}
\item There is a resonance in the $\nu_{eL}\leftrightarrows\nu_{\act R}$ channel for
\begin{equation}
V_{\text{CC}}+2V_{\text{NC}} = \frac{ \Delta{m}^{2} }{ 2 E_{\nu} } \, \cos2\vartheta_{12} \,. \label{F074}
\end{equation}
\item There is a resonance in the $\nu_{\act L}\leftrightarrows\nu_{eR}$ channel for
\begin{equation}
V_{\text{CC}}+2V_{\text{NC}} = - \frac{ \Delta{m}^{2} }{ 2 E_{\nu} } \, \cos2\vartheta_{12} \,. \label{F075}
\end{equation}
\end{enumerate}
The location of both resonances depend on both $N_{e}$ and $N_{n}$. If $\cos2\vartheta_{12}>0$, only the first resonance can
occur in normal matter, where $ N_{n} \simeq N_{e}/6 $. A realization of the second resonance requires a large neutron
number density, as that in a neutron star.

The neutrino spin oscillations in a transverse magnetic field with
a possible rotation of the field-strength vector in a plane
orthogonal to the neutrino-propagation direction (such rotating
fields may exist in the convective zone of the Sun) have been
considered in
\textcite{Vidal:1990fr,Smirnov:1991ia,Akhmedov:1993sh,Likhachev:1990ki}.
The effect of the magnetic-field rotation may substantially shift
the resonance point of neutrino oscillations.
Neutrino spin
oscillations in electromagnetic fields with other different
configurations, including a longitudinal magnetic field and the
field of an electromagnetic wave, were examined in
\textcite{Akhmedov:1988hd,Akhmedov:1990ng,Egorov:1999ah,Lobanov:2001ar,Dvornikov:2001ez,Dvornikov:2004en,Studenikin:2004bu,Studenikin:2004tv}
(see also Appendix~\ref{N001}).

It is possible to formulate a criterion \cite{Likhachev:1990ki}
for finding out if the neutrino spin and spin-flavor precession is
significant for given neutrino and background medium properties.
The probability of oscillatory transitions between two neutrino
states $\nu_{\afl L}\leftrightarrows\nu_{\bfl R}$ can be
expressed in terms of the elements of the effective Hamiltonian
matrices (\ref{F068}) and (\ref{F072}) as
\begin{equation}
P_{\nu_{\afl L} \leftrightarrows \nu_{\bfl R}}=\sin^{2} \vartheta_{\text{eff}} \sin^{2} \frac{x\pi}{L_{\text{eff}}},
\end{equation}
where
\begin{align}
\null & \null
\sin^{2} \vartheta_{\text{eff}}=
\frac{4\mathrm{H}^{2}_{\afl\bfl}}{4\mathrm{H}^{2}_{\afl\bfl}+(\mathrm{H}_{\bfl\bfl}-\mathrm{H}_{\afl\afl})^{2}}
,
\label{F076}
\\
\null & \null
L_{\text{eff}}=\frac {2\pi}{\sqrt
{4\mathrm{H}_{\afl\bfl}^{2}+(\mathrm{H}_{\bfl\bfl}-\mathrm{H}_{\afl\afl})^{2}}}
.
\label{F077}
\end{align}
The transition probability can be of order unity
if the following two conditions hold simultaneously:
1) the amplitude
of the transition probability must be sizable (at least
$\sin^{2} \vartheta_{\text{eff}} \gtrsim 1/2$);
2) the neutrino path length in a
medium with a magnetic field should be longer than half the effective
length of oscillations $L_{\text{eff}}$. In accordance with this criterion,
it is possible to introduce the critical strength of a magnetic field
$B_{\text{cr}}$ which determines the region of field values $B_{\perp}>
B_{\text{cr}}$ at which the probability amplitude is not small ($\sin^{2}
\vartheta_{\text{eff}} > 1/2$):
\begin{equation}\label{F078}
B_{\text{cr}}
=
\frac{1}{2\mgm_{\bfl\afl}}\sqrt
{(\mathrm{H}_{\bfl\bfl}-\mathrm{H}_{\afl\afl})^{2}}
.
\end{equation}

Consider, for instance, the case of
$\nu_{eL}\leftrightarrows\nu_{\act R}$
transitions of Majorana neutrinos.
From Eqs.~(\ref{F072}) and (\ref{F078}), it
follows \cite{Likhachev:1990ki} that
\begin{equation}
B_{\text{cr}}
=
\left|
\frac{1}{2\mgm_{\act e}}
\left(
\frac{\Delta{m}^{2}\cos2\vartheta_{12}}{2E_{\nu}}
-
\sqrt{2}G_{\text{F}} N_{\text{eff}}
\right)
\right|,
\label{F079}
\end{equation}
where $N_{\text{eff}}=N_{e}-N_{n}$.
For getting numerical estimates of $B_{\text{cr}}$ it is convenient
to rewrite Eq.~(\ref{F079}) in the following form:
\begin{align}
B_{\text{cr}}
\approx
43 \, \frac{\bmag}{\mgm_{\act e}}
\null & \null
\Bigg|
A
\left(\frac{\Delta{m}^{2}}{\text{eV}^{2}}\right)
\left(\frac{\text{MeV}}{E_{\nu}}\right)
\nonumber
\\
\null & \null
-
2.5\times 10^{-31}
\left(\frac{N_{\text{eff}}}{\text{cm}^{-3}}\right)
\Bigg|
\text{G}
.
\label{F080}
\end{align}

An interesting feature of the evolution equation~(\ref{F067}) in the case of
Majorana neutrinos is that the interplay of spin precession and flavor
oscillations can generate $\nu_{eL}\to\nu_{eR}$ transitions
\cite{Akhmedov:1991uk}. Since $\nu_{eR}$ interacts as right-handed Dirac
antineutrinos, it is often denoted by $\bar\nu_{eR}$, or only $\bar\nu_{e}$,
and called ``electron antineutrino''. This state can be detected through the
inverse $\beta$-decay reaction
\begin{equation}
\bar\nu_{e} + p \to n + e^{+} \,, \label{F081}
\end{equation}
having a threshold $ E_{\text{th}} = 1.8 \, \text{MeV} $.

The possibility of $\nu_{eL}\to\bar\nu_{eR}$ transitions
generated by spin-flavor precession of Majorana neutrinos is particularly interesting
for solar neutrinos, which experience matter effects in the
interior of the Sun in the presence of the solar magnetic field
(see \textcite{Pulido:1991fb,Shi:1992ek}).
Taking into account
the dominant $\nu_{e}\to\nu_{\act}$
transitions due to neutrino oscillations,
with $\nu_{\act}$ given by Eq.~(\ref{B074}),
the probability of solar $\nu_{eL}\to\bar\nu_{eR}$ transitions
is given by
\textcite{Akhmedov:2002mf}
\begin{align}
P_{\nu_{eL}\to\bar\nu_{eR}}
\simeq
\null & \null
1.8 \times 10^{-10}
\sin^2 2\vartheta_{12}
\nonumber
\\
\null & \null
\times
\left(
\frac{\mgm_{e\act}}{10^{-12}\,\bmag}
\,
\frac{B_{\perp}(0.05R_{\odot})}{10\,\text{kG}}
\right)^2
,
\label{F082}
\end{align}
where $\mgm_{e\act}$ is the transition magnetic moment in Eq.~(\ref{F073}),
$R_{\odot}$
is the radius of the Sun,
and the values of $\vartheta_{12}$ and $\vartheta_{23}$ are given in Tab.~\ref{B082}.

It is also possible that spin-flavor precession
occurs in the convective zone of the Sun,
where there can be random turbulent magnetic fields
\cite{Miranda:2003yh,Miranda:2004nz,Friedland:2005xh}.
In this case
\cite{Raffelt:2009mm},
\begin{align}
P_{\nu_{eL}\to\bar\nu_{eR}}
\approx
\null & \null
10^{-7} S^2
\left( \frac{\mgm_{e\act}}{10^{-12}\,\bmag} \right)^2
\left( \frac{B}{20\,\text{kG}} \right)^2
\nonumber
\\
\null & \null
\times
\left( \frac{3\times10^{4}\,\text{km}}{L_{\text{max}}} \right)^{p-1}
\left( \frac{8\times10^{-5}\,\text{eV}^2}{\Delta{m}^2_{\text{S}}} \right)^{p}
\nonumber
\\
\null & \null
\times
\left( \frac{E_{\nu}}{10\,\text{MeV}} \right)^{p}
\left( \frac{\cos^2\vartheta_{12}}{0.7} \right)^{p}
,
\label{F083}
\end{align}
where
$S$ is a factor of order unity depending on the spatial configuration of the magnetic field,
$B$ is the average strength of the magnetic field at the spatial scale $L_{\text{max}}$,
which is the largest scale of the turbulence,
$p$ is the power of the turbulence scaling,
$\Delta{m}^2_{\text{S}}$ is the solar neutrino squared-mass difference in Tab.~\ref{B082},
and
$E_{\nu}$ is the neutrino energy.
A possible value of $p$ is 5/3
\cite{Miranda:2003yh,Miranda:2004nz,Friedland:2005xh},
corresponding to Kolmogorov turbulence.
Conservative values for the other parameters are
$B=20\,\text{kG}$
and
$L_{\text{max}}=3\times10^{4}\,\text{km}$.

In 2002, the Super-Kamiokande Collaboration established for the
flux of solar $\bar\nu_{e}$'s a 90\% CL an upper limit of 0.8\%
of the Standard Solar Model (SSM) neutrino flux in the range of
energy from 8 to 20 MeV \cite{Gando:2002ub}
by taking as a reference the BP00 SSM prediction
$\phi_{^{8}\text{B}}^{\text{BP00}} = 5.05\times10^{6}\,\text{cm}^{-2}\,\text{s}^{-1}$
for the solar $^{8}\text{B}$ flux
\cite{Bahcall:2000nu}
and
assuming an undistorted $^{8}\text{B}$ spectrum for the $\bar\nu_{e}$'s.
This limit was
improved in 2003 by the KamLAND Collaboration
\cite{Eguchi:2003gg}
to
$ 2.8 \times 10^{-4} $
of the BP00 SSM prediction
at 90\% CL
by measuring
$\phi_{\bar\nu_{e}} < 370\,\text{cm}^{-2}\,\text{s}^{-1}$ (90\% CL)
in the energy range 8.3 -- 14.8 MeV,
which corresponds to
$\phi_{\bar\nu_{e}} < 1250\,\text{cm}^{-2}\,\text{s}^{-1}$ (90\% CL)
in the entire $^{8}\text{B}$ energy range
assuming an undistorted spectrum.

Recently,
the Borexino collaboration established the best limit on the probability of solar
$\nu_{eL}\to\bar\nu_{eR}$ transitions
\cite{Bellini:2010gn},
\begin{equation}
P_{\nu_{eL}\to\bar\nu_{eR}}
<
1.3 \times 10^{-4}
\qquad
\text{(90\% CL)}
,
\label{F084}
\end{equation}
by taking
as a reference
$\phi_{^{8}\text{B}}^{\text{SSM}} = 5.88\times10^{6}\,\text{cm}^{-2}\,\text{s}^{-1}$
\cite{Serenelli:2009yc}
and
assuming an undistorted $^{8}\text{B}$ spectrum for the $\bar\nu_{e}$'s.
They measured
$ \phi_{\bar\nu_{e}} < 320 \, \text{cm}^{-2} \, \text{s}^{-1} $ (90\% CL)
for $ E_{\bar\nu_{e}} > 7.3 \, \text{MeV} $,
which corresponds to
$ \phi_{\bar\nu_{e}} < 760 \, \text{cm}^{-2} \, \text{s}^{-1} $ (90\% CL)
in the entire $^{8}\text{B}$ energy range
assuming an undistorted spectrum

The implications of the limits
on the flux of solar $\bar\nu_{e}$'s on Earth
for the
spin-flavor precession of solar neutrinos have been studied in
several papers
\cite{Akhmedov:2002mf,Chauhan:2003wr,Miranda:2003yh,Miranda:2004nz,Balantekin:2004tk,Guzzo:2005rr,Friedland:2005xh,Yilmaz:2008vh},
taking into account the dominant $\nu_{e}\to\nu_{\mu},\nu_{\tau}$
transitions due to neutrino oscillations
(see Subsection~\ref{B035}).
Using Eqs.~(\ref{F082}) and (\ref{F084}),
we obtain
\begin{equation}
\mgm_{e\act}
\lesssim
1.3 \times 10^{-12}
\,
\frac{7\,\text{MG}}{B_{\perp}(0.05R_{\odot})}
\,
\bmag
,
\label{F085}
\end{equation}
with
$ 600 \, \text{G} \lesssim B_{\perp}(0.05R_{\odot}) \lesssim 7 \, \text{MG} $
\cite{Bellini:2010gn}.
In the case of spin-flavor precession in the convective zone of the Sun
with random turbulent magnetic fields,
Eqs.~(\ref{F083}) and (\ref{F084})
give,
assuming $p=5/3$,
\begin{equation}
\mgm_{e\act}
\lesssim
4 \times 10^{-11}
\,
S^{-1}
\,
\frac{20\,\text{kG}}{B}
\,
\left( \frac{L_{\text{max}}}{3\times10^{4}\,\text{km}} \right)^{1/3}
\,
\bmag
.
\label{F086}
\end{equation}

The spin-flavor precession mechanism was also considered \cite{Pulido:2005pt} in
order to describe time variations of solar-neutrino fluxes in
Gallium experiments.
The effect of a nonzero neutrino magnetic
moment is also of interest in connection with the analysis of
helioseismological observations \cite{Couvidat:2003ba}.

The idea that the neutrino magnetic moment may solve the problem of the explosion of core-collapse supernovae,
i.e. that the neutrino spin-flip transitions in a
magnetic field can provide an efficient mechanism of energy transfer
from a protoneutron star, was discussed in
\textcite{Fujikawa:1980yx,Dar:1987yv,Nussinov:1987zr,Goldman:1987fg,Lattimer:1988mf,Barbieri:1988nh,Voloshin:1988xu}.
The possibility of a
loss of up to half of the active left-handed neutrinos because of
their transition to sterile right-handed neutrinos in strong
magnetic fields at the boundary of the neutron star (the so-called
boundary effect) was considered in \textcite{Likhachev:1990ki}.

The possibility to observe the effects of
neutrino spin-flip transitions
in terrestrial measurements of the neutrino flux of a core-collapse supernova was studied in
\textcite{Ando:2003pj,Akhmedov:2003fu,Cuesta:2008te,Yoshida:2009ec,Yoshida:2011fc}.

Recently
\textcite{deGouvea:2012hg,deGouvea:2013zp}
studied the effects of spin-flavor precession on the evolution of neutrinos
with Majorana transition magnetic moments
inside the core of a supernova,
where the magnetic field can be as large as $10^{12} \, \text{G}$
at a radius of about $50 \, \text{km}$.
The high neutrino density in the protoneutron star induces neutrino-neutrino interactions
\cite{Notzold:1987ik}
that lead to collective neutrino flavor oscillations
(see \textcite{Duan:2009cd,Duan:2010bg,Volpe:2013kxa}).
This effect can swap the spectrum of different flavor neutrinos and antineutrinos emerging
from the supernova core
above a ``split'' energy.
\textcite{deGouvea:2012hg,deGouvea:2013zp}
studied the additional effects of spin-flavor precession
by considering a Hamiltonian of the type in Eq.~(\ref{F072})
with the addition of neutrino-neutrino interactions.
They found that there can be collective spin-flavor oscillations
in addition to the usual mass-generated collective neutrino oscillations,
which can lead to spectral swaps between neutrinos and antineutrinos\footnote{
In the traditional terminology,
although strictly speaking in the Majorana case there is no difference between a neutrino and an antineutrino,
it is common to call neutrinos the left-handed helicity states
and antineutrinos the right-handed helicity states,
which have the same weak interactions of the right-handed helicity states of Dirac antineutrinos.
}
for Majorana transition magnetic moments of the order of
$10^{-21} \, \bmag$.
These are extremely small values for the Majorana transition magnetic moments,
which are only two orders of magnitude larger than those predicted by the simplest extensions
of the Standard Model
(see Subsection~\ref{D024}, where it is explained that the Majorana transition magnetic moments are
expected to have the same order of magnitude (\ref{D019}) of the Dirac transition magnetic moments).
This may be the only potentially observable phenomenon sensitive to such small values
of the Majorana transition magnetic moments.

The neutrino spin (and spin-flavor) procession can be stimulated in the presence of moving matter when the
matter speed transverse to the neutrino propagation is not zero or when matter is polarized.
A detailed discussion of this phenomena can be found in \textcite{Studenikin:2004bu,Studenikin:2004tv}
(see also \textcite{Lobanov:2001ar}).
Note that these types of spin procession and the corresponding oscillations in matter occur without the presence of any electromagnetic field.

\subsection{Magnetic moment in a strong magnetic field}
\label{F087}

The discussion of the neutrino electromagnetic properties in Section~\ref{C001}
is based on the one-photon approximation of the response of a neutrino
to the presence of an electromagnetic field.
This approximation
is appropriate when the strength of the electromagnetic field is not too high.
In the case of a very strong electromagnetic field
one must take into account multiphoton contributions,
which can be effectively incorporated in the
neutrino form factors derived in Section~\ref{C001} by allowing the
form factors to depend on the strength of the external electromagnetic field.
In this Subsection we discuss the
dependence of the effective neutrino magnetic moments on the strength of
an external magnetic field,
which was investigated in
\textcite{Borisov:1985ha,Borisov:1987xm,Borisov:1988wy,Borisov:1989yw,Masood:1999qv}
through the calculation of the self-energy of a neutrino in the presence of
an arbitrary electromagnetic field.
In the following we generalize the results of
\textcite{Borisov:1985ha}
in order to take into account neutrino mixing.

The evaluation of the dependence of the neutrino magnetic moments on the
magnetic field is based on the Dirac-Schwinger equation for the wave function
$\Psi_{k}(x)$
of a neutrino with mass $m_{k}$:
\begin{equation}
\left(
i\partial_\mu \gamma^\mu-m_{k}
\right)
\Psi_{k}(x)
=
\int
M_{k}(x,x';\vet{B})
\,
\Psi_{k}(x')
\,
dx'
,
\end{equation}
where
$M_{k}(x',x;\vet{B})$ is the neutrino
mass operator in the presence of a magnetic field $\vet{B}$.
The diagonal matrix element calculated
on the mass shell ($p_{k}^{2}=m_{k}^{2}$) between the neutrino vacuum states gives the radiative
correction to the mass of the neutrino in the external field,
\begin{equation}\label{F088}
\Delta m_{k}
=
\frac{E_{k}}{m_{k}} \, \Delta E_{k}
.
\end{equation}
The shift of the neutrino energy due to the presence of the external field is given by
\begin{equation}\label{F089}
\Delta E_{k}(\vet{B})
=
\int
dx \, dx'
\,
\overline{\psi_{k}}(x)
\,
M_{k}(x,x';\vet{B})
\,
\psi_{k}(x')
,
\end{equation}
where $\psi_{k}(x) = \left( 2 E_{k} \right)^{-1/2} u(p_{k}) e^{- i p_{k} \cdot x}$
is the neutrino wave function in vacuum
with four-momentum
$p_{k}^{\mu} = (E_{k}, \vet{p})$
and energy
$E_{k} = \sqrt{ \vet{p}_{k}^2 + m_{k}^2}$.
The radiative correction $\Delta m_{k}$ to the neutrino mass in a constant electromagnetic field
described by the tensor $F_{\mu\nu}=\partial_{\mu}A_{\nu}-\partial_{\nu}A_{\mu}$
includes the Lorentz invariant
$s_{k}^{\mu}\widetilde{F}_{\mu\nu}p_{k}^{\nu}$ that depends on
$\widetilde{F}^{\mu\nu} = \frac{1}{2} \epsilon^{\mu\nu\alpha\beta} F_{\alpha\beta}$
and on the neutrino polarization vector
(see, for instance, \textcite{AkhBerQE1965})
\begin{equation}\label{F090}
s_{k}^{\mu}
=
\left(
\frac{\vec{S} \cdot \vec{p}}{m_{k}}
,
\,
\vec{S}
+
\frac{ \vec{p} \left( \vec{S} \cdot \vec{p} \right) }{ m_{k} \left(E_{k}+m_{k}\right) }
\right)
,
\end{equation}
where $\vec{S}$ is
the normalized neutrino spin vector in the rest frame.

The contribution to $\Delta m_{k}$
proportional to the Lorentz invariant
$s_{k}^{\mu} \widetilde{F}_{\mu\nu} p_{k}^{\nu}$
is due to the interaction of the neutrino magnetic moment with the external field.
Following \textcite{Ritus:1972ky},
for the real part of $\Delta m_{k}$ one gets
\begin{equation}\label{F091}
\operatorname{Re}\Delta m_{k}
=
\frac{\mgm_{k}}{m_{k}}
\,
s_{k}^{\mu}\widetilde{F}_{\mu\nu}p_{k}^{\nu}
,
\end{equation}
where
$\mgm_{k} = \mgm_{kk}$
are the diagonal magnetic moments of the massive neutrinos.
In the neutrino rest frame we obtain
\begin{equation}\label{F092}
\operatorname{Re}\Delta m_{k}
=
- \mgm_{k}
\left( \vec{B} \cdot \vec{S} \right)
.
\end{equation}
Using this equation one can extract
from $\Delta m_{k}$
the dependence of the effective magnetic moment
on the field strength $B = |\vet{B}|$.

In the framework of the minimal extension of the
Standard Model with right-handed neutrinos,
the virtual one-loop processes
$\nu_{k} \to e^{-}W^{+} \to \nu_{k}$,
$\nu_{k} \to \mu^{-}W^{+} \to \nu_{k}$ and
$\nu_{k} \to \tau^{-}W^{+} \to \nu_{k}$ contribute
to the mass operator
\begin{align}
M_{k}
\null & \null
(x,x';\vet{B})
=
-i
\,
\frac{g^2}{8}
\sum_{\afl=e,\mu,\tau}
|U_{\afl k}|^{2}
(1-\gamma^5)
\nonumber
\\
\null & \null
\times
\gamma_{\mu}
S_{\afl}(x,x';\vet{B})
\gamma_{\nu}
(1+\gamma^5)
D_{W}^{\mu\nu}(x,x';\vet{B})
,
\label{F093}
\end{align}
where $U_{\afl k}$ are the elements of neutrino mixing matrix, $S_{\afl}(x,x';\vet{B})$ and $D_{W}^{\mu \nu}(x,x';\vet{B})$ are the charged leptons and
$W$ boson propagators in the presence of the external magnetic field $\vet{B}$ and
$g$ is the $\text{SU}(2)_{L}$ weak-interaction coupling constant,
which is related to the Fermi coupling constant by
$G_{\text{F}}=\sqrt{2}g^2/8m^2_{W}$.
Neglecting terms proportional to $m_{k}^2 / m_{\afl}^2 \ll 1$
and considering a magnetic field
$B \ll B_{0}^{e} = m_{e}^2/\elechg \simeq 4.41 \times 10^{13} \, \text{G}$,
from a generalization of the results of
\textcite{Borisov:1985ha}
to the case of neutrino mixing
we obtain
\begin{equation}
\mgm_{k}(B)
=
\mgm_{k}(0)
\left[
1
+
\frac{4}{9}
\left( \frac{B}{B_{0}^{W}} \right)^2
\sum_{\afl=e,\mu,\tau}| U_{\afl k}|^{2}
\ln\frac{m_{W}^2}{m_{\afl}^2}
\right]
,
\label{F094}
\end{equation}
where
$B_{0}^{W} = m_{W}^2/\elechg \simeq 1.1 \times 10^{24} \, \text{G}$.
In this case the one-loop correction to the magnetic moment
given by the external magnetic field
is very small,
because $(B/B_{0}^{W})^2 \ll 10^{-22}$.

A significant difference of
$\mgm_{k}(B)$
from
$\mgm_{k}(0)$
is obtained when the
strength of the magnetic field approaches $B_{0}^{W}$.
For
$B_{0}^{W}-B_{0}^{e} \ll B \lesssim B_{0}^{W}$
we have
\cite{Borisov:1985ha}
\begin{equation}
\mgm_{k}(B)
=
\frac{2}{3}
\,
\mgm_{k}(0)
\,
\ln\left(\frac{B_{0}^{W}}{B_{0}^{W}-B}\right)
\sum_{\afl=e,\mu,\tau}| U_{\afl k}|^{2}
\frac{m_{W}^2}{m_{\ell}^2}
.
\end{equation}
The divergence of this expression for $B \to B_{0}^{W}$
must be treated with caution,
because
when the magnetic field $B$ is close to the critical value
$B_{0}^{W}$ the vacuum becomes unstable with respect to $W^{+}W^{-}$ pair production,
giving rise to $W$ bosons condensation
\cite{Nielsen:1978rm,Skalozub:1985ur,Skalozub:1986gw,Ambjorn:1988tm}.

Let us recall
that very strong fields are supposed to exist in some astrophysical domains.
For instance,
magnetic fields of the order of $10^{16} \, \text{G}$ or even up to $10^{18} \, \text{G}$ can be
produced in a supernova explosion or in the vicinity of magnetars,
as discussed by \textcite{Lai:2000at,Akiyama:2002xn,Mereghetti:2008je}.
For magnetar cores made of quark matter the interior magnetic field can reach values up to
about $10^{20} \, \text{G} $ \cite{Paulucci:2010uj}.
A more exotic possibility of superstrong magnetic fields is discussed in
\textcite{Ostriker:1986xc}, where it is shown that
magnetic fields stronger than $10^{30} \, \text{G}$ can be generated
in the vicinity of superconducting cosmic strings.

\textcite{Borisov:1985ha} calculated also the dependence of the
effective neutrino magnetic moment on the energy of neutrino.
In the case of a magnetic field which is not extremely strong
($B \ll B_{0}^{W}$),
a neutrino with transverse momentum
$p_{\perp} \gg m_{W}$
with respect to the magnetic field direction
and
\begin{equation}
\frac{B}{B_{0}^{\ell}}
\,
\frac{p_{\perp}}{m_{\ell}}
\gg
\left(
\frac{m_{W}}{m_{\ell}}
\right)^3
,
\label{F095}
\end{equation}
we have
\begin{align}
\null & \null
\mgm_{k}(B)
=
\frac{3^{5/6}\Gamma^{4}(1/3)}{20\pi}
\,
\mgm_{k}(0)
\nonumber
\\
\null & \null
\times
\sum_{\afl=e,\mu,\tau}
|U_{\afl k}|^{2}
\left(
\frac{B p_{\perp}}{B_{0}^{\ell} m_{\ell}}
\right)^{-2/3}
\left(
\frac{m_{W}}{m_{\ell}}
\right)^2
,
\label{F096}
\end{align}
where
$B_{0}^{\ell} = m_{\ell}^2 / \elechg$.
In this case,
the magnetic moment of a neutrino with very high energy
decreases to zero with the increase of the neutrino energy.

Let us finally remind the studies of the neutrino self-energy and electromagnetic vertex in matter without and with a magnetic field.
The neutrino self-energy and the electromagnetic vertex function in matter
without a magnetic field have been studied in \textcite{Notzold:1987ik,Nieves:1989xg,D'Olivo:1989cr}.
The vacuum dispersion relation in the presence of a constant magnetic field
has been studied by \textcite{Erdas:1990gy}.
Finite-temperature corrections to the neutrino self-energy in a
background medium without magnetic field have been calculated by \textcite{D'Olivo:1992vm}.
Those in the presence of an electromagnetic field
have been calculated in \textcite{Zhukovsky:1993kj,Esposito:1995db,Nieves:2003kw}.
The general expressions for the neutrino dispersion relation in a
magnetized plasma with wide ranges of temperature, chemical
potential and magnetic field strengths has been derived in
\textcite{Elmfors:1996gy,Elizalde:2000vz,Elizalde:2004mw}.
The one-loop thermal self-energy of a neutrino in an
arbitrary strong magnetic field has been calculated by \textcite{Erdas:1998uu,Erdas:2000iq}.
These calculations of the effective neutrino properties in a magnetized plasma
are useful for the study of the behavior of neutrinos in the early Universe.

\subsection{Beta decay of the neutron in a magnetic field}
\label{F097}

The first studies of neutrino
interactions in the presence of external electromagnetic fields
were performed by \textcite{Korovina:1964ivf,Ternov:1965vm},
who considered the $\beta$ decay $n \to p + e^{-} + \bar\nu_{e}$
of a polarized neutron in a magnetic field\footnote{
This process and the other URCA processes in Eq.~(\ref{F098}) are important for
the energy loss of stars
\cite{Gamow:1941pr}.
}.
It was shown that the differential rate of the process
exhibits resonance spikes which appear when the final electron energy
is equal to one of the allowed Landau energies in the
magnetic field.
It was also shown that the total rate depends on
the initial neutron polarization
and that the neutrino emission is asymmetric.
The
range of magnetic field strengths considered in these papers span
up to subcritical fields
$B \lesssim B_{0}^{e} = m_{e}^2/\elechg \simeq 4.41 \times 10^{13} \, \text{G}$.
It is worth to be noted that these studies were
performed before the discovery by \textcite{Hewish:1968bj} of
pulsars, where such strong magnetic fields are believed to exist.

In two papers by \textcite{Matese:1969zz,FassioCanuto:1970wk}, published a few years later, the
results of \textcite{Korovina:1964ivf,Ternov:1965vm} for the neutron decay rate in a magnetic
field were rederived, but there was no discussion of the
asymmetry in the neutrino emission.

Very strong magnetic fields are also supposed to exist in the
early Universe (see \textcite{Grasso:2000wj}).
As first discussed
by \textcite{Greenstein:1969xx,Matese:1970apj}, the
weak reaction rates of the URCA processes
\begin{equation} \label{F098}
n\to p + e^{-} +\bar\nu_e
,
\quad
\nu_e + n \leftrightarrows e^{-} + p
,
\quad
p + \bar\nu_e \leftrightarrows n + e^{+}
,
\end{equation}
which determine the conversions between neutrons and protons
and set the $n/p$ ratio in various environments, can be
significantly modified under the influence of magnetic fields.
This can be important for Big-Bang Nucleosynthesis and neutron star cooling
\cite{Cheng:1993ma}.

The aforementioned studies of neutrino interactions in the
presence of magnetic fields performed by
\textcite{Korovina:1964ivf,Ternov:1965vm,Matese:1969zz,Matese:1970apj,Greenstein:1969xx,FassioCanuto:1970wk}
gave birth to {\it neutrino astrophysics in magnetic fields}.

The $\beta$-decay process can be described by the well-known four-fermion
Lagrangian
\begin{equation}\label{F099}
\mathscr{L}=\frac{\widetilde{G}}{\sqrt{2}}\left[\overline\psi_{p}\gamma_{\mu}(1+g_{A} \gamma_5)
\psi_{n}\right]\left[\overline\psi_e\gamma^{\mu}(1+\gamma_5)\psi_{\nu}\right],
\end{equation}
where $\widetilde{G}=G_{\text{F}}\cos\theta_{\text{C}}$, $\theta_{\text{C}}$ is the Cabibbo angle, and
$g_{A} \simeq 1.27$ (see \textcite{PDG-2012}) is the axial coupling constants.
After standard calculations one can obtain the neutron decay rate
\begin{equation}
\Gamma
=
\sum_{\text{phase}\atop\text{space}}
|M|^{2}
\,
\delta(E_{n} - E_{p} - E_{e} - E_{\nu})
,
\label{F100}
\end{equation}
where the matrix element
\begin{equation}
M
=
\frac{\widetilde{G}}{\sqrt{2}}
\int d^4x
\left[\overline\psi_{p}\gamma_{\mu}
(1+g_{A} \gamma_5)
\psi_{n}\right]
\left[\overline\psi_e\gamma^{\mu}
(1+\gamma_5)\psi_{\nu}\right]
\label{F101}
\end{equation}
accounts for the influence of the magnetic field through the wave
functions of the electron and proton.
For the electron wave
function one has to use the exact solutions of the Dirac equation
in the magnetic field given in Appendix~\ref{P001} by
Eqs.~(\ref{P007}), (\ref{P010}), (\ref{P011}) and
(\ref{P012}). The wave function for a proton has similar form and
is given, for instance in \textcite{Studenikin:1989mm}. The
initial neutron and neutrino are supposed to be not directly
affected by the magnetic field and the plane waves are used for
these particles wave functions.

The argument of the $\delta$ function in (\ref{F100}), being
equated with zero, gives the law of energy conservation for the
particles in the process, that for the case of the neutron decay
at rest is
\begin{equation}\label{F102}
m_{n}
=
\sqrt{m_{e}^2 + 2eB N_{e} + {p_{e}^{3}}^2}
+
\sqrt{m_{p}^2 + 2eB N_{p} + {p_{p}^{3}}^2}
+
E_{\nu}
,
\end{equation}
where $N_{e}$ and $N_{p}$ are the numbers of Landau levels in the
magnetic field for the electron and proton.
The summation in
(\ref{F100}) is performed over the phase space of the final
particles:
$\vec{p}_{\nu}, p_{p}^2, p_{p}^3, N_{p}, s_{p}, p_{e}^2, p_{e}^3, N_{e}, s_{e}$,
where values $s_{e}, s_{p} = \pm 1$ denote the two possible spin states of the
the electron and proton.
For not very strong magnetic fields
$B<B_{\text{cr}} = (\Delta^2-m_{e}^2)/2\elechg \simeq 1.8 \times 10^{14} \, \text{G}$,
where $\Delta=m_{p}-m_{n}$,
the decay rate is
\begin{eqnarray}\label{F103}
\Gamma(&B&)=\frac{\Gamma(0)}{2}\int \sin\theta_{\nu}d\theta_{\nu}
\Bigg\{1+\frac{2(g_{A}^2+g_{A})}{1+3g_{A}^2}s_{n}\cos\theta_{\nu}
\nonumber\\&{-}&
4.9\frac{eB}{\Delta^2}\Big(\frac{g_{A}^2-1}{1+3g_{A}^2}\cos\theta_{\nu}+\frac{2(g_{A}^2-g_{A})}{1+3g_{A}^2}s_{n}\Big)\Bigg\},
\end{eqnarray}
where
$\theta_{\nu}$ is the angle between the neutrino propagation and the magnetic field vector
and
$\Gamma(0)$
is the decay rate of the neutron in the absence of the magnetic field,
given by
\begin{equation}
\Gamma(0)
=
0.47
\,
\frac{\widetilde{G}^2\Delta^5}{120\pi^3}(1+3g_{A}^2)
,
\end{equation}
where $s_{n}=\pm 1$ correspond to the neutron spin
polarization parallel or antiparallel to the magnetic field
vector.

From Eq.~(\ref{F103}) it follows that there is an
asymmetry in the spatial distribution of neutrinos.
This asymmetry
is due to the parity violation in weak interactions and it is
modified by the magnetic field presence.
In addition, as it is
also clear from (\ref{F103}) the average momentum of
antineutrinos on the magnetic field strength and the direction of
propagation with respect to the magnetic field vector.
That is why
we consider the total effect of the antineutrino spatial
distribution asymmetry as the neutrino electromagnetic properties
manifestation.

Note that the same asymmetry appears in case of much stronger
magnetic fields $B>B_{\text{cr}}$ as well as for other similar processes
(\ref{F098}). Recently the relativistic approach to the inverse
$\beta$ decay of a polarized neutron, $\nu_{e} + n \to p
+ e^{-}$, in a magnetic field has been developed by
\textcite{Shinkevich:2004ja} \footnote{This process is also
important for the neutrino transport inside the magnetized pulsar
and contribute to the kick velocities, as shown by
\textcite{Roulet:1997sw,Bhattacharya:2002aj,Duan:2004nc}.}. It was shown that in strong magnetic fields the
cross section can be highly anisotropic with respect to the neutrino
angle.
In the particular case of polarized neutrons, matter
becomes even transparent for neutrinos if neutrinos propagate
against the direction of neutrons polarization.

It was first claimed by
\textcite{Chugai:2005sal,Dorofeev:1985az,Dorofeev:1985saj,Zakhartsov:1985aj} that asymmetric neutrino
(antineutrino) emission in the direct URCA processes (\ref{F098})
during the first seconds after a magnetized massive star collapse
could provide explanations for the observed pulsar velocities.
As
shown by \textcite{Studenikin:1988sja}, in order to get a correct
prediction for the direction and value of the kick velocity of a
pulsar one has to account not only for the amount of neutrinos radiated in
the processes (\ref{F098}) but also for the fact that the
values of the average momentum of neutrinos propagating in the
opposite directions are not the same.
More detailed studies of the
neutrino asymmetry in relation to magnetized stars have been
performed by
\textcite{Leinson:1998yr,Goyal:1998nq,Lai:1998apj,Arras:1998mv,Gvozdev:1999md,Roulet:1997sw,Duan:2004nc,Kauts:2006rd}.

We recall also different other mechanisms for the asymmetry in the
neutrino emission from a magnetized pulsar studied by
\textcite{Kusenko:1996sr,BisnovatyiKogan:1997an,Akhmedov:1997qb,Lai:1998apj}.
For more complete references to the performed studies on the neutrino
mechanisms of the pulsar kicks see the introductions presented by
\textcite{Bhattacharya:2002aj,Shinkevich:2004ja}.
Presently there is no solid explanation of the observed pulsars kick velocities.
Thus, the origin of pulsar kicks is still an unsolved problem
(see, for instance, \textcite{Tamborra:2014aua}).
The phenomenon seems to be very complicated and is probably the result of
different mechanisms which are acting simultaneously.
One of these mechanisms can be the neutrino asymmetry considered in
this subsection.

\subsection{Neutrino pair production by an electron}
\label{F104}

It is well known that in the presence of external electromagnetic
fields particles interaction processes, that are forbidden in
vacuum, become possible.
One may consider the corresponding
processes of neutrinos interaction with real particles that could
only become possible under the influence of external
electromagnetic fields as manifestation of neutrinos
electromagnetic properties.

One of these processes is the
production of neutrino-antineutrino pair by an electron moving in
a constant magnetic field
\begin{equation} \label{F105}
e \to e + \nu_e + \bar\nu_e.
\end{equation}
Astrophysical significance of this process, termed the synchrotron
radiation of neutrinos, was discussed by
\textcite{Landstreet:1967pr}. Here it worth to be noted that the
possibility of $\nu {\bar {\nu}}$ emission by an electron through
the bremsstrahlung process on a nuclei
\begin{equation} \label{F106}
e + A \to e + A + \nu_e + \bar\nu_e
\end{equation}
was first discussed by \textcite{Pontecorvo:1959wb} who also
pointed out that for certain stages of a star evolution the
proposed mechanism of $\nu {\bar {\nu}}$ emission might be
important.

In vacuum, i.e. in the absence of the magnetic field, the process
(\ref{F105}) is obviously forbidden.
The dependence of the
rate of the process (\ref{F105}) on the magnetic field was
initially derived by \textcite{Baier:1966zz,Ritus:1969kk,Loskutov:1969aa} within the
local four fermion weak interaction model of Gell-Mann-Feynman.
In
the Weinberg-Salam model this process was considered by
\textcite{Ternov:1983hv,Ternov:1982qf}. In the low-energy
approximation of the model for the amplitude of the process
(\ref{F105}) we have used
\begin{equation}
M
=
-
\frac{G_{\text{F}}}{\sqrt{2}}
\overline{\psi'_e}
\gamma_{\mu}
( g_{V} + g_{A} \gamma_5 )
\psi_{e}
\overline{\psi}_{\nu_1}
\gamma^{\mu}
(1+\gamma_5)
\psi_{\nu_2}
,
\end{equation}
where
$\psi_{e}$ and $\psi'_e$ are the initial and final electron wave functions
and $\psi_{\nu_1}$ and $\psi_{\nu_2}$ are the two
neutrino wave functions.
In the case of the electron $\nu {\bar
{\nu}}$ pair emission
in Eq.~(\ref{F105}),
$g_{V} = \sin^2 \theta_{W} + 1/2$ and
$g_{A}=1/2$. The effect of a constant magnetic field
presence is accounted for by the wave functions of the initial and
final electron that are the exact solutions of the Dirac equation
in magnetic field given in Appendix~\ref{P001}. Performing
standard calculations accounting for the rotational symmetry of
the problem with respect to the magnetic field $\vec B$ oriented
along the $z$ axis one arrives to the rate given by
\textcite{Ternov:1983hv,Ternov:1982qf}
\begin{align}
\Gamma
=
\frac{G_{\text{F}}^2}{3(2\pi)^2}
\null & \null
\sum_{N}
\int_{|\vec{f}| \leq f_0} d^3f
\big[
f_0^2H_{00}
\nonumber
\\
\null & \null
-
\left(f_{0}^2 - |\vec{f}|^2\right)
\left(H_{00}-H_{11}-H_{22}-H_{33}\right)
\nonumber
\\
\null & \null
+|\vec{f}|^2
\left(H_{22}\sin^2 \theta + H_{33}\cos^2 \theta \right)
\nonumber
\\
\null & \null
-
2f_{0}|\vec{f}| \left( H_{20}\sin\theta +H_{30}\cos\theta \right)
\nonumber
\\
\null & \null
+
2 |\vec{f}|^2 H_{32} \cos\theta \sin\theta
\big]
,
\label{F107}
\end{align}
where the sum is performed over the Landau quantum number of the final
electron,
$f^{\mu} = (f_{0}, \vec{f}) = p_{\nu}^{\mu} + p_{\bar\nu}^{\mu} = p_{e}^{\mu} - p_{e}^{\prime\mu}$,
and
$\theta$ is the angle between $\vec{f}$ and $\vec{B}$.
The matrix elements
$H_{\alpha \beta}=j_\alpha j^\ast_\beta$ are determined by the
electron currents
\begin{align}
j_{\alpha}
=
\null & \null
\int dx \, dy \,
\overline{\psi'_e} \gamma_{\alpha} \left( g_{V} + g_{A} \gamma_5 \right) \psi_{e}
\nonumber
\\
\null & \null
\times
\exp\left\{-i\left[(\varkappa_{1} +\eta_{1})x+(\varkappa_{2} +\eta_{2})y\right]\right\}
,
\label{F108}
\end{align}
$\varkappa_{i}$ and $\eta_{i}$ are the corresponding neutrino and
antineutrino momenta components.
The functions $H_{\alpha \beta}$
are expressed in terms of quadratic combinations of Laguerre
functions which depend on the argument $\rho = |\vec{f}|^2 \sin^2 \theta /(2 \elechg B)$. In the case of the ultrarelativistic
electron energies the resulting expressions for the rate depend on
the electromagnetic field dynamical parameter
\begin{eqnarray}
\chi
=
\frac {e\sqrt{(F_{\mu \nu} p^\nu)^2}}{m^2_e}
=
\frac{B}{B_{0}^{e}} \frac{p_0}{m_{e}}
.
\end{eqnarray}
Integration in (\ref{F107}) can be performed analytically.
The final expressions for the
rate $\Gamma$ were obtained by \textcite{Ternov:1983hv,Ternov:1982qf}:
\begin{equation}\label{F109}
\Gamma=\frac {G_{\text{F}}^2 m^6_e \chi^{5}}{1152\sqrt{3} \pi^3 p_{0}}
\left[49g^2 + 437g^2_{A}\right],
\end{equation}
for $\chi\ll 1$ and
\begin{equation}\label{F110}
\Gamma
=
\frac{G_{\text{F}}^2 m^6_e \chi^{2}}{216 \pi^3 p_{0}}
\left(g^2_{V} + g^2_{A}\right)
\left[\ln\chi -C -\frac {\ln 3}{2} -\frac {5}{6}\right],
\end{equation}
for
$\chi\gg 1$,
where $C=0.577$ is the Euler constant.

From Eqs.~(\ref{F109}) and (\ref{F110}) one can see that rate is
governed by the value of the parameter $\chi$. It follows that the
rate is significantly dependent on the magnetic field strength and
the initial electron energy.
Therefore, for ultrarelativistic
energies and strong enough magnetic fields the $\nu {\bar {\nu}}$
synchrotron radiation by an electron can be important for
astrophysics.

As it has been demonstrated, for instance by
\textcite{Kaminker:1992su}, more consistent consideration of the
process $e\to e + \nu_e +\bar\nu_e$ appropriate for
astrophysical applications implies account for the presence of
background matter in addition to an external magnetic field.

\subsection{Neutrino pair production by a strong magnetic field}
\label{F111}

Over the years, starting from the observation of
\textcite{Klein:1929zz}, it has been known that the vacuum is not
stable under the influence of an external electric field.
\textcite{Schwinger:1951nm} has shown that
electron-positron pairs can be produced from the vacuum in the presence of
a strong electric field, with a strength that exceeds the critical
value $E_{\text{cr}}={m_{e}^2}/{e}$. It is also known that under the influence of
a homogeneous magnetic field the vacuum is stable,
because such a field does not produce work.
On the contrary, the presence of a strong inhomogeneous
magnetic field can produce an instability of the vacuum with respect
to neutral fermion-antifermion pair creation
if the fermion has a magnetic moment.

The interest in neutral particle-antiparticle pair creation from
the vacuum through the Pauli interaction
of a magnetic moment with external electromagnetic fields was raised by
\textcite{Lin:1999bb,Lee:2005aj,Lee:2006nc}.
In more recent papers
\textcite{Lee:2008zzy,Lee:2011rk}
discussed the vacuum instability in a strong magnetic field
due to neutrino-antineutrino pair production through the Pauli interaction.
However,
their results are questionable,
because they admit the creation of neutrino-antineutrino
pairs from the vacuum in a homogeneous magnetic field.

In a recent paper \textcite{Gavrilov:2012aw}
presented a nonperturbative calculation of neutrino-antineutrino pair
creation in a strong inhomogeneous magnetic field
in the framework of quantum field theory.
In particular,
they have shown that in specific cases (appropriate to
typical astrophysical applications) the problem can be technically
reduced to the problem of charged-particle creation by an
electric field.

Considering a generic neutrino $\nu$ with mass $m_{\nu}$
and magnetic moment $\mgm_{\nu}$,
the neutrino states in a magnetic field are
described by the Dirac-Pauli equation
\begin{equation}
\left\{
i\partial_\mu \gamma^\mu
-
m_{\nu}
+
\frac{\mgm_{\nu}}{2} \sigma_{\alpha\beta} F^{\alpha\beta}
\right\}
\Psi_{\nu}(x)
=0,
\end{equation}
where
$\Psi_{\nu}(x)$
is the neutrino wave function
and
$F^{\alpha\beta}$ is the electromagnetic field tensor.
\textcite{Gavrilov:2012aw}
have shown that
the energy spectrum of a neutrino that
interacts with an inhomogeneous magnetic field through
a magnetic moment consists of two branches separated by a gap.
Considering a magnetic field which is linearly growing on a given spatial interval $L$,
they demonstrated that the rate of pair creation
is determined by the gradient of the magnetic field.

A first condition
for neutrino-antineutrino pair production in a magnetic field $B$ is that
the magnetic energy must be enough to create a neutrino-antineutrino pair,
i.e.
$\mgm_{\nu} B > 2 m_{\nu}$.
Therefore,
the minimum value of the magnetic field for which neutrino-antineutrino pairs
are created is
\begin{equation}
B_{\text{cr}}
=
2
\,
\frac{ m_{\nu} }{ \mgm_{\nu} }
\simeq
3.4 \times 10^{8}
\left( \frac{m_{\nu}}{\text{eV}} \right)
\left( \frac{\bmag}{\mgm_{\nu}} \right)
\text{G}
.
\label{F112}
\end{equation}
Magnetic fields generated during a supernova explosion or in the vicinity of magnetars
can be of
the order of $10^{15} - 10^{16} \, \text{G}$ or even stronger,
up to about $10^{18} \, \text{G}$.
In this extreme case,
neutrino-antineutrino pair production
can occur for
${\mgm_{\nu}} \sim 10^{-12} \, \bmag$
and
$m_{\nu} \lesssim 10^{-2} \, \text{eV}$.
However,
it is also necessary to have a large gradient $B'$ of the magnetic field.
Considering a magnetic field which is linearly growing in a spatial interval $L$,
\textcite{Gavrilov:2012aw}
obtained the condition
$|\mu_{\nu} B'| \gtrsim m_{\nu}^2$,
which can be written as
$
|B'|
\gtrsim
m_{\nu}
B_{\text{cr}}
$.
Then,
for the maximum value $B_{\text{max}}$
of the magnetic field in the spatial interval $L$
we have the condition
\begin{equation}
|B_{\text{max}}|
\gtrsim
L
m_{\nu}
B_{\text{cr}}
.
\label{F113}
\end{equation}
Hence,
if the magnetic field is larger than $B_{\text{cr}}$
as required by the first condition above,
neutrino-antineutrino pair production can occur
if the size $L$ over which the magnetic field raises to such large values
is small enough:
\begin{align}
L
\lesssim
\null & \null
10^{-10}
\left( \frac{|B_{\text{max}}|}{B_{\text{cr}}} \right)
\left( \frac{\text{eV}}{m_{\nu}} \right)
\text{km}
\nonumber
\\
\sim
\null & \null
10^{-18}
\left( \frac{|B_{\text{max}}|}{\text{G}} \right)
\left( \frac{\text{eV}}{m_{\nu}} \right)^2
\left( \frac{\mgm_{\nu}}{\bmag} \right)
\text{km}
.
\label{F114}
\end{align}
Even considering the large values
$|B_{\text{max}}| \sim 10^{18} \, \text{G}$
and
${\mgm_{\nu}}\sim 10^{-12} \, \bmag$,
we need
$m_{\nu} \lesssim 10^{-6} \, \text{eV}$
in order to obtain
a distance of the order of a kilometer,
which may be appropriate for the spatial size of the magnetic field variations
in a supernova explosion or in the vicinity of magnetars.
Figure~\ref{B072}
shows that neutrino oscillation data allow one of the massive neutrinos
to be very light and even massless.
Hence,
there can be pair production of the lightest neutrino
in extreme astrophysical environments
if its mass is very small
and its magnetic moment is very large.
This is a condition which is contrary to the usual proportionality
between
the neutrino mass and the neutrino magnetic moment
and requires the intervention of powerful new physics
beyond the Standard Model,
as explained in Subsection~\ref{D068}.

\subsection{Energy quantization in rotating media}
\label{F115}

In Subsection~\ref{G012} we will discuss the possibility of nonzero neutrino
electric charge, that is predicted in a set of Standard Model
extensions.
If a neutrino is really a millicharged particle, in
the presence of a constant magnetic field it behaves in a way similar to an electron.
In particular, the energy of a
millicharged neutrino is quantized in a magnetic field (see Appendix~\ref{P001})
\begin{equation}\label{F116}
p^{\nu}_0=\sqrt{m_{\nu}^2 + p^2_3 + 2\chg_{\nu}BN_{\nu}},
\end{equation}
where $\chg_{\nu}$ is millicharge of the neutrino and $N_{\nu}=0,
1,2,\ldots $ is the Landau number of the millicharged neutrino
energy levels.
The corresponding radius of the neutrino classical
orbits in the magnetic field is given by \textcite{Balantsev:2010zw}
\begin{eqnarray}
\langle R^{\nu}_{B}\rangle=\sqrt{\frac{2N_{\nu}}{\chg_{\nu}B}}.
\end{eqnarray}
It is interesting to compare the radius of classical orbits in a
magnetic field of the millicharged neutrino, $\langle
R^{\nu}_{B}\rangle $, with that of the electron, $\langle
R^{e}_{B}\rangle $. If the relativistic electron and millicharged
neutrino are moving with the same energy in a constant magnetic
field then the ratio of orbits radiuses is equal to the inverse
ratio of electric charges
\begin{eqnarray}
\frac{\langle R^{\nu}_{B}\rangle }{\langle R^{e}_{B}\rangle
}=\frac{e}{\chg_{\nu}},
\end{eqnarray}
if for both particles the momentum components along the magnetic
field vector are zero.
From the obtained estimation for the ratio
of orbits radiuses, taking into account existing experimental
constraints on neutrino millicharge, we conclude that for the same
strength of the external magnetic field the motion of a charged
neutrino is much less localized as compared with an electron
motion.

The same method of wave equations exact solutions that is used in
studies of charged particles under the influence of external
electromagnetic fields (including millicharged neutrinos and
neutrinos with nonzero magnetic moment, see above discussions of
this Section and Appendix~\ref{P001}), as it has been
explicitly demonstrated by \textcite{Studenikin:2004dx,Studenikin:2008qk}, can be also used for investigations of
neutrinos moving in the background matter.
In particular, using
the method of exact solutions for a neutrino wave function in the
presence of matter it has been shown by \textcite{Grigoriev:2007zzc,Studenikin:2008qk} that the energy spectrum of a neutrino
moving in a rotating media is quantized.
This effect is very
similar to charged particles energy quantization in a magnetic
field.

The neutrino wave function exactly accounting for the neutrino
interaction with matter can be obtained by solving the modified
Dirac equation given by \textcite{Studenikin:2004dx} (see Appendix~\ref{P001}),
\begin{equation}\label{F117} \Big\{
i\gamma_{\mu}\partial^{\mu}-\frac{1}{2}
\gamma_{\mu}(1+\gamma_{5})f^{\mu}-m_{\nu} \Big\}\Psi(x)=0.
\end{equation}
In case an electron neutrino is propagating through a rotating
matter composed of neutrons then the matter potential, according
to \textcite{Balantsev:2009fd,Balantsev:2010zw}, is
\begin{equation}\label{F118}
f^{\mu} = -{G}(n,n \vec{v}), \ \ \vec{v}=(-\omega y,\omega x,0),
\end{equation}
where $\omega$ is the angular frequency of matter rotation around
the $z$ axis and $G=G_{\text{F}}/\sqrt{2}$. The neutrino energy spectrum
obtained by \textcite{Balantsev:2009fd,Balantsev:2010zw},
\begin{equation}\label{F119}
p_0=\sqrt{m_{\nu}^2+p_3^2+p_{\bot}^2}-Gn,
\end{equation}
contains the transverse momentum
\begin{equation}
p_{\bot} = 2\sqrt{NGn\omega}, \quad N = 0, 1, 2, \ldots
\end{equation}
that is quantized (see also \textcite{Grigoriev:2007zzc}). The quantum
number $N$ determines also the radius of classical orbits of
neutrino in rotating matter (it is supposed that $N\gg 1$ and
$p_3=0$),
\begin{equation}
R=\sqrt{\frac{N}{G n \omega}}.
\end{equation}
It is shown by \textcite{Studenikin:2008qk} that for low-energy
neutrinos it can be $R\sim R_{NS}=10 \ km$ that might be thought
to be of interest in applications for neutron stars.

It is interesting to note that within the quasiclassical approach
the neutrino binding on circular orbits is due to an effective
force that is orthogonal to the particle speed.
And an analogy
between a charged particle motion in a magnetic field and a
neutrino motion in a rotating matter can be established
(\textcite{Studenikin:2008qk}). It is possible to explain the
neutrino quasiclassical circular orbits as a result of action of
the attractive central force,
\begin{equation}
\vec{F}_m^{(\nu)}=q^{(\nu)}_m \vec{\beta} \times \vec{B}_m, \ \vec{B}_m= \vec{\nabla} \times \vec{A}_m, \ \vec{A}_m=n \vec{v},
\end{equation}
where the neutrino effective ``charge'' in matter (composed of
neutrons in the discussed case) is $q^{(\nu)}_m=-G$, whereas $\vec{B}_m$ and $\vec{A}_m$ play the roles of effective ``magnetic''
field and the correspondent ``vector potential''. Like the magnetic
part of the Lorentz force, $\vec{F}_m^{(\nu)}$ is orthogonal to
the speed $\vec{\beta}$ of the neutrino.

For the most general case the ``matter induced Lorentz force'' is given by
\begin{equation}\label{F120}
\vec{F}_m^{(\nu)}=q^{(\nu)}_m \vec{E}_m + q^{(\nu)}_m \vec{\beta}
\times \vec{B}_m,
\end{equation}
where the effective ``electric'' and ``magnetic'' fields are
respectively,
\begin{equation}\label{F121}
\vec{E}_m=-\vec{\nabla}n -\vec{v}\frac{\partial n}{\partial t} -
n\frac{\partial \vec{v}}{\partial t} ,
\end{equation}
and
\begin{equation}\label{F122}
\vec{B}_m=n \vec{\nabla} \times \vec{v}-\vec{v} \times {\vec
\nabla}n.
\end{equation}
The force acting on a neutrino, produced
by the first term of the effective ``electric'' field in the
neutron matter, was considered also by \textcite{Loeb:1989nb} and
the quasiclassical treatment of a neutrino motion in the electron
plasma was considered by \textcite{Mendonca:1998he}.

Note that while considering a neutrino effective electromagnetic
interactions with media an effective electric charge of the
neutrino has been introduced by
\textcite{Oraevsky:1986dt,Oraevsky:1987cu,Oraevsky:1994wb,Nieves:1993er,Mendonca:1998he,Bhattacharya:2001nm,Nieves:2003kw,Studenikin:2008qk}.

In the most general case the description of the millicharged
neutrino with anomalous magnetic moment motion in the presence of matter and
external electromagnetic fields can be obtained by solving the modified Dirac
equation
\begin{align}
\Big\{\gamma_{\mu}(p^{\mu}+\chg_{\nu}A^{\mu})
\null & \null
-
\frac{1}{2}
\gamma_{\mu}(1+\gamma_5)f^{\mu}
- \frac{i}{2} \mgm_{\nu} \sigma_{\mu\nu} F^{\mu\nu}
\nonumber
\\
\null & \null
- m_{\nu}
\Big\}
\Psi(x)
=
0,
\label{F123}
\end{align}
where $F^{\mu\nu}=\partial^{\mu}A^{\nu}-\partial^{\nu}A^{\mu}$ and $A^{\mu}$ is the
electromagnetic field potential, $\mgm_{\nu}$ is the neutrino anomalous magnetic
moment.
For several particular cases this equation can be solved exactly and the neutrino
wave functions and the corresponding energy spectra can be found
\cite{Grigoriev:2007zzc,Balantsev:2012ep,Balantsev:2013aya,Studenikin:2012vi}.
In particular, for a neutrino moving in a rotating matter with the potential
\begin{equation}
\label{F124}
f^{\mu}=-GN_{n}(1, -\epsilon y\omega, \epsilon x\omega, 0).
\end{equation}
and superimposed constant electric $\vec{E}$ and magnetic field $\vec{B}$, $\epsilon=\pm1$ corresponds to parallel and antiparallel directions of vectors $\vec{{\omega}}$ and $\vec{B}$,
for the neutrino energy spectrum we obtain
\begin{equation}
\label{F125} p_0=\sqrt{p_3^2+2N|2GN_{n}\omega-\epsilon
\chg_{\nu}B|+m_{\nu}^2}-GN_{n}-\chg_{\nu}\phi,
\end{equation}
where $\phi$ is the scalar potential of the electric field.
In this case the generalized effective Lorentz force introduced in
\textcite{Studenikin:2008qk} is
\begin{equation}
\label{F126}
\vec{F}_{eff}=\chg_{eff}\vec{E}_{eff}+\chg_{eff}
\left[\vec{\beta}\times\vec{B}_{eff}\right].
\end{equation}
Here $\vec{\beta}$ is the neutrino speed and
\begin{equation}
\label{F127}
\begin{aligned}
\chg_{eff}\vec{E}_{eff}&=\chg_m\vec{E}_m+\chg_{\nu}\vec{E},\\
\chg_{eff}\vec{B}_{eff}&=|\chg_mB_m+\epsilon \chg_{\nu} B|\vec{e}_z,\\
\end{aligned}
\end{equation}
where $\chg_m, \vec{B}_m, \vec{E}_m$ are the matter induced ``charge'',
``electric'' and ``magnetic'' fields correspondingly,
\begin{equation}
\label{F128}
\chg_m=-G, \quad
\vec{E}_m=-\vec{\nabla}N_{n}, \quad
\vec{B}_m=-2N_{n}\vec{{\omega}}.
\end{equation}
Note that the effective Lorentz force
(\ref{F126}), that directly follows from the exact form of the obtained
energy spectrum (\ref{F125}), is generated by both weak and electromagnetic
interactions.
The effect of the millicharged neutrino energy quantization
in a rotating magnetized matter was discussed in \textcite{Grigoriev:2007zzc,Studenikin:2008qk},
where is shown that the neutrino trapping in circular orbits exist due to
the neutrino millicharge interaction with the magnetic field and also due
to neutrino weak interaction with the rotating matter.

Under the influence of the effective
Lorentz force (\ref{F127}) the neutrino will move with acceleration given by \cite{Studenikin:2008qk}
\begin{equation}
\label{F129}
\vec{a}
=
\frac{1}{m_{\nu}}
\left(
G\vec{\nabla}N_{n}+\chg_{\nu}\vec{\nabla}\phi
+
\left| 2GN_{n}\omega-\epsilon\chg_{\nu}B \right|
\vec{\beta}\times\vec{e}_z
\right),
\end{equation}
where $\vec{e}_z$ is a unit vector in the direction of the
magnetic field and matter rotation.
The accelerated neutrino should produce
the electromagnetic radiation.
In the quasiclassical treatment the radiation
power of induced electromagnetic radiation is given by
\begin{equation}
\label{F130}
I_{LC\nu}=\frac{2\chg_{\nu}^2}{3}
\left(\frac{\vec{a}^2}{(1-|\vec{\beta}|^2)^2}+
\frac{(\vec{a}\cdot\vec{\beta})^2}{(1-|\vec{\beta}|^2)^3}\right).
\end{equation}
Such a mechanism of the neutrino electromagnetic
radiation due to the neutrino millicharge, that can be emitted in the
presence of the nonuniform rotating matter and electromagnetic fields,
is termed in \textcite{Studenikin:2012vi} the ``Light of (milli)Charged Neutrino'' ($LC\nu$).
It should be stressed, that the phenomenon exist even in the absence of the
electromagnetic fields, when the acceleration~(\ref{F129}) is
produced only due to the weak interactions of neutrinos with the
background particles.
So that the discussed mechanism is of a
different nature than that of the cyclotron radiation of a charged
particle in magnetic fields.

The $LC\nu$ mechanism manifests itself during the neutrino propagation from the central part of a rotating neutron star outwards through the crust.
The gradient of the matter density (the density variation along the neutrino path) gives the following contribution to the $LC\nu$ radiation power (see Eq.~(\ref{F129}))
\begin{equation}
\label{F131}
I_{LC\nu}
=
\frac{2\chg_{\nu}^2}{3m_{\nu}^2}
\left( G \vec{\nabla} N_{n} \right)^2
,
\end{equation}
and the effect of the matter rotation yields
\begin{equation}
\label{F132}
I_{LC\nu}
=
\frac{2\chg_{\nu}^2\gamma^2}{3m_{\nu}^2}
\left(
- \epsilon \chg_{\nu} B
+
2 G N_{n} \omega
\right)^2
,
\end{equation}
where $\gamma=(1-|\vec{\beta}|^2)^{-1/2}$.
The numerical estimations, that account for the $LC\nu$ power for the present limits on the neutrino millicharge and for a realistic gradient of a neutron star matter density $|G\vec{\nabla} N_{n}|\sim1 \text{eV}/1 \text{km}$ and the rotation frequency $\omega\sim2\pi\times10^3$ s$^{-1}$, show that the role of the $LC\nu$ in the explosion energetics is negligible with respect to the total energy of the collapse.
However, as discussed in Section~\ref{G001} \cite{Oraevsky:1994wb,Nieves:2003kw,Duan:2004nc},
in the presence of a dense plasma the induced neutrino effective electric charge can be reasonably large.
In addition, the phenomenon is of interest for astrophysics in light of the recently reported
measurement of ultra-high energy PeV neutrinos in the IceCube experiment \cite{Aartsen:2013bka,Aartsen:2013jdh,Aartsen:2014gkd}.

\section{Charge and anapole form factors}
\label{G001}

The magnetic and electric dipole moments are the most studied electromagnetic properties
in theoretical and experimental works,
but some attention has also been devoted to the possibility that
neutrinos have very small electric charges, usually called ``millicharges''.
Moreover,
even if neutrinos are exactly neutral,
they can have nonzero charge radii,
which can be probed in scattering experiments.
In Subsections~\ref{G012} and \ref{G043}
we review the theory of
electric charge and charge radius, respectively,
and we present the corresponding experimental limits.
In Subsection~\ref{G060}
we discuss the neutrino anapole moment,
which is the less known neutrino electromagnetic property.

\begin{figure}
\centering
\includegraphics*[bb=242 620 350 720, width=0.4\linewidth]{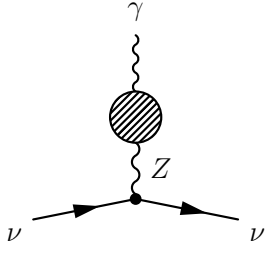}
\caption{\label{G002}
Contribution to the neutrino vertex function of $\gamma-Z$ self-energy.
Figure~\ref{G011} shows the diagrams contributing to the blob
at one loop
in the extended Standard Model with
right-handed neutrinos.}
\end{figure}

\begin{figure}
\begin{center}
\begin{tabular}{cc}
\subfigure[]
{\label{G003}
\includegraphics*[bb=240 644 353 710, width=0.3\linewidth]{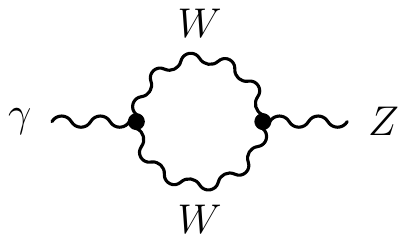}
}
&
\subfigure[]
{\label{G004}
\includegraphics*[bb=240 645 353 710, width=0.3\linewidth]{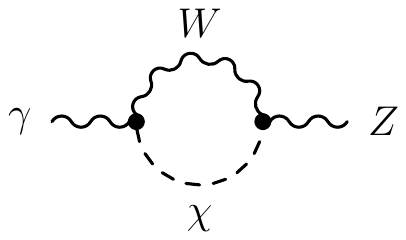}
}
\\
\subfigure[]
{\label{G005}
\includegraphics*[bb=240 673 353 720, width=0.3\linewidth]{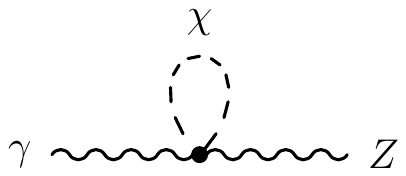}
}
&
\subfigure[]
{\label{G006}
\includegraphics*[bb=240 673 353 720, width=0.3\linewidth]{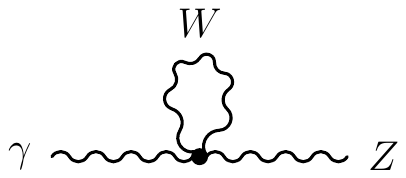}
}
\\
\subfigure[]
{\label{G007}
\includegraphics*[bb=240 644 353 711, width=0.3\linewidth]{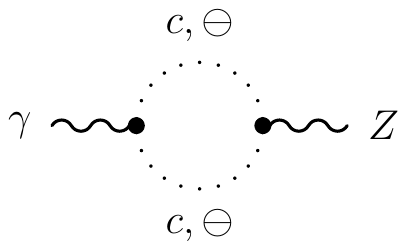}
}
&
\subfigure[]
{\label{G008}
\includegraphics*[bb=240 644 353 711, width=0.3\linewidth]{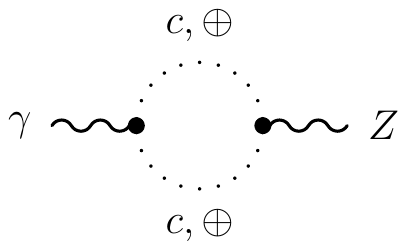}
}
\\
\subfigure[]
{\label{G009}
\includegraphics*[bb=240 645 353 710, width=0.3\linewidth]{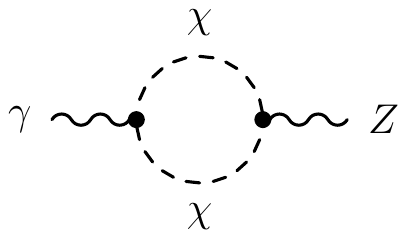}
}
&
\subfigure[]
{\label{G010}
\includegraphics*[bb=240 642 353 713, width=0.3\linewidth]{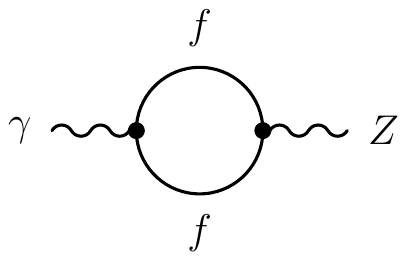}
}
\end{tabular}
\caption{ \label{G011}
$\gamma-Z$ self-energy diagrams contributing to the neutrino vertex function
at one loop
in the extended Standard Model with
right-handed neutrinos
\cite{Dvornikov:2003js,Dvornikov:2004sj}.
$f$ denotes a generic charged lepton
($e$, $\mu$, $\tau$)
or a quark
($u$, $c$, $t$, $d$, $s$, $b$).
$\chi$ is the unphysical would-be charged scalar boson.
The charge of ghosts $c$ is indicated by the symbols $\oplus$ and $\ominus$.
}
\end{center}
\end{figure}

\subsection{Neutrino electric charge}
\label{G012}

It is usually believed (see \textcite{Bernstein:1963qh}) that the neutrino
electric charge is exactly zero.
This is true in the Standard Model,
but in extensions of the Standard Model
neutrinos may be millicharged particles.

In the Standard Model of $\text{SU}(2)_L \times \text{U}(1)_Y$ electroweak interactions
the neutrality of neutrinos is a consequence of the quantization of electric charge
\cite{Geng:1988pr,Minahan:1989vd,Foot:1989fh,Babu:1989tq,Babu:1989ex}
(see also the earlier discussions in \textcite{Gross:1972pv,Bardeen:1972vi,Lee:1977tib}
and the reviews in \textcite{Foot:1990uf,Foot:1992ui}.
In the Standard Model
the electric charges of the particles are related to the
eigenvalue of the third component $I_{3}$ of the weak isospin $I$
and to the eigenvalue $Y$ of the hypercharge
by
\begin{equation}
Q = I_{3} + \frac{Y}{2}.
\label{G013}
\end{equation}
The hypercharges of the fermion multiplets are fixed by the requirement of cancellation of the triangle anomalies,
which is necessary for renormalizability.
For each generation,
let us denote with
$Y_{\Phi}$,
$Y_{L}$,
$Y_{Q}$,
$Y_{e}$,
$Y_{u}$,
$Y_{d}$
the hypercharges of
the Higgs doublet,
the left-handed lepton doublet,
the left-handed quark doublet,
the right-handed electron singlet,
the right-handed up-quark singlet,
the right-handed down-quark singlet,
respectively.
The electric charge can be defined in units of the charge of the Higgs field $\phi^{+}$
(see Table~\ref{B012})
by choosing $Y_{\Phi}=+1$.
Then,
the $\text{U}(1)_Y$ gauge invariance of the Yukawa couplings that generate the charged leptons and quarks masses
requires that
\begin{align}
\null & \null
Y_{e} = Y_{L} - 1
,
\label{G014}
\\
\null & \null
Y_{u} = Y_{Q} + 1
,
\label{G015}
\\
\null & \null
Y_{d} = Y_{Q} - 1
.
\label{G016}
\end{align}
Taking into account that quarks have three colors,
the values of $Y_{L}$ and $Y_{Q}$ are constrained by the cancellation of the
$\text{SU}(2)_L$
triangle anomaly by
\begin{equation}
Y_{Q} = - Y_{L} / 3
.
\label{G017}
\end{equation}
Finally,
the cancellation of the
$\text{U}(1)_Y$
triangle anomaly requires that
\begin{equation}
0
=
\text{Tr}\!\left[Y^3\right]
=
2 Y_{L}^3 + 6 Y_{Q}^3 - Y_{e}^3 - 3 \left( Y_{u}^3 + Y_{d}^3 \right)
,
\label{G018}
\end{equation}
where the right-handed fields enter with a minus sign.
Using Eqs.~(\ref{G014})--(\ref{G017}) in Eq.~(\ref{G018}),
we obtain
\begin{equation}
0
=
\text{Tr}\!\left[Y^3\right]
=
\left( Y_{L} + 1 \right)^3
\quad
\Longrightarrow
\quad
Y_{L} = -1
.
\label{G019}
\end{equation}
Therefore,
charge is quantized and from Eq.~(\ref{G013}) neutrinos are exactly neutral
(see also the explicit calculations in
\textcite{Bardeen:1972vi,Beg:1977xz,Marciano:1980pb,Sakakibara:1980hw,Lucio:1983mg,Lucio:1984jn,CabralRosetti:1999ad}).

This beautiful proof of charge quantization
is spoiled by the introduction of a right-handed $\text{SU}(2)_L$ singlet neutrino $\nu_{R}$
in order to have a Dirac neutrino mass.
Denoting with
$Y_{\nu}$
the hypercharge of $\nu_{R}$,
the $\text{U}(1)_Y$ gauge invariance of the Yukawa coupling that generates a Dirac neutrino mass
requires that
\begin{equation}
Y_{\nu} = Y_{L} + 1
.
\label{G020}
\end{equation}
Then, Eq.~(\ref{G019})
becomes
\begin{equation}
\text{Tr}\!\left[Y^3\right]
=
\left( Y_{L} + 1 \right)^3
-
\left( Y_{L} + 1 \right)^3
=
0
.
\label{G021}
\end{equation}
Therefore,
there is no $\text{U}(1)_Y$ triangle anomaly
for any value of $Y_{L}$,a right-handed $\text{SU}(2)_L$ singlet neutrino $\nu_{R}$
which remains unconstrained.
With the definition
$Y_{\nu} = 2 \varepsilon$,
using the relations in Eqs.~(\ref{G013})--(\ref{G017}) and (\ref{G020}),
we obtain
\begin{align}
Q_{\nu}
=
\null & \null
\varepsilon
,
\label{G022}
\\
Q_{e}
=
\null & \null
- 1 + \varepsilon
,
\label{G023}
\\
Q_{u}
=
\null & \null
2/3 - \varepsilon/3
,
\label{G024}
\\
Q_{d}
=
\null & \null
- 1/3 - \varepsilon/3
.
\label{G025}
\end{align}
For the proton and the neutron we have
\begin{equation}
Q_{p} = 1 - \varepsilon
,
\quad
Q_{n} = - \varepsilon
.
\label{G026}
\end{equation}
Hence,
the hydrogen atom is neutral,
but all the atoms with neutrons are not.
Obviously,
the limits on the non-neutrality of matter
\cite{Marinelli:1983nd,Bressi:2011pj}
imply that the value of $\varepsilon$ must be very small.
In this case,
neutrinos may be electrically millicharged particles
\cite{Minahan:1989vd,Foot:1989fh,Babu:1989tq,Babu:1989ex}
(see also the discussions in \textcite{Okun:1983vw,Shrock:1995bp}).

From Eqs.~(\ref{G022})--(\ref{G025})
one can see that the nonstandard hypercharge proportional to $\varepsilon$
is proportional to $\text{B}-\text{L}$,
where B and L are the baryon and lepton numbers.
With the introduction of the right-handed neutrino $\nu_{R}$
the $\text{U}(1)_{\text{B}-\text{L}}$
symmetry of the Standard Model becomes anomaly-free.
Adopting a notation similar to that used for the hypercharges,
in the Standard Model the $\text{U}(1)_{\text{B}-\text{L}}$ triangle anomaly is proportional to
\begin{align}
\text{Tr}\!\left[(\text{B}-\text{L})^3\right]
=
\null & \null
2 (\text{B}-\text{L})_{L}^3 + 6 (\text{B}-\text{L})_{Q}^3 - (\text{B}-\text{L})_{e}^3
\nonumber
\\
\null & \null
- 3 \left( (\text{B}-\text{L})_{u}^3 + (\text{B}-\text{L})_{d}^3 \right)
\nonumber
\\
=
\null & \null
-1
.
\label{G027}
\end{align}
Hence,
in the Standard Model
the $\text{U}(1)_{\text{B}-\text{L}}$
symmetry is not anomaly-free and cannot be gauged.
On the other hand,
with the introduction of $\nu_{R}$
which has
$(\text{B}-\text{L})_{\nu} = -1$
we obtain
$\text{Tr}\!\left[(\text{B}-\text{L})^3\right] = 0$.
In this case the $\text{U}(1)_{\text{B}-\text{L}}$
symmetry is anomaly-free and can be gauged.
Then,
there can be a mixing of the Standard Model hypercharge $Y_{\text{SM}}$
and $\text{B}-\text{L}$, which leads to the hypercharge
\begin{equation}
Y
=
Y_{\text{SM}} - 2 \varepsilon \left( \text{B}-\text{L} \right)
,
\label{G028}
\end{equation}
and the dequantized electric charges in Eqs.~(\ref{G022})--(\ref{G025}).
Hence,
the dequantization of the electric charge
is due to the appearance of an anomaly-free
$U(1)$ symmetry
which can be gauged and can mix with the standard hypercharge
\cite{Foot:1989fh,Babu:1989tq,Babu:1989ex}.
The addition of an anomaly-free $U(1)$ symmetry to the symmetries of the Lagrangian
is a general way to obtain charge dequantization
\cite{Holdom:1985ag}.

A well-known way to recover electric charge quantization in theories with
right-handed $\text{SU}(2)_L$ singlet neutrinos
is to consider grand unified theories (GUT)
in which there is no $U(1)$ symmetry
\cite{Pati:1974yy,Georgi:1974sy}.
However,
there is also the natural possibility to allow the right-handed neutrino to have a Majorana mass
\cite{Babu:1989tq,Babu:1989ex}.
In this case the gauge invariance of the Majorana mass term
$\nu_{R}^{T} \mathcal{C}^{\dagger} \nu_{R}$
requires that
$Y_{\nu}=0$
and,
from Eq.~(\ref{G020}),
$Y_{L}=-1$,
which gives
the same charge quantization as in the Standard Model.
This is consistent with the violation of the
$\text{U}(1)_{\text{B}-\text{L}}$
symmetry by the Majorana mass term,
which forbids the addition of the $\text{B}-\text{L}$ term to $Y_{\text{SM}}$ in Eq.~(\ref{G028}).

Until now in this Subsection we have considered only one generations,
but we know that there are three generations
and the Standard Model Lagrangian has four global $\text{U}(1)$ symmetries:
$\text{U}(1)_{\text{B}}$,
$\text{U}(1)_{\text{L}_{e}}$,
$\text{U}(1)_{\text{L}_{\mu}}$ and
$\text{U}(1)_{\text{L}_{\tau}}$,
associated with the conservation of
the baryon number B,
the electron lepton number $\text{L}_{e}$,
the muon lepton number $\text{L}_{\mu}$ and
the tau lepton number $\text{L}_{\tau}$.
It turns out that there is an infinite number of linear combinations of these $\text{U}(1)$ symmetries
which are anomaly-free
and lead to electric charge dequantization in the Standard Model with three generations
\cite{Foot:1990mn,Foot:1990uf,Foot:1992ui}.
Charge quantization can be recovered by introducing right-handed neutrinos
with Majorana mass terms which violate the conservation of all lepton numbers
\cite{Foot:1990mn,Foot:1990uf,Foot:1992ui,Sladkowski:1991gf}.

Some approximate constraints obtained with various assumptions
from reactor, accelerator and astrophysical data
are listed in Tab.~\ref{G029}
(see also
\textcite{Babu:1992sw,Raffelt:1996wa,Davidson:2000hf,PDG-2012}).

\begin{table*}
\renewcommand{\arraystretch}{1.2}
\begin{tabular}{lll}
Limit & Method & Reference\\
\hline
$|\chg_{\nu_{\tau}}| \lesssim 3 \times 10^{-4}\,\elechg$		&SLAC $e^{-}$ beam dump				&\textcite{Davidson:1991si}	\\
$|\chg_{\nu_{\tau}}| \lesssim 4 \times 10^{-4}\,\elechg$		&BEBC beam dump					&\textcite{Babu:1993yh}		\\
$|\chg_{\nu}| \lesssim 6 \times 10^{-14}\,\elechg$		&Solar cooling (plasmon decay)			&\textcite{Raffelt:1999gv}	\\
$|\chg_{\nu}| \lesssim 2 \times 10^{-14}\,\elechg$		&Red giant cooling (plasmon decay)		&\textcite{Raffelt:1999gv}	\\
$|\chg_{\nu_e}| \lesssim 3 \times 10^{-21}\,\elechg$	&Neutrality of matter				&\textcite{Raffelt:1999gv}	\\
$|\chg_{\nu_e}| \lesssim 3.7 \times 10^{-12}\,\elechg$	&Nuclear reactor				&\textcite{Gninenko:2006fi}	\\
$|\chg_{\nu_e}| \lesssim 1.5 \times 10^{-12}\,\elechg$	&Nuclear reactor				&\textcite{Studenikin:2013my}	\\
\hline
\end{tabular}
\caption{\label{G029}
Approximate limits for different neutrino effective charges.
The limits on $\chg_{\nu}$ apply to all flavors.
}
\end{table*}

The most severe experimental constraint on neutrino electric charges
is that on the effective electron neutrino charge
$\chg_{\nu_{e}}$,
which can be obtained from electric charge conservation in neutron beta decay
$n \to p + e^- + \bar\nu_{e}$,
from the experimental limits on the non-neutrality of matter
which constrain the sum of the proton and electron charges,
$\chg_{p} + \chg_{e}$,
and from the experimental limits on the neutron charge
$\chg_{n}$
\cite{Raffelt:1996wa,Raffelt:1999gv}.
Several experiments which measured the neutrality of matter give their results in terms of
\begin{equation}
\chg_{\text{mat}}
=
\frac{Z (\chg_{p} + \chg_{e}) + N \chg_{n}}{A}
,
\label{G030}
\end{equation}
where $A=Z+N$ is the atomic mass of the substance under study,
$Z$ is its atomic number and $N$ is its neutron number.
From electric charge conservation in neutron beta decay,
we have
\begin{equation}
\chg_{\nu_{e}}
=
\chg_{n} - (\chg_{p} + \chg_{e})
=
\frac{A}{Z} \left( \chg_{n} - \chg_{\text{mat}} \right)
.
\label{G031}
\end{equation}
The best recent bound on the non-neutrality of matter
\cite{Bressi:2011pj},
\begin{equation}
\chg_{\text{mat}}
=
(- 0.1 \pm 1.1) \times 10^{-21} \, \elechg
,
\label{G032}
\end{equation}
has been obtained with
$\text{S}\text{F}_{6}$,
which has
$A=146.06$
and
$Z=70$.
Using the independent measurement of the charge of the free neutron
\cite{Baumann:1988ue}
\begin{equation}
\chg_{n} = (- 0.4 \pm 1.1) \times 10^{-21} \, \elechg
,
\label{G033}
\end{equation}
we obtain
\begin{equation}
\chg_{\nu_{e}} = (-0.6 \pm 3.2) \times 10^{-21} \, \elechg
.
\end{equation}
This value is compatible with the neutrality of matter limit in Tab.~\ref{G029},
which has been derived \cite{Raffelt:1996wa,Raffelt:1999gv}
from the value of $\chg_{n}$ in Eq.~(\ref{G033})
and
$
\chg_{\text{mat}}
=
(0.8 \pm 0.8) \times 10^{-21} \, \elechg
$
\cite{Marinelli:1983nd}.

It is also interesting that
the effective charge of $\bar\nu_{e}$
can be constrained by the SN 1987A neutrino measurements
taking into account that
galactic and extragalactic magnetic field can lengthen the path of millicharged neutrinos
and requiring that neutrinos with different energies arrive
on Earth within the observed time interval of a few seconds
\cite{Barbiellini:1987zz}:
\begin{equation}
|\chg_{\nu_{e}}|
\lesssim
3.8 \times 10^{-12}
\frac{(E_{\nu}/10\,\text{MeV})}{(d/10\,\text{kpc}) (B/1\,\mu\text{G})}
\sqrt{
\frac{\Delta{t}/t}{\Delta{E_{\nu}}/E_{\nu}}
}
,
\label{G034}
\end{equation}
considering a magnetic field $B$ acting over a distance $d$ and the corresponding time $t=d/c$.
$E_{\nu} \approx 15\,\text{MeV}$ is the average neutrino energy,
$\Delta{E_{\nu}} \approx E_{\nu} / 2$ is the energy spread,
and
$\Delta{t} \approx 5 \, \text{s}$ is the arrival time interval.
\textcite{Barbiellini:1987zz} considered 2 cases:

\begin{enumerate}

\item
An intergalactic field $B \approx 10^{-3} \, \mu\text{G}$
acting over the whole path $d \simeq 50\,\text{kpc}$,
which corresponds to
$t \simeq 5 \times 10^{12} \, \text{s}$,
gives
\begin{equation}
|\chg_{\nu_{e}}|
\lesssim
2 \times 10^{-15} \, \elechg
.
\label{G035}
\end{equation}

\item
An galactic field $B \approx 1 \, \mu\text{G}$
acting over a distance $d \simeq 10\,\text{kpc}$,
which corresponds to
$t \simeq 1 \times 10^{12} \, \text{s}$,
gives
\begin{equation}
|\chg_{\nu_{e}}|
\lesssim
2 \times 10^{-17} \, \elechg
.
\label{G036}
\end{equation}

\end{enumerate}

The last two limits in Tab.~\ref{G029}
have been obtained
\cite{Gninenko:2006fi,Studenikin:2013my}
considering the results of reactor neutrino magnetic moment experiments
(see Sections~\ref{D034} and \ref{D064}).
The differential cross section of the
$\bar\nu_{e}$--$e^{-}$
elastic-scattering process due to a neutrino effective charge $\chg_{\nu_{e}}$
is given by
(see \textcite{Berestetskii:1979aa})
\begin{equation}
\left(\frac{d\sigma}{dT_{e}}\right)_{\text{charge}}
\simeq
2\pi\alpha
\frac{1}{m_{e}T_{e}^2}\chg_{\nu_{e}}^2
.
\label{G037}
\end{equation}
In reactor experiments the neutrino magnetic moment is searched by considering data with
$T_{e} \ll E_{\nu}$,
for which the ratio of the charge cross section (\ref{G037}) and the magnetic moment cross section in Eq.~(\ref{D041}),
for which we consider only the dominant part proportional to $1/T_{e}$,
is given by
\begin{equation}
R
=
\frac{ \left(d\sigma/dT_{e}\right)_{\text{charge}} }{ \left(d\sigma/dT_{e}\right)_{\text{mag}} }
\simeq
\frac{2m_e}{T_{e}}
\frac{ \left( \chg_{\nu_{e}} / \elechg \right)^2 }{ \left( \mgm_{\nu_{e}} / \bmag \right)^2 }
\label{G038}
\end{equation}
Considering an experiment which does not observe any effect of $\mgm_{\nu_{e}}$
and obtains a limit on $\mgm_{\nu_{e}}$,
it is possible to obtain,
following \textcite{Studenikin:2013my},
a bound on $\chg_{\nu_{e}}$ by demanding that
the effect of $\chg_{\nu_{e}}$ is smaller than that of $\mgm_{\nu_{e}}$,
i.e that $R \lesssim 1$:
\begin{equation}
\chg_{\nu_{e}}^{2}
\lesssim
\frac{T_{e}}{2m_e}
\left(\frac{\mgm_{\nu_{e}}}{\bmag}\right)^{2} \elechg^2
.
\label{G039}
\end{equation}
The last limit in Tab.~\ref{G029}
has been obtained from the 2012 results \cite{Beda:2012zz} of the GEMMA experiment,
considering $T_{e}$ at the experimental threshold of $2.8 \, \text{keV}$.

Let us finally note that
a strong limit on a generic neutrino electric charge
$\chg_{\nu}$
can be obtained by considering the influence of millicharged neutrinos on the rotation a magnetized star which is undergoing a core-collapse supernova explosion
(the neutrino star turning mechanism, $\nu$ST)
\cite{Studenikin:2012vi}.
During the supernova explosion,
the escaping millicharged neutrinos move along curved orbits inside the rotating magnetized star
and slow down the rotation of the star.
This mechanism could prevent the generation of a rapidly rotating pulsar in the supernova explosion.
Imposing that the frequency shift of a forming pulsar due to the neutrino star turning mechanism is less than a typical observed frequency of $0.1 \, \text{s}^{-1}$
and assuming a magnetic field of the order of $10^{14} \, \text{Gauss}$,
\textcite{Studenikin:2012vi}
obtained
\begin{equation}\label{G040}
|\chg_{\nu}| \lesssim 1.3 \times 10^{-19} \, \elechg
.
\end{equation}
Note that this limit is much stronger than the astrophysical limits in Tab.~\ref{G029}.

\begin{table*}
\renewcommand{\arraystretch}{1.2}
\begin{tabular}{llcll}
Method & Experiment & Limit [$\text{cm}^2$] & CL & Reference\\
\hline
\multirow{2}{*}{Reactor $\bar\nu_e$-$e^-$}
&Krasnoyarsk	&$|\langle{r_{\nu_e}^{2}}\rangle|<7.3\times10^{-32}$			&90\%	&\textcite{Vidyakin:1992nf}		\\
&TEXONO		&$-4.2\times10^{-32}<\langle{r_{\nu_e}^{2}}\rangle<6.6\times10^{-32}$	&90\%	&\textcite{Deniz:2009mu}$^\text{\ref{G041}}$\\
\hline
\multirow{2}{*}{Accelerator $\nu_e$-$e^-$}
&LAMPF		&$-7.12\times10^{-32}<\langle{r_{\nu_e}^{2}}\rangle<10.88\times10^{-32}$	&90\%	&\textcite{Allen:1992qe}\footnote{
\label{G041}
The published limits are half, because they use a convention which differs by a factor of 2 (see also \textcite{Hirsch:2002uv}).}\\
&LSND		&$-5.94\times10^{-32}<\langle{r_{\nu_e}^{2}}\rangle<8.28\times10^{-32}$		&90\%	&\textcite{Auerbach:2001wg}$^\text{\ref{G041}}$\\
\hline
\multirow{2}{*}{Accelerator $\nu_{\mu}$-$e^-$}
&BNL-E734	&$-4.22\times10^{-32}<\langle{r_{\nu_{\mu}}^{2}}\rangle<0.48\times10^{-32}$	&90\%	&\textcite{Ahrens:1990fp}$^\text{\ref{G041}}$\\
&CHARM-II	&$|\langle{r_{\nu_{\mu}}^{2}}\rangle|<1.2\times10^{-32}$				&90\%	&\textcite{Vilain:1994hm}$^\text{\ref{G041}}$\\
\hline
\end{tabular}
\caption{\label{G042}
Experimental limits for the electron neutrino charge radius.
}
\end{table*}

\subsection{Neutrino charge radius}
\label{G043}

Even if the electric charge of a neutrino is zero, the electric form factor
$\nff_{Q}(q^{2})$ can contain nontrivial information about the neutrino electric properties.
In fact,
a neutral particle can be characterized by a (real or virtual) superposition of two different
charge distributions of opposite signs,
which is described by a form factor $\nff_{Q}(q^{2})$ which is nonzero for $q^{2}\neq0$.

The neutrino charge radius is determined by the second
term in the expansion of the neutrino charge form factor $\nff_{Q}(q^{2})$
in series of powers of $q^{2}$:
\begin{equation}
\nff_{Q}(q^{2})
=
\nff_{Q}(0)
+
q^{2}
\left.\frac{d\nff_{Q}(q^{2})}{dq^{2}}\right|_{q^{2}=0}
+
\ldots
.
\label{G044}
\end{equation}
In the so-called ``Breit frame'', in which $q_0=0$,
the charge form factor $\nff_{Q}(q^{2})$
depends only on
$|\vec{q}|=\sqrt{-q^2}$
and can be interpreted as the Fourier transform of a spherically symmetric charge distribution
$\rho(r)$,
with
$r=|\vec{x}|$:
\begin{equation}
\nff_{Q}(q^{2})
=
\int
\rho(r) e^{-i\vec{q}\cdot\vec{x}}
d^3x
=
\int
\rho(r)
\frac{\sin(|\vec{q}|r)}{|\vec{q}|r}
d^3x
.
\label{G045}
\end{equation}
Deriving with respect to $q^2=-|\vec{q}|^2$, we obtain
\begin{equation}
\frac{d\nff_{Q}(q^{2})}{dq^{2}}
=
\int \rho(r) \frac{\sin(|\vec{q}|r) - |\vec{q}|r\cos(|\vec{q}|r)}{2q^{3}r} d^3x
,
\label{G046}
\end{equation}
and
\begin{equation}
\lim_{q^{2}\to0}
\frac{d\nff_{Q}(q^{2})}{dq^{2}}
=
\int \rho(r) \frac{r^2}{6} d^3x
=
\frac{\langle{r}^{2}\rangle}{6}
.
\label{G047}
\end{equation}
Therefore,
the squared neutrino charge radius is given by
\begin{equation}
\langle{r}^{2}\rangle
=
6
\,
\left.
\frac{d\nff_{Q}(q^{2})}{dq^{2}}
\right|_{q^2=0}
.
\label{G048}
\end{equation}
Note that
$\langle{r}^{2}\rangle$ can be negative,
because the charge density $\rho(r)$ is not a positively defined quantity.

As we have seen in Subsection~\ref{C089},
massless left-handed Weyl
neutrinos have the electromagnetic form factor in Eq.~(\ref{C094}).
This is the case of the Standard Model,
in which in addition neutrinos have zero electric charge,
$\nff_{Q}(0)=0$,
as explained at the beginning of Subsection~\ref{G012}.
Taking into account Eqs.~(\ref{G044}) and (\ref{G048}),
in the Standard Model the neutrino electromagnetic form factor
for small values of $q^2$ is given by
\begin{equation}
\nff(q^{2})
\simeq
\left(
\frac{\langle{r}^{2}\rangle}{6}
-
\anm
\right)
q^{2}
,
\label{G049}
\end{equation}
where
$\anm$
is the anapole moment.
Hence,
in the Standard Model the form factor
$\nff(q^{2})$
can be interpreted as a neutrino charge radius or as an anapole moment
(or as a combination of both).
In this section we consider the charge radius interpretation.
The equivalence between the
charge radius and anapole moment interpretations of
$\nff(q^{2})$
is further discussed in the following Subsection~\ref{G060}.

The Standard Model theory of the neutrino charge radius has a long history,
with some controversies which are shortly summarized in the following.

In one of the first studies \cite{Bardeen:1972vi}, it was
claimed that in the Standard Model and in the unitary gauge the
neutrino charge radius is ultraviolet-divergent and so it is not a
physical quantity.
A direct one-loop calculation
\cite{Dvornikov:2003js,Dvornikov:2004sj} of proper vertices
(Fig.~\ref{D008}) and $\gamma - Z$ self-energy
(Figs.~\ref{G002} and \ref{G011}) contributions to the neutrino charge radius
performed in a general $R_\xi $ gauge for a massive Dirac
neutrino gave also a divergent result.
However, it was shown
\cite{Lee:1973fw}, using the unitary gauge, that by including in
addition to the usual terms also contributions from diagrams of
the neutrino-lepton neutral-current scattering ($Z$ boson
diagrams), it is possible to obtain for the neutrino charge radius
a gauge-dependent but finite quantity.
Later on, it was also shown
\cite{Lee:1977tib} that in order to define the neutrino charge
radius as a physical quantity one has to consider additional box
diagrams
and that in combination with contributions from the
proper diagrams it is possible to obtain a finite and
gauge-independent value for the neutrino charge radius.
In this
way, the neutrino electroweak radius was defined by
\textcite{Lucio:1983mg,Lucio:1984jn} and an additional set of diagrams
that give contribution to its value was discussed by
\textcite{Degrassi:1989ip}. Finally,
\textcite{Bernabeu:2000hf,Bernabeu:2002nw,Bernabeu:2002pd}
introduced the neutrino electroweak radius as a physical observable.
In the corresponding calculations, performed in the
one-loop approximation including additional terms from the
$\gamma-Z$ boson mixing and the box diagrams involving $W$ and $Z$
bosons, the following gauge-invariant result for the neutrino
charge radius have been obtained:
\begin{equation}
\langle{r}_{\nu_{\afl}}^{2}\rangle_{\text{SM}}
=
\frac{G_{\text{F}}}{4\sqrt{2}\pi^{2}}
\left[
3-2\log\left(\frac{m_{\afl}^{2}}{m^{2}_{W}}\right)
\right]
,
\label{G050}
\end{equation}
where $m_{W}$ and $m_{\afl}$ are the $W$ boson and lepton masses
($\afl=e,\mu,\tau$).
This result, however, revived the discussion
\cite{Fujikawa:2003tz,Fujikawa:2003ww,Papavassiliou:2003rx,Bernabeu:2004jr}
on the definition of the neutrino charge radius.
Numerically, Eq.~(\ref{G050}) gives
\cite{Bernabeu:2000hf,Bernabeu:2002pd}
\begin{align}
\null & \null
\langle{r}_{\nu_{e}}^{2}\rangle_{\text{SM}}
=
4.1 \times 10^{-33} \, \text{cm}^{2}
,
\label{G051}
\\
\null & \null
\langle{r}_{\nu_{\mu}}^{2}\rangle_{\text{SM}}
=
2.4 \times 10^{-33} \, \text{cm}^{2}
,
\label{G052}
\\
\null & \null
\langle{r}_{\nu_{\tau}}^{2}\rangle_{\text{SM}}
=
1.5 \times 10^{-33} \, \text{cm}^{2}
.
\label{G053}
\end{align}
These value are of the same order of magnitude of the numerical estimation
$\langle{r}_{\nu_{\afl}}^{2}\rangle \approx 10^{-33} \, \text{cm}^{2}$
obtained by \textcite{Lucio:1984jn}.

The effects of new physics beyond the Standard
Model can contribute to the neutrino charge radius.
However,
\textcite{NovalesSanchez:2008tn} have shown that in the context of an effective electroweak Yang-Mills theory
the anomalous $WW\gamma$ vertex contribution to
the neutrino effective charge radius
is smaller than about
$10^{-34} \, \text{cm}^{2}$,
which is one order of magnitude smaller than the Standard Model values
in Eqs.~(\ref{G051})--(\ref{G053}).

The neutrino charge radius has an effect in the scattering of neutrinos
with charged particles.
The most useful process is the elastic scattering with electrons,
which has been discussed in Subsection~\ref{D034}
in connection with the searches of neutrino magnetic moments.
Since in the ultrarelativistic limit the charge form factor conserves the neutrino helicity
(see Appendix~\ref{K001}),
a neutrino charge radius contributes to the weak-interaction cross section
$(d\sigma/dT_{e})_{\text{SM}}$
of
$\nu_{\afl}$--$e^{-}$
elastic scattering through the following shift of the vector coupling constant $g_{V}^{\nu_{\afl}}$
\cite{Grau:1985cn,Degrassi:1989ip,Vogel:1989iv,Hagiwara:1994pw}:
\begin{equation}
g_{V}^{\nu_{\afl}}
\to
g_{V}^{\nu_{\afl}}
+
\frac{2}{3} m_{W}^{2}
\langle{r}_{\nu_{\afl}}^{2}\rangle
\sin^{2}\theta_{W}
.
\label{G054}
\end{equation}
Using this method,
experiments which measure neutrino-electron elastic scattering can
probe the neutrino charge radius.
Some experimental results are listed in Tab.~\ref{G042}.
In addition,
\textcite{Hirsch:2002uv}
obtained the following 90\% CL bounds on
$\langle{r_{\nu_{\mu}}^{2}}\rangle$
from a reanalysis of
CHARM-II \cite{Vilain:1994hm}
and
CCFR \cite{McFarland:1997wx}
data:
\begin{equation}
-0.52\times10^{-32}<\langle{r_{\nu_{\mu}}^{2}}\rangle<0.68\times10^{-32} \, \text{cm}^2
.
\label{G055}
\end{equation}
More recently,
\textcite{Barranco:2007ea}
obtained the following 90\% CL bounds on
$\langle{r_{\nu_e}^{2}}\rangle$
from a combined fit of all available
$\nu_{e}$--$e^-$
and
$\bar\nu_{e}$--$e^-$
data:
\begin{equation}
-0.26\times10^{-32}<\langle{r_{\nu_e}^{2}}\rangle<6.64\times10^{-32} \, \text{cm}^2
\label{G056}
\end{equation}

The single photon production process
$e^{+} + e^{-} \to \nu + \bar\nu + \gamma$
has been used to get bounds on the effective $\nu_{\tau}$ charge radius,
assuming a negligible contribution of the
$\nu_{e}$
and
$\nu_{\mu}$
charge radii
\cite{Altherr:1993hb,Tanimoto:2000am,Hirsch:2002uv}.
For Dirac neutrinos,
\textcite{Hirsch:2002uv}
obtained
\begin{equation}
-5.6\times10^{-32}<\langle{r_{\nu_{\tau}}^{2}}\rangle<6.2\times10^{-32} \, \text{cm}^2
.
\label{G057}
\end{equation}

Comparing the theoretical Standard Model values in Eqs.~(\ref{G051})--(\ref{G053})
with the experimental limits in Tab.~\ref{G042} and those in Eqs.~(\ref{G055})--(\ref{G057}),
one can see that they differ at most by one
order of magnitude.
Therefore, one may expect that the experimental
accuracy will soon reach the value needed to probe the
Standard Model predictions for the neutrino charge radii.
This will be an important test of the Standard Model calculation of the neutrino charge radii.
If the experimental value of a neutrino charge radius
is found to be different from the Standard Model prediction in
Eqs.~(\ref{G051})--(\ref{G053})
it will be necessary to clarify the precision of the theoretical calculation
in order to understand if the difference is due to
new physics beyond the Standard Model.

The neutrino charge radius has also some impact on astrophysical phenomena and on cosmology.
The limits on the cooling of the Sun and white dwarfs
due to the plasmon-decay process discussed in Subsection~\ref{E043}
induced by a neutrino charge radius
led \textcite{Dolgov:1981hv} to estimate
the respective limits
$|\langle{r}_{\nu}^{2}\rangle| \lesssim 10^{-28} \, \text{cm}^{2}$
and
$|\langle{r}_{\nu}^{2}\rangle| \lesssim 10^{-30} \, \text{cm}^{2}$
for all neutrino flavors.
From the cooling of red giants
\textcite{Altherr:1993hb}
inferred the limit
$|\langle{r}_{\nu}^{2}\rangle| \lesssim 4 \times 10^{-31} \, \text{cm}^{2}$.

If neutrinos are Dirac particles,
$e^{+}$--$e^{-}$
annihilations can produce right-handed neutrino-antineutrino pairs
through the coupling induced by a neutrino charge radius.
This process would affect primordial Big-Bang Nucleosynthesis
and the energy release of a core-collapse supernova.
From the measured $^4\text{He}$ yield in primordial Big-Bang Nucleosynthesis
\textcite{Grifols:1986ed}
obtained
\begin{equation}
|\langle{r}_{\nu}^{2}\rangle| \lesssim 7 \times 10^{-33} \, \text{cm}^{2}
,
\label{G058}
\end{equation}
and
from SN 1987A data
\textcite{Grifols:1989vi}
obtained
\begin{equation}
\langle{r}_{\nu}^{2}\rangle \lesssim 2 \times 10^{-33} \, \text{cm}^{2}
,
\label{G059}
\end{equation}
for all neutrino flavors.

\subsection{Neutrino anapole moment}
\label{G060}

The notion of an anapole moment for a Dirac particle was introduced by
\textcite{Zeldovich:1957zl}
after the discovery of parity violation.
The anapole form factor was not known before because it violates P.
Indeed,
taking into account that
\begin{equation}
A_{\mu}(x)
\xrightarrow{\;\text{P}\;}
A^{\mu}(x_{\text{P}})
,
\label{G061}
\end{equation}
P is conserved if
\begin{equation}
\Lambda_{\mu}(q)
\xrightarrow{\;\text{P}\;}
\Lambda^{\mu}(q)
.
\label{G062}
\end{equation}
Using the formulae in Appendix~\ref{I001},
one can find that
\begin{equation}
\Lambda_{\mu}(q)
\xrightarrow{\;\text{P}\;}
\gamma^{0}
\Lambda_{\mu}(q_{\text{P}})
\gamma^{0}
.
\label{G063}
\end{equation}
Using the form-factor expansion in Eq.~(\ref{C023}),
we obtain
\begin{align}
\Lambda_{\mu}(q)
\null & \null
\xrightarrow{\;\text{P}\;}
\nff_{Q}(q^{2}) \gamma^{\mu}
-
\nff_{M}(q^{2}) i \sigma^{\mu\nu} q_{\nu}
\nonumber
\\
\null & \null
-
\nff_{E}(q^{2}) \sigma^{\mu\nu} q_{\nu} \gamma_{5}
-
\nff_{A}(q^{2}) (q^{2} \gamma^{\mu} - q^{\mu} \slashed{q}) \gamma_{5}
.
\label{G064}
\end{align}
Hence, parity is violated by the electric and anapole moments.
Since the anapole moment conserves CP (and T, as a consequence of CPT symmetry),
as shown in Subsection~\ref{C002},
it follows that the anapole moment violates also C.

In order to understand the physical characteristics of the anapole moment,
we consider its effect in the interactions with external electromagnetic fields.
From the last term in Eq.~(\ref{F009})
one can see that the anapole moment describes an interaction with the current
which generates the external electromagnetic fields.

Using the method described in Appendix~\ref{M001},
we obtain the helicity-conserving potential
\begin{equation}
V_{h \to h}
=
- \anm \, h \frac{m}{E} \, s^{\mu} j_{\mu}
,
\label{G065}
\end{equation}
which is strongly suppressed for ultrarelativistic neutrinos.
In the nonrelativistic limit,
we obtain
\begin{equation}
V_{h \to h}^{\text{nr}}
\simeq
\vec{\anm} \cdot \vec{j}
,
\quad
\text{with}
\quad
\vec{\anm}
=
h \, \frac{\vec{p}}{|\vec{p}|} \, \anm
.
\label{G066}
\end{equation}
This is the anapole moment potential that
was introduced by
\textcite{Zeldovich:1957zl}.
It is proportional to the longitudinal component of the current.

Considering now the helicity-flipping potential,
as shown at the end of Appendix~\ref{M001},
we obtain
\begin{equation}
V_{-h \to h}
=
\anm \, \frac{m}{E} \, j_{\perp}
,
\label{G067}
\end{equation}
where
$j_{\perp}$ is the component of $\vec{j}$ orthogonal to $\vec{p}$.
For ultrarelativistic neutrinos,
the helicity-flipping potential is strongly suppressed,
but in the nonrelativistic limit we have
\begin{equation}
V_{-h \to h}^{\text{nr}}
\simeq
\anm \, j_{\perp}
=
|\vec{\anm} \times \vec{j}|
.
\label{G068}
\end{equation}
This potential corresponds to a classical torque
\cite{Zeldovich:1957zl}
which rotates the spin of the particle,
causing periodic changes of the helicity.

The anapole moment is a quantity which is difficult to
understand,
because it does not generate interactions with a free electromagnetic field,
but only contact interactions with the charge and current density which generates
an electromagnetic field.
A classical model which can help to visualize the behavior of
the anapole moment
has been given by \textcite{Zeldovich:1957zl}
(see also \textcite{Bukina:1998ae}).
In this model the anapole is represented by a current-carrying rigid toroidal solenoid.
The current generates a magnetic field only inside the toroidal solenoid.
Since the solenoid is rigid,
there is no external magnetic field which can act on the toroidal solenoid as a whole.
The only action on the toroidal solenoid can be generated by a current which passes through the solenoid
and interacts with the magnetic field inside.
For example,
the toroidal solenoid can be immersed in an electrolytic solution which fills also the space inside the solenoid.
If a current flows through the electrolytic solution,
it interacts with the magnetic field inside the solenoid
and generates a torque proportional to the sine of the angle between the
direction of the current and the axis of the toroid.
In this model the axis of the toroid corresponds to the direction of $\vet{\anm}$
in Eqs.~(\ref{G066}) and the torque corresponds to the
helicity-flipping potential in Eq.~(\ref{G068}).

The neutrino anapole moment contributes to the scattering of neutrinos
with charged particles.
In order to discuss its effects,
it is convenient to consider strictly neutral neutrinos with $\nff_{Q}(0)=0$
and define a reduced charge form factor
$\tilde{\nff}_{Q}(q^{2})$
such that
\begin{equation}
\nff_{Q}(q^{2}) = q^2 \, \tilde{\nff}_{Q}(q^{2})
.
\label{G069}
\end{equation}
Then,
from Eq.~(\ref{G048}),
apart from a factor $1/6$,
the reduced charge form factor at $q^2=0$ is just the squared neutrino charge radius:
\begin{equation}
\tilde{\nff}_{Q}(0) = \langle{r}^{2}\rangle / 6
.
\label{G070}
\end{equation}
Let us now consider the charge and anapole parts of the neutrino electromagnetic vertex function
in Eq.~(\ref{C043}),
which can be written as
\begin{equation}
\Lambda_{\mu}^{Q,A}(q)
=
\left( \gamma_{\mu} q^{2} - q_{\mu} \slashed{q} \right)
\left[
\tilde{\nff}_{Q}(q^{2})
+
\nff_{A}(q^{2}) \gamma_{5}
\right]
.
\label{G071}
\end{equation}
Since for ultrarelativistic neutrinos the effect of $\gamma_{5}$
is only a sign which depends on the helicity of the neutrino
(see Eq.~(\ref{K007})),
the phenomenology of neutrino anapole moments
is similar to that of neutrino charge radii.
Hence, the limits on the neutrino charge radii discussed in Subsection~\ref{G043}
apply also to the neutrino anapole moments multiplied by 6.

As we have discussed in the beginning of Subsection~\ref{G012},
in the Standard Model the neutrino electric charges are exactly zero.
Hence,
Eqs.~(\ref{G071}) applies to Standard Model
and can be further simplified taking into account that in the Standard Model neutrinos
are described by two-component massless left-handed Weyl spinors.
As discussed in Subsection~\ref{C089},
the $\gamma^{5}$ in Eq.~(\ref{G071})
becomes a minus sign,
leading to
\begin{equation}
\Lambda_{\text{SM}\mu}^{Q,A}(q)
=
\left( \gamma_{\mu} q^{2} - q_{\mu} \slashed{q} \right)
\nff^{\text{SM}}(q^{2})
,
\label{G072}
\end{equation}
with
\begin{equation}
\nff^{\text{SM}}(q^{2})
=
\tilde{\nff}_{Q}(q^{2})
-
\nff_{A}(q^{2})
\xrightarrow[q^2\to0]{}
\frac{\langle{r}^{2}\rangle}{6} - \anm
.
\label{G073}
\end{equation}
These equations correspond to Eqs.~(\ref{C093}) and (\ref{C094}) for $\nff_{Q}(0)=0$.
Hence,
in the Standard Model the neutrino charge radius and the anapole moment are not defined separately
and one can interpret arbitrarily $\nff^{\text{SM}}(0)$
as a charge radius or as an anapole moment.
This is the correct interpretation of the statement often found in the literature that in the Standard Model
$\anm = - \langle{r}^{2}\rangle / 6$.
Therefore, the Standard Model values for the neutrino charge radii in Eqs.~(\ref{G050})--(\ref{G053})
can be interpreted also as values of the corresponding neutrino anapole moments.

Some deep insight into an interpretation of the decompositions of the
vertex function (\ref{C023}) and the neutrino
form factors can be obtained in the framework of a multipole
expansions of the corresponding classical electromagnetic currents
\cite{Dubovik:1974dc,Dubovik:1984es,Dubovik:1996gx}. Since in this limit the
anapole form factor does not correspond to a certain multipole
distribution (that is why the term ``anapole'' was introduced by
\textcite{Zeldovich:1957zl}), the anapole moment has a quite
intricate classical analog.
Therefore,
\textcite{Dubovik:1996gx,Bukina:1998ae,Bukina:1998kw}
proposed to consider the toroidal dipole moment
as a characteristic of the neutrino
which is more convenient and transparent
than the anapole moment for the
description of T-invariant interactions with nonconservation of
the P and C symmetries.
In this case, the electromagnetic vertex
of a neutrino can be rewritten in the alternative multipole
(toroidal) parameterization
\begin{eqnarray}
\Lambda_{\mu}(q)= \nff_{Q}(q^{2})\gamma_{\mu}
&-& \nff_{M}(q^{2})i\sigma_{\mu\nu}q^\nu
+\nff_{E}(q^{2})\sigma_{\mu\nu}q^\nu\gamma_5
\nonumber\\ &+& i\nff_{T}(q^{2})
\epsilon_{\mu \nu \lambda \rho}
P^{\nu}q^{\lambda}\gamma^{\rho},
\label{G074}
\end{eqnarray}
where $\nff_{T}$ is the toroidal dipole form factor and
$P=p_{i}+p_{f}$. From the following identity
\begin{eqnarray}
\overline{u}_{f}(p_{f})\Bigl\{(m_{i}-m_{f})\sigma_{\mu\nu}q^\nu
+\left(q^{2}\gamma_{\mu}-\slashed{q}q_{\mu}\right)
\nonumber \\
- i\epsilon_{\mu\nu\lambda\rho}P^\nu q^\lambda
\gamma^\rho\gamma_5
\Bigr\}\gamma_5 u_{i}(p_{i}) = 0,
\label{G075}
\end{eqnarray}
it can be seen that the toroidal and anapole moments
coincide in the static limit when the masses of the initial and
final neutrino states are equal to each other, $m_{i}=m_{f}$
\cite{Bukina:1998kw}, i.e. the toroidal and anapole
parameterizations coincide in this case.

In some sense the toroidal parameterization has a more transparent
and clear physical interpretation, because it provides a
one-to-one correspondence between the multipole moments and the
corresponding form factors.
From the properties of each term in the
expression (\ref{G074}) for the vertex function under
C, P and T transformations, it follows that in the
Majorana case only the toroidal form factor survives
\cite{Zeldovich:1957zl,Kobzarev:1972ko} and the toroidal moment
of the Dirac neutrino is half of that in the Majorana case.

In one-loop calculations \cite{Dubovik:1996gx} of the toroidal
(and anapole) moment of a massive and a massless Majorana neutrino
(the diagrams in Figs.~\ref{D008}, \ref{G002} and \ref{G011} contribute)
it was shown that its value does not depend significantly on the
neutrino mass (through the ratios
$m^{2}_{\nu_{i}} / m^{2}_{W}$) and is of the order of
\begin{equation}
\nff_{T}(q^{2}=0)
\sim
e \times (10^{-33}-10^{-34}) \,
\text{cm}^{2},
\end{equation}
depending on the values of the quark masses that propagate in the
loop diagrams in Fig.~\ref{G011}.

Note that the toroidal form factors can contribute to the neutrino
vertex function in both the diagonal and off-diagonal cases.

The toroidal
(anapole) interactions of a Majorana as well as a Dirac neutrino
are expected to contribute to the total cross section of neutrino
elastic scattering off electrons, quarks and nuclei.
Because of the fact that the toroidal (anapole) interactions contribute to the
helicity preserving part of the scattering of neutrinos on
electrons, quarks and nuclei, its contribution to cross sections
are similar to those of the neutrino charge radius.
In principle,
these contributions can be probed and information about toroidal
moments can be extracted in low-energy scattering experiments in
the future.

Different effects of the neutrino toroidal moment are
discussed in \textcite{Ginzburg:1985gt,Dubovik:1996gx,Bukina:1998ae,Bukina:1998kw}.
In particular,
it has been shown that the neutrino toroidal
electromagnetic interactions can produce Cherenkov radiation of neutrinos
propagating in a medium.
\section{Summary and perspectives}
\label{H001}

In this review we discussed the theory and phenomenology
of neutrino electromagnetic properties and interactions.
We have seen that most of the theoretical and experimental research
has been devoted to the study of magnetic and electric dipole moments,
but there has been also some interest in the investigation of neutrino millicharges
and of the charge radii and anapole moments of neutrinos.

Unfortunately,
so far there is not any experimental indication in favor of neutrino electromagnetic interactions
and all neutrino electromagnetic properties are known to be small,
with rather stringent upper bounds obtained in laboratory experiments or from astrophysical observations.

The most accessible neutrino electromagnetic property may be the charge radius,
discussed in Subsection~\ref{G043},
for which the Standard Model gives a value which is only about one order of magnitude smaller
than the experimental upper bounds.
A measurement of a neutrino charge radius
at the level predicted by the Standard Model
would be another spectacular confirmation of the Standard Model,
after the recent discovery of the Higgs boson
(see \textcite{1312.5672}).
However, such a measurement would not give information on new physics beyond the Standard Model
unless the measured value is shown to be incompatible with the Standard Model value in a high-precision experiment.

The strongest current efforts to probe the physics beyond the Standard Model
by measuring neutrino electromagnetic properties
is the search for a neutrino magnetic moment effect
in reactor $\bar\nu_{e}$-$e^{-}$ scattering experiments.
The current upper bounds reviewed in Subsection~\ref{D064}
are more than eight orders of magnitude larger than the
prediction discussed in Subsection~\ref{D009}
of the Dirac neutrino magnetic moments in the
minimal extension of the
Standard Model with right-handed neutrinos.
Hence,
a discovery of a neutrino magnetic moment effect
in reactor $\bar\nu_{e}$-$e^{-}$ scattering experiments
would be a very exciting discovery of non-minimal new physics beyond the Standard Model.

In particular,
the GEMMA-II collaboration
expects to reach around the year 2017 a sensitivity to
$\mgm_{\nu_{e}} \approx 1 \times 10^{-11} \bmag$
in a new series of measurements at the Kalinin Nuclear Power Plant
with a doubled neutrino flux
obtained by reducing the distance between the reactor and the detector from 13.9 m to 10 m
and by reducing the energy threshold from 2.8 keV to 1.5 keV
\cite{Beda:2012zz,Beda:2013mta}.
The corresponding sensitivity to the neutrino electric millicharge discussed in Subsection~\ref{G012} will reach the level of
$|\chg_{\nu_{e}}|
\approx
3.7 \times 10^{-13} \, \elechg$ \cite{Studenikin:2013my}.

There is also a GEMMA-III project\footnote{Victor Brudanin and Vyacheslav Egorov, private communication.}
to further lower the energy threshold to about 350 eV,
which may allow the experimental collaboration to reach a sensitivity of
$\mgm_{\nu_{e}} \approx 9 \times 10^{-12} \bmag$.
The corresponding sensitivity to neutrino millicharge will be
$|\chg_{\nu_{e}}|
\approx
1.8 \times 10^{-13} \, \elechg$ \cite{Studenikin:2013my}.

An interesting possibility for exploring very small values of
$\mgm_{\nu_{e}}$ in $\bar\nu_{e}$-$e^{-}$
scattering experiments has been proposed by \textcite{Bernabeu:2004ay} on
the basis of the observation \cite{Segura:1993tu} that ``dynamical zeros'' induced
by a destructive interference between the left-handed and right-handed chiral couplings of the electron in the charged and neutral-current amplitudes
appear in the Standard Model contribution to
the scattering cross section.
It may be possible to enhance the sensitivity of an experiment to $\mgm_{\nu_{e}}$
by selecting recoil electrons contained in a forward narrow cone corresponding to a dynamical zero
(see Eq.~(\ref{D036})).

In the future experimental searches of neutrino electromagnetic properties
may be performed also with new neutrino sources,
as a tritium source
\cite{McLaughlin:2003yg},
a low-energy beta-beam
\cite{McLaughlin:2003yg,deGouvea:2006cb},
a stopped-pion neutrino source
\cite{Scholberg:2005qs},
or a neutrino factory
\cite{deGouvea:2006cb}.
Recently \textcite{Coloma:2014hka}
proposed to improve the existing limit on the electron neutrino magnetic moment
with a megacurie $^{51}\text{Cr}$ neutrino source
and a large liquid Xenon detector.

Neutrino electromagnetic interactions
could have important effects in astrophysical environments
and in the evolution of the Universe
and the current rapid advances of astrophysical and cosmological observations
may lead soon to the exciting discovery of nonstandard neutrino electromagnetic properties.
In particular,
future high-precision observations of supernova neutrino fluxes
may reveal the effects of collective spin-flavor oscillations
due to Majorana transition magnetic moments as small as
$10^{-21} \, \bmag$
\cite{deGouvea:2012hg,deGouvea:2013zp}.

Let us finally emphasize the importance of pursuing the experimental and theoretical studies
of electromagnetic neutrino interactions,
which could open a powerful window to new physics beyond the Standard Model.
\appendix
\section{Conventions, useful constants and formulae}
\label{I001}

In this Appendix we clarify the conventions and notation
used in the paper and
we list some useful physical constants and formulae.

We use natural units in which
$
c = \hbar = 1
$,
where
$c$ is the velocity of light
and
$\hbar$ is the reduced Planck constant.

The values
of the following physical constants
are taken from \textcite{PDG-2012}.
\begin{align}
&
\null\hspace{-0.3cm}
\text{Avogadro number:}
\nonumber
\\
&
N_{\text{A}} = 6.022\,141\,29\,(27) \times 10^{23} \, \text{mol}^{-1}
.
\label{I002}
\\
&
\null\hspace{-0.3cm}
\text{Bohr magneton ($\bmag \equiv e / 2 m_{e}$):}
\nonumber
\\
&
\bmag = 5.788\,381\,8066\,(38) \times 10^{-15} \, \text{MeV} \, \text{G}^{-1}
\nonumber
\\
&
\phantom{\bmag}
\simeq
0.296 \, \text{MeV}^{-1}
.
\label{I003}
\\
&
\null\hspace{-0.3cm}
\text{Electron mass:}
\nonumber
\\
&
m_{e} = 0.510\,998\,928\,(11) \, \text{MeV}
.
\label{I004}
\\
&
\null\hspace{-0.3cm}
\text{Conversion constant:}
\nonumber
\\
&
\hbar c = 1.973\,269\,718\,(44) \times 10^{-5} \, \text{eV} \, \text{cm}
.
\label{I005}
\\
&
\null\hspace{-0.3cm}
\text{Light velocity:}
\nonumber
\\
&
c = 299\,792\,458 \, \text{m} \, \text{s}^{-1}
.
\label{I006}
\\
&
\null\hspace{-0.3cm}
\text{Fermi constant:}
\nonumber
\\
&
G_{\text{F}} = 1.166\,378\,7\,(6) \times 10^{-5} \, \text{GeV}^{-2}
.
\label{I007}
\\
&
\null\hspace{-0.3cm}
\text{Fine-structure constant ($\alpha \equiv e^{2} / 4 \pi$) at $Q^2=0$:}
\nonumber
\\
&
\alpha^{-1} = 137.035\,999\,074\,(44)
.
\label{I008}
\\
&
\null\hspace{-0.3cm}
\text{Neutron mass:}
\nonumber
\\
&
m_{n} = 939.565\,379\,(21) \, \text{MeV}
.
\label{I009}
\\
&
\null\hspace{-0.3cm}
\text{Proton mass:}
\nonumber
\\
&
m_{p} = 938.272\,046\,(21) \, \text{MeV}
.
\label{I010}
\\
&
\null\hspace{-0.3cm}
\text{Muon mass:}
\nonumber
\\
&
m_{\mu} = 105.658\,3715\,(35) \, \text{MeV}
.
\label{I011}
\\
&
\null\hspace{-0.3cm}
\text{Planck constant, reduced:}
\nonumber
\\
&
\hbar = 6.582\,119\,28\,(15) \times 10^{-22} \, \text{MeV} \, \text{s}
.
\label{I012}
\\
&
\null\hspace{-0.3cm}
\text{Tau mass:}
\nonumber
\\
&
m_{\tau} = 1776.82\,(16) \, \text{MeV}
.
\label{I013}
\\
&
\null\hspace{-0.3cm}
\text{Weak mixing angle (on-shell: $\sin^{2} \theta_{W} \equiv 1 - m_{W}^2 / m_{Z}^2$):}
\nonumber
\\
&
\sin^{2} \theta_{W} = 0.222\,95\,(28)
.
\label{I014}
\end{align}

For Dirac $\gamma$ matrices
and related quantities we use the notation and conventions in \textcite{Giunti-Kim-2007}.
Curly and square brackets denote, respectively,
anticommutator and commutator.
For a four-vector $p^{\mu}$
we use the standard notation
$\slashed{p} \equiv p^{\mu} \gamma_{\mu}$,
with the metric tensor
$g^{\mu\nu}=g_{\mu\nu}=\text{diag}(1,-1,-1,-1)$
and
$p^{\mu}=(p^{0},p^{1},p^{2},p^{3})=(p^{0},\vet{p})$.

Dirac $\gamma$ matrices:
\begin{align}
\null & \null
\{ \gamma^{\mu} , \gamma^{\nu} \}
=
2 \, g^{\mu\nu}
,
\label{I015}
\\
\null & \null
\gamma^{0}
\,
\gamma^{\mu}
\,
\gamma^{0}
=
{\gamma^{\mu}}^{\dagger}
=
\gamma_{\mu}
.
\label{I016}
\end{align}

Definition and properties of $\gamma^{5}$:
\begin{align}
\null & \null
\gamma^{5}
\equiv
i
\,
\gamma^{0}
\,
\gamma^{1}
\,
\gamma^{2}
\,
\gamma^{3}
=
\gamma_{5}
,
\label{I017}
\\
\null & \null
\left\{
\gamma^{5}
,
\gamma^{\mu}
\right\}
=
0
,
\label{I018}
\\
\null & \null
\left(
\gamma^{5}
\right)^{2}
=
\openone
,
\label{I019}
\\
\null & \null
\left(
\gamma^{5}
\right)^{\dagger}
=
\gamma^{5}
,
\label{I020}
\\
\null & \null
\gamma^{\mu} \, \gamma^{5}
=
\frac{i}{6}
\,
\epsilon^{\mu\nu\rho\sigma} \, \gamma_{\nu} \, \gamma_{\rho} \, \gamma_{\sigma}
.
\label{I021}
\end{align}

Left-handed and right-handed chiral projectors:
\begin{equation}
P_{L}
=
\frac{1 - \gamma^{5}}{2}
,
\qquad
P_{R}
=
\frac{1 + \gamma^{5}}{2}
.
\label{I022}
\end{equation}
Chiral decomposition of a Dirac field $\psi$:
\begin{equation}
\psi
=
P_{L} \psi
+
P_{R} \psi
=
\psi_{L} + \psi_{R}
.
\label{I023}
\end{equation}

Definition and properties of $\sigma^{\mu\nu}$:
\begin{align}
\null & \null
\sigma^{\mu\nu}
\equiv
\frac{ i }{ 2 } \,
\left[
\gamma^{\mu}
,
\gamma^{\nu}
\right]
=
i \, \gamma^{\mu} \, \gamma^{\nu} - i \, g^{\mu\nu}
,
\label{I024}
\\
\null & \null
\left[ \gamma^{5} \, , \, \sigma^{\mu\nu} \right] = 0
,
\label{I025}
\\
\null & \null
\gamma^{0} \, \sigma^{\mu\nu} \, \gamma^{0}
=
(\sigma^{\mu\nu})^{\dagger}
=
\sigma_{\mu\nu}
,
\label{I026}
\\
\null & \null
\epsilon^{\mu\nu\alpha\beta} \sigma_{\alpha\beta}
=
- 2 i \sigma^{\mu\nu} \gamma^{5}
,
\label{I027}
\\
\null & \null
\epsilon^{\mu\nu\alpha\beta} \gamma_{\nu}
=
i \left(
g^{\mu\alpha} g^{\nu\beta}
-
g^{\mu\beta} g^{\nu\alpha}
\right)
\gamma_{\nu} \gamma^{5}
-
\gamma^{\mu} \sigma^{\alpha\beta} \gamma^{5}
.
\label{I028}
\end{align}

Definition and properties of $\vec{\Sigma}$:
\begin{align}
\null & \null
\Sigma^{k}
\equiv
\frac{1}{2}
\sum_{j,l}
\epsilon^{kjl} \, \sigma^{jl}
=
\gamma^{0} \, \gamma^{k} \, \gamma^{5}
,
\label{I029}
\\
\null & \null
[ \Sigma^{k} , \Sigma^{j} ] = 2 \, i \sum_{l} \epsilon^{kjl} \, \Sigma^{l}
,
\label{I030}
\\
\null & \null
\{ \Sigma^{k} , \Sigma^{j} \} = 2 \, \delta^{kj}
,
\label{I031}
\\
\null & \null
(\Sigma^{k})^{\dagger} = \Sigma^{k}
,
\label{I032}
\\
\null & \null
[ \Sigma^{k} , \gamma^{0} ]
=
[ \Sigma^{k} , \gamma^{5} ]
=
0
,
\label{I033}
\\
\null & \null
[ \Sigma^{k} , \gamma^{j} ] = 2 i \sum_{\ell} \epsilon^{kj\ell} \, \gamma^{\ell}
.
\label{I034}
\end{align}

Charge-conjugation matrix:
\begin{align}
\null & \null
\mathcal{C}
\,
\gamma_{\mu}^{T}
\,
\mathcal{C}^{-1}
=
- \gamma_{\mu}
,
\label{I035}
\\
\null & \null
\mathcal{C}^{\dagger} = \mathcal{C}^{-1}
,
\label{I036}
\\
\null & \null
\mathcal{C}^{T} = - \mathcal{C}
.
\label{I037}
\\
\null & \null
\mathcal{C}
\,
(\gamma^{5})^{T}
\,
\mathcal{C}^{-1}
=
\gamma^{5}
,
\label{I038}
\\
\null & \null
\mathcal{C}
\,
(\sigma^{\mu\nu})^{T}
\,
\mathcal{C}^{-1}
=
- \sigma^{\mu\nu}
.
\label{I039}
\end{align}

Traces of products of $\gamma$ matrices:
\begin{align}
\null & \null
\text{Tr}\!\left[ \gamma^{\alpha} \, \gamma^{\beta} \right]
=
4 \, g^{\alpha\beta}
,
\label{I040}
\\
\null & \null
\text{Tr}\!\big[ \gamma^{\alpha} \, \gamma^{\beta} \, \gamma^{\rho} \, \gamma^{\sigma} \big]
=
4 (
g^{\alpha\beta} g^{\rho\sigma}
-
g^{\alpha\rho} g^{\beta\sigma}
+
g^{\alpha\sigma} g^{\beta\rho}
)
,
\label{I041}
\\
\null & \null
\text{Tr}\!\left[ \gamma^{\alpha} \, \gamma^{\beta} \, \gamma^{\rho} \, \gamma^{\sigma} \, \gamma^{5} \right]
=
- 4 \, i \, \epsilon^{\alpha\beta\rho\sigma}
.
\label{I042}
\end{align}
The $\gamma$ matrices are traceless.
The trace of a product of an odd number of $\gamma$ matrices is zero.
$
\text{Tr}\!\left[ \gamma^{5} \right]
=
\text{Tr}\!\left[ \gamma^{\alpha} \, \gamma^{\beta} \, \gamma^{5} \right]
=
0
$.

Four-momentum and helicity eigenstate spinors:
\begin{align}
\null & \null
( \slashed{p} - m ) \, u^{(h)}(p) = 0
,
\label{I043}
\\
\null & \null
( \slashed{p} + m ) \, v^{(h)}(p) = 0
.
\label{I044}
\end{align}

Normalization:
\begin{align}
&
\overline{u^{(h)}}(p)
\,
u^{(h')}(p)
=
2 \, m \, \delta_{hh'}
,
\label{I045}
\\
&
\overline{v^{(h)}}(p)
\,
v^{(h')}(p)
=
- 2 \, m \, \delta_{hh'}
,
\label{I046}
\end{align}

Charge-conjugation relation:
\begin{align}
&
u^{(h)}(p)
=
\mathcal{C} \, \overline{v^{(h)}}^{T}(p)
\label{I047}
\\
&
v^{(h)}(p)
=
\mathcal{C} \, \overline{u^{(h)}}^{T}(p)
.
\label{I048}
\end{align}

Energy-projection matrices:
\begin{align}
\null & \null
\Lambda_{+}(p)
=
\frac{m + \slashed{p}}{2m}
=
\sum_{h=\pm1}
\frac{
u^{(h)}(p)
\,
\overline{u^{(h)}}(p)
}{2m}
,
\label{I049}
\\
\null & \null
\Lambda_{-}(p)
=
\frac{ m - \slashed{p} }{2m}
=
-
\sum_{h=\pm1}
\frac{
v^{(h)}(p)
\,
\overline{v^{(h)}}(p)
}{2m}
.
\label{I050}
\end{align}

Helicity-projection matrices
($[P_{h},\Lambda_{\pm}(p)] = 0$):
\begin{equation}
P_{h}
=
\frac{ 1 + h \, \gamma^{5} \, \slashed{s} }{ 2 }
,
\label{I051}
\end{equation}
for
$h=+1$ (right-handed)
and
$h=-1$ (left-handed),
with the polarization four-vector
\begin{equation}
s^{\mu}
=
\left(
\frac{|\vet{p}|}{m}
\, , \,
\frac{E}{m} \, \frac{ \vet{p} }{ |\vet{p}| }
\right)
,
\quad
s^{2} = -1
,
\quad
s \cdot p = 0
.
\label{I052}
\end{equation}
The helicity eigenstate spinors satisfy
\begin{align}
\null & \null
P_{h'}
u^{(h)}(p)
=
\frac{1}{2}
\left(
1
+
h'
\frac{\vet{p} \cdot \vet{\Sigma}}{|\vet{p}|}
\right)
u^{(h)}(p)
=
\delta_{h'h} u^{(h)}(p)
,
\label{I053}
\\
\null & \null
P_{h'}
v^{(h)}(p)
=
\frac{1}{2}
\left(
1
-
h'
\frac{\vet{p} \cdot \vet{\Sigma}}{|\vet{p}|}
\right)
v^{(h)}(p)
=
\delta_{h'h} v^{(h)}(p)
.
\label{I054}
\end{align}
Moreover,
since
$\gamma^{0} P_{h}^{\dagger} \gamma^{0} = P_{h}$,
we also have
\begin{equation}
\overline{u^{(h)}}(p) P_{h'} = \delta_{h'h} \overline{u^{(h)}}(p)
,
\quad
\overline{v^{(h)}}(p) P_{h'} = \delta_{h'h} \overline{v^{(h)}}(p)
.
\label{I055}
\end{equation}

Fourier expansion of a free Dirac field $\psi(x)$
(with $p^{0} = E = \sqrt{\vet{p}^{2} + m^{2}}$):
\begin{align}
\psi(x)
=
\int
\frac{ \text{d}^{3}p }{ (2\pi)^{3} \, 2E }
\null & \null
\sum_{h=\pm1}
\bigg[
a^{(h)}(p)
\,
u^{(h)}(p)
\,
e^{-ip{\cdot}x}
\nonumber
\\
\null & \null
+
b^{(h)\dagger}(p)
\,
v^{(h)}(p)
\,
e^{ip{\cdot}x}
\bigg]
.
\label{I067}
\end{align}
The operators
$a^{(h)}(p)$,
$a^{(h)\dagger}(p)$,
$b^{(h)}(p)$,
$b^{(h)\dagger}(p)$
anticommute,
except for
\begin{align}
\{
a^{(h)}(p)
,
a^{(h')\dagger}(p')
\}
=
\null & \null
\{
b^{(h)}(p)
,
b^{(h')\dagger}(p')
\}
\nonumber
\\
=
\null & \null
(2\pi)^{3} \, 2E \, \delta^{3}(\vet{p}-\vet{p}') \, \delta_{hh'}
.
\label{I068}
\end{align}

States describing a fermion $\fermion$
and an antifermion $\bar{\fermion}$
with four-momentum $p$ and helicity $h$
($|0\rangle$ is the vacuum, such that
$a^{(h)}(p) |0\rangle = 0$,
$b^{(h)}(p) |0\rangle = 0$
and
$\langle0|0\rangle = 1$;
$V$ is the total volume):
\begin{equation}
|\fermion(p,h)\rangle
=
a^{(h)\dagger}(p) \, |0\rangle
,
\quad
|\bar{\fermion}(p,h)\rangle
=
b^{(h)\dagger}(p) \, |0\rangle
,
\label{I069}
\end{equation}
\begin{equation}
\langle\fermion(p,h)|\fermion(p,h')\rangle
=
\langle\bar{\fermion}(p,h)|\bar{\fermion}(p,h')\rangle
=
2E \, V \, \delta_{hh'}
.
\label{I070}
\end{equation}

The Majorana condition (\ref{B028})
leads to the following Fourier expansion of a free Majorana field:
\begin{align}
\psi(x)
=
\int
\frac{ \text{d}^{3}p }{ (2\pi)^{3} \, 2E }
\null & \null
\sum_{h=\pm1}
\bigg[
a^{(h)}(p)
\,
u^{(h)}(p)
\,
e^{-ip{\cdot}x}
\nonumber
\\
\null & \null
+
a^{(h)\dagger}(p)
\,
v^{(h)}(p)
\,
e^{ip{\cdot}x}
\bigg]
.
\label{I071}
\end{align}

Gordon identities:
\begin{align}
\null & \null
\overline{u_{f}}(p_{f}) i \sigma^{\alpha\beta} (p_{f}-p_{i})_{\beta} u_{i}(p_{i})
\nonumber
\\
\null & \null
=
\overline{u_{f}}(p_{f})
\left[ \left( m_{f} + m_{i} \right) \gamma^{\alpha} - (p_{f}+p_{i})^{\alpha} \right]
u_{i}(p_{i})
,
\label{I072}
\\
\null & \null
\overline{u_{f}}(p_{f}) i \sigma^{\alpha\beta} (p_{f}+p_{i})_{\beta} u_{i}(p_{i})
\nonumber
\\
\null & \null
=
\overline{u_{f}}(p_{f})
\left[ \left( m_{f} - m_{i} \right) \gamma^{\alpha} - (p_{f}-p_{i})^{\alpha} \right]
u_{i}(p_{i})
,
\label{I073}
\\
\null & \null
\overline{u_{f}}(p_{f}) i \sigma^{\alpha\beta} (p_{f}-p_{i})_{\beta} \gamma^{5} u_{i}(p_{i})
\nonumber
\\
\null & \null
=
\overline{u_{f}}(p_{f})
\left[ \left( m_{f} - m_{i} \right) \gamma^{\alpha} - (p_{f}+p_{i})^{\alpha} \right]
\gamma^{5} u_{i}(p_{i})
,
\label{I074}
\\
\null & \null
\overline{u_{f}}(p_{f}) i \sigma^{\alpha\beta} (p_{f}+p_{i})_{\beta} \gamma^{5} u_{i}(p_{i})
\nonumber
\\
\null & \null
=
\overline{u_{f}}(p_{f})
\left[ \left( m_{f} + m_{i} \right) \gamma^{\alpha} - (p_{f}-p_{i})^{\alpha} \right]
\gamma^{5} u_{i}(p_{i})
.
\label{I075}
\end{align}

C, P, CP, T and CPT active transformations of a fermionic field $\psi(x)$
($x^{\mu}_{\text{P}}=x_{\mu}$
and
$x^{\mu}_{\text{T}}=-x_{\mu}$):
\begin{align}
\mathsf{U}_{\text{C}}
\psi(x)
\mathsf{U}_{\text{C}}^{\dagger}
=
\null & \null
\xi^{\text{C}} \mathcal{C} \overline{\psi}^{T}(x)
,
\label{I076}
\\
\mathsf{U}_{\text{P}}
\psi(x)
\mathsf{U}_{\text{P}}^{\dagger}
=
\null & \null
\xi^{\text{P}} \gamma^{0} \psi(x_{\text{P}})
,
\label{I077}
\\
\mathsf{U}_{\text{CP}}
\psi(x)
\mathsf{U}_{\text{CP}}^{\dagger}
=
\null & \null
\xi^{\text{CP}} \gamma^{0} \mathcal{C} \overline{\psi}^{T}(x_{\text{P}})
,
\label{I078}
\\
\mathsf{U}_{\text{T}}
\psi(x)
\mathsf{U}_{\text{T}}^{\dagger}
=
\null & \null
\xi^{\text{T}} \mathcal{C}^{\dagger} \gamma^{5} \psi(x_{\text{T}})
,
\label{I079}
\\
\mathsf{U}_{\text{CPT}}
\psi(x)
\mathsf{U}_{\text{CPT}}^{\dagger}
=
\null & \null
\xi^{\text{CPT}} \left[ \psi^{\dagger}(-x) \gamma^{5} \right]^{T}
.
\label{I080}
\end{align}
Here
$\xi^{\text{C}}$,
$\xi^{\text{P}}$,
$\xi^{\text{CP}}$,
$\xi^{\text{T}}$,
$\xi^{\text{CPT}}$ are phases such that
$\xi^{\text{CP}} = \xi^{\text{C}} \xi^{\text{P}}$,
$\xi^{\text{T}} = \pm \xi^{\text{CP}} \, \text{or} \, \pm i \xi^{\text{CP}}$, and
$\xi^{\text{CPT}} = \xi^{\text{T}} {\xi^{\text{CP}}}^{*} = \pm 1 \, \text{or} \, \pm i$.
The operators
$\mathsf{U}_{\text{C}}$, $\mathsf{U}_{\text{P}}$ and $\mathsf{U}_{\text{CP}}$
are unitary,
whereas
the operators
$\mathsf{U}_{\text{T}}$ and $\mathsf{U}_{\text{CPT}}$
are antiunitary.
An antiunitary operator $\mathsf{U}$
satisfies the standard relation $\mathsf{U}^{\dagger} = \mathsf{U}^{-1}$
of unitary operators,
but is antilinear,
i.e.
for a real number $z$
\begin{equation}
\mathsf{U} z \mathsf{U}^{\dagger} = z^{*}
,
\label{I081}
\end{equation}
and
if
$| p' \rangle = \mathsf{U} | p' \rangle$
we have
\begin{equation}
\langle p'_{1} | p'_{2} \rangle
=
\langle p_{1} | \mathsf{U}^{\dagger} \mathsf{U} | p_{2} \rangle^{*}
=
\langle p_{2} | p_{1} \rangle
.
\label{I082}
\end{equation}

Maxwell equations in the International System of Units (SI):
\begin{equation}
\partial_{\mu} F^{\mu\nu}(x) = j^{\nu}(x)
,
\qquad
\partial_{\mu} \widetilde{F}^{\mu\nu}(x) = 0
,
\label{I083}
\end{equation}
where
$j^{\mu}(x)=(\rho(x),\vec{j}(x))$
is the four-vector of the charge and current density
and
\begin{align}
\null & \null
F^{\mu\nu}(x)
=
\partial^{\mu} A^{\nu}(x)
-
\partial^{\nu} A^{\mu}(x)
,
\label{I084}
\\
\null & \null
\widetilde{F}^{\mu\nu}(x)
\equiv
\frac{1}{2} \epsilon^{\mu\nu\rho\sigma} F_{\rho\sigma}(x)
,
\label{I085}
\end{align}
where
$A^{\mu}(x)$ is the electromagnetic field.
The electromagnetic tensor $F^{\mu\nu}(x)$ contains the physical
electric field $\vec{E}(x)$ and magnetic field $\vec{B}(x)$:
\begin{equation}
E^{k}(x) = F^{k0}(x)
,
\qquad
B^{k}(x) = - \frac{1}{2} \sum_{j,\ell} \epsilon^{kj\ell} F^{j\ell}(x)
.
\label{I086}
\end{equation}
The Maxwell equations for the electromagnetic field are ($\square\equiv\partial_{\mu}\partial^{\mu}$):
\begin{equation}
\square A^{\mu}(x) - \partial^{\mu} \partial_{\nu} A^{\nu}(x) = j^{\mu}(x)
.
\label{I087}
\end{equation}
\section{Decomposition of $\Lambda_{\mu}$}
\label{J001}

\newcommand{\sif}{t}

In this Appendix
(see also \textcite{Nowakowski:2004cv})
we derive the general expression of
$\Lambda_{\mu}(p_{\text{i}},p_{\text{f}})$
in the matrix element
\begin{equation}
\langle \nu_{f}(p_{\text{f}}) |
j_{\mu}^{(\nu)}(0)
| \nu_{i}(p_{\text{i}}) \rangle
=
\overline{u_{f}}(p_{\text{f}})
\Lambda_{\mu}(p_{\text{i}},p_{\text{f}})
u_{i}(p_{\text{i}})
.
\label{J002}
\end{equation}
The initial and final massive neutrinos can be different,
but since they are considered as free particles they are on shell,
with four-momenta $p_{i}$ and $p_{f}$ such that
\begin{equation}
p_{i}^{2} = m_{i}^{2}
,
\qquad
p_{f}^{2} = m_{f}^{2}
,
\label{J003}
\end{equation}
We use the notation
\begin{equation}
q \equiv p_{i} - p_{f}
,
\qquad
\sif \equiv p_{i} + p_{f}
,
\label{J004}
\end{equation}
for which we have
\begin{equation}
q^{2} + \sif^{2} = 2 \left( m_{f}^{2} + m_{i}^{2} \right)
,
\qquad
q \cdot \sif = m_{i}^{2} - m_{f}^{2}
.
\label{J005}
\end{equation}

In general, $\Lambda^{\mu}(q,\sif)$ can be expanded as a linear combination of the 16 matrices
\begin{equation}
\openone
,
\qquad
\gamma^{\mu}
,
\qquad
\sigma^{\mu\nu} \equiv \frac{i}{2} \, [ \gamma^{\mu} , \gamma^{\nu} ]
,
\qquad
\gamma^{\mu} \gamma^{5}
,
\qquad
\gamma^{5}
,
\label{J006}
\end{equation}
which
form a basis in the vectorial space of
$4\times4$ matrices
(see \textcite{Giunti-Kim-2007}).
Since $\Lambda^{\mu}(q,\sif)$ carries a Lorentz index,
the coefficients of this expansion can depend on the available tensors:
the four-vectors
$q^{\mu}$, $\sif^{\mu}$,
the Lorentz-invariant metric tensor $g^{\mu\nu}$
and
the Lorentz-invariant antisymmetric tensor $\epsilon^{\mu\nu\alpha\beta}$.
Let us consider separately each term of the expansion:
\begin{enumerate}

\item
The $\openone$ term is a linear combination of the set
\begin{equation}
S(\openone)
=
\left\{
q^{\mu} \openone
,\,
\sif^{\mu} \openone
\right\}
.
\label{J007}
\end{equation}

\item
The $\gamma^{\mu}$ term is a linear combination of the set
\begin{equation}
S(\gamma^{\mu})
=
\left\{
\gamma^{\mu}
,\,
q^{\mu} \slashed{q}
,\,
q^{\mu} \slashed{\sif}
,\,
\sif^{\mu} \slashed{q}
,\,
\sif^{\mu} \slashed{\sif}
,\,
\epsilon^{\mu\nu\alpha\beta} \gamma_{\nu} q_{\alpha} \sif_{\beta}
\right\}
.
\label{J008}
\end{equation}

\item
The $\sigma^{\mu\nu}$ term is a linear combination of the set
\begin{align}
S(\sigma^{\mu\nu})
=
\null & \null
\left\{
\sigma^{\mu\nu} q_{\nu}
,\,
\sigma^{\mu\nu} \sif_{\nu}
,\,
q^{\mu} \sigma^{\alpha\beta} q_{\alpha} \sif_{\beta}
,\,
\sif^{\mu} \sigma^{\alpha\beta} q_{\alpha} \sif_{\beta}
,\,
\right.
\nonumber
\\
\null & \null
\phantom{\{}
\left.
\epsilon^{\mu\nu\alpha\beta} q_{\nu} \sigma_{\alpha\beta}
,\,
\epsilon^{\mu\nu\alpha\beta} \sif_{\nu} \sigma_{\alpha\beta}
,\,
\right.
\nonumber
\\
\null & \null
\phantom{\{}
\left.
\epsilon^{\mu\nu\alpha\beta} \sigma_{\nu\rho} q_{\alpha} \sif_{\beta} q^{\rho}
,\,
\epsilon^{\mu\nu\alpha\beta} \sigma_{\nu\rho} q_{\alpha} \sif_{\beta} \sif^{\rho}
\right\}
.
\label{J009}
\end{align}

\item
The $\gamma^{\mu}\gamma^{5}$ term is a linear combination of the set
\begin{align}
S(\gamma^{\mu}\gamma^{5})
=
\null & \null
\left\{
\gamma^{\mu} \gamma^{5}
,\,
q^{\mu} \slashed{q} \gamma^{5}
,\,
q^{\mu} \slashed{\sif} \gamma^{5}
,\,
\sif^{\mu} \slashed{q} \gamma^{5}
,\,
\sif^{\mu} \slashed{\sif} \gamma^{5}
,\,
\right.
\nonumber
\\
\null & \null
\phantom{\{}
\left.
\epsilon^{\mu\nu\alpha\beta} \gamma_{\nu} q_{\alpha} \sif_{\beta} \gamma^{5}
\right\}
.
\label{J010}
\end{align}

\item
The $\gamma^{5}$ term is a linear combination of the set
\begin{equation}
S(\gamma^{5})
=
\left\{
q^{\mu} \gamma^{5}
,\,
\sif^{\mu} \gamma^{5}
\right\}
.
\label{J011}
\end{equation}

\end{enumerate}

Several elements of the sets
(\ref{J007})--(\ref{J011})
can be expressed as linear combinations of others,
leading to only six independent elements.
It is convenient to choose the set of six independent elements as
\begin{equation}
q^{\mu} \openone
,\quad
q^{\mu} \gamma^{5}
,\quad
\gamma^{\mu}
,\quad
\gamma^{\mu} \gamma^{5}
,\quad
\sigma^{\mu\nu} q_{\nu}
,\quad
\epsilon^{\mu\nu\alpha\beta} q_{\nu} \sigma_{\alpha\beta}
.
\label{J012}
\end{equation}

We express all the elements in the sets (\ref{J007})--(\ref{J011})
in terms of the six elements in the set (\ref{J012})
using the equations in Appendix~\ref{I001}
as follows
(above each arrow we indicate the main equations used in the decomposition):

\begin{enumerate}

\item
From $S(\openone)$:
\begin{equation}
\sif^{\mu} \openone
\xrightarrow{\text{(\ref{I072})}}
\left\{
\gamma^{\mu}
,\,
\sigma^{\mu\nu} q_{\nu}
\right\}
.
\label{J013}
\end{equation}

\item
From $S(\gamma^{\mu})$:
\begin{align}
q^{\mu} \slashed{q}
\null & \null
\xrightarrow{\text{(\ref{I043})}}
\left\{
q^{\mu} \openone
\right\}
,
\label{J014}
\\
q^{\mu} \slashed{\sif}
\null & \null
\xrightarrow{\text{(\ref{I043})}}
\left\{
q^{\mu} \openone
\right\}
,
\label{J015}
\\
\sif^{\mu} \slashed{q}
\null & \null
\xrightarrow{\text{(\ref{I072})}}
\left\{
\gamma^{\mu}
,\,
\sigma^{\mu\nu} q_{\nu}
\right\}
,
\label{J016}
\\
\sif^{\mu} \slashed{\sif}
\null & \null
\xrightarrow{\text{(\ref{I072})}}
\left\{
\gamma^{\mu}
,\,
\sigma^{\mu\nu} q_{\nu}
\right\}
,
\label{J017}
\\
\epsilon^{\mu\nu\alpha\beta} \gamma_{\nu} q_{\alpha} \sif_{\beta}
\null & \null
\xrightarrow{\text{(\ref{I028})+(\ref{I074})}}
\left\{
q^{\mu} \gamma^{5}
,\,
\gamma^{\mu} \gamma^{5}
,\,
\epsilon^{\mu\nu\alpha\beta} q_{\nu} \sigma_{\alpha\beta}
\right\}
.
\label{J018}
\end{align}

\item
From $S(\sigma^{\mu\nu})$:
\begin{align}
\sigma^{\mu\nu} \sif_{\nu}
\null & \null
\xrightarrow{\text{(\ref{I073})}}
\left\{
q^{\mu} \openone
,\,
\gamma^{\mu}
\right\}
,
\label{J019}
\\
q^{\mu} \sigma^{\alpha\beta} q_{\alpha} \sif_{\beta}
\null & \null
\xrightarrow{\text{(\ref{I073})}}
\left\{
q^{\mu} \openone
\right\}
,
\label{J020}
\\
\sif^{\mu} \sigma^{\alpha\beta} q_{\alpha} \sif_{\beta}
\null & \null
\xrightarrow{\text{(\ref{I072})}}
\left\{
\gamma^{\mu}
,\,
\sigma^{\mu\nu} q_{\nu}
\right\}
,
\label{J021}
\\
\epsilon^{\mu\nu\alpha\beta} \sif_{\nu} \sigma_{\alpha\beta}
\null & \null
\xrightarrow{\text{(\ref{I075})}}
\left\{
q^{\mu} \gamma^{5}
,\,
\gamma^{\mu} \gamma^{5}
\right\}
,
\label{J022}
\\
\epsilon^{\mu\nu\alpha\beta} \sigma_{\nu\rho} q_{\alpha} \sif_{\beta} q^{\rho}
\null & \null
\xrightarrow{\text{(\ref{I028})+(\ref{I074})}}
\left\{
q^{\mu} \gamma^{5}
,\,
\gamma^{\mu} \gamma^{5}
,\,
\epsilon^{\mu\nu\alpha\beta} q_{\nu} \sigma_{\alpha\beta}
\right\}
,
\label{J023}
\\
\epsilon^{\mu\nu\alpha\beta} \sigma_{\nu\rho} q_{\alpha} \sif_{\beta} \sif^{\rho}
\null & \null
\xrightarrow{\text{(\ref{I028})+(\ref{I074})}}
\left\{
q^{\mu} \gamma^{5}
,\,
\gamma^{\mu} \gamma^{5}
,\,
\epsilon^{\mu\nu\alpha\beta} q_{\nu} \sigma_{\alpha\beta}
\right\}
.
\label{J024}
\end{align}

\item
From $S(\gamma^{\mu}\gamma^{5})$:
\begin{align}
q^{\mu} \slashed{q} \gamma^{5}
\null & \null
\xrightarrow{\text{(\ref{I043})}}
\left\{
q^{\mu} \gamma^{5}
\right\}
,
\label{J025}
\\
q^{\mu} \slashed{\sif} \gamma^{5}
\null & \null
\xrightarrow{\text{(\ref{I043})}}
\left\{
q^{\mu} \gamma^{5}
\right\}
,
\label{J026}
\\
\sif^{\mu} \slashed{q} \gamma^{5}
\null & \null
\xrightarrow{\text{(\ref{I074})}}
\left\{
\gamma^{\mu} \gamma^{5}
,\,
\epsilon^{\mu\nu\alpha\beta} q_{\nu} \sigma_{\alpha\beta}
\right\}
,
\label{J027}
\\
\sif^{\mu} \slashed{\sif} \gamma^{5}
\null & \null
\xrightarrow{\text{(\ref{I074})}}
\left\{
\gamma^{\mu} \gamma^{5}
,\,
\epsilon^{\mu\nu\alpha\beta} q_{\nu} \sigma_{\alpha\beta}
\right\}
,
\label{J028}
\\
\epsilon^{\mu\nu\alpha\beta} \gamma_{\nu} q_{\alpha} \sif_{\beta} \gamma^{5}
\null & \null
\xrightarrow{\text{(\ref{I028})+(\ref{I072})}}
\left\{
q^{\mu} \openone
,\,
\gamma^{\mu}
,\,
\sigma^{\mu\nu} q_{\nu}
\right\}
.
\label{J029}
\end{align}

\item
From $S(\gamma^{5})$:
\begin{equation}
\sif^{\mu} \gamma^{5}
\xrightarrow{\text{(\ref{I074})}}
\left\{
\gamma^{\mu} \gamma^{5}
,\,
\epsilon^{\mu\nu\alpha\beta} q_{\nu} \sigma_{\alpha\beta}
\right\}
.
\label{J030}
\end{equation}

\end{enumerate}

\section{Helicity and chirality}
\label{K001}

In this Appendix we derive the relation between helicity and chirality for ultrarelativistic neutrinos
and the corresponding helicity conservation properties of the different terms of the general expansion
of $\Lambda_{\mu}^{fi}(q)$ in Eq.~(\ref{C041}),
taking into account that
$\Lambda_{\mu}^{fi}(q)$
is sandwiched between $u$-spinors in the case of neutrinos
(Eq.~(\ref{C034}))
or
between $v$-spinors in the case of antineutrinos
(Eq.~(\ref{C051})).

With the help of Eqs.~(\ref{I043}) and (\ref{I044})
and using the definition (\ref{I052}) of the polarization four-vector
$s^{\mu}$, one can find that
\begin{align}
\null & \null
\slashed{s} \, u(p)
=
\left(
- \frac{m}{|\vet{p}|} \gamma^{0}
+ \frac{E}{|\vet{p}|}
\right)
u(p)
,
\label{K002}
\\
\null & \null
\slashed{s} \, v(p)
=
\left(
- \frac{m}{|\vet{p}|} \gamma^{0}
- \frac{E}{|\vet{p}|}
\right)
v(p)
.
\label{K003}
\end{align}
Therefore, in the ultrarelativistic limit $m \ll E$ we have
\begin{equation}
\slashed{s} \, u(p) \simeq u(p)
\quad
\text{and}
\quad
\slashed{s} \, v(p) \simeq - v(p)
,
\label{K004}
\end{equation}
and the helicity-projection matrices in Eq.~(\ref{I051})
have the same effect as the chirality projection matrices in Eq.~(\ref{I022}):
\begin{align}
\null & \null
P_{h} u^{(h)}(p)
\simeq
\frac{1+h\gamma^{5}}{2} u^{(h)}(p)
,
\label{K005}
\\
\null & \null
P_{h} v^{(h)}(p)
\simeq
\frac{1-h\gamma^{5}}{2} v^{(h)}(p)
.
\label{K006}
\end{align}
Then, Eqs.~(\ref{I053}) and (\ref{I054}) imply that
$u^{(h)}(p)$ and $v^{(h)}(p)$
are approximate eigenstates of $\gamma^{5}$:
\begin{align}
\null & \null
\gamma^{5} u^{(h)}(p)
\simeq
h u^{(h)}(p)
,
\label{K007}
\\
\null & \null
\gamma^{5} v^{(h)}(p)
\simeq
- h v^{(h)}(p)
.
\label{K008}
\end{align}
Hence,
in the ultrarelativistic limit we have
\begin{align}
\overline{u_{f}^{(h_{f})}}
\gamma^{\mu} \gamma^{5}
u_{i}^{(h_{i})}
\null & \null
\simeq
h_{i}
\overline{u_{f}^{(h_{f})}}
\gamma^{\mu}
u_{i}^{(h_{i})}
\nonumber
\\
\null & \null
\simeq
h_{f}
\overline{u_{f}^{(h_{f})}}
\gamma^{\mu}
u_{i}^{(h_{i})}
\propto
\delta_{h_{f}h_{i}}
,
\label{K009}
\\
\overline{u_{f}^{(h_{f})}}
\sigma^{\mu\nu} \gamma^{5}
u_{i}^{(h_{i})}
\null & \null
\simeq
h_{i}
\overline{u_{f}^{(h_{f})}}
\sigma^{\mu\nu}
u_{i}^{(h_{i})}
\nonumber
\\
\null & \null
\simeq
- h_{f}
\overline{u_{f}^{(h_{f})}}
\sigma^{\mu\nu}
u_{i}^{(h_{i})}
\propto
\delta_{-h_{f}h_{i}}
,
\label{K010}
\end{align}
and similar relations hold true in the case of $v$-spinors.
Of course, in the limit of massless neutrinos all the approximations above become
exact equalities.

Therefore,
in the ultrarelativistic limit the interactions generated by the charge and anapole form factors
conserve helicity,
whereas the interactions generated by the electric and magnetic dipole form factors
flip helicity.
In the same way, one can see that
the weak interactions generated by the charged current (\ref{B030})
or by the neutral current (\ref{B034}) conserve helicity
in the ultrarelativistic limit.
\section{Calculation of atomic ionization}
\label{L001}

In this Appendix we derive the neutrino magnetic moment contribution to the neutrino scattering on atomic targets.
The history and present status, including the corresponding references, of neutrino-atom collisions is given in \textcite{Kouzakov:2014lka}.

Consider the process where a neutrino with energy-momentum
$p_{\nu}=(E_{\nu}, \vet{p}_{\nu})$ scatters on an atom at
energy-momentum transfer $q=(T, \vec{q})$. In what follows the
recoil of atoms is neglected because of the reasonable assumption
$T\gg 2E_{\nu}^{2} / M$, $M$ is the nuclear mass.

The atomic target is supposed to be unpolarized and in its ground
state $|0\rangle$ with the corresponding energy $E_0$. It is
also supposed that $T\ll m_{e}$ and $\alpha Z \ll 1$, where $Z$ is
the nuclear charge and $\alpha$ is the fine-structure constant, so
that the initial and final electronic systems can be treated
nonrelativistically.
The neutrino states are described by the
Dirac spinors assuming $m_{\nu}=0$.

In the considered low-energy limit the neutrino magnetic moment
contribution to the electromagnetic vertex (\ref{C023}) can
be expressed in the following form
\begin{equation}
\label{L002}\Lambda_{\mu}=\frac{\mgm_{\nu}}{2m_{e}}\sigma_{\mu
\nu}q^{\nu}.
\end{equation}
Thus the magnetic moment interaction of a neutrino with the atomic
electrons is described by the Lagrangian
\begin{equation}\label{L003}
L_{int}=\frac{\mgm_{\nu}}{2m_e}{\bar \psi}(k')\sigma_{\alpha \beta}\psi (k)
q^{\alpha} A^{\beta},
\end{equation}
where the electromagnetic field $A_{\mu}=(A_0,\vec{A})$ of the
atomic electrons is $A_0(\vec q)=\sqrt{4 \pi \alpha} \, \rho( \vec
q)/ \vec{q}^{\,2}$, $\vec{A}(\vec q)=\sqrt{4 \pi \alpha} \, \vec{j}(
\vec q)/ \vec{q}^{\,2}$, where $\rho(\vec q)$ and $\vec{j}( \vec q)$
are the Fourier transforms of the electron number density and
current density operators, respectively,
\begin{equation} \rho(\vec{q)}= \sum_{a=1}^Z \exp(i \vec{q} \cdot \vec r_a)~,
\label{L004}
\end{equation}
\begin{align}
\vec{j}(\vec{q})
=
\null & \null
-\frac{i}{2m_e}\sum_{a=1}^Z
\left[
\exp(i {\vec q} \cdot \vec{r}_a)\frac{\partial}{\partial\vec{r}_a}
\right.
\nonumber
\\
\null & \null
\phantom{
-\frac{i}{2m_e}\sum_{a=1}^Z
}
\left.
+
\frac{\partial}{\partial\vec{r}_a}\exp(i{\vec{q}} \cdot \vec{r}_a)
\right]
,
\label{L005}
\end{align}
and the sums run over the positions $\vec r_a$ of all the $Z$
electrons in the atom.
The double differential cross section can
be presented as
\begin{equation}
\label{L006} \frac{d^{2}\sigma_{(\mu)}}{dTd{\vec
q}^{2}}=\left(\frac{d^{2}\sigma_{(\mu)}}{dTd{\vec
q}^{2}}\right)_\parallel+ \left(\frac{d^{2}\sigma_{(\mu)}}{dTd{\vec
q}^{2}}\right)_\perp,
\end{equation}
where

\begin{equation} {d^{2} \sigma_{(\mu)} \over dT \, d {\vec
q}^{2}}_\parallel = 4 \pi \, \alpha \, { \mgm_{\nu}^{2} \over \vec{q}^{2}}
.\left(1-\frac{T^{2}}{\vec{q}^{2}}\right)S(T,\vec{q}^{2}),
\label{L007}
\end{equation}
and
\begin{equation} {d^{2} \sigma_{(\mu)} \over dT \, d {\vec
q}^{2}}_\perp = 4 \pi \, \alpha \, { \mgm_{\nu}^{2} \over \vec{q}^{2}} \,
\left(1-\frac{\vec{q}^{2}}{4E_{\nu}^{2}}\right)R(T,\vec{q}^{2}),
\label{L008}
\end{equation}
where $S(T,\vec{q}^{2})$, also known as the dynamical structure
factor \cite{FanoARNS63}, and $R(T,\vec{q}^{2})$ are
\begin{equation}S(T,{\vec
q}^{2})=\sum_n \, \delta (T - E_n+E_0) \, \left | \langle n |
\rho(\vec{q}) | 0 \rangle \right |^{2}, \label{L009}
\end{equation}
\begin{equation} R(T,{\vec
q}^{2})=\sum_n \, \delta (T - E_n+E_0) \, \left | \langle n |
j_\perp(\vec{q}) | 0 \rangle \right |^{2}, \label{L010}
\end{equation}
with $j_\perp$ being the $\vec{j}$ component perpendicular to
$\vec{q}$ and parallel to the scattering plane, which is formed by
the incident and final neutrino momenta.
The sums in
Eqs.~(\ref{L009}) and~(\ref{L010}) run over all the states $| n
\rangle$ with energies $E_n$ of the electron system, with $|0
\rangle$ being the initial state.

The longitudinal term~(\ref{L007}) is associated with atomic
excitations induced by the force that the neutrino magnetic moment
exerts on electrons in the direction parallel to $\vec{q}$. The
transverse term~(\ref{L008}) corresponds to the exchange of a
virtual photon which is polarized as a real one, that is,
perpendicular to $\vec{q}$. It resembles a photoabsorption process
when $\vec{q}^{2} \to T$ and the virtual-photon four-momentum thus
approaches a physical value, $q^{2}\to 0$. Due to selections rules,
the longitudinal and transverse excitations do not interfere (see
\textcite{FanoARNS63} for details).

The factors $S(T,\vec{q}^{2})$ and $R(T,\vec{q}^{2})$ are related to
respectively the density-density $F(T,\vec{q}^{2})$ and
current-current $L(T,\vec{q}^{2})$ Green's functions
\begin{equation} S(T,{\vec
q}^{2})={1 \over \pi} \, {\rm Im}F(T,q^{2})~, \label{L011}
\end{equation}
\begin{equation} R(T,{\vec
q}^{2})={1 \over \pi} \, {\rm Im}L(T,\vec{q}^{2})~, \label{L012}
\end{equation}
where
\begin{align}
F(T,
\null & \null
\vec{q}^{2})
=
\sum_n {\left | \langle n | \rho(\vec{q}) | 0 \rangle \right
|^{2} \over T - E_n+E_0 - i \, \epsilon}
\nonumber
\\
\null & \null
=
\left \langle 0 \left
|\rho(- \vec{q}) \, {1 \over T-H+E_0- i \, \epsilon}\, \rho({\vec
q}) \right | 0 \right \rangle
,
\label{L013}
\\
L(T,
\null & \null
\vec{q}^{2})
=
\sum_n {\left | \langle n | j_\perp(\vec{q)} | 0 \rangle
\right |^{2} \over T - E_n+E_0 - i \, \epsilon}
\nonumber
\\
\null & \null
=
\left \langle
0 \left |j_\perp(- \vec{q)} \, {1 \over T-H+E_0- i \, \epsilon}\,
j_\perp(\vec{q)} \right | 0 \right \rangle
,
\label{L014}
\end{align}
$H$ being the Hamiltonian for the system of electrons.
For small
values of $\vec{q}$, in particular, such that $|\vec{q}|\sim T$,
only the lowest-order nonzero terms of the expansion of
Eqs.~(\ref{L011}) and (\ref{L012}) in powers of $\vec{q}^{2}$ are
of relevance (the so-called dipole approximation). In this case,
one has \cite{Kouzakov:2010tx}
\begin{equation}R(T,{\vec
q}^{2})=\frac{T^{2}}{\vec{q}^{2}}S(T,\vec{q}^{2}). \label{L015}
\end{equation}
Note that this ratio is much smaller than unity practically for
all $\vec{q}^{2}$ values involved in Eqs.~(\ref{L007})
and~(\ref{L008}). Thus, taking into account the foregoing
arguments, one might expect the transverse component to play a
minor role in Eq.~(\ref{L006}). The authors of
\textcite{Wong:2010pb}, however, came to the contrary
conclusion that this component dramatically enhances due to atomic
ionization when $T\sim\varepsilon_b$, where $\varepsilon_b$
is the electron binding energy.
The enhancement mechanism
proposed in \textcite{Wong:2010pb} is based on an analogy with
the photoionization process.
As mentioned above, when ${\vec
q}^{2}\to T^{2}$ the virtual-photon momentum approaches the physical
regime $q^{2}=0$. In this case, we have for the integrand in
Eq.~(\ref{L008})
\begin{equation}
\label{L016}\frac{R(T,\vec{q}^{2})}{\vec{q}^{2}}{\Bigg|}_{{\vec
q}^{2}\to T^{2}}=\frac{\sigma_\gamma(T)}{4\pi^{2}\alpha T},
\end{equation}
where $\sigma_\gamma(T)$ is the photoionization cross section at
the photon energy $T$ \cite{AkhBerQE1965}. The limiting
form~(\ref{L016}) was used in \textcite{Wong:2010pb} in
the whole integration interval.
Such a procedure is obviously
incorrect, for the integrand rapidly falls down as $\vec{q}^{2}$
ranges from $T^{2}$ up to $4E_{\nu}^{2}$, especially when ${\vec
q}^{2}\gtrsim r_a^{-2}$, where $r_a$ is a characteristic atomic size
(within the Thomas-Fermi model $r_a^{-1}=Z^{1/3}\alpha
m_{e}$ \cite{LanLif_QM_1977}). This fact reflects a strong departure
from the real-photon regime.
For this reason we can classify the
enhancement of the DCS determined in \textcite{Wong:2010pb} as
spurious.

Taking into account Eq.~(\ref{L015}), the experimentally
measured singe-differential inclusive cross section is, to a good
approximation, given by (see e.g. in
\textcite{Voloshin:2010vm,Kouzakov:2010tx,Kouzakov:2011vx})
\begin{equation} {d \sigma_{(\mu)} \over dT } = 4 \pi \, \alpha \, \mgm_{\nu}^{2}
.\int_{T^{2}}^{4E_{\nu}^{2}} \, S(T,\vec{q}^{2})\, {d\vec{q}^{2} \over
\vec{q}^{2}}~. \label{L017} \end{equation}
The standard electroweak contribution to the cross section can be
similarly expressed in terms of the same factor $S(T,{\vec
q}^{2})$ \cite{Voloshin:2010vm} as
\begin{eqnarray}{d \sigma_{EW} \over dT } &=& {G_{\text{F}}^{2} \over 4 \pi} \left ( 1+ 4
\sin^{2} \theta_{W} + 8 \, \sin^4 \theta_{W} \right
)\nonumber\\&{}&\times \int_{T^{2}}^{4E_{\nu}^{2}} \, S(T,\vec{q}^{2}) \,
d\vec{q}^{2} ~, \label{L018}
\end{eqnarray}
where the factor $S(T,\vec{q}^{2})$ is integrated over $\vec{q}^{2}$
with a unit weight, rather than $\vec{q}^{-2}$ as in
Eq.(\ref{L017}).

The kinematical limits for $\vec{q}^{2}$ in an actual neutrino
scattering are explicitly indicated in Eqs.(\ref{L017}) and
(\ref{L018}). At large $E_{\nu}$, typical for the reactor neutrinos,
the upper limit can in fact be extended to infinity, since in the
discussed here nonrelativistic limit the range of momenta $\sim
E_{\nu}$ is indistinguishable from infinity.
The lower limit can be
shifted to $\vec{q}^{2}=0$, since the contribution of the region of
$\vec{q}^{2} < T^{2}$ can be expressed in terms of the photoelectric
cross section \cite{Voloshin:2010vm} and is negligibly small (at
the level of below one percent in the considered range of $T$).
For this reason we henceforth discuss the momentum-transfer
integrals in Eqs.~(\ref{L017}) and~(\ref{L018}) running from ${\vec
q}^{2}=0$ to $\vec{q}^{2}=\infty$:
\begin{equation} I_1(T)=\int_0^{\infty} \, S(T,\vec{q}^{2})\, {d\vec{q}^{2} \over
\vec{q}^{2}}\label{L019},\end{equation}~
and
\begin{equation} I_2(T)=\int_0^{\infty} \, S(T,\vec{q}^{2})\, d\vec{q}^{2}~. \label{L020}\end{equation}

For a free electron, which is initially at rest, the
density-density correlator is the free particle Green's function
\begin{equation} F_{(FE)}(T,\vec{q}^{2})= \left ( T-{\vec{q}^{2} \over 2m_e} - i \,
\epsilon \right )^{-1}~ \label{L021} \end{equation}
so that the dynamical structure factor is given by
$S_{(FE)}(T,\vec{q}^{2})=\delta(T-\vec{q}^{2}/2m_e)$, and the discussed
here integrals are in the free-electron limit as follows:
\begin{equation} I_1^{(FE)}=\int_0^{\infty} \, S_{(FE)}(T,\vec{q}^{2})\, {d{\vec
q}^{2} \over \vec{q}^{2}} = {1 \over T}~,\label{L022}\end{equation}
\begin{equation} I_2^{(FE)}=\int_0^{\infty} \, S_{(FE)}(T,\vec{q}^{2})\, d{\vec
q}^{2} = 2 \, m_e~. \label{L023}
\end{equation}
Clearly, these expressions, when used in the formulas (\ref{L017})
and (\ref{L018}), result in the free-electron cross sections,
\begin{equation} {d \sigma_{(\mu)} \over dT }= 4 \pi \, \alpha \, \mgm_{\nu}^{2}
\left ( {1 \over T} - {1 \over E_{\nu} } \right ) ~ \label{L024}
\end{equation}
and
\begin{eqnarray} {d \sigma_{EW} \over dT }&=& {G_{\text{F}}^{2} \, m_e \over 2
\pi} \left ( 1+ 4 \, \sin^{2} \theta_{W} + 8 \, \sin^4 \theta_{W} \right
) \nonumber\\&{}&\times\left [ 1 + O \left ( {T \over E_{\nu}}
\right) \right ], \label{L025} \end{eqnarray}
correspondingly.

Now we consider neutrino scattering on an electron bound in an
atom following consideration of \textcite{Kouzakov:2011vx}. The
binding effects generally deform the density-density Green's
function, so that both the integrals (\ref{L019}) and
(\ref{L020}) are somewhat modified.
Namely, the binding effects
spread the free-electron $\delta$-peak in the dynamical structure
function at $\vec{q}^{2}=2 m_e T$ and also shift it by the scale of
characteristic electron momenta in the bound state.

We consider the scattering on just one electron.
The Hamiltonian
for the electron has the form $H=p^{2}/2m_e + V(r)$, and the
density-density Green's function from Eq.(\ref{L013}) can be
written as
\begin{align}
\null & \null
F(T,\vec{q}^{2})
=
\left \langle 0 \left | e^{-i \vec{q}
\cdot \vec{r}} \left [ T -H(\vec{p}, \vec{r}) + E_0 \right
]^{-1} e^{i \vec{q}
\cdot \vec{r}} \right | 0 \right \rangle
\nonumber
\\
\null & \null
=
\left \langle 0 \left | \left [ T -H(\vec{p} + \vec{q}, {\vec
r}) + E_0 \right ]^{-1} \right | 0 \right \rangle
\nonumber
\\
\null & \null
=
\left \langle 0 \left | \left [ T -{\vec{q}^{2} \over 2 m_e}- {\vec{p}
\cdot \vec{q} \over m_e}
- H(\vec{p}, \vec{r}) + E_0 \right ]^{-1} \right | 0 \right \rangle
,
\label{L026}
\end{align}
where the infinitesimal shift $T \to T - i \epsilon$ is implied.

Clearly, a nontrivial behavior of the latter expression in
Eq.(\ref{L026}) is generated by the presence of the operator $(\vec p
\cdot \vec{q})$ in the denominator, and the fact that it
does not commute with the Hamiltonian $H$. Thus an analytic
calculation of the Green's function as well as the dynamical
structure factor is feasible only in few specific problems.
Such a calculation for the
ionization from the $1s$, $2s$, and $2p$ hydrogen-like states
is presented in \textcite{Kouzakov:2011vx}.
In particular, it is shown that the deviation of the discussed integrals
(\ref{L019}) and (\ref{L020}) from their free values are very
small: the largest deviation is exactly at the ionization
threshold, where, for instance, each of the $1s$ integrals is
equal to the free-electron value multiplied by the factor $(1-7 \,
e^{-4}/3) \approx 0.957$~\footnote{It can be also noted that both
integrals are modified in exactly the same proportion, so that
their ratio is not affected at any $T$: $I_2(T)/I_1(T)= 2 m_e \, T$.
We find however that this exact proportionality is specific for
the ionization from the ground state in the Coulomb potential.}.
The same conclusion was also obtained in
\textcite{Kouzakov:2010tx} where the $1s$ case was examined
numerically.

The problem of calculating the integrals
(\ref{L019}) and (\ref{L020}) however can be solved in the
semiclassical limit, where one can neglect the noncommutativity of
the momentum $\vec{p}$ with the Hamiltonian, and rather treat this
operator as a number vector.
Taking also into account that
$(H-E_0) \, |0 \rangle =0$, one can then readily average the
latter expression in Eq.(\ref{L026}) over the directions of $\vec{q}$ and find the formula for the dynamical structure
factor:
\begin{align}
S(T,\vec{q}^{2})
=
\null & \null
{m \over 2 \, |\vec{p}|\, |\vec{q}|}
\left[
\theta \left( T- {\vec{q}^{2} \over 2m_e}+{|\vec{p}| \,|\vec{q}| \over m_e} \right)
\right.
\nonumber
\\
\null & \null
\left.
- \theta \left( T- {\vec{q}^{2} \over 2m_e}-{|\vec{p}| |\vec{q}| \over m_e} \right)
\right]
,
\label{L027}
\end{align}
where $\theta$ is the standard Heaviside step function.
The
expression in Eq.(\ref{L027}) is nonzero only in the range of
$|\vec{q}|$ satisfying the condition $-|\vec{p}|\, |\vec{q}|/m_e < T
- |\vec{q}|^{2}/2m_e < |\vec{p}|\, |\vec{q}|/m_e$, i.e. between the
(positive) roots of the binomials in the arguments of the step
functions: $|\vec{q}|_{min}=\sqrt{2m_e \, T + |\vec{p}|^{2}} - |{\vec
p}|$ and $|\vec{q}|_{max}=\sqrt{2m_e \, T + |\vec{p}|^{2}} + |{\vec
p}|$. One can notice that the previously mentioned `spread and
shift' of the peak in the dynamical structure function in this
limit corresponds to a flat pedestal between $|\vec{q}|_{min}$ and
$|\vec{q}|_{max}$. The calculation of the integrals (\ref{L019})
and (\ref{L020}) with the expression (\ref{L027}) is
straightforward, and yields the free-electron expressions
(\ref{L022}) and (\ref{L023}) for the discussed here
integrals in the semiclassical (WKB) limit~\footnote{The
appearance of the free-electron expressions here is not
surprising, since the equation (\ref{L027}) can be also viewed as
the one for scattering on an electron boosted to the momentum
$p$.}:
\begin{equation}
I_1^{(\text{WKB})}={1 \over T}
,
\qquad
I_2^{(\text{WKB})}=2 \, m_e
.
\label{L028} \end{equation}
The difference from the pure free-electron case however is in the
range of the energy transfer $T$. Namely, the expressions
(\ref{L028}) are applicable in this case only above the ionization
threshold, i.e. at $T \ge |E_0|$. Below the threshold the electron
becomes ``inactive''.
\section{Calculation of potentials}
\label{M001}

In this Appendix we describe the calculation of the potentials
in Eqs.~(\ref{F015}), (\ref{F017}), (\ref{F018}) and (\ref{G067}).
It is convenient to start by writing
the potential in Eq.~(\ref{F009})
(for $\chg=0$ and $j^{\mu}=0$)
as
\begin{equation}
V_{h_{i} \to h_{f}}
=
\frac{1}{4E}
\,
\text{Tr}
\left[
u^{(h_{i})}(p)
\overline{u^{(h_{f})}}(p)
\sigma_{\mu\nu} F^{\mu\nu}
\left(
\mgm
+ i
\elm
\gamma_{5}
\right)
\right]
.
\label{M002}
\end{equation}

For the helicity-conserving potential
$V_{h \to h}$,
we have
\begin{equation}
\dfrac{
u^{(h)}(p)
\,
\overline{u^{(h)}}(p)
}{2m}
=
\Lambda_{+}(p)
\,
P_{h}
,
\label{M003}
\end{equation}
with the energy and helicity projection operators
$\Lambda_{+}(p)$
and
$P_{h}$
given in Eqs.~(\ref{I049}) and (\ref{I051}).
Using the values of the traces of products of $\gamma$ matrices
given in Eqs.~(\ref{I040})--(\ref{I042}),
we obtain
\begin{equation}
V_{h \to h}
=
- \frac{h}{2E}
\left[
\mgm
\epsilon_{\alpha\beta\mu\nu}
p^{\alpha}
s^{\beta}
F^{\mu\nu}
-
2
\elm
F^{\mu\nu}
s_{\mu}
p_{\nu}
\right]
.
\label{M004}
\end{equation}
Then,
taking into account the expressions in Eq.~(\ref{I086})
for the electric and magnetic fields,
we obtain Eq.~(\ref{F015}).

In order to calculate the helicity-flipping potential
$V_{-h \to h}$,
we define the helicity-flipping matrix
\begin{equation}
F = \vec{\tau} \cdot \vec{\gamma} \, \gamma_{5}
,
\label{M005}
\end{equation}
where
$\vec{\tau}$ is an arbitrary unit vector orthogonal to $\vet{p}$,
i.e. such that
\begin{equation}
|\vec{\tau}|^{2} = 1
,
\qquad
\vec{\tau} \cdot \vet{p} = 0
.
\label{M006}
\end{equation}
One can check that
\begin{equation}
[F \,,\, \slashed{p}]
=
\{ F \,,\, \gamma_{5} \}
=
\{ F \,,\, \gamma^{5} \slashed{s} \}
=
0
,
\label{M007}
\end{equation}
and
\begin{equation}
F^2 = 1
,
\quad
\gamma^{0} F^{\dagger} \gamma^{0} = F
,
\quad
F P_{h}
=
P_{-h} F
.
\label{M008}
\end{equation}
Therefore,
we have
\begin{equation}
u^{(-h)}(p)
=
F
\,
u^{(h)}(p)
,
\label{M009}
\end{equation}
and
\begin{equation}
\dfrac{
u^{(-h)}(p)
\,
\overline{u^{(h)}}(p)
}{2 m}
=
F
\Lambda_{+}(p)
P_{h}
=
P_{-h}
\Lambda_{+}(p)
F
.
\label{M010}
\end{equation}

Plugging the expression (\ref{M010}) in Eq.~(\ref{M002}) for $h=h_{f}=-h_{i}$
and
using the values of the traces of products of $\gamma$ matrices
given in Eqs.~(\ref{I040})--(\ref{I042}),
we obtain
\begin{align}
V_{-h \to h}
\null & \null
=
-
\frac{\tau^{k}}{2E}
\Big[
\mgm
\left(
\epsilon^{k\alpha\mu\nu} p_{\alpha} F_{\mu\nu}
+
2 i m h F^{k\alpha} s_{\alpha}
\right)
\nonumber
\\
\null & \null
+
\elm
\left(
2 F^{k\alpha} p_{\alpha}
-
i m h \epsilon^{k\alpha\mu\nu} s_{\alpha} F_{\mu\nu}
\right)
\Big]
.
\label{M011}
\end{align}
Note that the $ih$ factors are correct in order to satisfy the hermiticity constraint in Eq.~(\ref{F004}).

Considering the expressions in Eq.~(\ref{I086})
for the electric and magnetic fields,
one can find
\begin{align}
V_{-h \to h}
=
\null & \null
\mgm
\Big(
- \vec{\tau} \cdot \vec{B}
- i h \vec{\tau} \cdot \frac{\vet{p} \times \vec{B}}{|\vet{p}|}
\nonumber
\\
\null & \null
\phantom{
\mgm
\Big(
}
+ \vec{\tau} \cdot \frac{\vet{p} \times \vec{E}}{E}
- i h \frac{|\vet{p}|}{E} \vec{\tau} \cdot \vec{E}
\Big)
\nonumber
\\
+
\null & \null
\elm
\Big(
- \vec{\tau} \cdot \vec{E}
- i h \vec{\tau} \cdot \frac{\vet{p} \times \vec{E}}{|\vet{p}|}
\nonumber
\\
\null & \null
\phantom{
\elm
\Big(
}
- \vec{\tau} \cdot \frac{\vet{p} \times \vec{B}}{E}
+ i h \frac{|\vet{p}|}{E} \vec{\tau} \cdot \vec{B}
\Big)
.
\label{M012}
\end{align}
Therefore,
only the components of the electric and magnetic fields orthogonal to $\vet{p}$
contribute to the helicity-flipping potential.

If we have only an electric or a magnetic field,
choosing $\vec{\tau}$ antiparallel to the component of the electric or magnetic field
orthogonal to $\vet{p}$,
we obtain the helicity-flipping potentials in Eqs.~(\ref{F017}) and (\ref{F018}).
In the general case of an electric and a magnetic field which are not parallel,
one must use the general equation (\ref{M012}),
which can be conveniently expressed in terms of the fields components.
Choosing, for example,
\begin{equation}
\vet{p} = (0,0,|\vet{p}|)
,
\qquad
\vec{\tau} = (-1,0,0)
,
\label{M013}
\end{equation}
we obtain
\begin{align}
V_{-h \to h}
=
\null & \null
\mgm
\left(
B^{1}
- i h B^{2}
+ \frac{|\vet{p}|}{E} \, E^{2}
+ i h \, \frac{|\vet{p}|}{E} \, E^{1}
\right)
\nonumber
\\
+
\null & \null
\elm
\left(
E^{1}
- i h E^{2}
- \frac{|\vet{p}|}{E} \, B^{2}
- i h \, \frac{|\vet{p}|}{E} B^{1}
\right)
,
\label{M014}
\end{align}
with
$\vec{E}=(E^{1},E^{2},E^{3})$
and
$\vec{B}=(B^{1},B^{2},B^{3})$.
The arbitrariness introduced by the choice of $\vec{\tau}$
is only apparent,
because the phase of
$V_{-h \to h}$
does not have physical effects.
For the physical absolute value of the potential we find
\begin{align}
|V_{-h \to h}|^2
=
\null & \null
\mgm^2
\left(
B_{\perp}^2
+
\frac{|\vet{p}|}{E} \, E_{\perp}^2
+
2
\,
\frac{\vet{p} \cdot \vec{B}_{\perp} \times \vec{E}_{\perp}}{E}
\right)
\nonumber
\\
+
\null & \null
\elm^2
\left(
E_{\perp}^2
+
\frac{|\vet{p}|}{E} \, B_{\perp}^2
-
2
\,
\frac{\vet{p} \cdot \vec{E}_{\perp} \times \vec{B}_{\perp}}{E}
\right)
\nonumber
\\
+
\null & \null
2 \, \mgm \, \elm
\,
\frac{m}{E}
\,
\vec{B}_{\perp} \cdot \vec{E}_{\perp}
,
\label{M015}
\end{align}
with
$\vec{E}_{\perp}=(0,E^{2},E^{3})$
and
$\vec{B}_{\perp}=(0,B^{2},B^{3})$.
Hence,
it is clear that
$|V_{-h \to h}|$
does not depend on the choice of $\vec{\tau}$.

The dependence
of the contribution
of the magnetic field on $\elm$
and that of the electric field on $\mgm$
are a consequence of the relativistic transformation of the electric and magnetic fields
and the fact that the classical electric and magnetic dipole moments
are defined for a nonrelativistic particle
through Eqs.~(\ref{F011}) and (\ref{F012}),
which establish the behavior of the nonrelativistic particle in electric and magnetic fields.
In fact,
for a nonrelativistic neutrino
($|\vet{p}| \ll E$)
we have
\begin{equation}
V_{-h \to h}^{\text{nr}}
\simeq
\mgm
\left(
B^{1}_{\text{rf}}
- i h B^{2}_{\text{rf}}
\right)
+
\elm
\left(
E^{1}_{\text{rf}}
- i h E^{2}_{\text{rf}}
\right)
,
\label{M016}
\end{equation}
where
the index ``rf'' indicates the rest frame of
the neutrino,
with the helicity defined as the projection of twice the spin on the $z$-axis.
One can see that for a nonrelativistic neutrino
the contribution of the magnetic field depends only on
the magnetic dipole moment $\mgm$
and
the contribution of the electric field depends only on
the electric dipole moment $\elm$.
Considering now a frame in which the neutrino is relativistic,
the components of the electric and magnetic fields are given by
\begin{align}
\null & \null
E^{1} = \gamma \left( E^{1}_{\text{rf}} + v B^{2}_{\text{rf}} \right)
,
\label{M017}
\\
\null & \null
E^{2} = \gamma \left( E^{2}_{\text{rf}} - v B^{1}_{\text{rf}} \right)
,
\label{M018}
\\
\null & \null
E^{3} = E^{3}_{\text{rf}}
,
\label{M019}
\\
\null & \null
B^{1} = \gamma \left( B^{1}_{\text{rf}} - v E^{2}_{\text{rf}} \right)
,
\label{M020}
\\
\null & \null
B^{2} = \gamma \left( B^{2}_{\text{rf}} + v E^{1}_{\text{rf}} \right)
,
\label{M021}
\\
\null & \null
B^{3} = B^{3}_{\text{rf}}
,
\label{M022}
\end{align}
with
$v=|\vet{p}|/E$
and
$\gamma=(1-v^{2})^{1/2}=E/m$.
Plugging these components of $\vec{E}$ and $\vec{B}$
in Eq.~(\ref{M014}),
we obtain
\begin{equation}
V_{-h \to h}
=
\mgm
\,
\frac{m}{E}
\left(
B^{1}_{\text{rf}}
- i h B^{2}_{\text{rf}}
\right)
+
\elm
\,
\frac{m}{E}
\left(
E^{1}_{\text{rf}}
- i h E^{2}_{\text{rf}}
\right)
,
\label{M023}
\end{equation}
Therefore,
the contribution of the magnetic field in the rest frame depends only on
the magnetic dipole moment $\mgm$
and
the contribution of the electric field in the rest depends only on
the electric dipole moment $\elm$.
If, for example,
there is no electric field in the rest frame,
$V_{-h \to h}$ depends only on the magnetic dipole moment $\mgm$.
In this case the coefficient of $\elm$
in Eq.~(\ref{M014}) vanishes because
the components of $\vec{E}$ and $\vec{B}$
are given by the relativistic transformations (\ref{M017})--(\ref{M022})
with
$\vec{E}_{\text{rf}}=0$
and the contribution of $\vec{E}$
in the coefficient of $\mgm$
is due to the same relativistic transformations.

Let us consider finally the contribution of the anapole moment
to the helicity-flipping potential.
From the last term in Eq.~(\ref{F009}),
using Eq.~(\ref{M010})
we obtain
\begin{equation}
V_{-h \to h}
=
\anm \, \frac{m}{E} \, \vec{j} \cdot \vec{\tau}
+
i \, \anm \, h \, \epsilon^{\alpha\beta\mu k} \, \frac{p_{\alpha}}{E}
\, s_{\beta} \, j_{\mu} \, \tau^{k}
.
\label{M024}
\end{equation}
Taking into account that $\vec{s}$ and $\vec{p}$ are parallel
(see Eq.~(\ref{I052})),
we can write this expression as
\begin{equation}
V_{-h \to h}
=
\anm \, \frac{m}{E} \, \vec{j} \cdot \vec{\tau}
+
i \, \frac{m}{E} \, \vec{\anm} \cdot \vec{j} \times \vec{\tau}
.
\label{M025}
\end{equation}
Then,
choosing $\vec{\tau}$ along the component of $\vec{j}$
orthogonal to $\vec{\anm}$
(which is defined in Eq.~(\ref{G066}) as parallel to $\vec{p}$),
the second term vanishes and we obtain Eq.~(\ref{G067}).
\section{Quasiclassical spin evolution in external fields}
\label{N001}

In this Appendix we show how the neutrino spin evolution can be described
in general case when the neutrino is subjected to
general types of non-derivative interactions with external fields
\cite{Dvornikov:2002rs} (see also \textcite{Bergmann:1999rz}).
Let the neutrino interactions are given by the Lagrangian
\begin{align}\label{N002}
-
\mathcal{L}
=
\null & \null
g_{s}s(x){\bar\nu}\nu
+
g_{p}{\pi}(x) {\bar\nu}\gamma^{5}\nu
+
g_{v}V^{\mu}(x){\bar\nu}\gamma_{\mu}\nu
\nonumber
\\
\null & \null
+
g_{a}A^{\mu}(x){\bar\nu}\gamma_{\mu}\gamma^{5}\nu
+
{{g_{t}}\over{2}}T^{\mu\nu}{\bar\nu}\sigma_{\mu\nu}\nu
\nonumber
\\
\null & \null
+
{{g^{\prime}_{t}}\over{2}} \Pi^{\mu\nu}{\bar\nu}\sigma_{\mu\nu}\gamma_{5}\nu
,
\end{align}
where
$s$,
$\pi$,
$V^{\mu}=(V^{0}, \vec{V})$,
$A^{\mu}=(A^{0}, \vec{A})$,
$T_{\mu\nu}=(\vec{a}, \vec{b})$,
$\Pi_{\mu\nu}=(\vec{c}, \vec{d})$ are scalar,
pseudoscalar, vector, axial-vector, tensor and pseudotensor fields,
respectively.
This Lagrangian accounts for a wide set of non-derivative neutrino
interactions with external fields.
We introduce the neutrino spin operator in a
usual form,
\begin{equation}\label{N003}
\vec{O}=\gamma_{0}\vec{\Sigma}-\gamma_{5}{\vec{p}\over{p_0^{\nu}}}-
\gamma_{0}
{{\vec{p}(\vec{p}\vec{\Sigma})}\over{p_0^{\nu}(p_0^{\nu}+m_{\nu})}},
\end{equation}
where $\vec{\Sigma}=\gamma^{0}\gamma^{5}\vec{\gamma}$.
Its overage over the neutrino stationary states gives the neutrino three-dimensional
spin vector
\begin{equation}\label{N004}
<\vec{O}>=\vec{S}
\end{equation}
that determines the four-dimensional spin vector given by Eq.~(\ref{F090}) in
Section~\ref{F001}. The corresponding spin evolution equation is obtained
in \textcite{Dvornikov:2002rs},
\begin{equation}\label{N005}
\begin{array}{c} \displaystyle {{d\vec{S} \over dt}}=
2g_{a}\left\{ A^{0}[\vec{S}
\times\vec{\beta}]- {{(\vec{A}\vec{\beta})[\vec{S} \times{\vec
\beta}]}\over{1+{\gamma}^{-1}}} - {1 \over \gamma}[\vec{S}\times\vec{A}]
\right\}
\\ \displaystyle +2g_{t}\left\{ [\vec{S}\times\vec{b}]-
{{(\vec{\beta}\vec{b})[\vec{S}\times\vec{\beta}]}\over{1+{\gamma}^{-1}}} +
[\vec{S}\times[\vec{a}\times\vec{\beta}]] \right\} \\
\displaystyle + 2ig^{\prime}_{t}\left\{ [\vec{\zeta}_{\nu}\times\vec{c}]- {{({\vec
\beta}\vec{c})[\vec{S}\times\vec{\beta}]}\over{1+{\gamma}^{-1}}}- [{\vec
S}\times[\vec{d}\times\vec{\beta}]] \right\},
\end{array}
\end{equation}
where $\gamma = p^{0}_{\nu} / m_{\nu}$ and $\vec \beta$ is the neutrino speed.
This is a rather general equation for the neutrino spin evolution
that can be also used for the description of neutrino spin
oscillations in different environments, such as moving and
polarized matter with external electromagnetic fields (see
\textcite{Studenikin:2004bu,Studenikin:2007zz}).
The $SL\nu$ in gravitational fields has been studied
(see \textcite{Grigoriev:2004bm}) on the basis of a neutrino
spin evolution equation
(\ref{N005}).

The Lorentz invariant form of Eq.~(\ref{N005}) can be obtained
using the four-dimensional spin vector $S^{\mu}$ which is determined
by the three-dimensional spin vector
$\vec{S}$
in accordance with the relation:
\begin{equation}
S^{\mu}=
\left(
{{(\vec{S}\vec{p})}\over{m_{\nu}}},
\vec{S}+
{{\vec{p}(\vec{S}\vec{p})}
\over{m_{\nu}(m_{\nu}+p^0_{\nu})}}
\right)
.
\label{N006}
\end{equation}
Thus, we get the Lorentz invariant form for the neutrino spin
$S^{\mu}$ evolution
equation accounting for the general interactions with external
fields
\begin{eqnarray}\label{N007}{{dS^{\mu}}\over{d\tau}}=&{}&2g_{t}
(T^{\mu\nu}S_{\nu}-
u^{\mu}T^{\lambda\rho}u_{\lambda}S_{\rho})+\nonumber\\&{}&
2ig^{\prime}_{t}
({\tilde \Pi}^{\mu\nu}S_{\nu}-
u^{\mu}{\tilde \Pi}^{\lambda\rho}u_{\lambda}S_{\rho})
+2g_{a}G^{\mu\nu}S_{\nu},
\end{eqnarray}
where
$G^{\mu\nu}=\epsilon^{\mu\nu\alpha\beta}A_{\alpha}u_{\beta}$,
$u^{\mu} = (1,\vec{\beta}) E_{\nu} / m_{\nu}$,
${\tilde \Pi}^{\mu\nu}=\epsilon^{\mu\nu\alpha\beta}\Pi_{\alpha\beta}/2$.
The tensor $G_{\mu\nu}$ can be expressed through two vectors
$G_{\mu\nu}=(-\vec{P},\vec{M})$ which are presented in the form,
\begin{equation}
\vec{M}=\gamma(A^{0}\vec{\beta}-\vec{A}), \ \ \
\vec{P}=-\gamma[\vec{\beta}\times\vec{A}].
\end{equation}
The derivation in the left-handed side of this equation is taken over the
neutrino proper time $\tau =\gamma^{-1}t$, where $t$
is the time in the laboratory frame of reference.

Note that Eq.~(\ref{N007}) can be considered as the generalized
Bargmann-Michel-Telegdi equation and that it was used in
\textcite{Egorov:1999ah,Lobanov:2001ar} for description of the neutrino spin and flavor
oscillations in arbitrary electromagnetic fields.
Some general aspects of
the neutrino spin dynamics in case of
non-minimal couplings with an external magnetic field
was studied in \textcite{Bernardini:2006cy}.
\section{Spin precession in moving matter}
\label{O001}

An approach based on the generalized Bargmann-Michel-Telegdi equation can be used for derivation of an impact of matter motion and polarization on the neutrino spin (and spin-flavor) evolution.
Consider, as an example, an electron neutrino spin procession in the case when neutrinos with the Standard Model interaction are propagating through moving and polarized matter composed of electrons (electron gas) in the presence of an electromagnetic field given by the electromagnetic-field tensor $F_{\mu \nu}=({\vec E}, {\vec B})$. As discussed in \textcite{Studenikin:2004bu} (see also \textcite{Egorov:1999ah,Lobanov:2001ar}) the evolution of the
three-di\-men\-sio\-nal neutrino spin vector $\vec S $ is given by
\begin{equation}\label{O002} {d\vec S
\over dt}={2\mgm\over \gamma} \Big[ {\vec S \times ({\vec
B_0}+\vec M_0)} \Big],
\end{equation}
where the magnetic field $\vec{B_0}$ in the neutrino rest frame is determined by the transversal
and longitudinal (with respect to the neutrino motion) magnetic and electric field components in the
laboratory frame,
\begin{equation}
\vec B_0=\gamma\Big(\vec B_{\perp} +{1 \over \gamma} \vec
B_{\parallel} + \sqrt{1-\gamma^{-2}} \Big[\vec E_{\perp} \times
\frac{\vec \beta}{\beta} \Big]\Big).
\end{equation}
The matter term $\vec M_0$ in Eq. (\ref{O002}) is also composed of the transversal $\vec {M}{_{0_{\parallel}}}$
and longitudinal  $\vec {M_{0_{\perp}}}$ parts,
\begin{equation}
\vec {M_0}=\vec {M}{_{0_{\parallel}}}+\vec {M_{0_{\perp}}},
\label{O003}
\end{equation}
\begin{equation}
\begin{array}{c}
\displaystyle \vec {M}_{0_{\parallel}}=\gamma\vec\beta{n_{0} \over
\sqrt {1- v_{e}^{2}}}\left\{ \rho^{(1)}_{e}\left(1-{{\vec v}_e
\vec\beta \over {1- {\gamma^{-2}}}} \right)\right. \\-
\displaystyle\rho^{(2)}_{e}\left. \left(\vec\zeta_{e}\vec\beta
\sqrt{1-v^2_e}+ {(\vec \zeta_{e}{\vec v}_e)(\vec\beta{\vec v}_e)
\over 1+\sqrt{1-v^2_e} }\right){1 \over {1- {\gamma^{-2}}}}
\right\}, \label{O004}
\end{array}
\end{equation}
\begin{equation}\label{O005}
\begin{array}{c}
\displaystyle \vec {M}_{0_{\perp}}=-\frac{n_{0}}{\sqrt {1-
v_{e}^{2}}}\Bigg\{ \vec{v}_{e_{\perp}}\Big(
\rho^{(1)}_{e}+\rho^{(2)}_{e}\frac
{(\vec{\zeta}_{e}{\vec{v}_e})} {1+\sqrt{1-v^2_e}}\Big) \\+
\displaystyle {\vec{\zeta}_{e_{\perp}}}\rho^{(2)}_{e}\sqrt{1-v^2_e}\Bigg\}.
\end{array}
\end{equation}
Here $n_0=n_{e}\sqrt {1-v^{2}_{e}}$ is the invariant number density of
matter given in the reference frame for which the total speed of
matter is zero.
The vectors $\vec v_e$, and $\vec \zeta_e \
(0\leqslant |\vec \zeta_e |^2 \leqslant 1)$ denote, respectively,
the speed of the reference frame in which the mean momentum of
matter (electrons) is zero, and the mean value of the polarization
vector of the background electrons in the above mentioned
reference frame.
The coefficients $\rho^{(1,2)}_e$ are calculated
if the neutrino Lagrangian is given, and within the extended
standard model supplied with $SU(2)$-singlet right-handed neutrino
$\nu_{R}$,
\begin{equation}\label{O006}
\rho^{(1)}_e={\tilde{G}_F \over {2\sqrt{2}\mu }}\,, \qquad
\rho^{(2)}_e =-{G_F \over {2\sqrt{2}\mu}}\,,
\end{equation}
where $\tilde{G}_{F}={G}_{F}(1+4\sin^2 \theta _W).$
For the probability of the neutrino spin oscillations in the adiabatic approximation we get from
Eqs. (\ref{O004}) and (\ref{O005})
\begin{equation}\label{O007}
P_{\nu_L \rightarrow \nu_R} (x)=\sin^{2} 2\theta_\textmd{eff}
\sin^{2}{\pi x \over L_\textmd{eff}},\end{equation}
\begin{equation}
sin^{2} 2\theta_\textmd{eff}={E^2_\textmd{eff} \over
{E^{2}_\textmd{eff}+\Delta^{2}_\textmd{eff}}}, \ \ \
L_\textmd{eff}={2\pi \over
\sqrt{E^{2}_\textmd{eff}+\Delta^{2}_\textmd{eff}}},
\end{equation}
where
\begin{equation}\label{O008}
E_\textmd{eff}=\mgm \Big|{\vec B}_{\perp} + {1\over
\gamma}{\vec M}_{0\perp} \Big|,
\end{equation}
\begin{equation}\label{O009}
\Delta^{2}_\textmd{eff}={\mgm \over \gamma}\Big|{\vec
M}_{0\parallel}+{\vec B}_{0\parallel} \Big|.
\end{equation}
It follows that even without presence of an electromagnetic field,
${\vec B}_{\perp}={\vec B}_{0\parallel}=0$,
neutrino spin (or spin-flavor) oscillations can be induced in the presence of matter
when the transverse matter term ${\vec M}_{0\perp}$ is not zero.
This possibility is realized
in the case of nonzero transversal matter velocity or polarization.
A detailed discussion of this phenomenon can be found in \textcite{Studenikin:2004bu,Studenikin:2004tv}.

\section{Wave functions in magnetic field and matter}
\label{P001}

In this Appendix we derive exact solutions of the Dirac equation
for two cases: 1) for an electron in a constant magnetic field and
2) for a neutrino in presence of matter.
The electron wave
function in magnetic field is used in Section~\ref{F001}, in
calculations a neutrino mass-operator, of the beta decay of a
neutron and $\nu {\bar \nu}$ synchrotron radiation by an electron
in magnetic field.
The neutrino wave function in matter is used in
Subsection~\ref{E049} in studies of the spin light of neutrino in matter.

Following \textcite{Balantsev:2010cy} and \textcite{Balantsev:2010zw},
we derive two exact solutions for the Dirac equation for two
considered cases starting with general equation for a charged
particle wave functions that accounts both for the presence of a
magnetic field and matter.
In the case of the standard model
interaction of an electron neutrino and electron with matter
composed of neutrons, the modified Dirac equations as is given by
\textcite{Studenikin:2004dx}, \textcite{Studenikin:2006jv} and
\textcite{Studenikin:2008qk}
\begin{equation}\label{P002}
\Big\{ i\gamma_{\mu}(\partial^{\mu}+\chg_{\afl} A^{\mu})-\frac{1}{2}
\gamma_{\mu}(c_{\afl}+\gamma_{5})\widetilde{f}^{\mu}-m_{\afl}
\Big\}\Psi^{(l)}(\vec{r}, t)=0,
\end{equation}
where for the case of the electron $m_{\afl}=m_e$,
$c_{\afl}=c_e=1-4\sin^{2}\theta_{W}$ and $\chg_{\afl}=-\elechg$. For neutrinos $m_{\afl}=m_\nu$,
$c_{\afl}=c_{\nu}=1$ and $\chg_{\afl}=\chg_{\nu}$ is the possible neutrino
millicharge \cite{Balantsev:2010cy} (see Subsection~\ref{G012}). For
unpolarized and not moving matter $\widetilde{f}^{\mu}=
G(n,0,0,0)$, $n$ is the matter number density,
$G=\frac{G_{\text{F}}}{\sqrt{2}}$, and for the magnetic field potential
$A^{\mu}=(0,0,Bx,0)$.

Equation~(\ref{P002}) can be rewritten in the
Hamiltonian form
\begin{equation}\label{P003}
\mathrm{i}\frac{\partial}{\partial
t}\Psi(\vec{r}, t)=\hat{H}\Psi(\vec{r}, t),
\end{equation}
\begin{eqnarray}\hat{H}=\gamma^0\vec{\gamma}(\vec{p}+\chg_{\afl}\vec{A})+m_{\afl}\gamma^0+
\frac {Gn}{2}(c_{\afl}+\gamma^5).
\end{eqnarray}

The solution of Eq.~(\ref{P003}) due to
symmetries can be sought in the form
\begin{eqnarray}\label{P004}
\Psi^{(l)}(\vec{r}, t) = e^{-i p_0 t + i p_2y +i p_3z}
\begin{pmatrix}
\psi_1(x)\\
\psi_2(x)\\
\psi_3(x)\\
\psi_4(x)\end{pmatrix}.
\end{eqnarray}
Substituting (\ref{P004}) into (\ref{P003}) and
introducing the increasing and decreasing operators,
\begin{equation}\label{P005}
\hat{a}=\frac{1}{\sqrt{2}}(\eta+\frac{d}{d\eta}),\qquad
\hat{a}^{+}=\frac{1}{\sqrt{2}}(\eta-\frac{d}{d\eta}),
\end{equation}
where
\begin{equation} \eta =x \sqrt{\gamma_{\afl}} +\frac{p_2}{\sqrt
{\gamma_{\afl}}}, \ \ \gamma_{\afl}=\chg_{\afl} B,
\end{equation}
we arrive at a system of linear equations for the particle wave
function components:
\begin{align}
\null & \null
(\tilde{p}_0-m_{\afl})\psi_1+i\sqrt{2\chg_{\afl} B}\hat{a}\psi_{4}-\left(p_3-\frac{Gn}{2}\right)\psi_3=0
,
\nonumber
\\
\null & \null
(\tilde{p}_0-m_{\afl})\psi_2-i\sqrt{2\chg_{\afl} B}\hat{a}^{+}\psi_{3}+\left(p_3+\frac{Gn}{2}\right)\psi_4=0
,
\nonumber
\\
\null & \null
(\tilde{p}_0+m_{\afl})\psi_3+i\sqrt{2\chg_{\afl} B}\hat{a}\psi_{2}-\left(p_3-\frac{Gn}{2}\right)\psi_1=0
,
\nonumber
\\
\null & \null
(\tilde{p}_0+m_{\afl})\psi_4-i\sqrt{2\chg_{\afl}B}\hat{a}^{+}\psi_{1}+\left(p_3+\frac{Gn}{2}\right)\psi_2=0
,
\label{P006}
\end{align}
where $\tilde{p}_0 = p_0+ G n c_{\afl} / 2$.

The exact solution for Dirac equation for a particles wave
function in presence of the magnetic field and matter can be
written in the form
\begin{eqnarray}\label{P007}
\Psi^{(l)}(\vec{r}, t)= \frac{1}{L}e^{-i p_0 t + i p_2y +i p_3z}
\begin{pmatrix}
C_1 U_{N-1}(\eta)\\
iC_2 U_{N}(\eta)\\
C_3 U_{N-1}(\eta)\\
iC_4 U_{N}(\eta)
\end{pmatrix},
\end{eqnarray}
where $U_{N}(\eta)$ are Hermite functions (see, for instance, \textcite{Balantsev:2010zw}).
Then the system of equations for the coefficients $C_i$,
with $i=1,2,3,4$, is
\begin{align}
\null & \null
(\tilde{p}_0-m_{\afl})C_1-\sqrt{2\chg_{\afl}B N}C_4-\left((p_3-\frac{Gn}{2}\right)C_3=0
,
\nonumber
\\
\null & \null
(\tilde{p}_0-m_{\afl})C_2-\sqrt{2\chg_{\afl}B N}C_3+\left((p_3+\frac{Gn}{2}\right)C_4=0
,
\nonumber
\\
\null & \null
(\tilde{p}_0+m_{\afl})C_3-\sqrt{2\chg_{\afl}B N}C_2-\left((p_3-\frac{Gn}{2}\right)C_1=0
,
\nonumber
\\
\null & \null
(\tilde{p}_0+m_{\afl})C_4-\sqrt{2\chg_{\afl}B N}C_1+\left((p_3+\frac{Gn}{2}\right)C_2=0
.
\label{P008}
\end{align}
From (\ref{P008}) we get the energy spectrum
\begin{equation}\label{P009}
p_0=\varepsilon\sqrt{m_{\afl}^2+\left({m_{\afl}}T^0-\frac{Gn}{2}\right)^2}+\frac{Gn}{2}c_{\afl},
\end{equation}
where \begin{equation}\label{P010}
T^{0} = \frac
{s}{m_{\afl}}\sqrt{p_3^2+2\chg_{\afl}BN},
\end{equation}
$s=\pm 1$, are the eigenvalues of the longitudinal spin
polarization operator
$\hat{T}^0 = \vec\sigma \cdot (\vec{p}+\elechg\vec{A}) / m_{\afl}$
\cite{SokolovTernov:1968}
and $\varepsilon=\pm 1$ is the sign of the energy.
Note that in the presence of the matter potential proportional to
$Gn$ the transverse spin polarization operator does not commute
with the Hamiltonian (\ref{P003}), that is a
consequence of $\gamma^5$ presence in (\ref{P002}).
Taking into account the normalization condition $\sum_{i} C_i^2=1$
the solution of the system (\ref{P008}) is
\begin{align}
\null & \null
C_1=\frac{1}{2}\sqrt{1+\frac{m_{\afl}}{p_0-\frac{Gn}{2}c_{\afl}}}\sqrt{1+\frac{p_3}{m_{\afl}T^0}}
,
\nonumber
\\
\null & \null
C_2=\frac{s}{2}\sqrt{1+\frac{m_{\afl}}{p_0-\frac{Gn}{2}c_{\afl}}}\sqrt{1-\frac{p_3}{m_{\afl}T^0}}
,
\nonumber
\\
\null & \null
C_3=\frac{s\varepsilon \eta}{2}\sqrt{1-\frac{m_{\afl}}{p_0-\frac{Gn}{2}c_{\afl}}}\sqrt{1+\frac{p_3}{m_{\afl}T^0}}
,
\nonumber
\\
\null & \null
C_4=\frac{\varepsilon \eta}{2}\sqrt{1-\frac{m_{\afl}}{p_0-\frac{Gn}{2}c_{\afl}}}\sqrt{1-\frac{p_3}{m_{\afl}T^0}}
,
\label{P011}
\end{align}
where $\eta=$sign$\big(p-s\frac{Gn}{2}\big)$.

From the obtained exact solution of Dirac equation
(\ref{P002}) for a charged massive particle moving
in a magnetic field and matter, that is given by Eqs.~(\ref{P007}), (\ref{P009}), and
(\ref{P011}) it is easy to get the solution for the
electron wave function in the magnetic field by ``switching off''
the matter term $Gn\rightarrow 0$ and the corresponding proper
choice for other values: $m_{\afl}=m_e$,
$c_{\afl}=c_e=1-4\sin^{2}\theta_{W}$ and $\chg_{\afl}=-\elechg$. In particular,
from (\ref{P009}) one obtains the well known
energy spectrum for the electron in a constant magnetic field
(see \textcite{SokolovTernov:1968})
\begin{equation}\label{P012}
p_0=\sqrt{m_e^2 + p^2_3 + 2{\elechg}BN}
\end{equation}
where $N = 0, 1, 2, \ldots $ is the Landau number of the energy levels.

From the obtained general solution for the wave function given by
Eqs.~(\ref{P007}), (\ref{P009}), and
(\ref{P011}) it is also possible to get the wave function
for a neutrino moving in matter by ``switching off'' the magnetic
field strength by considering the wave function in the limit
$\chg_{\nu}B\rightarrow 0$. Of course, the corresponding choice of
values, $m_{\afl}=m_\nu$, $c_{\afl}=c_{\nu}=1$ and $\chg_{\afl}=\chg_{\nu}$, should be
done.
When the magnetic field is ``switching off'', the maximal
number of Landau levels $N_{\text{max}}$ is increasing to infinity,
however the product $\chg_{\nu}BN=\gamma_{\nu}N$ remains constant
\begin{equation}
\lim_{\gamma_{\nu}\rightarrow 0, N\rightarrow \infty }
2\gamma_{\nu}N=p_\bot^2.
\end{equation}
Accounting also the asymptotic behavior of Hermite functions,
\begin{equation}
\lim_{\gamma_{\nu}\rightarrow 0, N\rightarrow \infty } U_N
(\eta)\sim e^{ip_1 x},
\end{equation}
we arrive to the neutrino wave function in matter obtained by
\textcite{Studenikin:2004dx}
\begin{align}
\Psi_{\varepsilon}(\vec{r},t)
=
\null & \null
\frac{e^{-i(p_0^{\varepsilon}t-\vec{p}\vec{r})}}{2L^{\frac{3}{2}}}
\nonumber
\\
\null & \null
\times
\begin{pmatrix}
\sqrt{
\left( 1+ \frac{m_{\nu}}{p_0^{\varepsilon}-\frac{Gn}{2}} \right)
\left( 1 + s \frac{p_{3}}{p} \right)
}
\\
s
e^{i\delta}
\sqrt{
\left( 1 + \frac{m_{\nu}}{p_0^{\varepsilon}-\frac{Gn}{2}} \right)
\left( 1 - s \frac{p_{3}}{p} \right)
}
\\
s
\varepsilon
\eta
\sqrt{
\left( 1 - \frac{m_{\nu}}{p_0^{\varepsilon}-\frac{Gn}{2}} \right)
\left( 1 + s \frac{p_{3}}{p} \right)
}
\\
s
\varepsilon
\eta
e^{i\delta}
\sqrt{
\left( 1 - \frac{m_{\nu}}{p_0^{\varepsilon}-\frac{Gn}{2}} \right)
\left( 1 - s \frac{p_{3}}{p} \right)
}
\end{pmatrix}
,
\label{P013}
\end{align}
where $\delta=\arctan{p_2/p_1}$ and the neutrino energy
\begin{equation}\label{P014}
p_{0}^{\varepsilon}=\varepsilon \eta {\sqrt{m_{\nu}^2+\Big(p-s\frac{Gn}{2}\Big)^{2}}
+\frac{Gn}{2}}.
\end{equation}
In the limit of vanishing density of matter, when
$n\rightarrow 0$, the wave function (\ref{P013})
transforms to the vacuum solution of the Dirac equation.
The
values $s=\pm 1$ specify the two neutrino helicity states,
$\nu_{+}$ and  $\nu_{-}$.

\begin{acknowledgments}
We thank Vadim Bednyakov, Samoil Bilenky, Victor Brudanin, Vyacheslav Egorov, Konstantin Kouzakov, Alexander Starostin and Alexey Ternov
for fruitful discussions.
The work of A. Studenikin was partially supported by the Russian Basic Research Foundation grants N. 14-22-03043\_ofi\_m and 15-52-53112.
The work of C. Giunti was partially supported by the research grant {\sl Theoretical Astroparticle Physics} number 2012CPPYP7
under the program PRIN 2012 funded by the Ministero dell'Istruzione, Universit\`a e della Ricerca (MIUR).
\end{acknowledgments}

%

\end{document}